\pdfoutput=1
\documentclass[12pt,a4paper,twoside,openright]{book}

\usepackage{xcolor}

\usepackage[T1]{fontenc}

\usepackage[utf8]{inputenc}
\usepackage[polish,english]{babel}

\usepackage[
linktocpage=true,
  colorlinks   = true,    
  urlcolor     = blue,    
  linkcolor    = blue,    
  citecolor    = red      
]{hyperref}

\let\orgautoref\autoref

\providecommand{\Autoref}
        {\def\equationautorefname{Equation}%
         \def\figureautorefname{Figure}%
         \def\subfigureautorefname{Figure}%
         \def\chapterautorefname{Chapter}%
         \def\sectionautorefname{Section}%
         \def\subsectionautorefname{Section}%
         \def\subsubsectionautorefname{Section}%
         \def\appendixautorefname{Appendix}%
         \def\Itemautorefname{Item}%
         \def\tableautorefname{Table}%
         \orgautoref}

\renewcommand{\autoref}
        {\def\equationautorefname{Eq.}%
         \def\figureautorefname{Fig.}%
         \def\subfigureautorefname{Fig.}%
         \def\chapterautorefname{Chap.}%
         \def\sectionautorefname{Sec.}%
         \def\subsectionautorefname{Sec.}%
         \def\subsubsectionautorefname{Sec.}%
         \def\appendixautorefname{App.}%
         \def\Itemautorefname{item}%
         \def\tableautorefname{Tab.}%
         \orgautoref}

\usepackage{emptypage}
\newcommand{\dilog} {\text{Li}_2}

\usepackage{subfig}

\usepackage[cache=true,frozencache=true]{minted}

\usepackage[sort&compress,square,numbers]{natbib}
\usepackage{mathtools}

\usepackage[left=2cm,right=2cm,top=2cm,bottom=4cm]{geometry}

\usepackage{fancyhdr}
\usepackage{booktabs}
\usepackage{amsmath}
\DeclareMathOperator{\arctanh}{arctanh}
\usepackage{amssymb}
\usepackage{graphicx}
\usepackage{lipsum}
\usepackage{slashed}
\usepackage[export]{adjustbox}

\usepackage{gnuplot-lua-tikz}

\usepackage{multirow}

\usepackage{acro}
\acsetup{first-style=short,list-style=longtable}

\DeclareAcronym{LEP}{
  short = LEP,
  long  = Large Electron–Positron Collider,
  class = abbrev
}
\DeclareAcronym{BMP}{
  short = BMP,
  long  = Benchmark Point,
  class = abbrev
}
\DeclareAcronym{ATLAS}{
  short = ATLAS,
  long  = A Toroidal LHC Apparatus,
  class = abbrev
}
\DeclareAcronym{CMS}{
  short = CMS,
  long  = Compact Muon Solenoid,
  class = abbrev
}
\DeclareAcronym{LHC}{
  short = LHC,
  long  = Large Hadron Collider,
  class = abbrev
}
\DeclareAcronym{MSSM}{
  short = MSSM,
  long  = Minimal Supersymmetric Standard Model,
  class = abbrev
}
\DeclareAcronym{MRSSM}{
  short = MRSSM,
  long  = Minimal R-symmetric Supersymmetric Standard Model,
  class = abbrev
}
\DeclareAcronym{SM}{
  short = SM,
  long  = Standard Model (of Elementary Interactions),
  class = abbrev
}
\DeclareAcronym{SUSY}{
  short = SUSY,
  long  = Supersymmetry,
  class = abbrev
}
\DeclareAcronym{SSB}{
  short = SSB,
  long  = Soft SUSY Breaking,
  class = abbrev
}
\DeclareAcronym{CKM}{
  short = CKM,
  long  = Cabibbo-Kobayashi-Maskawa,
  class = abbrev
}
\DeclareAcronym{EW}{
  short = EW,
  long  = Electroweak,
  class = abbrev
}
\DeclareAcronym{QCD}{
  short = QCD,
  long  = Quantum Chromodynamics,
  class = abbrev
}
\DeclareAcronym{SQCD}{
  short = SQCD,
  long  = Supersymmetric Quantum Chromodynamics,
  class = abbrev
}
\DeclareAcronym{EWPO}{
  short = EWPO,
  long  = Electroweak Precision Observables,
  class = abbrev
}
\DeclareAcronym{EWSB}{
  short = EWSB,
  long  = Electroweak Symmetry Breaking,
  class = abbrev
}
\DeclareAcronym{LSP}{
  short = LSP,
  long  = Lightest Supersymmetric Particle,
  class = abbrev
}
\DeclareAcronym{COM}{
  short = COM,
  long  = Center of Mass,
  class = abbrev
}
\DeclareAcronym{LO}{
  short = LO,
  long  = Leading Order,
  class = abbrev
}
\DeclareAcronym{NLO}{
  short = NLO,
  long  = Next-to-Leading Order,
  class = abbrev
}
\DeclareAcronym{PDF}{
  short = PDF,
  long  = Parton Distribution Function(s),
  class = abbrev
}
\DeclareAcronym{SPA}{
  short = SPA,
  long  = Supersymmetry Parameter Analysis,
  class = abbrev
}
\DeclareAcronym{VEV}{
  short = VEV,
  long  = Vacuum Expectation Value,
  class = abbrev
}
\DeclareAcronym{1PI}{
  short = 1PI,
  long  = One-Particle Irreducible,
  class = abbrev
}
\DeclareAcronym{BSM}{
  short = BSM,
  long  = (physics) Beyond the Standard Model,
  class = abbrev
}
\DeclareAcronym{UV}{
  short = UV,
  long  = Ultraviolet,
  class = abbrev
}
\DeclareAcronym{IR}{
  short = IR,
  long  = Infrared,
  class = abbrev
}
\usepackage[toc,numberedsection,nonumberlist]{glossaries}
\setglossarysection{section}

\makeglossaries

\newglossaryentry{M^D_S}{
  name={\ensuremath{M^D_S}},
  description={Dirac  mass}
}
\newglossaryentry{M^D_O}{
  name={\ensuremath{M^D_O}},
  description={Dirac gluino mass}
}

\usepackage{tikz}
\usepackage{tikz-3dplot}
\usetikzlibrary{shapes,arrows}

\usepackage{color, colortbl}

\usepackage[nottoc,numbib]{tocbibind}

\usepackage{caption}
\usepackage{feynmp}
\usepackage{tabularx}

\usepackage{longtable}
\usepackage{placeins} 

\makeatletter  
\renewcommand\maketitle{%
  \begin{titlepage}%
    \let\footnotesize\small
    \let\footnoterule\relax
    \begin{center}%
     \par \vspace*{2.25cm} 
         {\fontsize{20pt}{24pt}\selectfont\textbf{\@title}\par}
      \vspace{1cm}
      {\fontsize{17pt}{22pt}\selectfont\textsl{\@author}\par}
         \fontsize{12pt}{14pt}\selectfont
      \vspace{2cm}
      \begin{figure}[h!]
        \centering
        \includegraphics[scale=1.2]{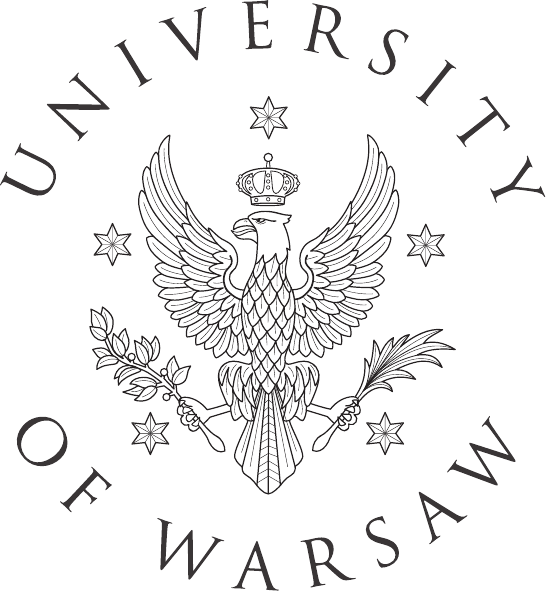}
       \end{figure}
       \vspace{2cm}
      \begin{flushright}
        \begin{tabular}{r}
          Doctoral dissertation prepared under the supervision of\\[2pt]
          \bfseries prof. dr hab. Jan Henryk Kalinowski \\
          at the Institute of Theoretical Physics, Faculty of Physics, University of Warsaw\\[2pt]
        \end{tabular}
      \end{flushright}
      \vspace{20mm plus .1fill}
      {\fontsize{14pt}{18pt} September 2016}
    \end{center}
 \end{titlepage}%
}
\makeatother   

\usepackage[explicit]{titlesec}

\definecolor{myrulecolor}{RGB}{150,20,0}

\begin{document}

\frontmatter
\title{Analysis of the R-symmetric supersymmetric models including quantum corrections}
\author{Wojciech Michał Kotlarski}
\maketitle

\titleformat{name=\chapter}[display]
  {\normalfont\scshape\Huge}
  {\hspace*{100pt}#1}
  {-15pt}
  {\hspace*{-110pt}{\color{myrulecolor}\rule{\dimexpr\textwidth+80pt\relax}{3pt}}\Huge} 
  \titleformat{name=\chapter,numberless}[display]
  {\normalfont\scshape\Huge}
  {\hspace*{180pt}#1}
  {-15pt}
  {\hspace*{-110pt}{\color{myrulecolor}\rule{\dimexpr\textwidth+80pt\relax}{3pt}}\Huge} 
  
\titlespacing*{\chapter}{0pt}{0pt}{30pt}

\chapter*{\center Acknowledgments}
\thispagestyle{empty}
\pagenumbering{gobble}
\addcontentsline{toc}{chapter}{Acknowledgments}
First and foremost, I would like to thank professor Jan Kalinowski for his help, support and encouragement.
I'm also grateful to people at the Institute for Nuclear and Particle Physics of the Technische Universität Dresden.
To professor Dominik Stöckinger for knowing the answer to every random QFT question I had and to the people I had the luck of sharing the office with there: Adriano, Josua, Markus, Philip, Patrik, Sebastian, Tom and Tomek. 
I'm indebted to dr Tania Robens, without whom I'd have never go there in the first place. 
Special thanks go to Philip and Tom for showing me what's what in Germany.

Last but not least, I would like to thank my fellow PhD students at the University of Warsaw: Bogusia, Marek, Ola, Olga and, especially, my office mates Mateusz and Mateusz. 

This work was partially supported by the German DFG grant STO 876/4-1.
\chapter*{\center Abstract}
\thispagestyle{empty}
\addcontentsline{toc}{chapter}{Abstract}

In this thesis the Minimal R-symmetric Supersymmetric Standard Model (MRSSM) is analyzed at the quantum level.
The MRSSM presents a viable alternative to the most commonly considered supersymmetric extensions of the Standard Model like the MSSM or the NMSSM.
It is based on the assumption that the $U(1)_R$ symmetry of the $\mathcal{N}=1$ SUSY algebra is left unbroken at the weak scale.
This assumption leads to a model with a distinct phenomenology, containing an extended Higgs sector, Dirac gauginos and color octet scalars.

The thesis consists of two parts. 
First one treats about the electroweak sector of the model.
Among others, it identifies the parameter region allowed by the electroweak precision observables.
Since the MRSSM contains an $SU(2)_L$-triplet with a non-zero vacuum expectation value the emphasis is put on the calculation of the $W$ boson mass.
To that end, a full one-loop calculation of $m_W$ augmented with the leading two-loop SM result is presented.
This allows to identify the region of parameter space consistent with the measured value of $m_W$.
The region is then checked against the measurement of the Higgs boson mass (where it is assumed that the lightest MRSSM Higgs state is SM-like).
For this, the full one-loop and leading two-loop corrections to the Higgs boson mass in the MRSSM are calculated.
Devised benchmark points, consistent with both of these observables, are shown to fulfill also a number of additional experimental constraints like properties of the Higgs boson(s), $b$-physics observables and vacuum stability.
Correlating all of these observables allows to put bounds on the parameters of the model.

Second part of the thesis treats about the strongly interacting scalar sector. 
First, NLO QCD corrections to the production of scalar gluon (sgluon) pair at the LHC are calculated.
The calculation of the virtual part and details of the zero-momentum subtraction scheme are given.
Large emphasis is put on the treatment of the infrared/collinear divergences which are dealt with using the two-cut phase space slicing method.
Their cancelation between real and virtual contributions is also explicitly checked.
A set of $K$-factors for a selected sgluon masses at 13 and 14 TeV LHC is presented.
This calculation is applied to constrain the sgluon mass using 2015 data set from Run 2 of the LHC.
To that end, a same-sign lepton search by ATLAS is recasted for the case of the production of the sgluon pair decaying to $t\bar{t}$ pairs.
The analysis is reproduced with the help of shower Monte Carlo softwares and the  program performing a fast detector response simulation.
Before applying this analysis to the sgluon pair production, it is validated on the selection of background processes showing reasonable agreement with the ATLAS one.
Its application to the sgluon signal gives (to my knowledge) first limits on the sgluon mass from the 13 TeV data.
The analysis shows that already using 3.2/fb of integrated luminosity the exclusion limits from Run 2 are competitive with the 8 TeV ones.

\addcontentsline{toc}{chapter}{Contents} 
\addcontentsline{toc}{chapter}{Preface}

\tableofcontents

\chapter*{\center Preface \label{sec:preface}}
\thispagestyle{empty}
\setcounter{page}{0}
\pagenumbering{gobble}
\setcounter{page}{0}

This thesis is partly based on the following publications \cite{Diessner:2014ksa,Diessner:2015yna,Diessner:2015iln,Kotlarski:2016zhv}:

\begin{itemize}
  \item[1] Philip~Diessner, Jan~Kalinowski, Wojciech~Kotlarski and Dominik~Stöckinger, \textit{Higgs boson mass and electroweak observables in the MRSSM}, JHEP 1412 (2014) 124
  \item[2] Philip~Diessner, Jan~Kalinowski, Wojciech~Kotlarski and Dominik~Stöckinger,
  \textit{Two-loop correction to the Higgs boson mass in the MRSSM},
  Adv. High Energy Phys. 2015 (2015) 760729
  \item[3] Philip~Diessner, Jan~Kalinowski, Wojciech~Kotlarski and Dominik~Stöckinger,
  \textit{Exploring the Higgs sector of the MRSSM with a light scalar},
  JHEP 1603 (2016) 007 
  \item[4] Wojciech Kotlarski,
  \textit{Sgluons in the same-sign lepton searches},
  arXiv:1608.00915 [hep-ph]
\end{itemize}
and conference proceedings \cite{Kotlarski:2011zz,Kotlarski:2011zza,Kotlarski:2013lja,Diessner:2015bna,Diessner:2015mml}:
\begin{itemize}
  \item[1] Wojciech~Kotlarski and Jan~Kalinowski,
            \textit{Scalar gluons at the LHC},
            Acta Phys.\ Polon.\ B 42 (2011) 2485
  \item[2]  Wojciech~Kotlarski,
            \textit{Searching for octet scalars in the t anti-t channel at the early stage of the LHC},
            Acta Phys.\ Polon.\ B 42 (2011) 1457
  \item[3]  Wojciech~Kotlarski, Artur~Kalinowski and Jan~Kalinowski,
            \textit{Searching for Sgluons in the Same-sign Leptons Final State at the LHC},
            Acta Phys.\ Polon.\ B 44 (2013) no.11,  2149
  \item[4]  Philip~Diessner and Wojciech~Kotlarski,
            \textit{Higgs and the electroweak precision observables in the MRSSM},
            PoS CORFU 2014 (2015) 079
  \item[5]  Philip~Diessner and Wojciech~Kotlarski,
            \textit{Meeting Higgs and Electroweak Precision Observables with R-symmetric SUSY},
            Acta Phys.\ Polon.\ B 46 (2015) no.11,  2179
\end{itemize}
The work concerning the electroweak sector of the MRSSM, published in Refs.~\cite{Diessner:2014ksa,Diessner:2015yna,Diessner:2015iln}, was done in collaboration with Jan Kalinowski from the University of Warsaw and Philip Diessner and Dominik Stöckinger from the TU Dresden.
The parts presented in this thesis were prepared, unless stated otherwise, by me. 

\pagenumbering{arabic}
\setcounter{page}{1}

\titleformat{\chapter}[display]
  {\normalfont\scshape\Huge}
  {\vspace{0cm}\hspace*{-1.1cm}\thechapter.~#1}
  {-15pt}
  {\hspace*{-110pt}{\color{myrulecolor}\rule{\dimexpr\textwidth+80pt\relax}{3pt}}\Huge}
\titleformat{name=\chapter,numberless}[display]
  {\normalfont\scshape\Huge}
  {\hspace*{-70pt}#1}
  {-15pt}
  {\hspace*{-110pt}{\color{myrulecolor}\rule{\dimexpr\textwidth+80pt\relax}{3pt}}\Huge}

\mainmatter

\chapter{Introduction \label{sec:intro}}

\section{Introduction}
The work that will be presented in this thesis was done while the most powerful particle accelerator to date, the Large Hadron Collider (LHC), was gathering its first data.
On July 4th, 2012 LHC experiments \ac{ATLAS} and \ac{CMS} announced the discovery of a new \textit{Higgs boson like} resonance with a mass of around 125 GeV.
Since then, additional data confirmed beyond any doubt that this was indeed the long sought Higgs boson.
With this single achievement LHC completed the particle spectrum of the Standard Model (\ac{SM}).
To a general sadness of particle physics community, no other (elementary) resonances where found, though.

\begin{figure}[]
  \centering
  \includegraphics[width=0.7\textwidth]{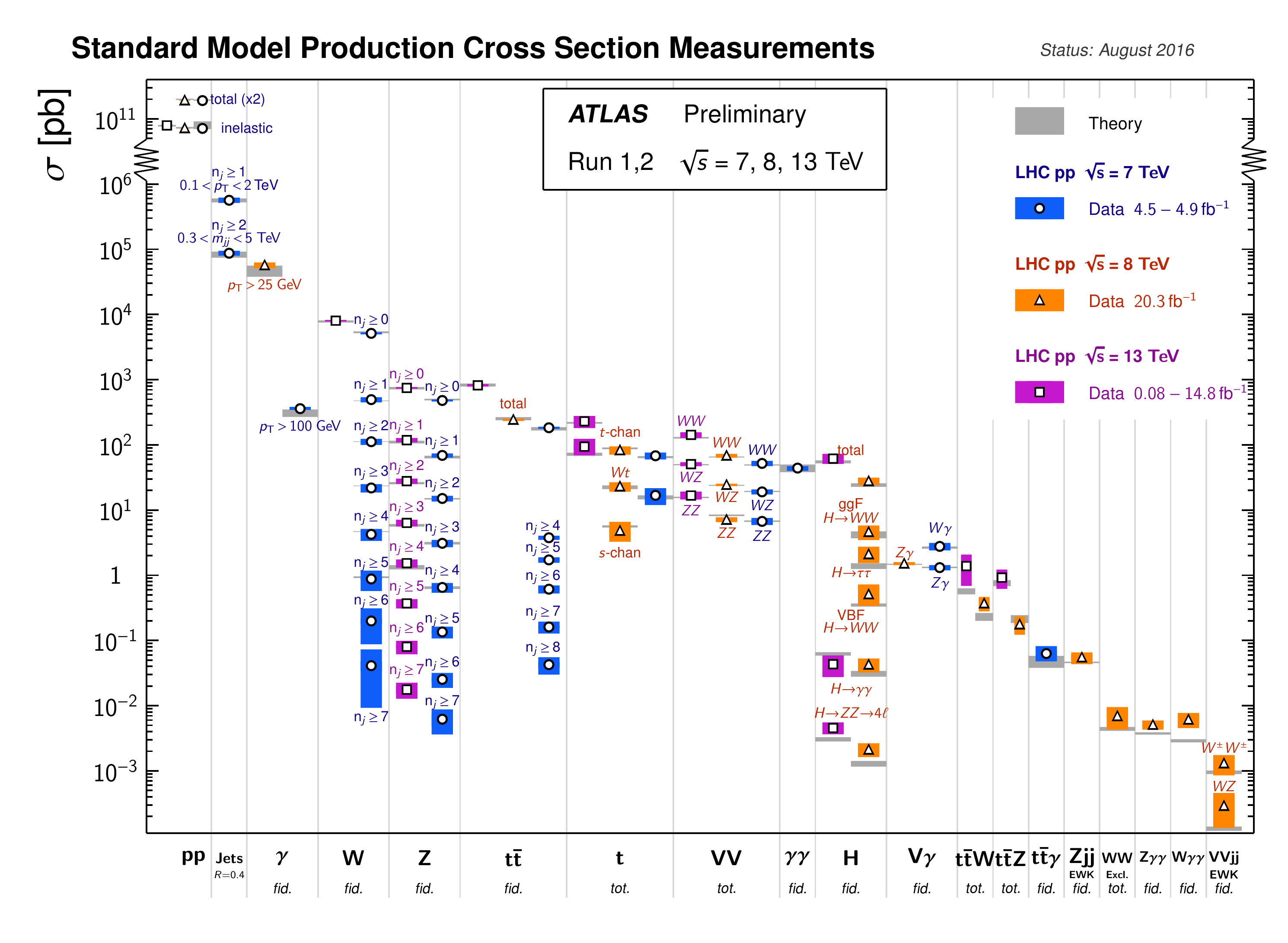}
  \caption{Standard Model 'status' circa ICHEP2016. \label{fig:sm_xsection}}
\end{figure}

The SM turns out to work extremely well also at LHC Run 2 energy frontier.
It correctly describes LHC results over the range of 14 orders of magnitude, as shown for example in \autoref{fig:sm_xsection}.\footnote{Figure available at \url{https://atlas.web.cern.ch/Atlas/GROUPS/PHYSICS/CombinedSummaryPlots/SM/ATLAS_b_SMSummary_FiducialXsect/ATLAS_b_SMSummary_FiducialXsect.pdf} [accessed 7.9.2016].}
As strange as it seems, particle physicists are not happy with that. 
The reason is that SM features certain experimental and theoretical shortcomings. 
On the experimental side phenomena like the presence of a dark matter, the muon anomalous magnetic moment discrepancy or neutrino masses call for its extension.
On the theoretical side, the issue of fine-tuning, lack of gauge coupling unification or not enough CP-violation to generate visible matter anti-matter asymmetry in the universe are puzzling.
A great hope of the physics community was that already first years of the LHC operating would point to a possible solution to those problem, commonly believed to be Beyond the Standard Model physics (BSM).

Among the possible candidates for the BSM physics, supersymmetry (\ac{SUSY}) is probably the most studied one.
This is, in part, because SUSY is arguably a beautiful theory. 
It is built upon our conviction that on a fundamental level world should be maximally symmetric.
Although theoretically very appealing, so far we weren't been able to confirm its existence.
Moreover, more and more stringent limits on particles of its minimal realization, called the Minimal Supersymmetric Standard Model (\ac{MSSM}), mean that one of the strongest arguments in its favor, naturalness, gets weakened.
Despite the sentiment of the physics community to the MSSM, SUSY does not have to be realized in this way.
In light of the null result on the MSSM search at the \ac{LHC}, it became important to consider its possible non-minimal realizations.

This realization generated a lot of interest in the non-minimal supersymmetric models.
They range from simple extensions of the MSSM like the NMSSM to models with Dirac gauginos and beyond.  
Among these models also appeared an old idea of Fayet, Salam and Strathdee of models with an unbroken R-symmetry, on which this thesis is focused.

Although R-symmetry is uniquely specified by the SUSY algebra, depending on fields' R-charge assignments, different R-symmetric models can be constructed.
One possible way, called in the literature the Minimal R-symmetric Supersymmetric Standard Model (\ac{MRSSM})~\cite{Kribs:2007ac}, was shown to be especially interesting.
In Ref.~\cite{Kribs:2007ac} it was shown that the MRSSM ameliorates the flavor problem of the MSSM, even in the case of anarchic flavor structure of the sfermion mass matrices.\footnote{This argument works in the so-called mass insertion approximation. See \autoref{sec:MRSSM} for the comment about a general case.}

The urgent open question that remained was whether MRSSM can also accommodate then newly discovered Higgs boson with a mass of $125$ GeV?
The answer to this question was not obvious as by construction MRSSM does not allow for left-right stop squark mixing, a feature that is often invoked in the MSSM as the source of an additional contributions to the Higgs boson mass in scenarios with light stop squarks.
Also, the MRSSM features an $SU(2)_L$ triplet with a non-zero vacuum expectation value (\ac{VEV}) which could spoil the electroweak precision observables (\ac{EWPO}) by directly contributing to the Peskin-Takeuchi $T$ parameter at the tree level.
These two questions form the core of the electroweak part of this thesis contained in \autoref{sec:ew} -- \autoref{sec:mhmw}.

In \autoref{sec:sgluon_analytic} -- \autoref{sec:sgluon_mc} the question of  how can the MRSSM be confirmed experimentally at the LHC is addressed.
In hadron colliders, like the LHC, it is best to focus first on the production of strongly interacting particles.
A non-typical (compared to the MSSM) feature of the MRSSM is the presence of a color octet scalars (sgluon) whose R-charge is the same as the one of the SM particles.
It therefore may be produced singly and decay without a company of the lightest  supersymmetric particle (\ac{LSP}).
Moreover, one of the octets might be naturally light.
In case of a favorable mass hierarchy of the MRSSM particles, it may decay exclusively to $t\bar{t}$ pairs.
Its pair production would give a quite uncommon in the SM $t\bar{t}t\bar{t}$ signature.
To study it, the NLO QCD corrections to the sgluon pair production are presented and the available experimental searches of this kind of topology are used to directly constrain the sgluon mass. 

\section{Structure of the thesis}

The thesis is structured as follows. 
In \Autoref{sec:SUSY} the basics of supersymmetry together with the construction of the \ac{MSSM} are recapitulated.
The chapter introduces SUSY algebra and the notion of superspace and superfields.
The rules for the construction of a SUSY invariant actions are given and applied in the construction of the MSSM Lagrangian.
The phenomenological status of the MSSM is also briefly discussed.

\Autoref{sec:MRSSM} then describes R-symmetry and goes into details of a construction of the Minimal R-symmetric Supersymmetric Standard Model.
It describes the field content of the model and its Lagrangian, presents the renormalization group equations for the gauge couplings and minimization conditions for its tree-level Higgs potential (tadpoles equations).
Explicit forms of selected mass matrices and Feynman rules are postponed to Appendices.
The discussion of the MRSSM finishes with a presentation of the benchmark points used in later chapters and with the description of the computational setup used for the numerical analysis of the model.
 
\Autoref{sec:ew} discusses the calculation of the $W$ boson mass.
It re-derives $m_W$ master formula of Degrassi, Fanchiotti and Sirlin \cite{Degrassi:1990tu} and discusses the role of $SU(2)_L$ triplet VEV in this calculation.
It then focusses on the calculation of vertex and box corrections to the muon decay constant in the MRSSM.
Chapter concludes with a short discussion of the numerical results, with the main numerical analysis postponed until \Autoref{sec:mhmw}.

\Autoref{sec:higgs_chapter} analyzes the Higgs sector of the MRSSM.
The Higgs boson mass is calculated at the full one-loop level including leading two-loop contributions in the effective potential approximation.
The chapter introduces the notion of the effective potential and gives an explicit expression for the MRSSM specific contributions.
Short discussion of the gauge dependence of the two-loop contribution is also given.

\Autoref{sec:mhmw} is devoted to the numerical study of results obtained in \autoref{sec:ew} and \ref{sec:higgs_chapter}.
The parameter region in agreement with both measured $W$ and Higgs boson masses is identified.
Moreover, benchmark points presented in \autoref{sec:MRSSM} are justified and also shown to pass other phenomenological constraints like $b$-physics observables and vacuum stability.
That chapter ends the discussion of the \ac{EW} sector of the model.

\Autoref{sec:sgluon_analytic} begins the part of the thesis devoted to the strong interactions.
It contains the calculation of the one-loop QCD corrections to the sgluon pair production at the \ac{LHC} using a simplified phenomenological model.
The chapter goes into details of renormalization and treatment of infrared/collinear divergences using the two-cut phase space slicing method.
After a thorough validation of the results and a cross-check with an independent calculation, a set of cross sections for 13 and 14 TeV LHC and selected sgluons masses is presented.

Those results are then applied in \Autoref{sec:sgluon_mc} to discus collider phenomenology of the sgluon pair production.
Under a reasonable assumptions about the mass hierarchy of the MRSSM particle spectrum the signature for pseudoscalar sgluon pair production is $t\bar{t}t\bar{t}$.
This kind of topology was already searched for by ATLAS and CMS in Run 2 data but not in the context of the sgluon production.
Therefore \Autoref{sec:sgluon_mc} recasts current experimental limits from search of same-sign lepton signature by ATLAS for the case of sgluon pair production, deriving lower limit on their mass.

The work ends with a summary in \autoref{sec:conclusions}.
Appendices contain supplementary materials: list of abbreviations and symbols, MRSSM mass matrices and Feynman rules, kinematic integrals and the derivation of sgluon decay widths.
\chapter{Supersymmetry \label{sec:SUSY}}

The Standard Model (SM) together with neutrino masses still provides an accurate description of all experimental data.\footnote{I do not mention dark matter here since its particle interpretation is still open to debate.}
The SM by itself is a consistent, renormalizable quantum field theory - valid to (almost) arbitrary energy scales.
The problem appears if there is any "physics beyond it". 
The (squared) Higgs boson mass is then quadratically sensitive to the scale of this physics.
This is known as the hierarchy problem.
Stabilization of the EW scale would then require precise cancelation between model parameters known as fine-tuning.
Even if there is no genuine new physics, effects of the gravity (which is not included in the SM) might also destabilize the EW scale.
Several solutions have been proposed to solve (or postpone) the hierarchy problem. To name few, these include:
\begin{itemize}
  \item \textbf{Technicolor} - Since there is no hierarchy problem for fermions (as they are protected by the chiral symmetry) one solution is to postulate that there are no elementary scalars. 
  In technicolor models the (effective) scalar degrees of freedom appear then as bound states of technifermions at the TeV scale.
  \item \textbf{Extra dimensions} - It is possible that the gravity becomes strong already at the TeV scale and hence there is no large hierarchy between gravity scale and the EW one.
  This occurs in models with extra spatial dimensions.
  \item \textbf{Little Higgs models} - In this models quadratic contributions to the Higgs boson mass appear at the two-loop level reducing the severity of fine-tuning by an additional loop factor.
  \item \textbf{Supersymmetry} (SUSY) - In the presence of the same number of fermionic and bosonic degrees of freedom with properly chosen couplings only wave-functions renormalize and they can depend at most logarithmically on the new physics scale  (the so-called non-renormalization theorem 
      \cite{Grisaru:1979wc,Seiberg:1993vc}). 
\end{itemize}
Among these models the specific appeal of supersymmetry comes as it solves many problems at once, from providing a dark matter candidate to gauge coupling unification and more.

The notion of supersymmetry originates from the string theory where it is realized on a $2 d$ world sheet.
The first spacetime SUSY theories were created by Wess and Zumino~\cite{Wess:1974tw} and by Salam and Strathdee~\cite{Salam:1974yz}.
The realization that SUSY might not suffer from the fine-tuning problem gave it a lot of interest which in the end led to the formulation of the first phenomenologically viable SUSY model, the Minimal Supersymmetric Standard Model (MSSM), in the early '80s.
Therefore, before going into details of the construction of the R-symmetric supersymmetric model it is instructive to remind basics of SUSY and the MSSM.\footnote{This chapter is based on Refs.~\cite{Drees:2004jm,Baer:2006rs,Martin:1997ns}.}

The chapter is structured as follows. 
First the formal aspects of supersymmetry are introduced.
To that end, the next section (\autoref{sec:super_poincare_algebra}) introduces the supersymmetric extension of the Poincar\'e algebra.
In \autoref{sec:superspace_and_superfields} the formalism of superspace and superfields is described.
\Autoref{sec:supersymmetric_lagrangian} then gives rules for constructing explicitly supersymmetric actions.
After that, in \autoref{sec:mssm}, the minimal supersymmetric extension of the SM is discussed.
The chapter ends with a brief comment about the validity of the MSSM in the light of new experimental data and possible non-minimal realizations of SUSY.

\section{Super-Poincaré algebra \label{sec:super_poincare_algebra}}
Theoretical efforts in the '60s to embed spacetime symmetries into a larger symmetry group culminated in the Coleman-Mandula no-go theorem \cite{Coleman:1967ad}.
Coleman and Mandula demonstrated that, abiding a set of reasonable physical assumptions, it is not possible to extend the Poincar\'e symmetry in a non-trivial way by using bosonic charges.
Golfand and Likhtman in \cite{Golfand:1971iw} proposed, following earlier works on SUSY on a $2d$ world sheet, a realization of SUSY in physical, 1+3 dimensional Minkowski space as a graded Lie algebra.
The case of spinorial charges, not covered by the Coleman-Mandula no-go theorem, was worked out by Haag, \L{}opusza{\'n}ski and Sohnius \cite{Haag:1974qh}.
They showed that, for a single spinorial charge $Q$, the most generic extension of the Poincaré algebra $\{P^\mu,M^{\mu \nu}\}$ is given by the so-called Super-Poincaré algebra\footnote{Only relations involving spinorial generators $Q$ are written below.}
\begin{align}
    \{Q_\alpha, Q_\beta \}&  = \{ \bar{Q}_{\dot{\alpha}},
\bar{Q}_{\dot{\beta}}\} = 0,\label{storm-hawk-1}\\
[ Q_\alpha, P_\mu ] & = [ \bar{Q}_{\dot{\alpha}}, P_\mu ] = 0 \\
    \{Q_\alpha, \bar{Q}_{\dot{\alpha}}\}&  = 2 \sigma^\mu_{\alpha
\dot{\alpha}} P_\mu,\label{ben-10}\\
    [M_{\mu \nu}, Q_{\alpha} ] & =  \imath {\left( \sigma_{\mu \nu}
\right)_{\alpha}}^\beta Q_\beta,\label{storm-hawk-2}\\
    [M_{\mu \nu}, \bar{Q}^{\dot{\alpha}} ] & =  \imath {\left( \bar{\sigma}_{\mu
\nu} \right)^{\dot{\alpha}}}_{\dot{\beta}}
\bar{Q}^{\dot{\beta}},\label{storm-hawk-3}
\end{align}
where $\sigma^{\mu \nu}\equiv\frac{1}{4}\left ( \sigma^\mu \bar{\sigma}^\nu - \sigma^\nu \bar{\sigma}^\mu\right)$, $\bar{\sigma}^{\mu \nu}\equiv\frac{1}{4}\left ( \bar{\sigma}^\mu \sigma^\nu - \bar{\sigma}^\nu \sigma^\mu\right)$ and $\sigma^\mu \equiv (I, \vec \sigma)$, $\bar{\sigma}^\mu \equiv (I, -\vec \sigma)$ with $\vec\sigma$ being Pauli matrices and $\mu,\nu=0,1,2,3$, $\mu < \nu$. 

Phenomenologically interesting are mainly $\mathcal{N}=1$ supersymmetries as extended SUSY theories require, among others, vector-like (and not chiral) representations of gauge groups.

\section{Superspace and superfields \label{sec:superspace_and_superfields}}
Superfields, introduced by Salam and Strathdee \cite{Salam:1974yz}, provide a convenient tool in constructing manifestly supersymmetric theories.
This is done by extending Minkowski space by four fermionic coordinates
\begin{equation}
  x^\mu \to x^\mu, \theta^\alpha, \bar{\theta}_{\dot{\alpha}},
\end{equation}
with indices $\alpha, \dot{\alpha} = 1, 2$ and $\theta$ of dimension [mass]$^{-1/2}$. 
The SUSY generators $Q$ are then represented as differential operators on the superspace
\begin{align}
  Q_\alpha &= -\imath \left(\partial_\alpha + \imath \sigma^\mu_{\alpha \dot{\beta}} \bar \theta^{\dot{\beta}} \partial_\mu \right) \\
  \bar{Q}_{\dot{\alpha}} &= \imath \left(\partial_{\dot \alpha} + \imath \theta^\beta \sigma^\mu_{\beta \dot \alpha} \partial_\mu  \right)
\end{align}

A function of superspace coordinates $\Phi(x^\mu, \theta^\alpha, \bar{\theta}^{\dot{\alpha}})$ is called a superfield. 
An expansion of $\Phi$ as a power series in $\theta$ has the following general form
\begin{multline}
\Phi(x, \theta, \bar{\theta} ) = f(x) + \sqrt{2} \theta \psi(x) + \sqrt{2}  \bar{\theta}
\bar{\chi}(x) + \theta \theta m(x) + \bar{\theta}\bar{\theta} n(x)
+\\  \theta \sigma^\mu \bar{\theta} A_\mu(x) + \theta \theta
\bar{\theta} \bar{\lambda}(x) + \bar{\theta} \bar{\theta} \theta
\rho(x) + \frac{1}{2}\theta \theta \bar{\theta} \bar{\theta} D(x)\,,
\end{multline}
where for convenience (and without loss of generality) one introduces the numerical factors of $\sqrt{2}$ and 1/2.
Under the transformation of the Poincar\'e group, $A_\mu$  transforms as a vector, $\psi$, $\bar \chi$, $\rho$, $\bar \lambda$ as Weyl spinors and $f,\,m,\,n,\,d$ are scalars.
This SUSY representation turns out to be reducible.
To obtain irreducible representations of the supersymmetry algebra, called supermultiplets, it is convenient to utilize SUSY covariant derivatives 
\begin{align}
D_\alpha & \equiv  \frac{\partial}{\partial \theta^\alpha} - \imath \sigma^\mu _{\alpha \dot{\alpha}} \bar{\theta}^{\dot{\alpha}} \partial_\mu , \label{eq:czerwony1}\\
\bar{D}_{\dot{\alpha}} & \equiv  - \frac{\partial}{\partial \bar{\theta}^{\dot{\alpha}}} + \imath \theta^{\alpha} \sigma^\mu_{\alpha \dot{\alpha}} \partial_\mu \label{eq:czerwony2},
\end{align}
fulfilling
\begin{align}
\label{eq:susy_gen_com}
\left \{ D_\alpha, Q_\beta \right \} = \left \{ D_\alpha,
\bar{Q}_{\dot{\beta}} \right \} = \left \{ \bar{D}_{\dot{\alpha}},
Q_\beta \right \} = \left \{ \bar{D}_{\dot{\alpha}},
\bar{Q}_{\dot{\beta}} \right \} = 0.
\end{align}
A left-chiral superfield is defined as  
\begin{align}
  \label{eq:left_chiral_condition}
  \bar{D}_{\dot{\alpha}} \Phi &\overset{!}{=} 0 
\end{align}
(analogously, superfields fulfilling $D_\alpha \Phi^\dagger \overset{!}{=} 0$ are called right-chiral). 
It follows from \autoref{eq:susy_gen_com} that a (right)left-chiral stays (right)left-chiral under the SUSY transformation.

The form a superfield $\Phi$ fulfilling \autoref{eq:left_chiral_condition} can be easily obtained by observing that superspace coordinates $y^\mu \equiv x^\mu + \imath \theta \sigma^\mu \bar{\theta}$ and $\theta$ are annihilated by $\bar{D}_{\dot{\alpha}}$, $\bar{D}_{\dot{\alpha}} (x^\mu + \imath \theta \sigma^\mu \bar{\theta}) = 0,\,\bar{D}_{\dot{\alpha}} (\theta) =0$. 

An (off-shell) left-chiral field must therefore have the form
\begin{align}
\label{eq:chiral_superfield}
\Phi(y^\mu, \theta)  = & \phi(y^\mu) + \sqrt{2} \theta \psi(y^\mu) + \theta\theta F(y^\mu)
 =  \phi(x^\mu) + \sqrt{2} \theta \psi(x^\mu) + \theta \theta F(x^\mu) \\
 & +
\imath \partial_\mu \phi (x^\mu) \theta \sigma^\mu \bar{\theta} +
      -\frac{\imath}{\sqrt{2}} \theta \theta \partial_\mu \psi(x^\mu) 
\sigma^\mu \bar{\theta} - \frac{1}{4} \partial_\mu \partial^\mu \phi(x^\mu)
\theta \theta \bar{\theta}  \bar{\theta}, \nonumber
\end{align}

A vector superfields fulfills condition $V\overset{!}{=}V^\dagger$. 
Most general form of a vector superfield is given by
\begin{align}
V(x^\mu, \theta, \bar{\theta}) = & C(x^\mu) + \imath \theta
\chi(x^\mu) - \imath \bar{\theta} \bar{\chi}(x^\mu) + \frac{1}{2}
\imath \theta \theta [ M(x^\mu) + \imath N(x^\mu) ] \\
& - \frac{1}{2}
\imath \bar{\theta} \bar{\theta} [ M(x^\mu) - \imath N(x^\mu) ] +
\theta \sigma^\mu \bar{\theta} A_\mu (x^\mu) + \imath \theta \theta
\bar{\theta} \left [\bar{\lambda}(x^\mu) + \frac{\imath}{2}
\bar{\sigma}^\mu \partial_\mu \chi(x^\mu) \right ] \\ 
&- \imath
\bar{\theta} \bar{\theta} \theta \left [ \lambda(x^\mu) +
\frac{\imath}{2} \sigma^\mu \partial_\mu \bar{\chi}(x^\mu) \right ]
+ \frac{1}{2} \theta \theta \bar{\theta} \bar{\theta} \left [
D(x^\mu) - \frac{1}{2} \partial_\mu \partial^\mu C(x^\mu) \right ],
\end{align}
where $C,M,N,D$ are real scalar fields, $\lambda,\chi$ are Weyl spinors and $A_\mu$ is a real vector field. 
Some components might be removed by the supergauge transformation giving the vector multiplet in the Wess-Zumino gauge \cite{Wess:1974jb}
\begin{align}
  \label{eq:wz_supermultiplet}
    V_{WZ}^a = \theta \sigma^\mu \bar{\theta} A^a_\mu(x^\mu) + \imath \theta
\theta \bar{\theta} \bar{\lambda}^a(x^\mu) - \imath \bar{\theta} \bar{\theta}
\theta \lambda^a(x^\mu) + \frac{1}{2} \theta \theta \bar{\theta} \bar{\theta}
D^a(x^\mu).
\end{align}
It should be denoted that the Wess-Zumino gauge is non-supersymmetric (although one can always go back to Wess-Zumino form of a vector supermultiplet by using again a super-gauge transformation after a SUSY one). 
\section{Supersymmetric Lagrangian \label{sec:supersymmetric_lagrangian}}
Using the superfield formalism, construction of explicitly supersymmetric actions follows simple rules.
Since $F$ and $D$ terms transform as full derivatives, they are candidates for Lagrangian densities.
For a given chiral superfield $\Phi$ of \autoref{eq:chiral_superfield} the $F$ term is obtained as
\begin{equation}
  \label{eq:fterm}
  F = \left.\int d^2 \theta \, \Phi \right|_{\bar \theta = 0}.
\end{equation} 
Analogously, for a vector superfield $V$ the $D$-term is obtained as
\begin{equation}
  \label{eq:dterm}
  \frac{1}{2} D = \int d^4 \theta \, V .
\end{equation}

\subsection{Lagrangian for interacting chiral superfields}
Using the observation that the product of left-chiral superfields is a left-chiral superfield and that the product $\Phi^\dagger \Phi$ is a vector superfield the Lagrangian for a theory with only chiral superfields is given by
\begin{equation}
\label{eq:pure_chiral}
\mathcal{L} = \int d^4 \theta \, \Phi^\dagger \Phi + \left ( \int
d^2 \theta \, W[\Phi_i] + h.c.\right ),
\end{equation}
where $W[\Phi_i]$ is a polynomial in superfields $\Phi_i$.
For renormalizable theories the degree of $W$ is at most 3.

\subsection{Lagrangian for free vector superfields}
Kinetic terms for vector superfields are constructed with the help of spinor superfields $W$ defined for the non-abelian gauge group as
\begin{align}
    W_\alpha =& -\frac{1}{4} \bar{D}^{\dot{\beta}} \bar{D}_{\dot{\beta}} e^{-V} D_\alpha e^V,\\
    \overline{W}^{\dot \alpha} =& -\frac{1}{4} D^{\beta} D_{\beta} e^{V} \bar D^{\dot \alpha} e^{-V} ,
\end{align}
where $D$s are defined in \autoref{eq:czerwony1} and \autoref{eq:czerwony2} and $V \equiv 2 g V_a T^a$ with $g$ being a gauge coupling and $T$ denoting generators of the gauge group representation.
Since $\bar{D}_{\dot{\alpha}} W_\alpha = D_{\alpha} \bar W_{\dot \alpha} =0$ the $W_\alpha$ ($\bar W_{\dot \alpha}$) is a left (right) chiral field. 
The $F$-term of a product $W_\alpha W^\alpha$ (or its conjugate) is therefore a candidate for supersymmetric action
\begin{equation}
  \label{eq:pure_vector}
  \mathcal{L} = \left. \frac{1}{4} \int d^2\theta\, W^{a,\alpha} W^a_\alpha \right|_{\bar \theta=0} + h.c.,
\end{equation}
with the normalization to correctly reproduce canonic kinetic terms for the gauge bosons.

\subsection{Lagrangian for Super-Yang-Mills theory}
Lagrangian for a Super-Yang-Mills theory with matter interacting with non-abelian gauge bosons is a simple generalization of Lagrangians in \autoref{eq:pure_chiral} and \autoref{eq:pure_vector}. 
For a gauge group with coupling $g$ and a matter $\Phi$ in the gauge group representation generated by $T$ it reads
\begin{equation}
\label{eq:full_susy_lagrangian}
\mathcal{L} = \left. \frac{1}{4} \int d^2\theta\, W^{a,\alpha} W^a_\alpha \right|_{\bar \theta=0} + h.c. 
+ \int d^4 \theta\,  \Phi^\dagger e^{2 g  V_a T^a} \Phi + \left ( \left.\int
d^2 \theta \, W[ \Phi_i] \right|_{\bar \theta=0} + h.c.\right ).
\end{equation}

\section{SUSY breaking \label{sec:susy_breaking}}
Since experiments do not detect signals of supersymmetry, SUSY must be broken in realistic models.
As every symmetry, it can be broken either explicitly or spontaneously, but due to the SUSY mass sum rules, the latter one cannot happen in the visible sector. 
It can happen in the sector mostly decoupled from the visible one and communicated to it, though.
The breaking might occur through $F$-terms (O'Raifeartaigh \cite{O'Raifeartaigh:1975pr}), $D$-terms (Fayet-Iliopoulos \cite{Fayet:1974jb}) or the combination of both mechanisms.
Communicated to the visible sector through, for example, gravity or gauge interactions spontaneous SUSY breaking will look like an explicit one.
Girardello and Grisaru \cite{Girardello:1981wz} classified which explicit SUSY breaking terms do not introducing quadratic sensitivity to the high scale of the theory - the so-called soft-SUSY breaking (\ac{SSB}) terms.
This allows to talk about SUSY phenomenology without the knowledge of its exact breaking mechanism.
The Lagrangian of \autoref{eq:full_susy_lagrangian} can therefore be supplemented by the following one
\begin{align}
\label{eq:soft_breaking_terms}
\mathcal{L}_{\text{soft}} = &
\left ( - \frac{1}{2} M \lambda^a \lambda^a - M^D \lambda^a \psi_a - \frac{1}{6} A_{ijk}
\phi_i \phi_j \phi_k - \frac{1}{2} B_{\mu,ij} \phi_i \phi_j - t_i \phi_i
+ c.c. \right )  \\ 
& - \left(m^2\right)^{i}_j \phi^{j*} \phi_i \nonumber,
\end{align}
where the terms are, in order:  Majorana and Dirac gaugino masses, holomorphic tri- and bilinear scalar couplings, tadpole terms and scalar masses.
The second term does not exist in the MSSM but it will be crucial in the construction of the MRSSM.

\section{Minimal Supersymmetric Standard Model \label{sec:mssm}}
Having discussed theoretical basics of SUSY it is instructive to examine the construction of the simplest phenomenologically viable SUSY model -- the MSSM -- before discussing non-minimal models like the MRSSM, which will be done in the next chapter.

The model is based on the SM gauge group and extends its particle content by the minimal number of new fields (hence the name 'minimal').
Following subsections will describe MSSM field content, its Lagrangian and provide a short discussion of its experimental status.
The last part will serve as a motivation for the discussion of non-minimal SUSY models.

\subsection{Field content}
\Autoref{tab:mssm_field_content} shows the field content of the MSSM.
It lists superfields, their matter parity, (gauge) quantum numbers and names of their components.
The model consist of supersymmetrized matter content and two $SU(2)_L$ Higgs doublets which are needed due to holomorphicity of the superpotential $W$ and for anomaly cancelation.
Fermionic particles of the SM are part of chiral superfields together with their superpartners dubbed sfermions.
Higgs bosons also reside in chiral superfiels, with their fermionic partners called higgsinos.
Following the discussion in \autoref{sec:superspace_and_superfields} gauge fields reside in the vector supermultiplets, with fermionic partners of gauge bosons called gauginos.\footnote{The general naming convention is that scalar partners of the SM particles get prefix \textit{s} while fermionic ones get suffix \textit{inos}.}
\begin{table}
\begin{center} 
\begin{tabular}{ccccc} 
chiral superfield & matter parity & spin 0 & spin \(\frac{1}{2}\) & \(U(1)_Y \otimes\, SU(2)_L \otimes\, SU(3)_C\)\\
\midrule 
\(\hat{q}\) & -1 & \(\tilde{q}\) & \(q\) & \(\left(\frac{1}{6},{\mathbf 2},{\mathbf 3} \right) \)\\ 
\(\hat{d}\) & -1 & \(\tilde{d}_R^*\) & \(d_R^*\) & \(\left(\frac{1}{3},{\mathbf 1},\bar{\mathbf {3}}\right) \) \\
\(\hat{u}\) & -1 & \(\tilde{u}_R^*\) & \(u_R^*\) & \(\left(-\frac{2}{3},{\mathbf 1},\bar{\mathbf {3}}\right) \) \\
\(\hat{H}_d\) & 1 & \(H_d\) & \(\tilde{H}_d\) & \(\left(-\frac{1}{2},{{\mathbf 2}},{\mathbf 1}\right) \) \\
\(\hat{H}_u\) & 1 & \(H_u\) & \(\tilde{H}_u\) & \(\left(\frac{1}{2},{\mathbf 2},{\mathbf 1} \right) \) \\
\(\hat{l}\) & -1 & \(\tilde{l}\) & \(l\) & \(\left(-\frac{1}{2},{\mathbf 2},{\mathbf 1}\right) \) \\ 
\(\hat{e}\) & -1 & \(\tilde{e}_R^*\) & \(e_R^*\) & \(\left(1,{\mathbf 1},{\mathbf 1}\right) \)
\\
\\
vector superfield & matter parity & spin \(\frac{1}{2}\) & spin 1 & \(U(1)_Y \otimes\, SU(2)_L \otimes\, SU(3)_C\)\\
\midrule 
  $\hat B$ & 1 & $\tilde B^0$ & $B^0$ & \(\left(1,{\mathbf 1},{\mathbf 1}\right) \)\\
$\hat W$ & 1 & $\tilde W^\pm$, $\tilde W^0$ & $\tilde W^\pm$, $W^0$ & \(\left(1,{\mathbf 3},{\mathbf 1}\right) \)\\
$\hat g$ & 1 & $\tilde g$ & $g$ & \(\left(0,{\mathbf 1},{\mathbf 8}\right) \)
\end{tabular}
\end{center} 
\caption{Field content of the MSSM. No generation indices for (s)fermions are written.}
\label{tab:mssm_field_content}
\end{table}

\subsection{Superpotential}
Knowing the field content one can write superpotential for the model.
The superpotential $W$ of \autoref{eq:full_susy_lagrangian} for the MSSM then reads
\begin{align}
W = & \mu \,\hat{H}_u \cdot \hat{H}_d\, - Y_d \,\hat{d}\,\hat{q}\cdot\hat{H}_d\,- Y_e \,\hat{e}\,\hat{l}\cdot\hat{H}_d\, +Y_u\,\hat{u}\,\hat{q}\cdot\hat{H}_u\, .
\label{eq:mssm_superpot}
 \end{align}
where $A \cdot B \equiv A_i \epsilon_{ij} B_j$ with the Levi-Civita symbol $\epsilon_{12} = 1$.

\subsection{Matter parity and R-parity}
The superpotential in \autoref{eq:mssm_superpot} is not the most general, gauge invariant expression that can be written using fields in \autoref{tab:mssm_field_content}.
In general, MSSM allows also for the following Baryon ($B$) and Lepton ($L$) number violating terms to appear in the superpotential
\begin{eqnarray}
W_{\Delta L=1} & = & \frac{1}{2} \lambda^{ijk} \hat{L}_i \hat{L}_j
\bar{\hat{e}}_k + \lambda '^{ijk} \hat{L}_i \hat{Q}_j \bar{\hat{d}}_k + \mu '
\hat{L}_i \cdot \hat{H}_u \label{eq:knd1} ,\\
W_{\Delta B =1} & = & \frac{1}{2} \lambda ''^{ijk} \bar{\hat{u}}_i
\bar{\hat{d}}_j \bar{\hat{d}}_k \label{eq:knd2}.
\end{eqnarray} 
These terms generate unobserved effects like proton decay and are therefore highly constrained.
They can be eliminated by enforcing a so-called matter parity
\begin{equation}
\label{eq:matter_parity}
  M_p = (-1)^{3(B-L)}
\end{equation}
acting multiplicatively on the superfields (see \autoref{tab:mssm_field_content} for $M_P$ assignment).
Since an entire supermultiplet has the same matter parity this symmetry commutes with SUSY generators.
The matter parity is equivalent (if theory conserves angular momentum) to the R-parity defined as
\begin{equation}
\label{eq:r_parity}
R_p = (-1)^{3(B-L)+2S} = (-1)^{(3B-L)+2S}
\end{equation}
which acts on particles (not supermultiplets) and where $S$ is the spin of the particle.

The building assumption of the MSSM is therefore that there is a discrete $Z_2$ symmetry called R-parity $R_p$.
One assigns $R_p = 1$ to particles of the SM (including all Higgses) and $R_p = -1$ for sparticles (which is consistent with the $M_p$ assignment in \autoref{tab:mssm_field_content}).
As a consequence of R-parity the lightest supersymmetric particle (LSP) is stable.
If neutral, LSP is a candidate for dark matter particle.

\subsection{Soft-breaking terms in the MSSM}
Following the discussion in \autoref{sec:susy_breaking}, realistic model must be supplemented by a set of soft-breaking terms consistent with the gauge symmetries of the model.
Using \autoref{eq:soft_breaking_terms}, for the field content of  \autoref{tab:mssm_field_content}, the soft-breaking terms are
\begin{itemize}
  \item gaugino masses
  \begin{equation}
  \label{eq:mssm_gaugino_masses}
    \mathcal{L}_{\text{SSB}} \ni
  -\frac{1}{2} M_B \tilde{B} \tilde{B}   
  -\frac{1}{2} M_W \tilde{W}_i \tilde{W}_i 
  -\frac{1}{2} M_O \tilde{g} \tilde{g} + h.c.
  \end{equation}
  \item scalar masses
  \begin{align}
  \label{eq:mssm_ssb_scalar_masses}
  \mathcal{L}_{\text{SSB}} \ni & - m_{H_d}^2 \left(  |H_d^0|^2 + |H_d^-|^2 \right)
  - m_{H_u}^2 \left( |H_u^0|^2 + |H_u^+|^2 \right) \\
  & - \tilde{d}^*_{L,i} m_{q,{i j}}^{2} \tilde{d}_{L,j} 
   - \tilde{d}^*_{R,i} m_{d,{i j}}^{2} \tilde{d}_{R,j} \nonumber \\
  & - \tilde{u}^*_{L,i} m_{q,i j}^{2} \tilde{u}_{L,j}
  - \tilde{u}^*_{R,i} m_{u,{i j}}^{2} \tilde{u}_{R,{j}} \nonumber \\
 & - \tilde{e}^*_{L,{i}} m_{l,{i j}}^{2} \tilde{e}_{L,    {j}} +\tilde{e}^*_{R,{i}} m_{e,{i j}}^{2} \tilde{e}_{R,{j}} \nonumber \\
 & - \tilde{\nu}^*_{L,{i}} m_{l,{i j}}^{2} \tilde{\nu}_{L,{j}} \nonumber
  \end{align}
  \item holomorphic bilinear terms
  \begin{equation}
  \label{eq:mssm_holomorhic_bilinear}
  \mathcal{L}_{\text{SSB}} \ni B_{\mu} \left( H_d^0 H_u^0  - H_d^- H_u^+ \right)
  \end{equation}
  \item trilinear terms
  \begin{align}
\mathcal{L}_{\text{SSB}} \ni & \, 
  H_d^0 \tilde{d}^*_{R,i} \tilde{d}_{L,j} A_{d,{i j}} - H_d^- \tilde{d}^*_{R,i} \tilde{u}_{L,j} A_{d,{i j}} 
 +H_d^0 \tilde{e}^*_{R,{i}} \tilde{e}_{L,{j}} A_{e,{i j}}  \\ 
 & - H_d^- \tilde{e}^*_{R,{i}} \tilde{\nu}_{L,{j}} A_{e,{i j}} - H_u^+ \tilde{u}^*_{R,i} \tilde{d}_{L,j} A_{u,{i j}} + H_u^0 \tilde{u}^*_{R,i}  \tilde{u}_{L,j} A_{u,{i j}} + h.c. \nonumber
\end{align}
\end{itemize}

\begin{figure}
  \centering
  \hspace{-1.5cm}
  \includegraphics[width=\textwidth]{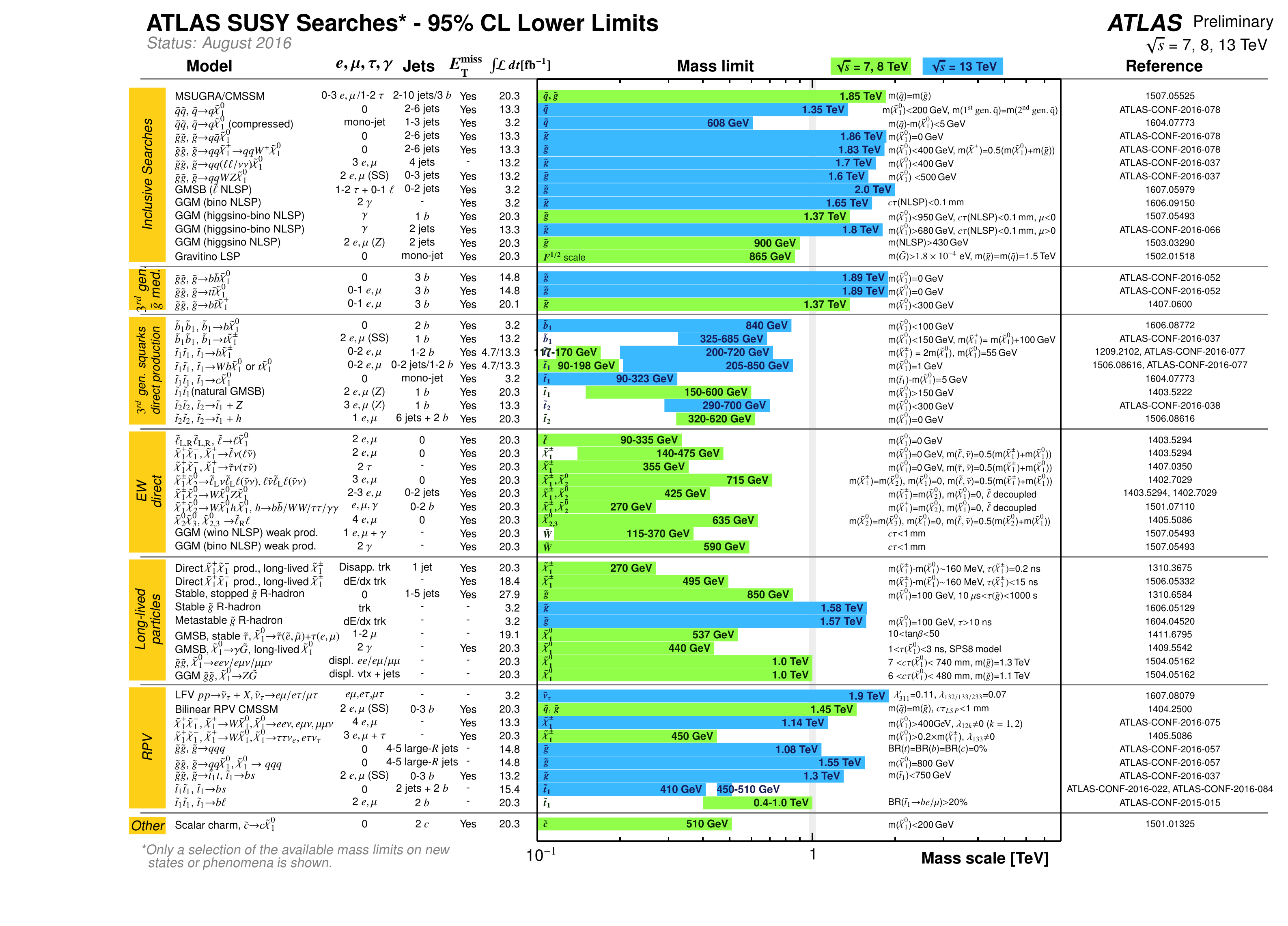}
  \caption{Selection of ATLAS SUSY mass limits.
  \label{fig:susy_exclusion} 
  }
\end{figure}

\subsection{Physical states}
After supplementing MSSM with the soft SUSY breaking terms listed in the previous subsection, the model becomes rather complicated.
CP-conserving MSSM has (including SM ones) 124 free parameters \cite{Dimopoulos:1995ju}.
This number also includes new parameters of the Higgs sector not present in the SM as after the EW symmetry breaking both of the Higgs boson doublets receive vacuum expectation values
\begin{align}
  H_u = & 
  \left( 
  \begin{matrix}
    H_u^+ \\
    \frac{1}{\sqrt{2}} \left(v_u + \Re H_u^0 +\imath \Im H_u^0 \right)
  \end{matrix} 
  \right)\\
  H_d = & 
  \left( 
  \begin{matrix}
    \frac{1}{\sqrt 2}(v_d + \Re H_d^0 + \imath \Im H_d^0) \\
    H_d^-
  \end{matrix} 
  \right)
\end{align} 
It is customary to exchange those VEVs for parameters $\beta$ and $v$ defined as $\tan \beta \equiv v_u/v_d$ and $v \equiv \sqrt{v_u^2 + v_d^2} \approx 246$ GeV.
On the other hand, in contrast to the SM, the quartic Higgs coupling in the MSSM is not a free parameter.

To summarize, the spectrum of physical states of the model consists of the SM particles but with 2 CP-even Higgs bosons, CP-odd one, 2 Dirac charginos, 4 Majorana neutralinos, Majorana gluino, squarks and sleptons.
This list will be stacked up against the MRSSM particle spectrum in the next chapter.

\section{Going beyond}
Supersymmetry was searched for intensively during Run 1 of the LHC.
Gathered data was interpreted mainly in the context of the MSSM (either directly in the constrained MSSM or within simplified models with the MSSM particle content).
\Autoref{fig:susy_exclusion} shows a selection of exclusion limits as for ICHEP 2016, that is including full Run I data together with the first Run II results.\footnote{Figure available at \url{https://atlas.web.cern.ch/Atlas/GROUPS/PHYSICS/CombinedSummaryPlots/SUSY/ATLAS\_SUSY\_Summary/ATLAS\_SUSY\_Summary.png} [accessed 7.9.2016].}
In many cases, the exclusions reach almost to 2 TeV.
Large mass splitting in the supermultiplets reintroduces a fine-tuning problem.
A folklore is that this splitting should not be bigger than a few TeV.
If this is the case, we are slowly approaching a point where MSSM becomes a problem, not a solution.
The caveat, as already mentioned in \Autoref{sec:intro}, is that these results are model dependent.
It has been shown in Ref.~\cite{Heikinheimo:2011fk} that in models with Dirac gluinos squark-gluino exclusion limits are strongly altered.
Similar effect has been shown in Ref.~\cite{Diessner:2015iln} for EW-gauginos.
It is therefore important not to identify SUSY with the MSSM, remembering that SUSY might be realized in a different way.
With this in mind, the MRSSM presented in the next chapter might provide a solution to certain shortcomings of the MSSM.

\chapter{Minimal R-symmetric Supersymmetric Standard Model \label{sec:MRSSM}}

The concept of R-symmetry is roughly as old as the supersymmetry itself \cite{Salam:1974xa,Fayet:1974pd}.
Beside being a symmetry of the SUSY algebra, R-symmetry is also intimately connected with the SUSY breaking.
Nelson and Seiberg showed that, under certain assumptions, spontaneously broken R-symmetry is a sufficient condition for SUSY breaking \cite{Nelson:1993nf}.
This does not exclude the possibility of existing of non-supersymmetric metastable vacua with an unbroken R-symmetry \cite{Intriligator:2007py,Intriligator:2006dd,Amigo:2008rc}, though. 
Since both gauge and matter kinetic terms are automatically R-invariant, the question arrises if R-symmetry could also be a symmetry of the superpotential and soft breaking terms.
Since for phenomenologically viable models this requires, as will be shown in this chapter, increasing the particle content of the model beyond just simply supersymmetrizing the SM - this road was not pursued in the '70s.
Instead most studies of the last 40 years were focused on the MSSM where R-symmetry is explicitly broken by, for example, Majorana gaugino masses.
The failure to discover the MSSM, together with its certain theoretical shortcomings, revived the interest in R-symmetry as a guiding principle in constructing low energy SUSY models.
This chapter is therefore devoted to describing such, phenomenologically viable, R-symmetric SUSY models.

The chapter is structured as follows.
In the next section theoretical aspects of R-symmetry are discussed.
After that, \Autoref{sec:mrssm_sec} describes the structure of the MRSSM.
Following the same outline as \autoref{sec:mssm} it lists MRSSM field content (\autoref{sec:field_content}), superpotential (\autoref{sec:mrssm_superpotential}), soft breaking terms (\autoref{sec:mrssm_ssb}) and physical states of the model (\Autoref{sec:particle_spectrum}).
\Autoref{sec:mrssm_rges} comments on the difference of one-loop renormalization group equations for gauge coupling between MSSM and MRSSM.
Also, MRSSM tree-level tadpole equations are given in \autoref{sec:mrssm_tadpoles}.
The discussion of the structure of the MRSSM finishes in \autoref{sec:mrssm_bmps} with a presentation of 3 benchmark points which will be used later on in confirming the viability of the model.
The chapter finishes with a description of the computational setup used for numerical analysis.

\section{R-symmetry}

As mentioned in the introduction, the super-Poincar\'e algebra described in \autoref{sec:super_poincare_algebra} has an additional $U(1)$ symmetry called R-invariance or R-symmetry (denoted $U(1)_R$).
Commutation relations of its generator $R$ with super-Poincar\'e generators are given by
\begin{align}
  [P_\mu, R ] & = [M_{\mu \nu}, R ] = 0\\
  \label{eq:scooby_doo1}
  [ Q_\alpha, R ] & = Q_\alpha\\
    \label{eq:scooby_doo2}
  [\bar{Q}^{\dot \alpha}, R] &=  -\bar{Q}^{\dot \alpha}
\end{align}
From \Autoref{eq:scooby_doo1} and \autoref{eq:scooby_doo2}, using the Baker–Campbell–Hausdorff, it follows that
\begin{align}
  R: Q_\alpha \to e^{\imath \varphi R} \, Q_\alpha e^{-\imath \varphi R} = e^{-\imath \varphi} Q_\alpha,
\end{align}
meaning that the R-charge of $Q$ is $-1$ (and analogously $+1$ for $\bar Q$).
The R-charge is also defined for an entire superfield
\begin{equation}
  R: \Phi(x, \theta, \bar \theta) \to \Phi'(x, e^{\imath \varphi} \theta, e^{-\imath \varphi} \bar \theta) = e^{\imath \varphi R_\Phi} \Phi(x, \theta, \bar \theta) .
\end{equation}
Reality of a vector superfield requires that $R(\hat V) = 0$.
From \autoref{eq:wz_supermultiplet} it then follows that
\begin{equation}
  R(V_\mu) = 0, \hspace{1cm} 
  R(\lambda) = - R(\bar \lambda) = 1, \hspace{1cm} 
  R(D) = 0
\end{equation}
For a chiral superfield $\Phi$ with R-charge $R_\Phi$, using \autoref{eq:chiral_superfield}, one gets
\begin{equation}
  R(\phi) = R_\Phi, \hspace{1cm} 
  R(\psi) = R_\Phi - 1, \hspace{1cm} 
  R(F) = R_\Phi - 2
\end{equation}
The R-charges of various objects appearing in the construction of SUSY Lagrangians are summarized in \autoref{tab:r_charge_assignment}.
It should be stressed again that there is no freedom in the choice of R-charge of the vector superfield.
Freedom in the constructions of an R-symmetric model, apart from the choice the field content with their gauge representations, is therefore only in the choice of R-charges of chiral superfields.
With the constraint that their charges must aggregate to the R-charge of the superpotential which is 2.

Before continuing to the construction of the MRSSM it is instructive to relate R-symmetry to previously considered matter and R parities. 
Technically, a matter parity in \autoref{eq:matter_parity} is a discrete subgroup of R-symmetry with $\varphi = \pi$
\begin{equation}
  M_p = e^{\imath \pi R} = (-1)^R,
\end{equation}
where one identifies R-charge with $B-L$.
Since for $\varphi = \pi$ a variable $\theta$ transforms as $\theta \to -\theta$ then assigning R-charges to reproduce matter parity of the MSSM according to the above equation means that the corresponding component fields of supermultiplets will transform under R-symmetry as dictated by $R_p$ of \autoref{eq:r_parity}.

\begin{table}
  \centering
  \begin{tabular}{c|cccccccccc}
    & $\theta$ & $\bar \theta$ & $d^2 \theta$ & $d^2 \bar \theta$ & $Q$ & $\bar Q$\\
    \hline
    R-charge & +1 & -1 & -2 & +2 & -1 & +1
  \end{tabular}
  \\
  \vspace{1cm} 
  \begin{tabular}{c||cccccc|ccc}
    & $D_\alpha$ & $W_\alpha$ & $A_\mu$ & $\lambda_\alpha$ & $D$ & $W$ & $\phi$ & $\psi_\alpha$ & $F$\\
    \hline
    R-charge & -1 & +1 & 0 & +1 & 0 & 2 & $R_\Phi$ & $R_\Phi - 1$ & $R_\Phi-2$
  \end{tabular}
  \caption{R-charges of different objects appearing in the construction of the SUSY Lagrangian.
  The only freedom is in the assignment of R-charges $R_\Phi$ to chiral supermultiplets.
  \label{tab:r_charge_assignment}
  }
\end{table}

\section{Minimal R-symmetric Supersymmetric Standard Model \label{sec:mrssm_sec} }
A successful construction of low energy SUSY model preserving above mentioned R-symmetry was presented in \cite{Kribs:2007ac}.
The model, dubbed The Minimal R-symmetric Supersymmetric Standard Model (MRSSM) was formulated as a means of solving the flavor problem of the MSSM.\footnote{This conclusion was later weaken by Ref.~\cite{Dudas:2013gga} where it was shown not to hold beyond the mass insertion approximation.}
To preserve R-symmetry its particle content has to be extended beyond what is present in MSSM.
Looking at \autoref{tab:r_charge_assignment} this is obviously necessary if one wants to assign R-charge 0 to SM fields.
Since the R-charge of superpotential $W$ must be 2, the MSSM-like $\mu$-terms of the form $\mu \hat H_u \hat H_d$ will not work as it has R-charge 0.
Also, gaugino masses, which have a general form of $\lambda \lambda$ have R-charge 2 and as such cannot appear in the soft SUSY breaking Lagrangian.
Following subsections will therefore describe how MRSSM construction circumvents those problems.\footnote{For possible alternative realization see Ref.~\cite{Frugiuele:2011mh}.}

\subsection{Field content \label{sec:field_content}}
In the MRSSM one introduces 5 new superfields: $U(1)_Y$, $SU(2)_L$, $SU(3)_C$ gauge adjoints called $\hat{S}$, $\hat{T}$, $\hat{O}$, respectively, and two $SU(2)_L$ doublets: $\hat{R}_u$, $\hat{R}_d$.
The R-charges of adjoints are 0 to allow them to mix with gauginos while R-charges of R-Higges are 2 (hence the name). 
Table \ref{tab:mrssm_field_content} summarizes the field content of the MRSSM.
For every chiral or vector supermultiplet it states its R-charge, gauge quantum numbers as well as names of its component fields.
This should be compared with the analogous \autoref{tab:mssm_field_content} done for the MSSM.
The gauge part is identical as in the MSSM and is shown in \autoref{tab:mssm_field_content} only for convenience.
\begin{table}
\begin{center} 
\begin{tabular}{ccccc} 

chiral superfield & R-charge & spin 0 & spin \(\frac{1}{2}\) & \(U(1)_Y \otimes\, SU(2)_L \otimes\, SU(3)_C\)\\
\midrule 
\(\hat{q}\) & 1 & \(\tilde{q}\) & \(q\) & \(\left(\frac{1}{6},{\mathbf 2},{\mathbf 3} \right) \)\\ 
\(\hat{l}\) & 1 & \(\tilde{l}\) & \(l\) & \(\left(-\frac{1}{2},{\mathbf 2},{\mathbf 1}\right) \) \\ 
\(\hat{H}_d\) & 0 & \(H_d\) & \(\tilde{H}_d\) & \(\left(-\frac{1}{2},{\mathbf 2},{\mathbf 1}\right) \) \\
\(\hat{H}_u\) & 0 & \(H_u\) & \(\tilde{H}_u\) & \(\left(\frac{1}{2},{\mathbf 2},{\mathbf 1} \right) \) \\
\(\hat{d}\) & 1 & \(\tilde{d}_R^*\) & \(d_R^*\) & \(\left(\frac{1}{3},{\mathbf 1},{\mathbf \overline{3}}\right) \) \\
\(\hat{u}\) & 1 & \(\tilde{u}_R^*\) & \(u_R^*\) & \(\left(-\frac{2}{3},{\mathbf 1},{\mathbf \overline{3}}\right) \) \\
\(\hat{e}\) & 1 & \(\tilde{e}_R^*\) & \(e_R^*\) & \(\left(1,{\mathbf 1},{\mathbf 1}\right) \) \\
\midrule 
\(\hat{S}\) & 0 & \(S\) & \(\tilde{S}\) & \(\left(0,{\mathbf 1},{\mathbf 1}\right) \) \\
\(\hat{T}\) & 0 & \(T\) & \(\tilde{T}\) & \(\left(0,{\mathbf 3},{\mathbf 1}\right) \) \\
\(\hat{O}\) & 0 & \(O\) & \(\tilde{O}\) & \(\left(0,{\mathbf 1},{\mathbf 8}\right) \) \\
\(\hat{R}_d\) & 2 & \(R_u\) & \(\tilde{R}_d\) & \(\left(\frac{1}{2},{\mathbf 2},{\mathbf 1}\right) \) \\
\(\hat{R}_u\) & 2 & \(R_d\) & \(\tilde{R}_u\) & \(\left(-\frac{1}{2},{\mathbf 2},{\mathbf 1}\right) \)
\\
\\
vector superfield & R-charge & spin \(\frac{1}{2}\) & spin 1 & \(U(1)_Y \otimes\, SU(2)_L \otimes\, SU(3)_C\)\\
\midrule 
  $\hat B$ & 0 & $\tilde B^0$ & $B^0$ & \(\left(1,{\mathbf 1},{\mathbf 1}\right) \)\\
$\hat W$ & 0 & $\tilde W^\pm$, $\tilde W^0$ & $\tilde W^\pm$, $W^0$ & \(\left(1,{\mathbf 3},{\mathbf 1}\right) \)\\
$\hat g$ & 0 & $\tilde g$ & $g$ & \(\left(0,{\mathbf 1},{\mathbf 8}\right) \)
\end{tabular} 
\end{center} 
\caption{Field content of the Minimal R-symmetric Supersymmetric Standard Model (MRSSM). R-charges in the second column correspond to the entire superfield. The table should be compared with \autoref{tab:mssm_field_content} for the MSSM. }
\label{tab:mrssm_field_content}
\end{table}

\subsection{Superpotential \label{sec:mrssm_superpotential}}
With the field content in \autoref{tab:mrssm_field_content} the MRSSM superpotential is given by
\begin{align}
\label{eq:mrssm_superpot}
W = & \mu_d\,\hat{R}_d \cdot \hat{H}_d\,+\mu_u\,\hat{R}_u\cdot\hat{H}_u  \\
 & - Y_d \,\hat{d}\,\hat{q}\cdot\hat{H}_d\,- Y_e \,\hat{e}\,\hat{l}\cdot\hat{H}_d\, +Y_u\,\hat{u}\,\hat{q}\cdot\hat{H}_u  \nonumber\\
& + \Lambda_d\,\hat{R}_d\cdot \hat{T}\,\hat{H}_d\,+\Lambda_u\,\hat{R}_u\cdot\hat{T}\,\hat{H}_u\, +\lambda_d\,\hat{S}\,\hat{R}_d\cdot\hat{H}_d\,+\lambda_u\,\hat{S}\,\hat{R}_u\cdot\hat{H}_u \nonumber ,
\end{align}
where $A \cdot B \equiv A_i \epsilon_{ij} B_j$ with the Levi-Civita symbol, $\epsilon_{12} = 1$.
This superpotential consists of 3 parts
\begin{itemize}
  \item first row of \autoref{eq:mrssm_superpot} is the MRSSM's replacement for the conventional MSSM $\mu$-term, necessary to generate Higgsino masses
  \item second row contains Yukawa interactions, identical to those in the MSSM superpotential in \autoref{eq:mssm_superpot}
  \item third row contains interactions between singlet/triplet and Higgses, similar in their structure and impact to the Yukawas
 
\end{itemize}
\subsection{Soft breaking terms \label{sec:mrssm_ssb} }
As in the MSSM, supersymmetry of the MRSSM is broken by supplementing it with a set of soft SUSY breaking terms.
Decomposing the $SU(2)_L$ triplet field $\hat T$ as
\begin{equation}
  \hat{T} \equiv \left ( 
  \begin{matrix}
    \hat T^0/\sqrt{2} & \hat T^+\\
    \hat T^- & - \hat T^0/\sqrt{2}
  \end{matrix}
  \right )
\end{equation}
the soft SUSU breaking terms are
\begin{itemize}
\item scalar masses (cf. \autoref{eq:mssm_ssb_scalar_masses})
  \begin{align}
  \label{eq:scalar_masses}
  \mathcal{L}_{\text{SSB}} \ni & - m_{H_d}^2 \left(  |H_d^0|^2 + |H_d^-|^2 \right)
  - m_{H_u}^2 \left( |H_u^0|^2 + |H_u^+|^2 \right) \\
  & - \tilde{d}^*_{L,i} m_{q,{i j}}^{2} \tilde{d}_{L,j} 
   - \tilde{d}^*_{R,i} m_{d,{i j}}^{2} \tilde{d}_{R,j} \nonumber \\
  & - \tilde{u}^*_{L,i} m_{q,i j}^{2} \tilde{u}_{L,j}
  - \tilde{u}^*_{R,i} m_{u,{i j}}^{2} \tilde{u}_{R,{j}} \nonumber \\
 & - \tilde{e}^*_{L,{i}} m_{l,{i j}}^{2} \tilde{e}_{L,    {j}} +\tilde{e}^*_{R,{i}} m_{e,{i j}}^{2} \tilde{e}_{R,{j}} \nonumber \\
 & - \tilde{\nu}^*_{L,{i}} m_{l,{i j}}^{2} \tilde{\nu}_{L,{j}} \nonumber \\
 & - m_{R_d}^2 (|R_d^0|^2 + |R_d^+|)^2 
  - m_{R_u}^2 (|R_u^0|^2 + |R_u^-|)^2 \nonumber\\
  & - m_S^2 |S|^2 - m_T^2 (|T^0|^2 + |T^+|^2 + |T^-|^2 ) - m_O^2 |O|^2 \nonumber
  \end{align}
  \item Dirac gaugino masses (cf. \autoref{eq:mssm_gaugino_masses})\\
  The Dirac nature of the MRSSM fermions is a prevalent feature of the model.
Their Dirac SSB masses arise in models with $D$-term SUSY breaking from a hidden $U(1)'$ \cite{Fox:2002bu} as
\begin{equation}
\label{eq:wujcio_dobra_rada}
  \int d^2 \theta \sqrt{2} \frac{W'_\alpha W^\alpha_i \hat A_i}{M} ,
\end{equation}
where $W'$,$W$ are field-strength spinorial fields for hidden $U(1)'$ and SM gauge group, respectively, and $\hat A$ is a chiral multiplet in the adjoint representation of the same SM gauge group.
Once the superfield $W'_\alpha$  with $R  = 1$ develops a vacuum $D$-term $\langle W'_\alpha \rangle=D’\theta_\alpha$ this expression will generate Dirac mass terms with mass $M^D \equiv D'/M$.
In the MRSSM those are
  \begin{equation}
  \label{eq:dirac_masses}
    \mathcal{L}_{\text{SSB}} \ni
  - M_B^D \tilde{B} \tilde{S}   
  - M_W^D \left( \tilde T^+  \tilde{W}^- + \tilde T^-  \tilde{W}^+ + \tilde{T}^0 \tilde{W}^3\right)
  - M_O^D \tilde{g} \tilde{O} + h.c.
  \end{equation}
    where $\tilde{W}^\pm \equiv (\tilde{W}^1 \mp \imath \tilde{W}^2)/\sqrt{2}$.
  Following the discussion in \autoref{sec:susy_breaking} Dirac gaugino masses are also soft (in fact they are even supersoft \cite{Fox:2002bu}).
  \Autoref{eq:wujcio_dobra_rada} also generates mass splitting between real and imaginary parts of scalar gauge adjoints through a $D$ term contribution
  \begin{align}
  \label{eq:dirac_splitting}
    \mathcal{L}_{\text{SSB}} \ni - (M^D_S)^2 (S + S^*)^2 - (M^D_T)^2 \sum_{i=1}^3 (T^i + T^{i,*})^2 - (M^D_O)^2 (O + O^*)^2 
  \end{align}
  \item holomorphic bilinear terms (cf. \autoref{eq:mssm_holomorhic_bilinear})
  \begin{equation}
  \label{eq:bmu_sign_convention}
  \mathcal{L}_{\text{SSB}} \ni B_{\mu} \left( H_d^0 H_u^0  - H_d^- H_u^+ \right) 
  \end{equation}
  The peculiar feature of the R-symmetry is that although the MSSM $\mu \hat H_u \hat H_d$ term in the superpotential is not allowed, the corresponding soft breaking term $B_\mu H_u H_d$ in the SSB Lagrangian is.
  Conversely, the $B_{\mu_d} R_d H_d$ and $B_{\mu_u} R_u H_u$ terms, being counterparts of the superpotential terms $\mu_d\hat{R}_d \hat{H}_d$ and $\mu_u\hat{R}_u \hat{H}_u$ would have R-charge 2 and are therefore forbidden here.
  The $R_d R_u$ term is also forbidden, as it has R-charge 4.
  \end{itemize}

\subsection{Particle spectrum \label{sec:particle_spectrum} }
Having shown the complete Lagrangian of the MRSSM, this subsection lists the physical states of the model after SUSY and EW symmetry breaking.
Since the analysis of the Higgs sector is one of the pillars of this thesis its good to start from it.

The Higgs sector of the MRSSM consists of 4 fields, $H_u^0$, $H_d^0$, $S$, $T^0$,
whose scalar components acquire VEVs after the electroweak symmetry breaking (\ac{EWSB}) as
\begin{align}
  \label{eq:gustav1}
  H_d^0 = & \frac{1}{\sqrt{2}} \left( v_d + \phi_d + \imath \sigma_d \right)\\
  \label{eq:gustav2}
  H_u^0 = & \frac{1}{\sqrt{2}} \left( v_u + \phi_u + \imath \sigma_u \right)\\
  \label{eq:gustav3}
  T^0 = & \frac{1}{\sqrt{2}} \left( v_T + \phi_T + \imath \sigma_T \right)\\
  \label{eq:gustav4}
  S = & \frac{1}{\sqrt{2}} \left( v_S + \phi_S + \imath \sigma_S \right)
\end{align}
and which will mix to form physical mass eigenstates.
Since MRSSM is CP-conserving, states $\phi_d, \phi_u, \phi_T, \phi_S$ and $\sigma_d, \sigma_u, \sigma_T, \sigma_S$ mix separatelly.
This gives rise to four physical CP-even Higgs bosons and 3 physical CP-odd ones (+ one Goldstone boson).
The mass matrices for scalar and pseudoscalar Higgses are shown in the Appendix, in \autoref{sec:higgs_mass_matrix} and \autoref{sec:pseudohiggs_mass_matrix}, respectively.
Similarly, charged components of the $SU(2)_L$ doublets $H_u$, $H_d$ and $SU(2)_L$ triplet $T$ mix to form 3 physical charged Higgs bosons + one Goldstone boson. 

Scalar component of the $R=2$ $SU(2)_L$ doublets $\hat R_u, \hat R_d$ form the so-called R-Higgses \cite{Choi:2010an}.
While the neutral ones mix giving the mass matrix in \autoref{eq:r_higgs_mass_matrix}, the charged ones do not, with masses given by \autoref{eq:rum-higgs_mass} and \autoref{eq:rdp-higgs_mass}.
It should be noted that neutral R-Higgses are complex scalars as masses of the scalar and pseudoscalar parts do not split.

The fermionic components of the $H_u$, $H_d$, $S$ and $T$ superfields form, together with gauginos and R-Higgsinos, set of fermions called charginos and neutralinos.
There are two important distinctions here compared to the MSSM. 
First, there are two sets of charginos, denoted $\tilde \chi^+$ and $\tilde \rho^-$. 
Their mass matrices are shown in \autoref{sec:chargino_mass_matrix}.
As for the neutralinos, there are 4 of them like in the MSSM but they are of a Dirac nature in the MRSSM.\footnote{The Dirac nature of neutralinos has an important consequences if they are to be the dark matter particles \cite{Buckley:2013sca,Diessner:2015iln}.}
Their mass matrix is shown in \autoref{sec:chargino_mass_matrix}
For the gluino, defining a Dirac bi-spinor $\tilde{g}_D$ in the Weyl representation as  
\begin{equation}
\tilde g_D \equiv \left (
  \begin{matrix}
    \tilde g_\alpha \\
    \overline{\tilde O^{\dot \alpha}}
  \end{matrix}
\right ),
\end{equation}
the Dirac gluino mass term in the \autoref{eq:dirac_masses} can be concisely written (reintroducing spinor indices) as
\begin{equation}
 \mathcal{L}_{\text{SSB}} \ni -M_O^D \tilde{O}^\alpha \tilde{g}_\alpha  + h.c. 
  = -M_O^D \left(\tilde{O}^\alpha \tilde{g}_\alpha + \overline{\tilde{g}_{\dot \alpha}} \overline{\tilde{O}^{\dot\alpha}} \right) = -M_O^D \tilde{g}_D^\dagger \gamma^0 \tilde{g}_D = -M_O^D \overline{\tilde{g}_D} \tilde{g}_D
\end{equation}


As will be shown during the discussion of the strongly interacting sector of the model in \autoref{sec:sgluon_mc}, the mass splitting in \autoref{eq:dirac_splitting} has especially important consequences for the color octet scalars.
Masses of the CP-even ($O_S$) and CP-odd ($O_A$) components of the sgluon field $O \equiv (O_S + \imath O_A)/\sqrt{2}$ are split as
\begin{align}
  \label{eq:sgluons_mass_splitting1}
  m_{O_S}^2 =& m_O^2 + 4 (M_O^D)^2 ,\\
  \label{eq:sgluons_mass_splitting2}
    m_{O_A}^2 =& m_O^2 .
\end{align}
Since the gluino mass parameter $M^D_O$ phenomenologically must be larger than around 1 TeV, scalar sgluon will be relatively heavy.
The pseudoscalar one can be quite light though.
This will be exploited in \autoref{sec:sgluon_mc}.

Before finishing the discussion of the particle spectrum of the model it is important to highlight the difference of the sfermion sectors between MSSM and MRSSM.
The left-right sfermion mixing in the MSSM is generated by the superpotential $\mu \hat H_u \hat H_d$ term.
Since such term is not present in the MRSSM, no mixing is present either.
This has an important consequences for the Higgs boson mass, as will be discussed in \autoref{sec:higgs_chapter}.

\subsection{Comment on the renormazalization group running \label{sec:mrssm_rges}}
It is important to comment on the impact of the additional fields charged under the gauge group that were added to form the MRSSM on its $\beta$ functions, defined as
\begin{equation}
 \beta_g \equiv \mu \frac{\partial g}{\partial \mu} = \frac{g^3}{16\pi^2} b_g .
\end{equation}
The one-loop coefficients $b$ can be read off from the general formula for the Yang-Mills theory coupled to $\mathcal{F}$ fermions $\psi_i$ in representations $R_i$ and $S$ complex scalars in representations $R_a$ as
\begin{equation}
  b = -\frac{11}{3} C_2(G) + \frac{2}{3} \sum_{i=1}^F T(R_i) + \frac{1}{3} \sum_{a=1}^S T(R_a),
\end{equation}
where $C_2 (G)$ is the quadratic Casimir of gauge group $G$ and $T(R)$ is the  structure constant of the representation $R$.
The MRSSM differs from the MSSM by a presence of 5 chiral superfields. 
Each chiral superfield contributes $\Delta b = - T(R_i)$ since both components of the chiral superfield are in the same representation $R_i$.

Since the $T(\text{Adj}(SU(N))) = N$, the color-octet superfield $\hat O$ gives
\begin{equation}
 b_{g_s}^{\text{1-loop}} = b_{g_s}^{\text{1-loop},\text{MSSM}} + 3 = 0  .
\end{equation}
Analogously, the $SU(2)_L$ adjoint $\hat T$ and doublets $\hat R_u$, $\hat R_d$ give
\begin{equation}
b_{g_2}^{\text{1-loop}} = b_{g_2}^{\text{1-loop},\text{MSSM}} + 3 = 4  
\end{equation}
and the $\hat R_u$, $\hat R_d$ contribute identical to the $H_u$ and $H_d$
\begin{align}
  b_{g_1}^{\text{1-loop}} & = b_{g_1}^{\text{1-loop, MSSM}} + \frac{3}{5} = \frac{36}{5}
\end{align}
It turns out that at the one-loop level QCD coupling gets frozen at the scale where MRSSM is matched to the SM.
Moreover, since two-loop $\beta_{g_s} >0$, the asymptotic freedom of QCD is lost.
This is not a problem from the points of view of low-energy theory which is considered here as the QCD Landau pole lies far beyond the Planck scale.

To construct a complete high scale theory the model would therefore had to be extended by additional fields which would alter the running of the gauge couplings.
This kind of constructions might be realized in gauge-mediated SUSY breaking.

The full set of two-loop RGEs derived by \texttt{SARAH}~\cite{Staub:2010jh} is used in the matching the MRSSM parameters between the SUSY and EWSB scales, as will be explained in \autoref{sec:mrssm_computational_setup}.

\subsection{Tadpole equations \label{sec:mrssm_tadpoles}}
For completeness, this subsection gives tree-level minimization conditions for the scalar potential.
Using the decomposition in \autoref{eq:gustav1} - \autoref{eq:gustav4} the minimum  of the potential is determined by the set of following four tadpole equations
\begin{align}
\label{eq:tadpole_equation1}
0 = & v \cos \beta  \left(4 m_Z^2 \cos 2 \beta  - 8 g_1 \gls{M^D_S} v_S + 8 g_2
   M^D_T v_T+4 \lambda_D^2 v_S^2+ 4 \sqrt{2} \lambda_D v_S
   (\Lambda_D v_T + 2 \mu_D)\right .\\
   \nonumber
   & \left. +2 (\Lambda_D v_T + 2 \mu_D)^2+8
  m_{H_d}^2 \right) \\
  \label{eq:tadpole_equation2}
  0 = & v \sin \beta \left(- 4 m_Z^2 \cos 2 \beta + 8 g_1 M^D_S v_S - 8 g_2
   M^D_T v_T+4 \lambda_U^2 v_S^2-  4 \sqrt{2} \lambda_U
   \Lambda_U v_S v_T \right. \\
   \nonumber
   & \left.+ 8 \sqrt{2} \lambda_U \mu_u v_S + 2
   \Lambda_U^2 v_T^2-8 \Lambda_U \mu_u v_T + 8 m_{H_u}^2 + 8 \mu_u^2 \right)\\
  \label{eq:tadpole_equation3}
   0 = & v_T \left[ 4 (M_T^D)^2 + m_T^2 \right] + \left[ 2 g_2 M^D_T + \Lambda_d ( \sqrt{2} \lambda_d v_S + \Lambda_d v_T + 2 \mu_d ) \right] v^2 \cos^2 \beta  \\
   & - \frac{1}{4} \left[2 g_2 M^D_T + \Lambda_u (\sqrt{2} \lambda_u v_S - \Lambda_u v_T + 2 \mu_u ) \right] v^2 \sin^2 \beta \nonumber
   \\
   \label{eq:tadpole_equation4}
   0 = & v_S \left[4 (M^D_S)^2 + m_S^2\right] + \frac{1}{4} \left[- 2 g_1 M^D_S + \lambda_d (2 \lambda_d v_S + \sqrt{2} \Lambda_d v_T + 2 \sqrt{2} \mu_d) \right] v^2 \cos^2 \beta\\
   & + \frac{1}{4} \left[ 2 g_1 M^D_S + \lambda_u (2 \lambda_u v_S + 2\sqrt{2} \mu_u - \sqrt{2}\Lambda_u v_T)) \right] v^2 \sin^2 \beta \nonumber
   \end{align}
   where $m_Z^2 \equiv \frac{1}{4} (g_1^2 + g_2^2 ) v^2$ is square of the tree-level $Z$-boson mass.
   Those equations are then solved analytically for $m_{H_d}^2$, $m_{H_u}^2$, $v_S$ and $v_T$.
   The advantage of this approach over solving them for $m_S^2$ and $m_T^2$ is that by controlling these parameters directly one avoids  accidental tachyonic states in Higgs mass matrices.

\subsection{Benchmark points \label{sec:mrssm_bmps}}
Having described the structure of the MRSSM this section presents benchmark points which will be used to test phenomenological validity of the presented model.
These points are called BMP1, BMP2 and BMP3 and are presented in \autoref{tab:BMP}.
The upper part of the table presents parameters which were set differently for every BMP.
Middle part lists parameters whose values were common for all BMPs.
Sfermion mass matrices were chosen to be diagonal, with parameters for the third generation different than for the first two in case of squark matrices.
Sfermion masses are (generation) universal.
These parameters are defined following the \ac{SPA} convention \cite{AguilarSaavedra:2005pw}, meaning, among others, that they are defined in the $\overline{\text{DR}}$ scheme at the scale of 1 TeV.
Lower section of the table shows parameters calculated using one-loop corrected tadpole equations (cf. \autoref{eq:tadpole_equation1} - \autoref{eq:tadpole_equation4}).
Together with SM inputs this completely specifies benchmark points.

These benchmark points are characterized by widely different values of $\tan \beta$, the Yukawa-like couplings $\lambda$, $\Lambda$ of the order of the $top$-quark Yukawa and scalar soft masses of around 1 TeV (with the notable exception of the triplet soft-mass, which must be large to give small triplet VEV).
The numerical values of these parameters are quite generic, in that no special fine-tuning is needed.

Next chapters will compute a selection of physical predictions for these benchmark points, most importantly Higgs and $W$ boson masses.
These points were of course chosen such, as to be in agreement with them, i.e. were chosen based on the calculations done in the following chapters.
Their placement here is an editorial decision.

\begin{table}[t]
\begin{center}
\begin{tabular}{lrrr}
&BMP1&BMP2&BMP3\\
\hline
$\tan\beta$  &  3  &  10  &  40\\
$B_\mu$      &  $500^2$  &  $300^2$  &  $200^2$\\
$\lambda_d$, $\lambda_u$&   $1.0,-0.8$ &  $1.1,-1.1$  &   $0.15,-0.15$\\
$\Lambda_d$, $\Lambda_u$&  $-1.0,-1.11$ &  $-1.0,-0.85$ & $-1.0,-1.03$\\
$M_B^D$&$600$&$1000$&$250$\\
$m_{R_u}^2$ & $2000^2$ & $1000^2$ & $1000^2$\\
\midrule
$\mu_d$, $\mu_u$&\multicolumn{3}{c}{$400,400$}\\
$M_W^D$&\multicolumn{3}{c}{$500$}\\
$M_O^D$&\multicolumn{3}{c}{$1500$}\\
$m_T^2$, $m_S^2$, $m_O^2$&\multicolumn{3}{c}{$3000^2$, $2000^2$, $1000^2$}\\
$\left(m_{u}^2\right)_{11,22}$, $\left(m_{u}^2\right)_{33}$ &\multicolumn{3}{c}{$2500^2,1000^2$}\\
$\left(m_{d}^2\right)_{11,22}$, $\left(m_{d}^2\right)_{33}$ &\multicolumn{3}{c}{$2500^2$, $1000^2$}\\
$\left(m_{u}^2\right)_{11,22}$, $\left(m_{u}^2\right)_{33}$ &\multicolumn{3}{c}{$2500^2$, $1000^2$}\\
$m_l^2$, $m_e^2$&\multicolumn{3}{c}{$1000^2$}\\
$m_{R_d}^2$&\multicolumn{3}{c}{$700^2$}\\
\midrule
$v_S$       & $5.27$    & $1.3$ &$-0.14$\\
$v_T$       & $-0.267$  & $-0.19$ &$-0.34$\\
$m_{H_d}^2$ & $674^2$   &$761^2$ &$1158^2$ \\
$m_{H_u}^2$ & $-498^2$  &$-544^2$ & $-543^2$ \\
\end{tabular}
\end{center}
\caption{
  \label{tab:BMP}
  Definition of benchmark points. 
  Dimensionful parameters are given in GeV or GeV${}^2$, as appropriate. 
  The first part gives input parameters that are specific for each point, while the second part gives input parameters common for all points. 
  Sfermion mass matrices are diagonal.
  The last part shows parameters derived from loop-corrected tadpole equations.
}
\end{table}

\section{Computational setup \label{sec:mrssm_computational_setup}}

\begin{figure}
  \centering
  \includegraphics[width=\textwidth]{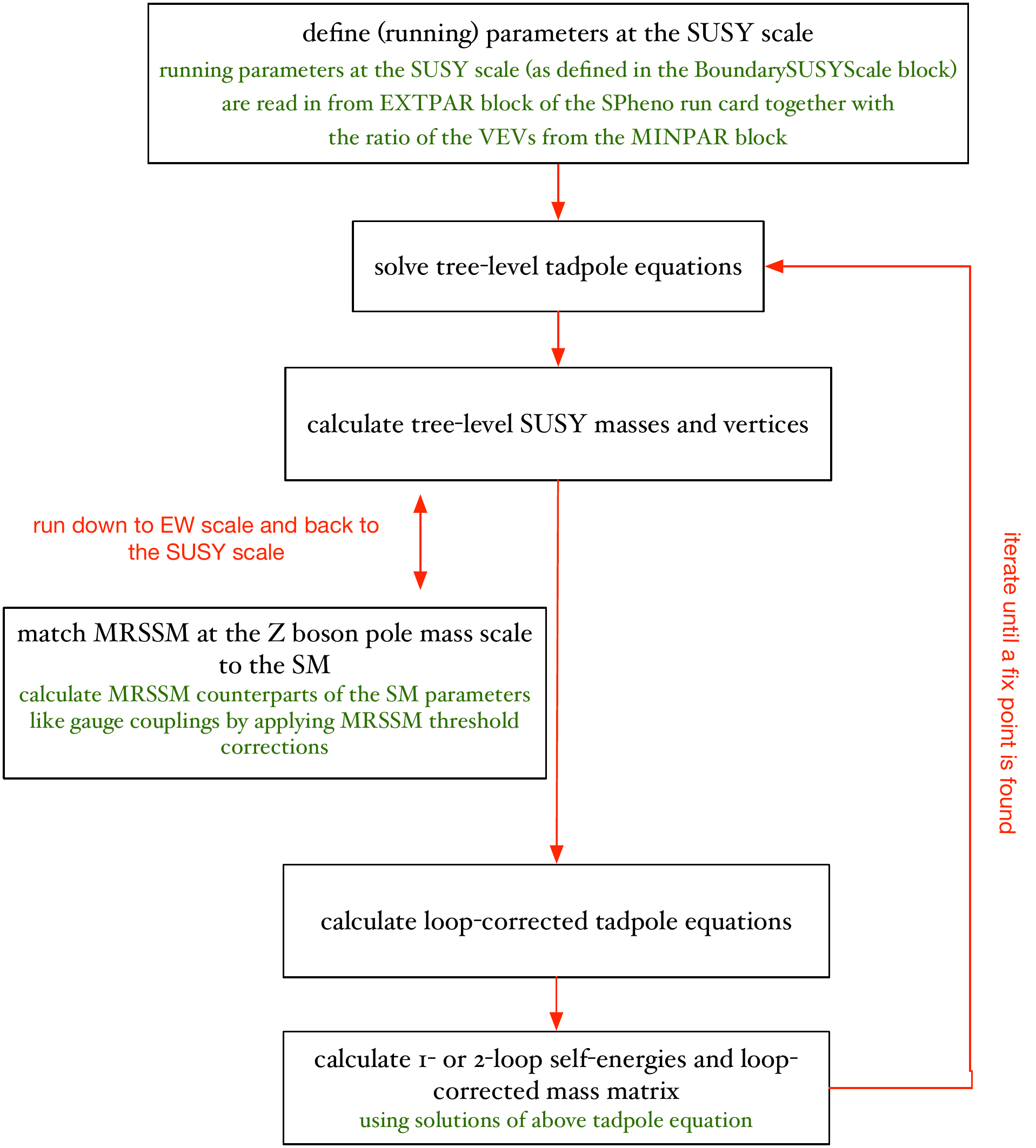}
  \caption{
    Schematic flow of the \texttt{SARAH} generated \texttt{SPheno} code (see Ref.~\cite{Staub:2015kfa} to learn more). 
    See \autoref{sec:sarah_mathematica} for the definition of \texttt{EXTPAR} and \texttt{BoundarySUSYScale} blocks.
    \label{fig:sarah_framework_flowchart}
    }
\end{figure}

Before  delving into the actual calculations it is important to outline how they will be organized.
The computational setup is based on the \texttt{SARAH}~\cite{Staub:2010jh, Staub:2012pb,Staub:2013tta} and \texttt{SPheno}~\cite{Porod:2003um,Porod:2011nf} framework.\footnote{Throughout this thesis a \textit{v.4.8.6} of \texttt{SARAH} and \textit{v.3.3.8} of \texttt{SPheno} are used.}
\texttt{SARAH} is a code sometimes referred to as a spectrum-generator generator.
Based on the analytic expressions created by \textit{Mathematica} it generates \texttt{SPheno}-like spectrum generator.
\Autoref{sec:sarah_mathematica} shows the steering file controlling this generation, based on the model file \texttt{MRSSM.m} distributed with \texttt{SARAH}, together with some useful in code comments.\footnote{The original \texttt{SPheno.m} steering file for the MRSSM distributed with \texttt{SARAH} generates a low energy model i.e., among others, without any RGE running.}
The numerical values in \Autoref{tab:BMP} are used as the boundary conditions at the scale of 1 TeV.
This requires setting flag 2 in the \texttt{SPhenoInput} block to 1 (see \autoref{sec:spheno_card} for the description of the \texttt{SPheno} input card.)
These parameters are then run down to the scale of $Z$ boson mass using two-loop renormalization group equations and matched to the SM as described in the next chapter.
The procedure is iterated until an (approximate) fix point is found.
\autoref{fig:sarah_framework_flowchart} shows the general flow of the program.
The input to the entire calculation (both input and technical parameters) is specified in \autoref{sec:spheno_card}.

\section{Conclusions}
This chapter described in detail the philosophy and the construction of the Minimal R-symmetric Supersymmetric Standard Model.
Starting with the principle of preserving the R-symmetry at the low scale, the phenomenologically viable model was constructed.
The necessary extensions which had to be made compared to MSSM lead to a model with an interesting phenomenology, different from what is usually considered in SUSY studies.
The role of next chapters will therefore be to confront this model with available experimental data, especially against the ever increasing set of LHC measurements.
\chapter{$W$ boson mass in the MRSSM \label{sec:ew} }

\begin{figure}[h!]
  \centering
  \includegraphics[width=0.55\textwidth]{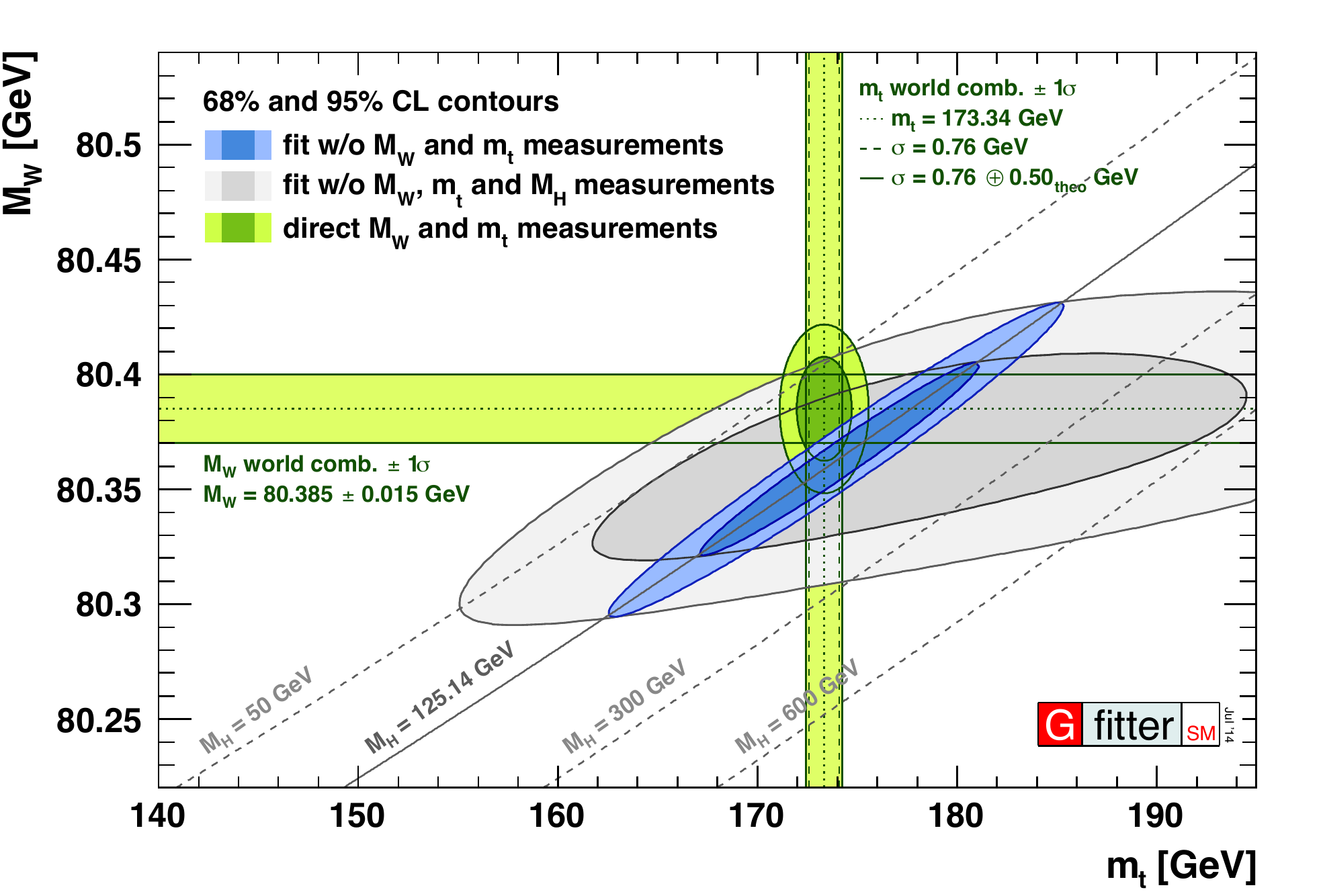}
  \caption{Contours of 68\% and 95\% confidence level obtained from scans of fits with fixed variable pairs $m_W$ vs. $m_t$. Plot taken from Ref.~\cite{Baak:2014ora}.}
  \label{fig:gfitter}
\end{figure}

In literature, the expression 'electroweak precision observables' (EWPO) usually refers to a multitude of EW observables measured either at low energies (like the muon decay constant) or around the $Z$-pole (like the widths of the gauge bosons). 
Historically, EWPO played a crucial role in constraining values of masses of the top quark and the Higgs boson before their actual discovery~\cite{Sirlin:2012mh}.  
In 1994, a year before the discovery of the top quark by the Tevatron \cite{Abe:1995hr}, a global EW analysis led to an indirect determination of its mass as \cite{Pietrzyk:1994rx}
\begin{equation}
  m_t = 177 \pm 11^{+18}_{-19} \text{ GeV}.
\end{equation}
Since Higgs boson mass $m_h$ was unknown at that time, the central value of the above prediction corresponds to $m_h = 300$ GeV with the second error describing the change of the top mass with variation of $m_h$ between 60 GeV and 1 TeV while the first error is experimental.
Unfortunately, EWPO are much less constraining for the Higgs boson mass, since they depend on it only logarithmically. 
This can be seen in \autoref{fig:gfitter}, which shows the result of a recent EW fit, done already after the discovery of the Higgs boson. 
It is nevertheless clear that even without discovering $W, h$ and $t$ one could say quite a lot about their masses (grey region).
This shows the importance of EW observables in indirectly constraining models.

Nowadays, since masses of $W, h$ and $t$ are well known, EWPO can be used to constrain models of physics beyond the SM.
One of those observables is the mass of the $W$ boson.
If one takes as know best measured observables such as $Z$-boson mass $m_Z$, muon decay constant $G_\mu$ and the electromagnetic coupling constant in the $\overline{\text{MS}}$ scheme at the $m_Z$ scale, then $m_W$ becomes a calculable quantity which can be compared with experiment.
This observable is of special interest for the MRSSM since, as was explained in the previous chapter, contrary to the models with a custodial $SU(2)$, in the MRSSM $m_W$ receives additional tree-level contribution from the VEV of the $SU(2)_L$ triplet $T$.

The chapter is structured as follows. Before discussing the full one-loop calculation of the $W$ boson mass next section reminds the notion of $STU$ parameters since, as will be argued in \autoref{eq:mw_numerical_results}, the full MRSSM contribution is well approximated by just the $T$ parameters.
In \autoref{eq:mw_in_mrssm_full_results} the one-loop MRSSM specific contributions are discussed.
That section describes how muon decay constant is extracted from experiment and how it is related to the $W$ boson mass.
The general setup of the calculation is also explained.
Since, as was mentioned in the Preface to this thesis, part of the calculations was done in collaboration with Philip Diessner, Jan Kalinowski and Dominik Stöckinger, \Autoref{sec:mrssm_deltaVB} presents only details of the $\delta_{VB}$ calculation which was performed by me.
For the discussion of oblique corrections, see original work \cite{Diessner:2014ksa} or Philip Diessner's thesis \cite{philips_phd}.
However, when discussing the W boson mass in later chapters, all corrections will be included.
 
\FloatBarrier
\section{Peskin-Takeuchi $STU$ parameters}

Before starting a full calculation of $m_W$, it is instructive to discuss the so-called $STU$ parameters~\cite{Peskin:1990zt,Peskin:1991sw,Altarelli:1990zd,Kennedy:1990cx,Marciano:1990ji}.
\begin{figure}
  \centering
  \includegraphics[width=0.55\textwidth]{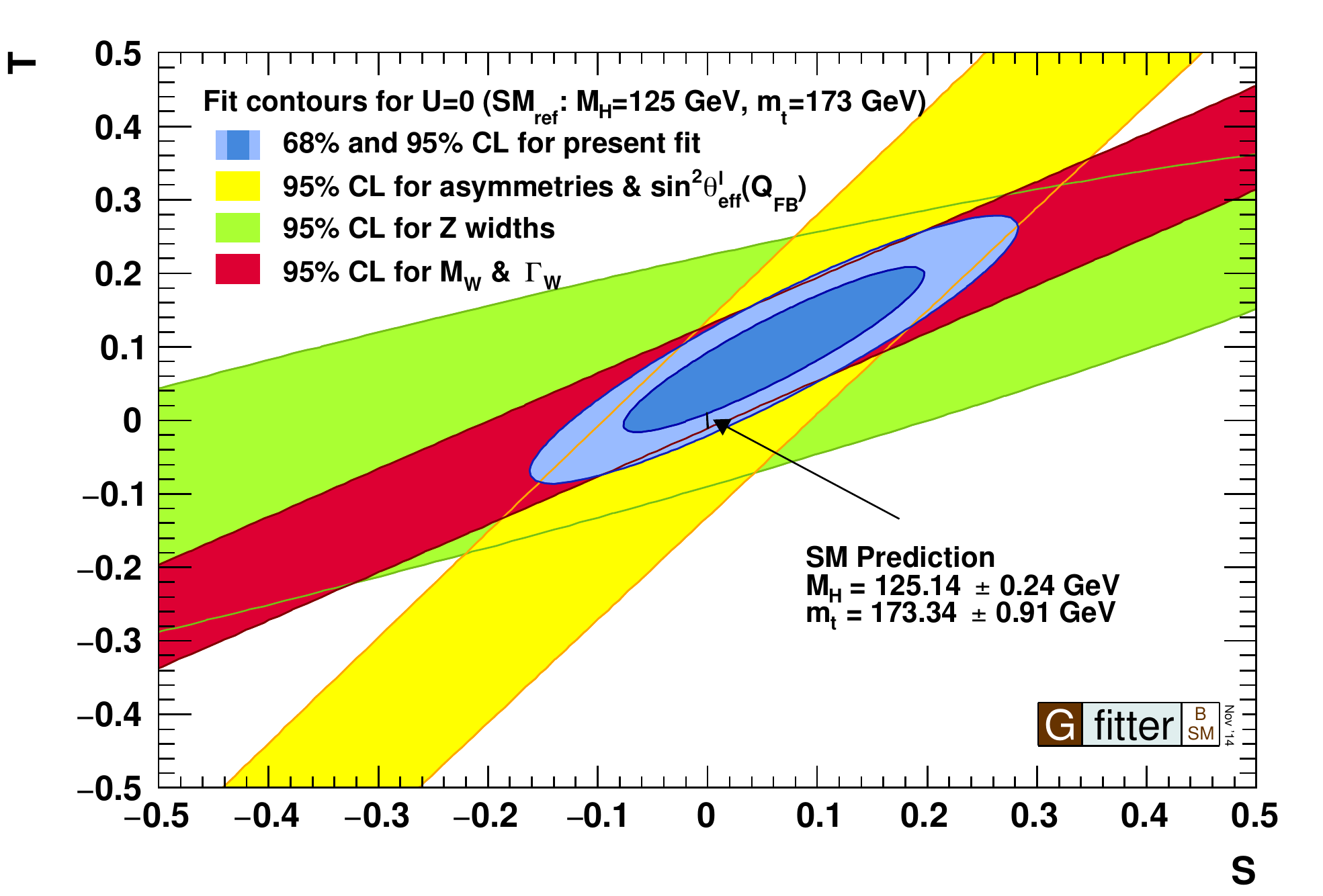}
  \caption{Allowed region in the $S-T$ plane for $U=0$. Plot taken from Ref.~\cite{Baak:2014ora}.}
  \label{fig:gfitter_stu}
\end{figure}
Under the assumptions that
\begin{enumerate}
  \item there are no new gauge bosons
  \item \ac{BSM} coupling to light fermions is suppressed
  \item the scale $m$ of BSM is large compared to $m_Z$\footnote{This parametrization can be extended for the case of $m \gtrsim m_Z$ by introducing 3 additional parameters, called $VWX$ \cite{Maksymyk:1993zm}.}
\end{enumerate}
most new physics effects will be captured by corrections to gauge bosons self-energies. 
The $STU$ parameters are therefore defined as certain combinations of BSM contributions to self-energies $\Pi^{\text{new}}_{ZZ}$, $\Pi^{\text{new}}_{Z\gamma}$, $\Pi^{\text{new}}_{\gamma\gamma}$, $\Pi^{\text{new}}_{WW}$ as
\begin{eqnarray}
  \frac{\alpha}{4 c_w^2 s_w^2} S & \equiv & \frac{\Pi^{\text{BSM}}_{ZZ} (m_Z^2) - \Pi^{\text{BSM}}_{ZZ} (0)}{m_Z^2} - \frac{c_w^2 - s^2_w}{c_w s_w} \frac{\Pi^{\text{BSM}}_{Z\gamma} (m_Z^2)}{m_Z^2} - \frac{\Pi^{\text{BSM}}_{\gamma\gamma} (m_Z^2)}{m_Z^2}\\
  \alpha T &\equiv& \frac{\Pi^{\text{BSM}}_{WW} (0)}{m_W^2} - \frac{\Pi^{\text{BSM}}_{ZZ} (0)}{m_Z^2} \label{eq:T-parameter_definition}\\
  \frac{\alpha}{4 s_w^2} (S+U) &\equiv & \frac{\Pi^{\text{BSM}}_{WW} (m_W^2) - \Pi^{\text{BSM}}_{WW} (0)}{m_W^2} - \frac{c_w}{s_w} \frac{\Pi^{\text{BSM}}_{Z\gamma} (m_Z^2)}{m_Z^2} - \frac{\Pi^{\text{BSM}}_{\gamma\gamma} (m_Z^2)}{m_Z^2}
\end{eqnarray}
where $\alpha \equiv \alpha_e (m_Z)$ and $c_w$ is the weak-mixing angle. 
Obviously, $S$, $T$ and $U$ are 0 in the SM by definition.  
As an example, the approximate expression for the shift of $W$ boson mass due to BSM contributions is \cite{Maksymyk:1993zm}
\begin{equation}
  m_W^2 = (m_W^{\text{SM}})^2 \left (1 - \frac{\alpha S}{2(c_w^2 - s_w^2)} + \frac{c_w^2 \alpha T}{c_w^2 - s_w^2} + \frac{\alpha U}{4 s_w^2}\right)
  \label{eq:mw_from_stu}
\end{equation}
\Autoref{fig:gfitter_stu} shows the result of the fit done by the \texttt{Gfitter} group~\cite{Baak:2014ora} with allowed region in the $S-T$ plane for  $U=0$.\footnote{In the language of effective field theory $U$-parameter is proportional to the Wilson coefficient of a dimension 8 operator and thus is often assumed to be 0.}

In the MRSSM, which contains an $SU(2)_L$-triplet with non-zero VEV $v_T$, one gets contribution to the $T$ parameter already at the tree level, modifying \autoref{eq:mw_from_stu} to
\begin{equation}
  m_W^2 \approx (m_W^{\text{SM}})^2 \left (1 + \frac{4 v_T^2}{v^2} - \frac{\alpha S}{2(c_w^2 - s_w^2)} +  \frac{c_w^2 \alpha T}{c_w^2 - s_w^2} + \frac{\alpha U}{4 s_w^2}\right).
  \label{eq:mw_from_stu_and_vT}
\end{equation}
The $|v_T|$ is strongly constrained by the EWPO to be $\lesssim 4$ GeV.
Solving the tadpole equation in \autoref{eq:tadpole_equation3} for $v_T$ one obtains
\begin{equation}
v_T = \frac{v^2}{4} \cdot \frac{\sin^2 \beta \left(2 g_2 M^D_T + \sqrt{2} \lambda_u \Lambda_u v_S + 2 \Lambda_u \mu_u \right) - \cos^2 \beta \left(2 g_2 M^D_T + \sqrt{2} \lambda_d \Lambda_d v_S + 2 \Lambda_d \mu_d \right)}{ m_T^2 + 4 (M^D_T)^2 + \frac{v^2}{4} \left(\Lambda_d^2 \cos^2 \beta + \Lambda_u^2 \sin^2\beta \right)} .
\label{eq:mT2_tadpole}
\end{equation}
If $|v_T|$ is to be small, the denominator must be large unless one introduces large cancelations in the numerator.
The pragmatic approach is therefore to assume that $|m_T| \gtrsim 2$ TeV, with $v_T$ being a solution of the tadpole equations as was done for \ac{BMP}s of \autoref{tab:BMP}. 
The advantage of this solution over choosing the $v_T$ as input is that in the latter case $m_T^2$ often becomes negative.

As will be shown later, in the MRSSM the most important contribution to the $m_W$ comes from the $T$ parameters.
In some cases the contribution to the $T$ parameter has an especially simple form.
Such is the case of the R-Higgs sector where one gets
\begin{multline}
T \approx \frac{1}{16\pi\hat{s}_W^2 \hat{m}_W^2} \left \{ -\sin^2 2 \theta_R F_{0} \left (m_{R_1}^2, m_{R_2}^2 \right ) + 
\cos^2 \theta_R \left [ F_0 \left ( m_{R_1}^2, m_{R^+_1}^2 \right ) + (1 \rightarrow 2)
\right ] 
\right . \\+ \left . \sin^2 \theta_R \left [ F_0 \left ( m_{R_1}^2, m_{R^+_2}^2 \right ) + 
(1 \leftrightarrow 2)
\right ] \right \} ,
\label{eq:drhorh}
\end{multline}
In the above equation $\theta_R$ parametrizes rotation diagonalizing neutral R-Higgs mass matrix in \autoref{eq:r_higgs_mass_matrix} and
\begin{equation}
F_0 (x, y) = x + y + \frac{2 x y}{x-y} \log \frac{y}{x} .
\end{equation}
Also, contributions proportional to the $\left ( v_T/v \right )^2$ were neglected.
The \autoref{eq:drhorh} has a structure similar to the stop-sbottom one~\cite{Drees:1990dx}.
Since 
\[
\lim_{y \to x} F_0 (x, y) = 0,
\]
in the case of no-mixing in the R-Higgs sector ($\theta_R = 0$) the contribution is proportional to the mass splittings in the R-Higgs $SU(2)_L$ doublets as expected.

Reference \cite{Diessner:2014ksa} contains approximate expressions also for  contributions from other sectors of the model.
Unfortunately, due to the non-trivial nature of the mixing matrices, they are much more complicated than \autoref{eq:drhorh} and can be derived only in suitable limits.
Hence only the numerical results for the benchmark points of \autoref{tab:BMP}, but with values of $\Lambda_u$ -1.2 (BMP1), -1.0 (BMP2) and -1.15 (BMP3), are given here\footnote{These values of $\Lambda_u$ are tailored to give $\approx 125$ GeV Higgs boson mass at the one-loop level and are the original BMPs of Ref.~\cite{Diessner:2014ksa}. See Ref.~\cite{philips_phd} for a thorough discussion of the $STU$ parameters in the MRSSM. }
\begin{center}
\begin{tabular}{c|ccc}
  & $S$ & $T$ & $U$ \\
  \hline
  BMP1 & 0.0097 & 0.090 & 0.00067 \\
  BMP2 & 0.0092 & 0.091 & 0.00065 \\
  BMP3 & 0.0032 & 0.085 & 0.0010
\end{tabular}
\end{center}
These numbers agree with experimental measurements which are $S = -0.03 \pm 0.1$, $T= 0.01 \pm 0.12$, $U = 0.05 \pm 0.1$ \cite{Agashe:2014kda}.
As mentioned before, the advantage of $STU$ parameters is that all EWPO can be approximated through them.
They do not capture the process specific corrections though.
From the viewpoint of constraining MRSSM through $m_W$ measurement it is therefore worthwhile to perform a full one-loop calculation of that observable.

\section{$W$ boson mass in the MRSSM at full one-loop level \label{eq:mw_in_mrssm_full_results}}
In the SM, EW gauge sector can be described at the tree level in terms of only 3 parameters: $g_1$, $g_2$ and $v$.
The same is true for the MSSM with $v^2 \equiv v_u^2 + v_d^2$.
In the MRSSM a forth parameter has to be chosen, for example the $SU(2)_L$ triplet VEV $v_T$. 
Beyond the tree level, these parameters have to be interpreted as the renormalization scale dependent quantities, which I denote with a caret.
Following arguments in \cite{Chankowski:2006hs}, $\hat g_1$, $\hat g_2$ and $\hat v$ are traded in favor of physical quantities $m_Z$ (physical $Z$-boson mass), $G_\mu$ (muon decay constant) and the $\overline{\text{MS}}$ electromagnetic coupling constant at the $Z$ pole  $\hat{\alpha}^{\overline{\text{MS}},\text{SM}}(m_Z)$.
As was explained in \autoref{sec:mrssm_tadpoles}, $v_T$ is exchanged in favor of other $\overline{\text{DR}}$ parameters of the model through the tadpole equation.

At this point all model parameters are specified - either as a direct  $\overline{\text{DR}}$ quantities or through their matching to respective SM counterparts.
This allows to proceed with the calculation of the $W$ boson mass. 

\subsection{Master formula} 
The starting point of the calculation of the W boson mass is its relation to the muon decay constant $G_\mu$.
The muon decay can be described in terms of an (effective) four-fermion operator as
\begin{equation}
  \mathcal{L}_{\text{eff}} = - \frac{G_\mu}{\sqrt{2}} \left[\bar \psi_{\nu_\mu} \gamma^\mu (1-\gamma_5)  \psi_\mu \right] \left[\bar \psi_{e} \gamma^\mu (1-\gamma_5)  \psi_{\nu_e}\right],
\end{equation}

\begin{eqnarray}
\label{eq:muon_measurment}
  \frac{1}{\tau_\mu} = \frac{G_\mu^2 m_\mu^5}{192 \pi^3} F\left( \frac{m_e^2}{m_\mu^2} \right) \left( 1 + \frac{3}{5} \frac{m_\mu^2}{m_W^2} \right) \left[ 1 + H_1 \left( \frac{m_e^2}{m_\mu^2} \right) \frac{\hat{\alpha}(m_\mu)}{\pi} + H_2 \left( \frac{m_e^2}{m_\mu^2} \right) \frac{\hat{\alpha}^2(m_\mu)}{\pi^2} \right],
\end{eqnarray} 
where
\begin{align}
  F(\rho) = & 1 - 8\rho + 8 \rho^3 - \rho^4 - 12 \rho^2 \ln \rho = 0.99981295, \\
  H_1(\rho) = & \frac{25}{8} - \frac{\pi^2}{2} - (9 +4\pi^2+12\ln \rho)\rho + 16\pi^3 \rho^{3/2}+ \mathcal{O}(\rho^2) = -1.80793~\text{\cite{NIR1989184}}, \\
  H_2(\rho) = & \frac{156815}{5184} - \frac{518}{81} \pi^2 - \frac{895}{36} \zeta(3) + \frac{67}{720} \pi^4 + \frac{53}{6} \pi^2 \ln 2 \\
  & - (0.042 \pm 0.002)_{\text{had.}} - \frac{5}{4}\pi^2 \sqrt{\rho} + \mathcal{O}(\rho) = 6.64~\text{\cite{Pak:2008qt}},\\
  \hat \alpha^{-1}(m_\mu) = & \alpha^{-1} + \frac{1}{3\pi} \ln \frac{m_e^2}{m_\mu^2} + \mathcal{O}\left( \frac{m_e^2}{m_\mu^2} \right)  = 135.901, 
\end{align}
for $m_e = 0.510998928 \pm 0.000000011$ and $m_\mu = 105.6583715 \pm 0.0000035$  MeV.
$F$ is a phase space correction due to the non-zero mass of an electron and $H_1$ and $H_2$ capture higher order QED corrections in the Fermi theory.
With the muon lifetime of $\tau_\mu = (2.1969811 \pm 0.0000022) \cdot 10^{-6}$~\textit{s}, this corresponds to $G_\mu = 1.1663787(6) \times 10^{-5} \text{ GeV}^{-2}$ \cite{Agashe:2014kda}. 

In the full theory, like the MRSSM, the muon decay constant is not an independent parameter but a function of more fundamental Lagrangian parameters
\begin{equation}
  \frac{G_\mu}{\sqrt{2}} = 
  \frac{\pi \hat{\alpha}}{2 \hat{s}^2_{W}
 m_{W}^2} (1 + \Delta \hat{r}_W ) 
\label{w-mass-master-formula}
\end{equation}
where $\hat{\alpha}$ and $\hat{s}_W^2 \equiv {\hat{g}_1^2}/({\hat{g}_1^2 + \hat{g}_2^2})$ are $\overline{\text{DR}}$-renormalized running MRSSM electromagnetic coupling constant and sine of the weak mixing angle, respectively.
The first-order correction from $W$ boson propagator, term $3/5 \, m_\mu^2/m_W^2$ present in \autoref{eq:muon_measurment}, is not taken into account as it is numerically negligible. 
The denominator $(1-\Delta \hat{r}_W)$ contains quantum corrections from the W boson self-energy, process dependent box- and vertex-type contributions and counterterms. 
It also properly resums leading two-loop SM corrections as shown in \cite{Degrassi:1990tu}.

One can eliminate the $\hat s_W$ in terms of the $\hat \rho$ parameter, connecting the on-shell and $\overline{\text{DR}}$ renormalization schemes, defined as
\begin{equation}
\hat{\rho} =  \frac{m_W^2}{m_Z^2 \hat{c}_W^2}, 
\label{Definitionrho}
\end{equation}
to obtain the master formula for the $W$ boson mass
\begin{equation}
m_W^2 = \frac{1}{2} m_Z^2 \hat{\rho} \left [ 1 + \sqrt{1
- \frac{4 \pi \hat{\alpha}}{\sqrt{2} G_\mu m_Z^2 \hat{\rho}
(1-\Delta \hat{r}_W)}}\;\right ] \label{w-mass-master-formula2}
\end{equation}
Hence, to calculate $m_W$, one needs to compute the quantities $\hat\alpha$,
$\hat\rho$, and $\Delta\hat{r}_W$ which depend on the entire particle content of the model. 

\subsection{Computational framework}

The computation of the quantities $\hat\alpha$, $\hat\rho$, $\Delta\hat{r}_W$ proceeds as follows. 
The $\overline{\text{DR}}$ running electromagnetic coupling $\hat\alpha$ in the MRSSM can be obtained from the known running SM 5-flavor coupling $\hat{\alpha}^{\overline{\text{MS}},\text{SM}}$ by matching them at the scale of $m_Z$. 
This requires adding MRSSM threshold corrections and the finite counterterm which converts from $\overline{\text{MS}}$ to $\overline{\text{DR}}$. 
These are
\begin{align}
\frac{2 \pi}{\alpha} \Delta \hat{\alpha}^{\overline{\text{DR}},\text{MRSSM}}(m_Z)
& =  \frac{1}{3}
- \sum_{i=1}^6 \left(\frac{1}{3} \log \frac{m_{\tilde{l}_i^\pm}}{m_Z} 
+
 \frac{1}{9} \log \frac{m_{\tilde{d}_i}}{m_Z} 
+ \frac{4}{9} \log \frac{m_{\tilde{u}_i}}{m_Z}
\right )
\notag\\
&- \sum_{i=1}^3\left( \frac{4}{3} \log \frac{m_{l^\pm_i}}{m_Z}
+
\frac{4}{9} \log \frac{m_{d_i}}{m_Z}
+  \frac{16}{9} \log \frac{m_{u_i}}{m_Z}
\right )
\notag\\
&- \sum_{i=1}^3 \frac{1}{3} \log \frac{m_{H_i^\pm}}{m_Z} - \sum_{i=1}^2 \frac{1}{3}  \log \frac{m_{R^\pm_i}}{m_Z} 
- \sum_{i=1}^2 
\frac{4}{3} \left (
 \log \frac{m_{\chi^\pm_i}}{m_Z}
+  \log \frac{m_{\rho^\pm_i}}{m_Z}
\right ),
\end{align}
where $\alpha$ is the electromagnetic coupling   in the Thomson limit. 
In case of benchmark points in \autoref{tab:BMP} this expression is always negative, reducing the value of the running coupling
\begin{equation}
\hat{\alpha}(m_Z) = \frac{\hat{\alpha}^{\overline{\text{MS}},\text{SM}}(m_Z)}{1-\Delta \hat{\alpha}^{\overline{\text{DR}},\text{MRSSM}}(m_Z)} \leq \hat{\alpha}^{\overline{\text{MS}},\text{SM}}(m_Z)
\end{equation}
to approximately 
\begin{equation}
\hat{\alpha}^{-1}(m_Z) \approx 132 \, .
\end{equation}

Large corrections to the $W$ boson mass originate from the $\hat\rho$ parameter defined in \autoref{Definitionrho}. 
In the SM, the
dominant contributions arise from top/bottom loop; in the MRSSM there
are not only loop contributions but already a tree-level
contribution due to the presence of the Higgs triplet with a VEV
$v_T$ as already pointed out in \autoref{eq:mw_from_stu_and_vT}.
This is used to define the tree-level shift $\Delta\hat\rho_{\text{tree}}$ using
\begin{equation}
  \hat{\rho}_{\text{tree}} = \frac{\hat{m}^2_W}{\hat{m}^2_Z \hat{c}_W^2}\equiv1+\Delta\hat\rho_{\text{tree}}
 =  1 + \frac{4 v_T^2}{v^2}. \label{eq:tree-rho}
\end{equation}
Here $\hat{m}_{V}$ ($V=W,Z$) are the tree-level $\overline{\text{DR}}$ masses related to the pole masses $m_V$ by
\begin{equation}
  \label{eq:physical_mass_definition}
  \hat{m}_V^2 = m_V^2 + \Re(\hat{\Pi}_{VV}^T (m_V^2)),
\end{equation}
where $\hat{\Pi}_{VV}^T$ denotes the finite part of the respective transverse vector boson self energy. 
The loop contributions to $\hat{\rho}$ are given by
\begin{eqnarray}
\frac{\hat\rho}{\hat{\rho}_\text{tree}}\equiv\frac{1}{1-\Delta\hat\rho}
 &=& \frac{m_W^2}{m_Z^2 \hat \rho_{\text{tree}} \hat c^2_W} = \frac{\hat m_Z^2/m_Z^2}{\hat m_W^2/m_W^2} = 
 \frac{1+ \frac{\Re(\Pi_{ZZ}^T
 (m_Z^2))}{m_Z^2}}{1+ \frac{\Re(\Pi_{WW}^T(m_W^2))}{m_W^2}},
\end{eqnarray}
where \autoref{eq:physical_mass_definition} and \ref{eq:tree-rho} were used.
The full $\hat\rho$ can then be approximated by 
\begin{equation}
\label{eq:rho_definition}
\hat\rho \approx \frac{1}{1-\Delta\hat\rho_{\text{tree}}-\Delta\hat\rho},
\end{equation}
neglecting products of the form $\Delta\hat\rho_{\text{tree}}\Delta\hat\rho$ which are numerically negligible.

The remaining quantity $\Delta\hat{r}_W$ can then be written
as \cite{Degrassi:1990tu}
\begin{align}
\Delta\hat{r}_W &= \Delta\hat\rho(1-\Delta\hat{r})+\Delta\hat{r},
\\
\label{eq:delta_r-definition}
\Delta\hat{r}
&= \Re\left(\frac{1}{1-\Delta\hat{\rho}}\frac{\hat\Pi_{WW}^T(0)}{m_W^2}
-\frac{\hat\Pi_{ZZ}^T(m_Z^2)}{m_Z^2}\right)
+\frac{1}{1-\Delta\hat{\rho}}\delta_{VB}.
\end{align}
The term $\delta_{VB}$ contains vertex and box diagram contributions to muon decay and is calculated explicitly in the next section.
It is worth noting that in these equations only the loop contributions to $\hat\rho$ appear. 
This way of writing the contributions and the master formula \ref{w-mass-master-formula} automatically resums leading reducible two-loop contributions.
Inclusion of further leading irreducible SM-like two-loop contributions according to \cite{Degrassi:1990tu,Fanchiotti:1992tu,Pierce:1996zz} is discussed in \autoref{sec:mw_1and2-loop_sm}. 

The numerical calculation of $W$ mass is part of the $\texttt{BoundaryEW}$ subroutine of \texttt{SPheno} and follows the setup outlined in this subsection.

\subsection{MRSSM vertex and box corrections \label{sec:mrssm_deltaVB}}

The $\delta_{VB}$ is defined as
\begin{equation}
  \delta_{VB} = 2\cdot\frac{\sqrt{2}}{g_2} \delta V + \frac{2 \hat c_w^2 m_Z^2}{g_2^2}  B + \frac{1}{2}\delta Z^e_L + \frac{1}{2} \delta Z^\mu_L + \frac{1}{2}\delta Z^{\nu_e}_L + \frac{1}{2} \delta Z^{\nu_\mu}_L,
\label{eq:deltaVB}
\end{equation}
where $\delta V$ and $B$ are vertex- and box-corrections to the muon decay and $\delta Z^i_L$ stands for the external wave function renormalization of the lepton $i$.
The calculation of those quantities is done with the help of \texttt{SARAH}~\cite{Staub:2009bi} generated \texttt{FeynArts}~\cite{Hahn:2000kx} model, \texttt{FeynArts} and \texttt{FormCalc}~\cite{Hahn:1998yk}.
Passarino-Veltman loop functions \cite{Passarino:1978jh} $B_0, B_1, C_0, C_{00}, D_{00}$ appearing in the final result follow the \texttt{LoopTools}~\cite{Hahn:1998yk} convention while the mixing matrices $U,V, N$ are defined in \autoref{eq:diag_rot1}, \ref{eq:diag_rot2} and \ref{eq:diag_rot3}.

\subsubsection{External wave functions renormalization}
\begin{figure}
\centering
\includegraphics[width=0.34\textwidth]{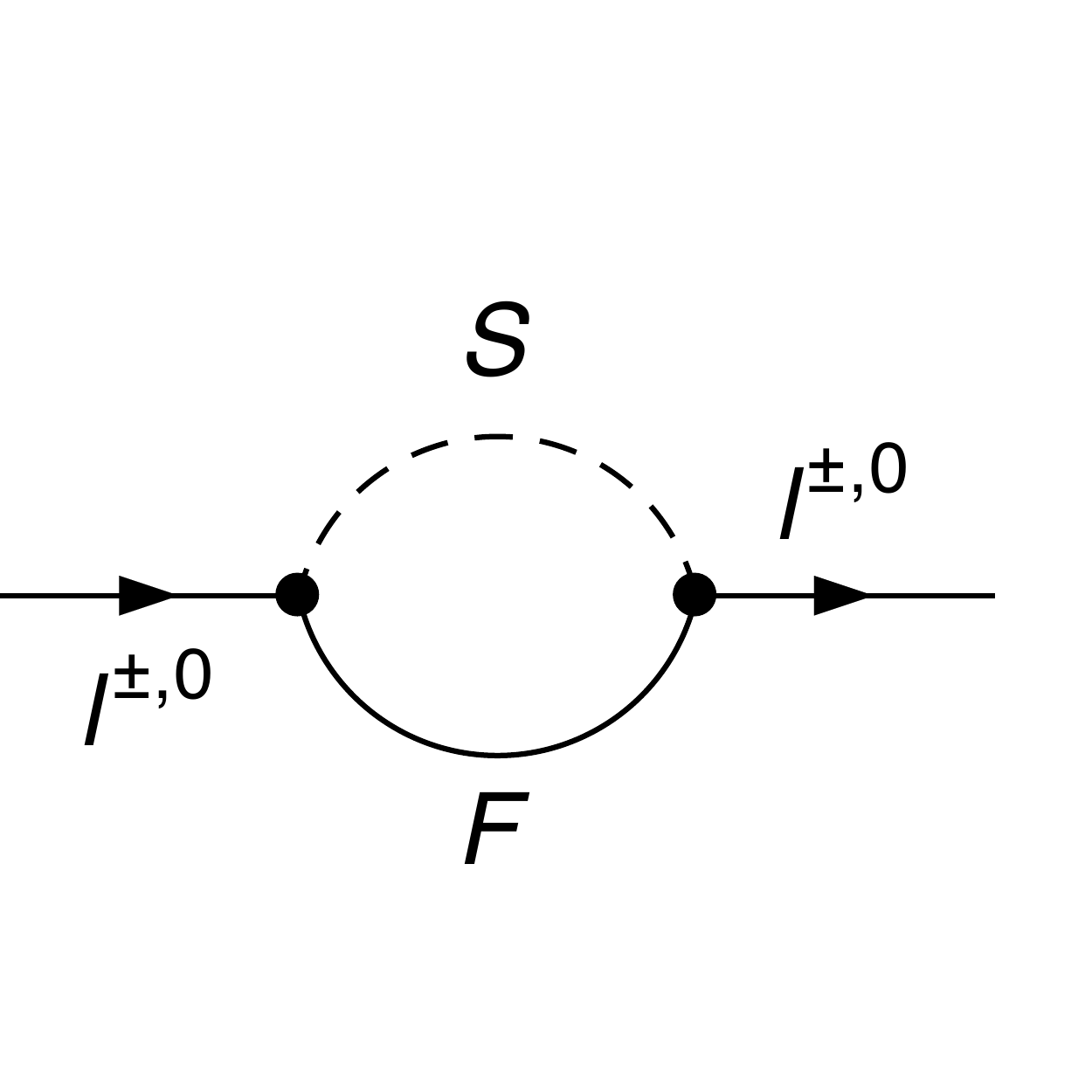}
\caption{Generic form of a diagram giving the non SM-like contribution to the external wave function renormalization.}
\label{external-wave-function}
\end{figure}

A generic diagram contributing to the renormalization of the external lepton wave functions is shown in \autoref{external-wave-function}. 
Decomposing the general fermion self-energy $-\imath \Sigma_{ff}$ as
\begin{equation}
\Sigma_{ff} = \Sigma^L_{ff} \slashed{p} P_L + \Sigma^R_{ff} \slashed{p} P_R + \Sigma^M_{ff} m_f,
\end{equation}
where $P_{L,R} \equiv (1 \mp \gamma^5)/2$ and $m_f$ is the fermion mass, one finds, in the flavor conserving limit and assuming that leptons are massless, for the left-handed projector
\begin{eqnarray}
\label{eq:cosiek1}
\Sigma_{\mu \mu}^{L,\text{MRSSM-SM}} &=& \frac{g_2^2}{16 \pi^2} \sum_{i=1}^2 \left | V^{1}_{i1} \right|^2 B_1 \left ( m_{\chi^\pm_i}^2, m_{\tilde{\nu}_\mu}^2 \right )  \\
&& + \frac{1}{16 \pi^2} \sum_{i=1}^4 \left | \frac{g_2 N^1_{i2} + g_1 N^1_{i1} }{\sqrt{2}} \right |^2 B_1 \left ( m_{\chi_i}^2, m_{\tilde{\mu}_L}^2 \right ), \nonumber \\
\label{eq:cosiek2}
\Sigma_{\nu_\mu \nu_\mu}^{L,\text{MRSSM-SM}} &=& \frac{g_2^2}{16 \pi^2} \sum_{i=1}^2 \left | U^{2}_{i1} \right|^2 B_1 \left (m_{\rho^\pm_i}^2, m_{{\tilde{\mu}_L}}^2 \right )  \\
&& + \frac{1}{16 \pi^2} \sum_{i=1}^4 \left | \frac{g_2 N^1_{i2} - g_1 N^1_{i1} }{\sqrt{2}} \right |^2 B_1 \left (m_{\chi_i}^2, m_{\tilde{\nu}_\mu}^2 \right ) \nonumber,
\end{eqnarray}
where the superscript MRSSM-SM is to remind that only contributions not present in the SM were written.
Since it was assumed that leptons are massless, Eqs. \ref{eq:cosiek1} and \ref{eq:cosiek2} hold also for leptons of the first generation. 
External wave functions are renormalized in the on-shell scheme, i.e.
\begin{equation}
 \delta Z^L_i = \Sigma_{i i}^{L,\text{MRSSM}} .
\end{equation}

\subsubsection{Vertex correction}
\begin{figure}
\centering
\includegraphics[width=0.8\textwidth]{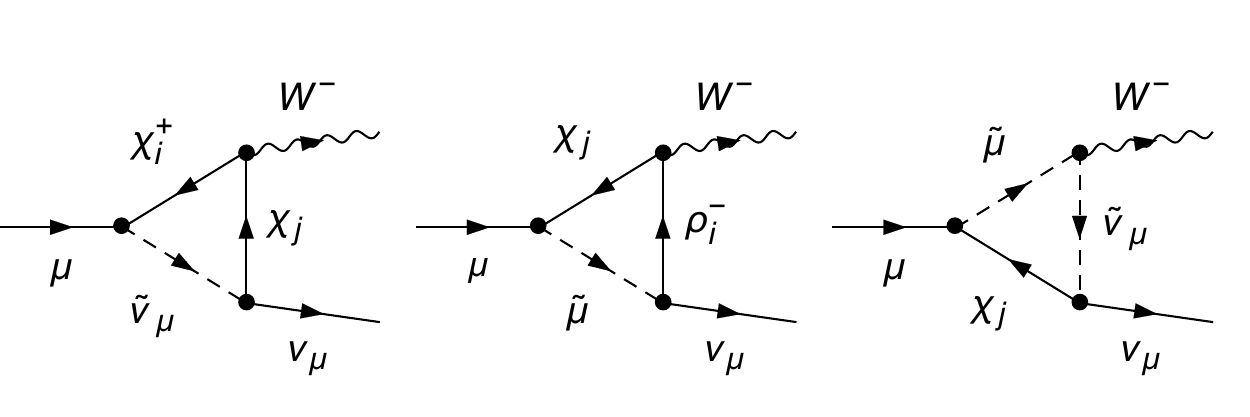}
\caption{Non-SM corrections to the $\mu^- \to W^- \nu_\mu$ decay vertex. Diagrams proportional to the muon Yukawa coupling are not shown. \label{fig:vertex-diagrams}}
\end{figure}

\autoref{fig:vertex-diagrams} contains non-SM corrections to the muon decay vertex. The corresponding analytic expression for the amplitude reads
%
\begin{eqnarray}
-\imath \delta V &=& 
-\frac{\imath g_2}{16 \sqrt{2} \pi^2} \sum_{i=1}^2 \sum_{j=1}^4 g_2 V^{1*}_{i1} \left ( \frac{g_2 N^1_{j2} - g_1 N^1_{j1} }{\sqrt{2}} \right ) \left \{ \left ( \sqrt{2} N^{1*}_{j2} V^1_{i1} - N^{1*}_{j3} V^1_{i2} \right )\right .\\
&&  \left [m_{\tilde{\nu}_\mu}^2 C_{0} \left ( m_{\chi^\pm_i}^2, m_{\chi_j}^2, m_{\tilde{\nu}_\mu}^2 \right ) + B_0 \left (m_{\chi^\pm_i}^2, m_{\chi_j}^2 \right ) - 2 C_{00} \left ( m_{\chi^\pm_i}^2, m_{\chi_j}^2, m_{\tilde{\nu}_\mu}^2 \right ) \right ] \nonumber \\ 
 && -  \left . m_{\chi^\pm_i}^2  m_{\chi_j}^2 \left (\sqrt{2} N^2_{j2} U^{1*}_{i1} + N^2_{j3} U^{1*}_{i2} \right ) C_{0} \left ( m_{\chi^\pm_i}^2, m_{\chi_j}^2, m_{\tilde{\nu}_\mu}^2 \right ) \right \}  \nonumber\\ 
&& - \frac{\imath g_2}{16 \sqrt{2} \pi^2} \sum_{i=1}^2 \sum_{j=1}^4 g_2 U^{2}_{i1} \left ( \frac{g_2 N^{1*}_{j2} + g_1 N^{1*}_{j1}}{\sqrt{2}} \right ) \left \{ \left ( \sqrt{2} N^1_{j2} U^{2*}_{i1} + N^1_{j3} U^{2*}_{i2} \right )\right . \nonumber\\
&&  \left [m_{\tilde{\mu}}^2 C_{0} \left ( m_{\rho^\pm_i}^2, m_{\chi_j}^2, m_{\tilde{\mu}}^2 \right ) + B_0 \left (m_{\rho^\pm_i}^2, m_{\chi_j}^2 \right ) - 2 C_{00} \left ( m_{\rho^\pm_i}^2, m_{\chi_j}^2, m_{\tilde{\mu}}^2 \right ) \right ] 
   \nonumber\\ 
 && - \left . \left . m_{\rho^\pm_i}^2  m_{\chi_j}^2 \left (\sqrt{2} N^2_{j2} U^{1*}_{i1} - N^{2*}_{j4} V^2_{i2} \right )\right ] C_{0} \left ( m_{\rho^\pm_i}^2, m_{\chi_j}^2, m_{\tilde{\mu}}^2 \right ) \right \}  \nonumber\\
 && + \frac{\imath g_2}{8 \sqrt{2} \pi^2} \sum_{i=1}^4 \left ( \frac{g_2 N^1_{j2} - g_1 N^1_{j1} }{\sqrt{2}} \right ) \left ( \frac{g_2 N^{1*}_{j2} + g_1 N^{1*}_{j1} }{\sqrt{2}} \right ) C_{00} \left ( m_{\chi_i}^2, m_{\tilde{\mu}}^2, m_{\tilde{\nu}_\mu}^2 \right ) \nonumber .
\end{eqnarray}
Although diagrams from \autoref{fig:vertex-diagrams} have the same analytic expression as in the MSSM, due to the absorption of $\tilde{W}^\pm$ into $\chi^+$ and $\rho^-$, respectively,  diagram 2 is independent in magnitude from diagram 1, as it depends on the  mass of a different particle.
\begin{figure}
\centering
\includegraphics[width=0.525\textwidth]{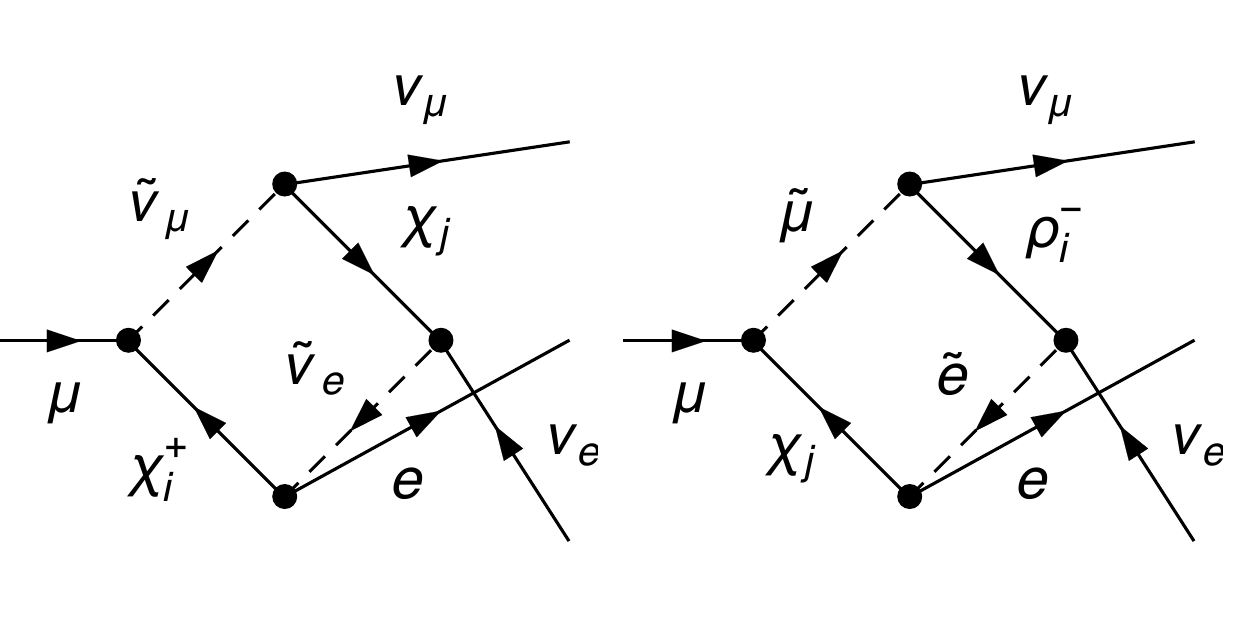}
\caption{Non-SM box contributions to $\mu^- \to \nu_\mu e^- \bar{\nu}_e$ in the MRSSM. Diagrams proportional to lepton's Yukawa or vanishing in the flavor-conserving limit are not shown. \label{box-diagrams}}
\end{figure}
\subsubsection{Box correction}
\Autoref{box-diagrams} contains the most relevant box-type contributions to muon decay in the MRSSM. Those diagrams are both \ac{UV} and \ac{IR} finite. The expression for them, after factorizing out the spinor structure (cf. \autoref{eq:deltaVB}), reads 
%
%
\begin{multline}
-\imath B  =  -\frac{\imath}{16 \pi^2} \sum_{i=1}^2 \sum_{j=1}^4  \left |g_2^2 V^1_{i1} \left (\frac{g_1 N^1_{j1} - g_2 N^1_{j2}}{\sqrt{2}} \right )\right |^2 D_{00} \left ( m_{\chi^\pm_i}^2, m_{\chi_j}^2, m_{\tilde{\nu}_\mu}^2, m_{\tilde{\nu}_e}^2 \right ) \\
 - \frac{\imath}{16 \pi^2} \sum_{i=1}^2 \sum_{j=1}^4  \left |g_2^2 U^2_{i1} \left (\frac{g_1 N^1_{j1} + g_2 N^1_{j2}}{\sqrt{2}} \right )\right |^2 D_{00} \left ( m_{\rho^\pm_i}^2, m_{\chi_j}^2, m_{\tilde{\mu}}^2, m_{\tilde{e}}^2 \right ) .
\end{multline}
The structure of this correction is different from the one in the MSSM due to the Dirac nature of MRSSM's neutralinos and the fact that $\tilde{W}^+$ and $\tilde{W}^-$ are parts of two different types of charginos. 
This forbids the existence of two additional MSSM-like diagrams with mass-term $\tilde{W}^+-\tilde{W}^-$ mixing. 
Although distinct from the MSSM, for benchmark points under consideration, the contribution from the box-correction is below 1 MeV.

\section{One- and two-loop SM corrections \label{sec:mw_1and2-loop_sm}}

The MRSSM result from the previous section needs to be supplemented by the one-loop SM contribution
\begin{align}
\Delta \hat r_{\text{SM}}^{\text{1-loop}} = & \hat \rho \frac{\hat \alpha}{4\pi \hat s_w^2} \left[6 + \left( \frac{7}{2} - \frac{5}{2} s_w^2 - \hat s_w^2 \left( 5-\frac{3}{2} \frac{c_w^2}{\hat c_w^2}\right)\right) \frac{\ln c_w^2}{s_w^2} \right] .
\end{align}
Also, for numerical studies done in later chapters, additional two-loop $\alpha \alpha_s$ corrections to $\Delta \hat r$ and $\Delta \hat \rho$ according to \cite{PhysRevD.48.307,Pierce:1996zz} are implemented 
\begin{align}
\Delta \hat r_{\text{SM}}^{\text{2-loop}} = &\frac{\hat \alpha \alpha_s}{4\pi^2} \left(  2.145\frac{m_t^2}{m_Z^2} + 0.575 \ln \frac{m_t}{m_Z} - 0.224 - 0.144 \frac{m_Z^2}{m_t^2} \right)  \\
& - \frac{1}{3} x_t^2 \left(\frac{Z^H_{12}}{\sin \beta}\right)^2 \rho^{(2)} \left(\frac{m_h}{m_t}\right) (1-\Delta \hat r) \hat \rho \nonumber \\
\Delta \hat \rho_{\text{SM}}^{\text{2-loop}} = & \frac{\hat \alpha \alpha_s }{4\pi^2 \hat s_w^2} \left( -2.145\frac{m_t^2}{m_W^2} + 1.262 \ln \frac{m_t}{m_Z} -2.24- 0.85 \frac{m_Z^2}{m_t^2}  \right) \\
& + \frac{1}{3} x_t^2 \left(\frac{Z^H_{12}}{\sin \beta}\right)^2 \rho^{(2)} \left(\frac{m_h}{m_t}\right), \nonumber
\end{align}
where $x_t \equiv 3 G_\mu m_t^2/(8\sqrt{2}\pi^2)$ and for $r \leq 1.9$ the $\rho^{(2)}(r)$ can be approximated as \cite{Fleischer:1993ub}
\begin{align}
  \rho^{(2)}(r) \equiv & 19 - \frac{33}{2} r + \frac{43}{12} r^2 + \frac{7}{120} r^3 - \pi \sqrt{r} \left(  4 - \frac{3}{2} r + \frac{3}{32} r^2  + \frac{1}{256} r^3 \right ) \\
  & - \pi^2 \left(2 - 2 r + \frac{1}{2} r^2 \right) - \ln r \left ( 3r -  \frac{1}{2} r^2 \right) \nonumber
\end{align}
In the above formulas it is assumed that the lightest Higgs boson of mass $m_h$ is mainly composed of $H_u$ gauge eigenstate which, in the convention of this work, is the second state in the Higgs boson mass matrix (hence the presence of the $Z_{12}^H$ element of the mixing matrix).

\section{Numerical results \label{eq:mw_numerical_results}}
\begin{figure}
\centering
  \includegraphics[width=0.45\textwidth]{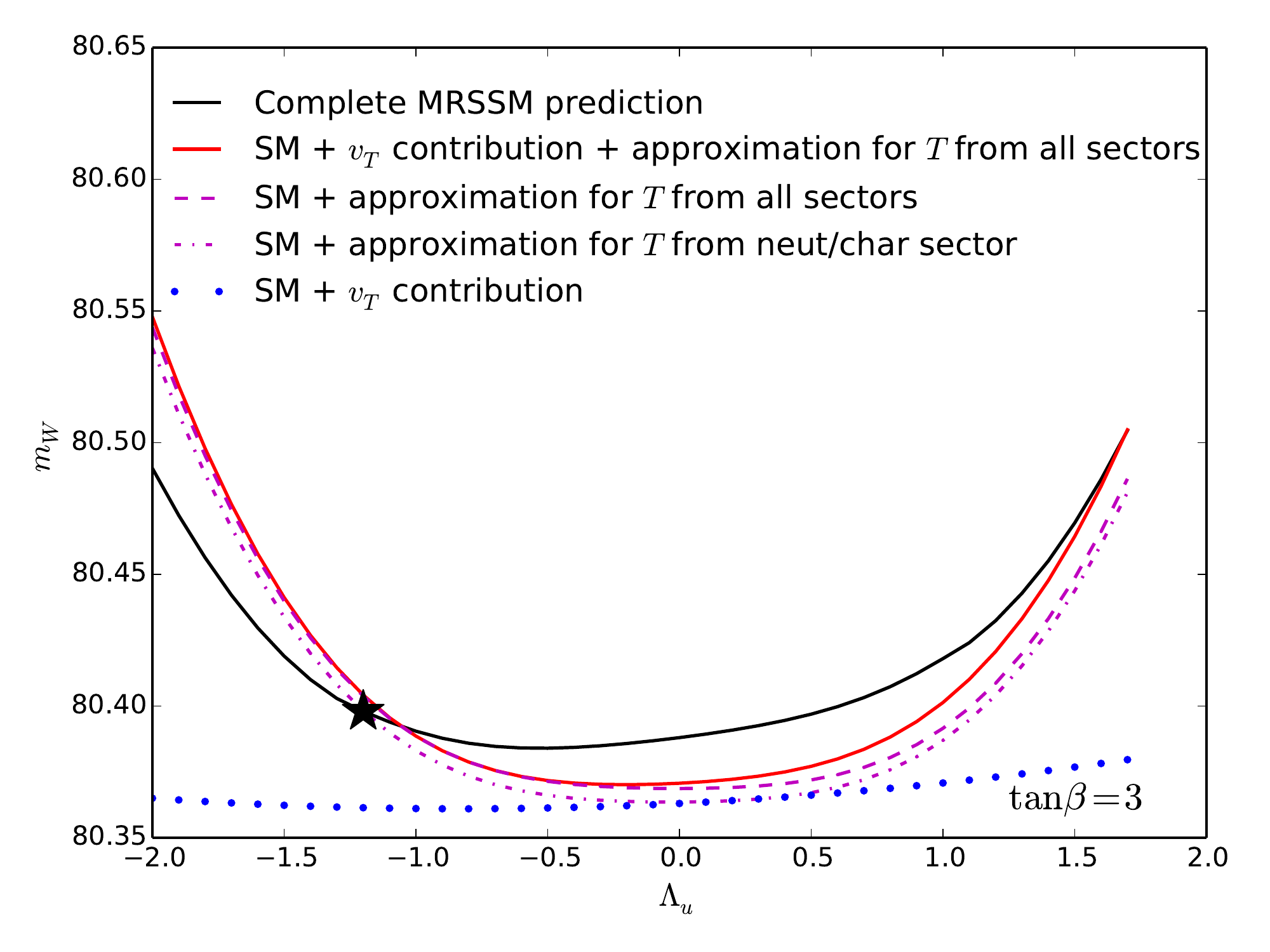}
  \includegraphics[width=0.45\textwidth]{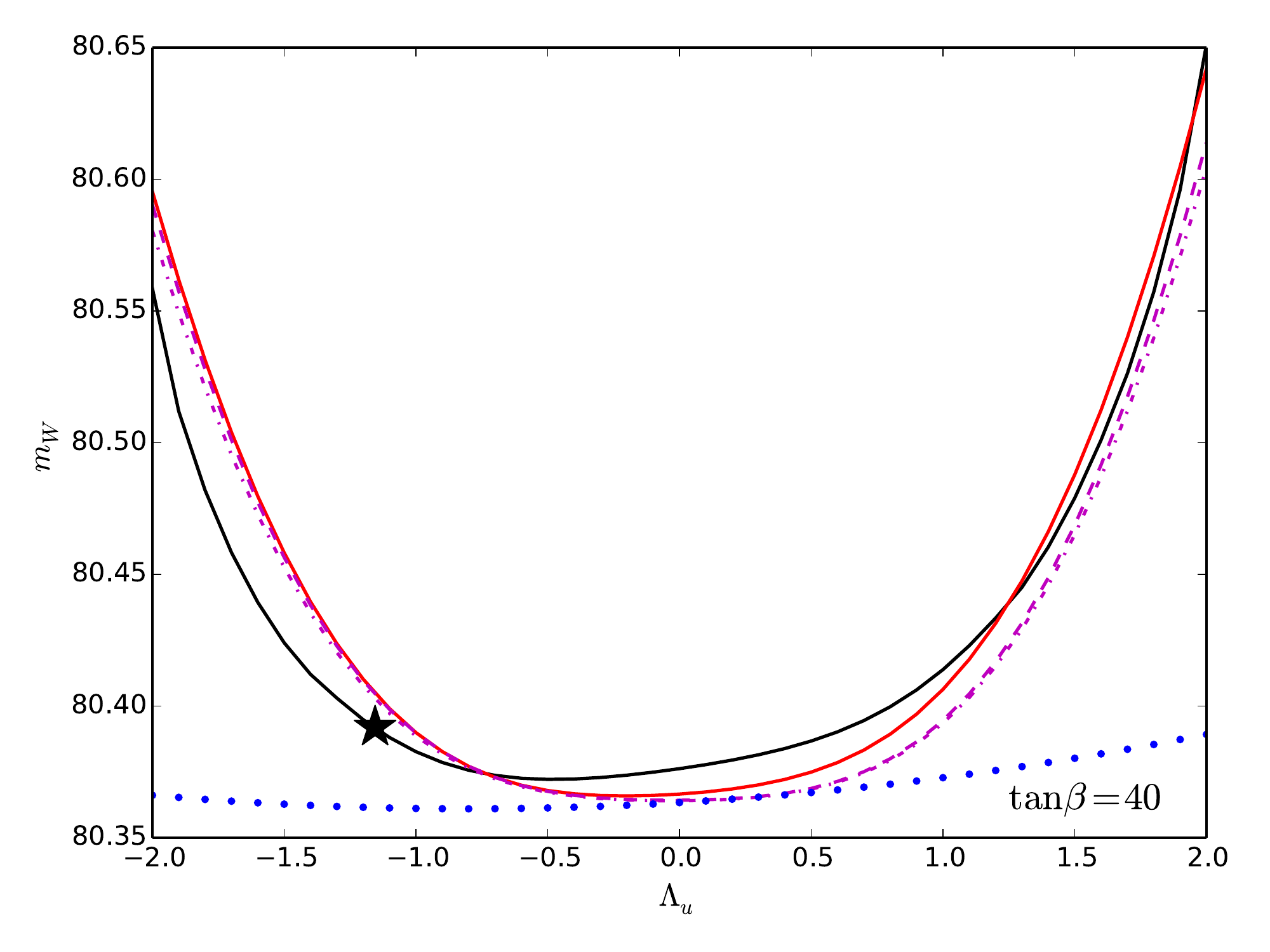}
\caption{
  Comparison of the mass of the W boson depending on $\Lambda_u$, calculated using full MRSSM contributions and different approximations for the $T$ parameter (see Refs.~\cite{Diessner:2014ksa,philips_phd} for details). 
  The black stars mark the corresponding benchmark points of Ref.~\cite{Diessner:2014ksa}.
  Plots prepared by P. Diessner.
}
\label{img:mwlam}
\end{figure}

Before discussing numerical results of the full one-loop calculation, it is instructive to look at its approximation based on \autoref{eq:mw_from_stu_and_vT}.
Full discussion can be found in \cite{Diessner:2014ksa,philips_phd}, here I only briefly state that the full result turns out to be well approximated by just the $T$ parameter, which in turn is dominated by the neutralino/chargino contribution.
This can be seen in \autoref{img:mwlam}, where \autoref{eq:mw_from_stu} with different approximations of the $T$ parameter, is used. 
The complete MRSSM result (i.e. including vertex and box corrections) is well approximated by the SM two-loop calculation supplemented with a $T$ parameter contribution.
Moreover, the $T$ parameter is dominated by chargino/neutralino contributions.
From this it is clear that the MRSSM $\Delta \hat r$ contribution is small.
This is proved in \Autoref{fig:mw_numerical_results}.
As before, the numerical analysis was performed with the help of \texttt{SPheno}.
Treatment of $\delta_{VB}$ in \texttt{SPheno} is controlled by the flag 58 of \texttt{SPhenoInput} block. 
Setting it to 2 excludes BSM contributions, while setting it to 0 sets $\delta_{VB} = 0$ altogether.
The first column of \autoref{fig:mw_numerical_results} shows predictions for $m_W$ without any $\delta_{VB}$ contribution, second after including one-loop SM contribution, third after adding genuine MRSSM effects and forth after adding partial two-loop SM results. 
It is clear from \autoref{fig:mw_numerical_results} that the dominant contribution to $\delta_{VB}$ comes from the SM, with MRSSM one being negligible.
This confirms the aforementioned agreement seen in \autoref{img:mwlam}.
\begin{table}
  \centering
  \begin{tabular}{ccccc}
    & $\delta_{VB} = 0$ & one-loop SM & MRSSM & two-loop SM \\
    \hline
    BMP1 & 80.586 & 80.413 & 80.415 & 80.396\\
    BMP2 & 80.570 & 80.397 & 80.398 & 80.382\\
    BMP3 & 80.574 & 80.401 & 80.402 & 80.386
  \end{tabular}
  \caption{Impact of different contributions to $\delta_{VB}$ on $m_W$ ($m_W$ given in GeV). 
  Contributions are stacked from left to right.
  \label{fig:mw_numerical_results} }
\end{table}
\section{Conclusions}
In this chapter it was shown, that MRSSM may be consistent with the measurement of the $W$ boson mass.
To this end, the chapter described how MRSSM is matched to the SM and how $m_W$ is calculated. 
It also gave explicit expressions for MRSSM contributions to $\delta_{VB}$.
Numerically, it turns out that $\delta_{VB}$ is dominated by one- and two-loop SM contributions, with negligible MRSSM effect.
As such, the $m_W$ dependence on MRSSM model parameters comes mainly from universal self-energy corrections, which in turn are well described by the $T$ parameter.

The full constraining power of the $m_W$ as an observables will be discussed in \autoref{sec:mwmh_conclusions}, for now it is important to note two things.
Due to necessarily small triplet VEV $v_T$ the connected with it through tadpole equation $m_T$ will generally be large, of the order of $2-3$ TeV (unless one arranges special cancelation between model parameters).
Second, chosen benchmark points of \autoref{tab:BMP} are in agreement with the measured $m_W$ within one experimental standard deviation.

\chapter{Higgs boson mass in the MRSSM \label{sec:higgs_chapter}}

On July 4, 2012 ATLAS and CMS experiments announced the discovery of a Higgs boson, together with the first measurements of its properties.
By the end of Run 1, its mass was determined to be $125.09 \pm 0.21 \text{ (stat.) } \pm 0.11 \text{ (syst.)}$ GeV \cite{Aad:2015zhl}.
With that precision, the Higgs boson mass became an electroweak precision observable, much like  $m_W$, and as such any viable model of BSM physics needs now to accommodate it.
To this end, this chapter is devoted to the calculation of full one-loop and leading two-loop corrections to its mass in the MRSSM.\footnote{Since the SUSY scale is not greater than a few TeV a fixed order calculation is used. 
See Ref.~\cite{Athron:2016fuq} for the discussion of the impact of an all order resummation on the Higgs boson mass in the MRSSM, important for SUSY scales $\gtrsim 10$ TeV.}

The chapter is structured as follows. 
The next section gives tree-level values of the lightest Higgs boson mass for benchmark points of \autoref{tab:BMP}, explaining the necessity of calculation of higher order corrections.
After that, in \autoref{sec:mh_1loop}, the complete, momentum dependent, one-loop corrections to its mass are introduced, while \autoref{sec:mh_2loop} describes effective potential based calculation of two-loop Higgs boson mass in the gauge-less limit.
To this end, \autoref{sec:mh_2loop} introduces the notion of effective potential together with the discussion of its limitations and made approximations.
It also gives the explicit expression for MRSSM specific contribution.
Chapter ends with the comparison of results at the tree, one-loop and two-loop levels for BMPs of \autoref{tab:BMP}, as well as a function of the most important superpotential parameters.

\section{Higgs boson mass at the tree level}
As discussed in \autoref{sec:particle_spectrum}, in MRSSM physical CP-even Higgs bosons are mixtures of real parts of 4 gauge eigenstates: scalar fields $H_u^0$, $H_d^0$, $S$ and $T^0$.
The full tree-level mass matrix is given in \autoref{sec:higgs_mass_matrix}.
The matrix is taken at the EW minimum, determined from \autoref{eq:tadpole_equation1} - \autoref{eq:tadpole_equation4}, and written in term of 3 2-by-2 blocks: MSSM-like one, singlet-triplet one and a mixing block.
The mixing between the MSSM and the singlet-triplet blocks results in the lightest Higgs boson mass which is smaller in the MRSSM compared to analogous parameter point in the MSSM (under the assumptions used in this thesis, that $m_S^2, m_T^2$ are large).\footnote{See also Ref.~\cite{Diessner:2015iln} for the discussion of an alternative scenario, in which the SM-like Higgs boson is the second-to-lightest one.
}
The values of tree-level Higgs bosons masses for 3 benchmark points from \autoref{tab:BMP} are given in \autoref{tab:higgs}, and as a function of superpotential parameters $\lambda_{u,d}$, $\Lambda_{u,d}$ at the end of the chapter, in \autoref{fig:mh_1d}.
The smallest tree-level mass occurs, as expected, for BMP with the smallest value of $\tan \beta$, giving $\approx 70$ GeV. 

Clearly, substantial loop corrections are needed to bring predicted Higgs boson mass into agreement with measurements.

\section{One-loop corrections to Higgs boson mass\label{sec:mh_1loop}}

One-loop corrected Higgs boson masses are obtained as a solution in $p^2$ of the equation
\begin{equation}
  \text{determinant}[p^2 \cdot \mathbf{1} - \Re( m_H^2(p^2)) ] = 0,
\end{equation}
where $\mathbf{1}$ is a 4-by-4 identity matrix and
\begin{equation}
  \label{eq:one_loop_formula}
  m_H^2 (p^2) = (\hat m_H^\text{tree})^2 - \hat \Pi^{\text{1-loop}}(p^2) .
\end{equation}
where $\hat m_H^\text{tree}$ and $\hat \Pi^{\text{1-loop}}(p^2)$ are the $\overline{\text{DR}}$~\cite{Jack:1994rk} tree and renormalized one-loop self-energy matrices, respectively.
The \Autoref{eq:one_loop_formula} cannot be solved analytically due to a complicated structure of one-loop corrections. 
Therefore, this calculation was embed in the numerical \texttt{SARAH} - \texttt{SPheno} framework which was then adapted to our purposes.
The general setup for the calculation is the same as illustrated in \autoref{fig:sarah_framework_flowchart} for the case of calculation of $W$ boson mass.


Whether one-loop masses of particles other than $m_W$ are calculated is controlled by flag 55 of \texttt{SphenoInput} block (see \autoref{sec:spheno_card}).
In \texttt{SPheno} there is an important distinction between the calculation of one-loop corrections to particle masses, and inverting these relation to obtain tree-level masses from on-shell masses which are used as input ( so for example pole $m_Z$).
The first one is done in the \texttt{OneLoopMasses} subroutine using loop-corrected tadpole equations while the latter is done in the \texttt{BoundaryEW} routine using tree-level tadpoles (as explained in the previous chapter).

Impact of different contributions to loop-corrected masses can be studied in two ways. 
One is generating \texttt{Spheno} code with \texttt{FlagLoopContributions = True} option (see \autoref{sec:sarah_mathematica}) which includes switches to exclude loop contributions from selected particle(s).
These switches are global, that is they influence calculation of every loop quantity in \texttt{SPheno}, and as such they are affecting how the MRSSM is matched to SM in \texttt{BoundaryEW}. 
Therefore including or not given sectors does also change tree-level value of Higgs boson mass through threshold corrections to $\overline{\text{DR}}$ parameters.

Alternatively, one can comment out in the source code contributions to tadpoles  and one-loop Higgs boson self-energies from given particles (one also has to disable calculation of two-loop contributions, more on them in the next section). 
Calculation of self-energies for particle $X$ is done in \texttt{SPheno} inside \texttt{Pi1Loop}$X$ subroutine while the loop-corrected tadpoles are calculated in \texttt{OneLoopTadpoleshh} subroutine.
The loop-corrected masses of all particles will change, but the tree-level parameters used in the calculation of self-energies will remain constant (apart for those extracted from tadpole equations).

Here, the latter option is used.
\Autoref{tab:higgs} analyzes contributions to the lightest Higgs boson mass sector by sector.
Since the benchmark points were chosen such as to give the correct $\approx125$ GeV Higgs boson mass at the two-loop level, final numbers in \Autoref{tab:higgs} are still of by a few GeV.  
This is discussed in the next section.
\begin{table}
\centering 
\begin{tabular}{l|cccccc}
 & tree-level & \begin{tabular}{@{}c@{}}+\\gauge \\ ghost \\ Higgs\end{tabular} & \begin{tabular}{@{}c@{}}+\\quark/squark\end{tabular} & \begin{tabular}{@{}c@{}}+\\chargino\\ neutralino\end{tabular} & \begin{tabular}{@{}c@{}}+\\R-Higgses \end{tabular}& all\\
\hline
BMP1 & 70.0 & 78.5 & 105.7 & 115.1 & 120.3 & 120.3\\
BMP2 & 86.6 & 93.2 & 116.5 & 119.9 & 120.5 & 120.5\\
BMP3 & 90.8 & 91.0 & 114.8 & 120.4 & 120.7 & 120.7 
\end{tabular}
\caption{
  The lightest Higgs boson mass (in GeV) for the benchmark points: tree-level value and after adding one-loop contributions sector by sector of the MRSSM. 
  \label{tab:higgs}
 }
\end{table}

\FloatBarrier
\section{Two-loop corrections to Higgs boson mass\label{sec:mh_2loop}}
At the time of writing, most advanced results for Higgs boson mass in the MSSM are based on momentum dependent two-loop calculation done for selected sectors of the MSSM \cite{Borowka:2014wla}.
Unfortunately, up until recently, calculations in other SUSY models were much less advanced.
Since \cite{Goodsell:2014bna} an automatized tool for calculating Higgs bosons masses based on two-loop effective potential in the gauge-less limit is available for broad range of SUSY models. 
In Ref.~\cite{Diessner:2015yna} we analyzed impact of two-loop corrections on the Higgs boson mass in the MRSSM, where I have derived expression for the MRSSM-specific contributions.

In this sections I summarize the method of calculating corrections to the mass matrix based on the effective potential approximation and calculate ${\cal O}(\alpha_t\alpha_s)$ corrections that involve Dirac gluino or scalar sgluon (since if parameter $M_O^D$ is real, a pseudoscalar part does not contribute).

Physical masses come from the poles of the propagator matrix as in the one-loop case, see \autoref{eq:one_loop_formula}, where this time
\begin{equation}
  m_H^2 = (m_H^\text{tree})^2 - \Pi^{\text{1-loop}}(p^2) - \Pi^{\text{2-loop}}(p^2).
\end{equation}
Since the calculation of momentum dependent two-loop self energies $\Pi^{\text{2-loop}}(p^2)$ is computationally challenging, one can rewrite the above equation as
\begin{equation}
  m_H^2 (p^2) = (m_H^\text{tree})^2 - \Pi^{\text{1-loop}}(p^2) - \Pi^{\text{2-loop}}(0) - ( \Pi^{\text{2-loop}}(p^2) - \Pi^{\text{2-loop}}(0) ),
\end{equation}
Since the last term is subdominant, one can approximate $m_H^2 (p^2)$ as
\begin{equation}
  m_H^2 (p^2) \approx (m_H^\text{tree})^2 - \Pi^{\text{1-loop}}(p^2) - \Pi^{\text{2-loop}}(0).
\end{equation}
The self-energy in the zero-momentum approximation $\Pi^{\text{2-loop}}(0)$ can be the calculated using the effective potential.
This method was employed in the calculation of two-loop corrections to Higgs boson mass in the MSSM in Refs.~\cite{Zhang:1998bm,Hempfling:1993qq}.

\subsection{Definition of the effective potential}
I start with recalling the definition of generating functional $Z$ (see for example \cite{Schwartz:2013pla}),
\begin{equation}
  Z[J] = \int \mathcal{D}\phi \exp\left\{\imath S[\phi]  + \imath \int d^4 x J(x) \phi(x)\right\},
\end{equation}
for action $S$ depending on field $\phi$. 
One can also define the functional $W[J] \equiv - \imath \ln Z[J]$ which, contrary to $Z[J]$, generates only connected diagrams. 
The (quantum) effective action $\Gamma$ is then defined as the Legendre transform of $W[J]$,
\begin{equation}
  \label{eq:eff_action_definition}
  \Gamma [\phi]  = W[J_\phi] - \int d^4 x J_\phi (x) \phi (x) ,
\end{equation}
where $J_\phi$ is an implicit function of $\phi$ as a solution of the equation
\begin{equation}
  \label{eq:phi_sol}
  \phi(x) \equiv \frac{\partial W[J_\phi]}{\partial J_\phi(x)}.
\end{equation}
meaning that $J_\phi$ is the current needed for the expectation value of $\hat \phi$ operator to be\footnote{This justifies often used name 'classical field'.}
\begin{equation}
  \label{eq:eff_action_definition3}
  \phi(x) = \langle J | \hat \phi | J \rangle,
\end{equation}
Differentiating \autoref{eq:eff_action_definition} with respect to $\phi$ gives
\begin{equation}
  \label{eq:eff_action_definition2}
  \frac{\partial \Gamma[\phi]}{\partial \phi(x)} = \int dy \left( \frac{\partial W[J_\phi]}{\partial J_\phi(y)} \frac{\partial J_\phi(y)}{\partial \phi(x)} -\frac{\partial J_\phi(y)}{\partial \phi(x)} \phi(y) \right) - J_\phi (x) = - J_\phi (x),
\end{equation}
which means that the field configuration $\phi$ for which a solution of \autoref{eq:phi_sol} is $J_\phi = 0$ extremizes the effective potential. 
Conversely, from \autoref{eq:eff_action_definition3}, $\phi(x)$ corresponding to $J_\phi = 0$ is given by the VEV of the field operator. 
This means, that \textit{the effective action has a minimum in a true vacuum state of the theory}.

One can expand effective action in terms of derivatives of $\phi$ as
\begin{equation}
  \Gamma[\phi] = \int d^4 x \left [ -V_{eff}(\phi) + \text{derivative and non-local terms} \right ].
\end{equation}
For a constant field $\phi$ this gives
\begin{equation}
  \Gamma [\phi = const.] = - V_{eff} (\phi) \int d^4 x = - \text{4-volume} \cdot V_{eff} (\phi).
\end{equation}
Since the ground state of the theory must be translationally invariant not to break momentum conservation, one can search for it minimizing $\Gamma[\phi]$ instead of $V_{eff} (\phi)$.

Effective potential can be calculated in a few ways. 
One, used in this work, is by expanding the Lagrangian around a stationary point $\phi_{b}$ as $\phi \to \phi_b +\phi$, $\phi_{b}$ being the solution of equations of motion in the presence of current $J_{\phi}=0$, and doing the functional integral over $\phi$.
It can be shown that this new background effective action $\Gamma_b$ fulfills the relation
\begin{equation}
  \Gamma_b [\phi, \phi_b] = \Gamma[\phi+\phi_b]
\end{equation}
For $\phi = 0$ one gets $\Gamma_b [0, \phi_b] = \Gamma[\phi_b]$. 
Calculation of $\Gamma_b [0, \phi_b]$ means to calculate diagrams with no external $\phi$ legs (i.e. only $\phi$ in the loops) in the presence of a background field $\phi_b$.
The background field can then be treated either perturbatively or, in the case of simple backgrounds like a constant field and more than one-loop, exactly (that is operating with $\phi_b$ dependent masses and vertices, see Appendix A.1 of \cite{PhysRevD.7.1888}).
The latter method is used here.

\subsection{Comment on gauge-dependence of the mass corrections from effective potential}
Before continuing with the discussion of the two-loop corrections, it is instructive to comment on the gauge dependence of the effective potential.
This dependence comes both from the fact that effective action generates only one-particle irreducible (\ac{1PI}) graphs and from the fact that the effective potential will give only off-shell Higgs boson self energies $\Sigma_{h_i h_j}(0)$.
It turns out though, that the $\alpha_s$ corrections will be gauge-independent.
Heuristically, this can be understood as follows: 
all two-loop vacuum diagrams proportional to  $\alpha_s$ are 1PI; moreover, QCD does not 'know' what is the Higgs boson mass, hence it cannot be sensitive to whether it is on- or off-shell.
The formal proof of this follows the derivation of the Nielsen identity \cite{Nielsen:1975fs} using the notion of extended BRS symmetry \cite{Piguet:1984mv,KlubergStern:1974rs,KlubergStern:1974xv,KlubergStern:1975hc} as done in \cite{DelCima:1999gg}.
Using the extended BRS symmetry $s$ (with the BRS transformation for the gauge parameter $\xi$) being
\begin{equation}
  s \xi = \chi,  \qquad s \chi = 0,
\end{equation}
where $\chi$ is a Grassmann variable with a Faddeev-Popov charge +1 gives the extended Slavnov-Taylor identity (with an implicit sum over $i$)
\begin{align}
  0 = S(\Gamma) = & \chi \frac{\partial \Gamma}{\partial \xi} + \int d^4 x \frac{\partial \Gamma}{\partial Y_i} \frac{\partial \Gamma}{\partial \varphi^i} .
\end{align}
In the above equation $Y_i$ are sources and $\varphi^i$ are particle fields.
Differentiating with respect to $\chi$ at $\chi=0$ yields 
\begin{align}
  \frac{\partial \Gamma}{\partial \xi} = -\int d^4 x \left. \Gamma_{\chi Y_i} \Gamma_{\varphi^i} + \Gamma_{Y_i} \Gamma_{\chi \varphi^i } \right|_{\chi=0},
\end{align}
where a shorthand notation $\Gamma_x \equiv \partial \Gamma/ \partial x$ was introduced.
Differentiating twice more with respect to Higgs fields $\varphi$ gives
\begin{align}
  \frac{\partial \Gamma_{\phi \phi}}{\partial \xi} = - \int d^4 x & 
  \, \Gamma_{\phi \phi \chi Y_i } \Gamma_{\varphi^i} +
  \Gamma_{\phi \chi Y_i  } \Gamma_{\phi \varphi^i}+
  \Gamma_{\chi Y_i } \Gamma_{\phi \phi \varphi^i} +
  \Gamma_{\phi \phi Y_i } \Gamma_{ \chi \varphi^i} \\
  & \left. +
    \Gamma_{\phi Y_i } \Gamma_{\phi \chi \varphi^i} +
      \Gamma_{Y_i} \Gamma_{\phi \phi \chi \varphi^i  }
  \right|_{\chi=0},
\end{align}
where indices on $\phi$ were suppressed.
Since $Y_i$ couples to gauge transformation, for $\alpha_s$ corrections $Y_i$ and $\varphi_i$ will be a color-charged.
All terms on the right hand side are products of two derivates of effective actions, each derivative containing only one differentiation with respect to strongly interacting particle.
No such 1PI diagrams can exist due to color conservation.
This proves that the calculation is QCD gauge invariant.

\subsection{Sgluon and gluino contributions to the effective potential}

In this section the effective potential method is applied to the calculation of two-loop corrections to Higgs boson mass of  ${\cal O}(\alpha_t\alpha_s)$ that involve Dirac gluino or sgluon. 
At the one-loop level, the Higgs boson mass depends on Dirac gluino and sgluon only through threshold corrections to quark Yukawas and this dependence is small. 
Therefore, dependence on the Dirac soft mass $\gls{M^D_O}$ and sgluons soft mass $m_O$ is a genuine two-loop effect.
Since both $Y_t$ and $g_s$ are of the $\mathcal{O}(1)$, one can expect a sizable contribution.
In contrast, superpotential parameters $\Lambda_i, \lambda_j$, which are also of the $\mathcal{O}(1)$, appear already at the one-loop level and therefore one expects that their impact at two loops should be moderate, hence the focus is first on the sgluons and gluino contributions.
 This statement is checked later on by an explicit numerical calculation, see  \autoref{fig:mh_1d}.

\begin{figure}
  \begin{center}
  \subfloat[]{\includegraphics[width=0.24\textwidth]{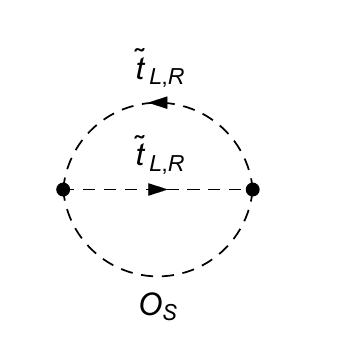}}
  \subfloat[]{\includegraphics[width=0.48\textwidth]{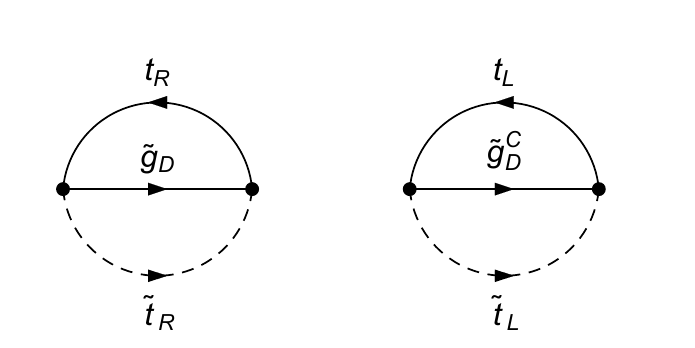}}
  \end{center}
  \caption{Two-loop corrections to the Higgs boson mass corresponding to \autoref{eq:two_loop_eff_pot}, depending on the Dirac mass $M_O^D$ and the soft-sgluon mass $m_O^2$.}
  \label{fig:2L_diagrams}
\end{figure}

\Autoref{fig:2L_diagrams} shows all relevant Feynman diagrams with vertices and Feynman rules as depicted in \autoref{sec:sqcd_feynman_rules}. 
The analytic expression for these contribution, based on loop integrals from Ref.~\cite{Martin:2001vx}, reads

\begin{align}
\label{eq:two_loop_eff_pot}
V^{(2)}_{eff} =& 
  8 \left(\frac{g_s M_O^D}{16\pi^2}\right)^2 \sum_{i=L,R} f_{SSS} (m^2_{\tilde{t}_i}, m^2_{\tilde{t}_i}, m^2_{O_S} ) \\
  & + 
  8 \left(\frac{g_s}{16\pi^2}\right)^2 \sum_{i=L,R} f_{FFS} (m_t^2, m_{\tilde t_i}^2, m_{\tilde g_D}^2 ) \nonumber
\end{align}
where
\begin{eqnarray}
  f_{FFS} &=& J(x,y) - J(x,z) - J(y,z) + (x + y - z) I(x,y,z) \\
  f_{SSS} &=& -I(x,y,z)
\end{eqnarray}
with 
\begin{eqnarray}
  J(x,y) &=& x y ( \overline{\ln} \, x - 1) ( \overline{\ln} \, y - 1) \\
  I(x,y,z) &=& \frac{1}{2}(x-y-z) \overline{\ln} \, y \, \overline{\ln} \, z +
  \frac{1}{2}(y-x-z) \overline{\ln} \, x \, \overline{\ln} \, z +
  \frac{1}{2}(z-x-y) \overline{\ln} \, x \, \overline{\ln} \, y \\
  && + 2 \, x \, \overline{\ln} \, x + 2 \, y \, \overline{\ln} \, y + 2 \, z \, \overline{\ln} \, z - \frac{5}{2} (x+y+z) - \frac{1}{2} \xi (x, y, z)
\end{eqnarray}
and $\overline{\ln} \, x = \ln (x/Q^2)$. For $x, y \leq z$
\begin{eqnarray}
  \xi(x,y,z) &=& R \cdot \left [ 2 \ln \left (\frac{z+x-y-R}{2 z} \right) \cdot 
  \ln \left (\frac{z+y-x-R}{2 z} \right) \right. \\
  && -\ln \left (\frac{x}{z} \right)\cdot \ln \left (\frac{y}{z} \right) -2 \text{Li}_2 \left ( \frac{z+x-y-R}{2 z} \right ) \\
  && \left. -2 \text{Li}_2 \left ( \frac{z+y-x-R}{2 z} \right ) + \frac{\pi^3}{3} \right ]
\end{eqnarray}
and $R = (x^2 + y^2 + z^2 - 2 x y - 2 x z - 2 y z)^{1/2}$.
Coefficients of \autoref{eq:two_loop_eff_pot} can be obtained from Ref.~\cite{Martin:2001vx} by applying translation rules from real to complex fields. Many such rules can be found in Ref.~\cite{Goodsell:2014bna}; an additional rule needed here for the case of a Lagrangian $\mathcal L = -c \Phi_1 |\Phi_2|^2$, where $\Phi_1,c \in \mathbb{R}, \Phi_2 \in \mathbb{C}$, is 
$  V_{\text{SSS}} = \textstyle{\frac{1}{2}} |c|^2 f_{SSS} (m_1^2, m_2^2, m_2^2)$.

\begin{figure}
  \centering
  \includegraphics[width=0.15\textwidth]{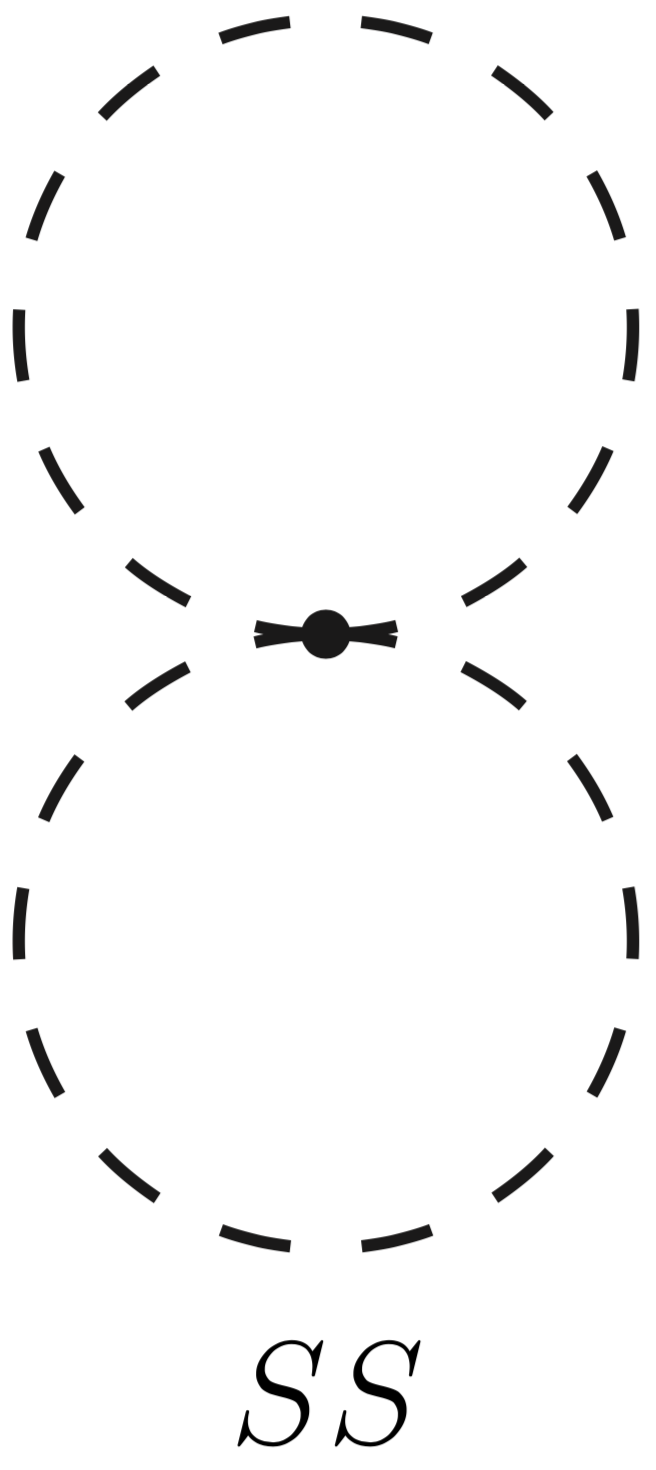}
  \hspace{0.25cm}
  \raisebox{+1.1cm}{\includegraphics[width=0.15\textwidth]{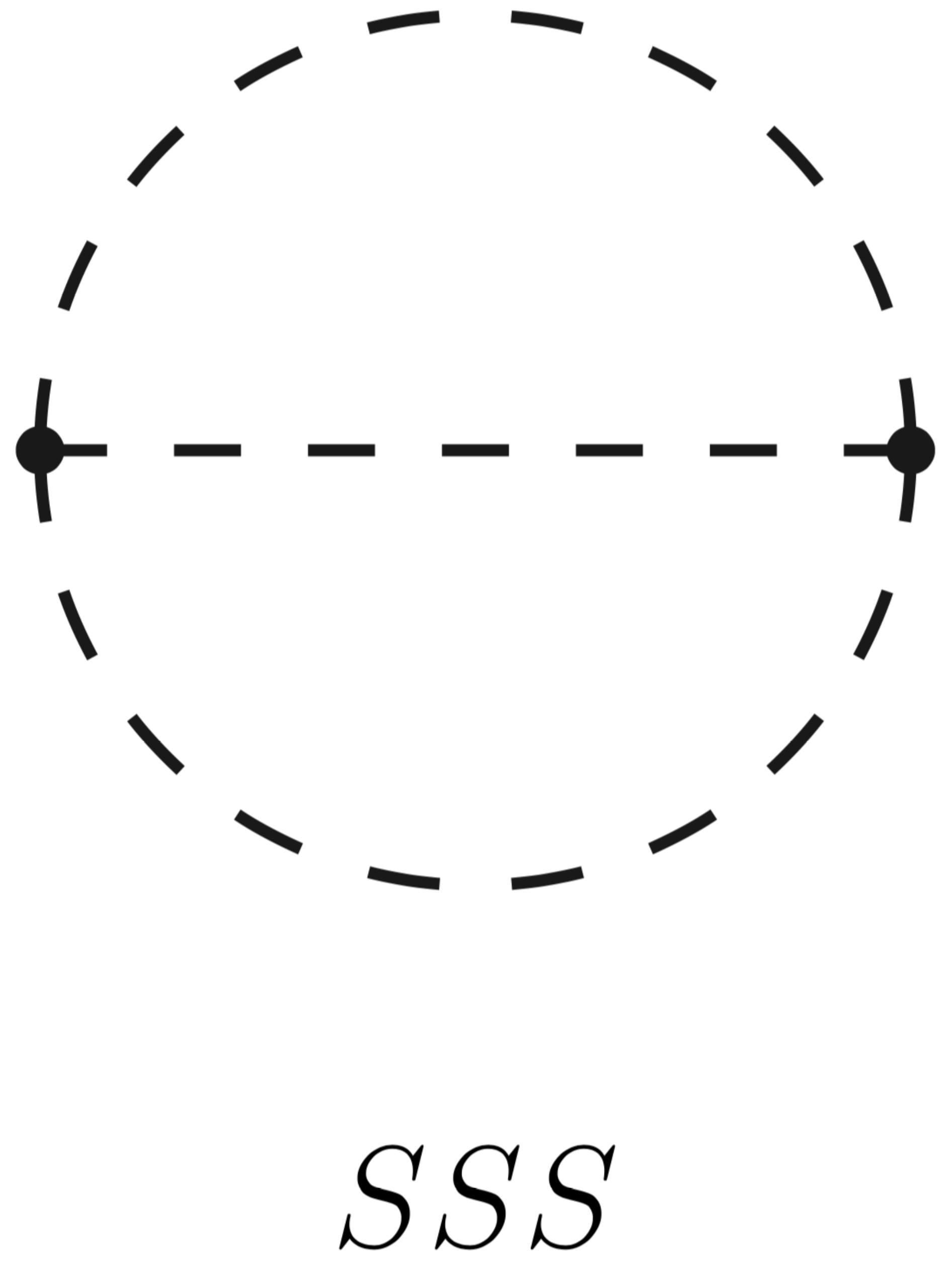}}
  \hspace{0.25cm}
  \raisebox{+1.1cm}{\includegraphics[width=0.15\textwidth]{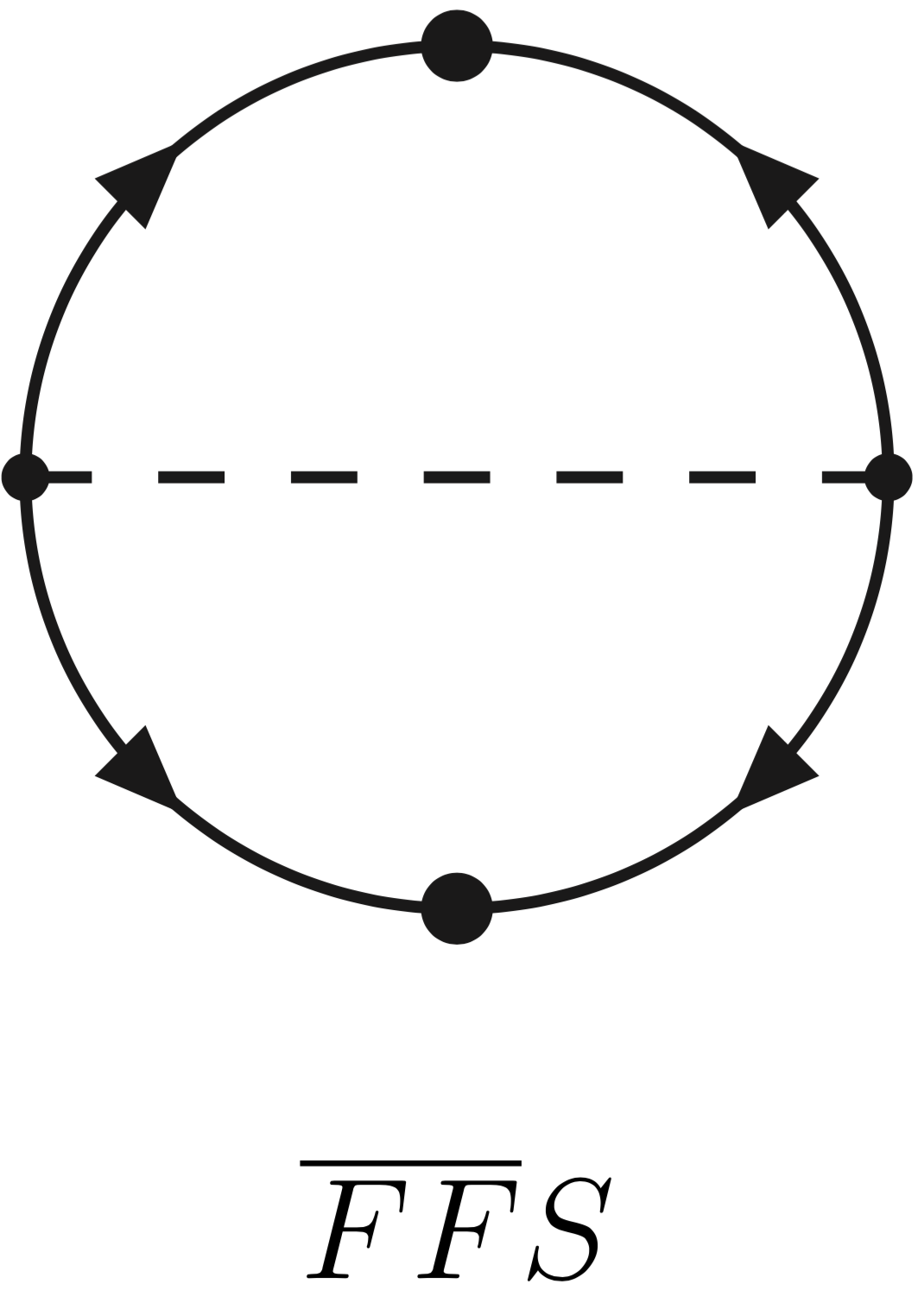}}
  \caption{Naming convention for the two-loop vacuum diagrams.
  Dashed lines denote scalars, solid ones fermions.
  Large dots between clashing arrows on $\bar F \bar F S$ topology denote mass insertion.
  Diagrams come from Ref.~\cite{Martin:2001vx}.
    \label{fig:2l_topologies}
  }
\end{figure}

The structure of the supersymmetric-QCD (SQCD) corrections in \autoref{eq:two_loop_eff_pot} is markedly different from the MSSM case. 
An important difference to the MSSM is that contributions with fermion mass insertions, corresponding to $\overline{FF}S$-type diagram (see \autoref{fig:2l_topologies} for the naming convention of topologies), are not present in the MRSSM. 
Such contributions vanish due to the lack of L-R mixing between squarks.
On the other hand, MRSSM features $SSS$-type diagram with the sgluon field.
As discussed in \autoref{sec:particle_spectrum}, after the SUSY breaking, the mass of the pseudoscalar part of the complex sgluons field $O$, decomposed as $O = \frac{1}{\sqrt{2}}(O_S + \imath O_A)$, is equal to the soft-breaking parameter  $m_{O_A}^2 = m_O^2$, while the mass of the scalar part is shifted by the $D$-term contribution $m_{O_S}^2 = m_O^2 + 4 (M_D^O)^2$.\footnote{In Ref.~\cite{Diessner:2014ksa} a simplifying assumption was made that masses of the scalar and pseudoscalar components of (complex) sgluon field were equal, since it was unimportant for that analysis.}
Thus, this diagram also depends on Dirac gluino mass $M_O^D$.
In the CP-conserving MRSSM considered in this work there is no \textit{SS}-type diagram with sgluon and squarks.

\subsection{Higgs mass from the effective potential}

Following the calculation of the two-loop contributions to the effective potential $V^{(2)}_{eff}$ in the last section, the two-loop correction to the $\phi_u \phi_u$ Higgs mass matrix element in the zero-momentum approximation is given by\footnote{As pointed out in \cite{Goodsell:2014bna}, in \texttt{SARAH} the two-loop tadpole contributions are included directly in vacuum minimization condition and not in \autoref{eq:2}.}
\begin{equation}
  \Pi^{\text{2-loop}}_{\phi_u \phi_u}(0) = \left ( \frac{\partial^2 }{\partial v_u \partial v_u} - \frac{1}{v_u} \frac{\partial }{\partial v_u} \right ) V^{(2)}_{eff} .
  \label{eq:2}
\end{equation}
The effective potential $V^{(2)}_{eff}$ depends on $v_u$ through stop masses, which in the gauge-less limit approach:
\begin{eqnarray}
  m_{\tilde t_L \tilde t_L}^2 &\to& m_q^2 + \frac{1}{2} Y_t^2 v_u^2 ,   \\
  m_{\tilde t_R \tilde t_R}^2 &\to& m_u^2 + \frac{1}{2} Y_t^2 v_u^2 ,
\end{eqnarray}
where $m_q$ and $m_u$ are soft squark masses and $Y_t$ is top quark Yukawa coupling.

For large $\tan \beta$ corrections to the Higgs mass matrix of order $\mathcal O (\alpha_b \alpha_s)$ cannot be neglected any more. 
But since they contribute only to $\phi_d \phi_d $ matrix element, their impact  on the mass of the lightest Higgs, which stems mainly from the $\phi_u \phi_u$ element, is minimal.
Although the expression \autoref{eq:two_loop_eff_pot} can be differentiated analytically, the result is quite lengthy and not given here.
\begin{figure}
  \centering
    \subfloat[]{\includegraphics[width=0.4\textwidth]{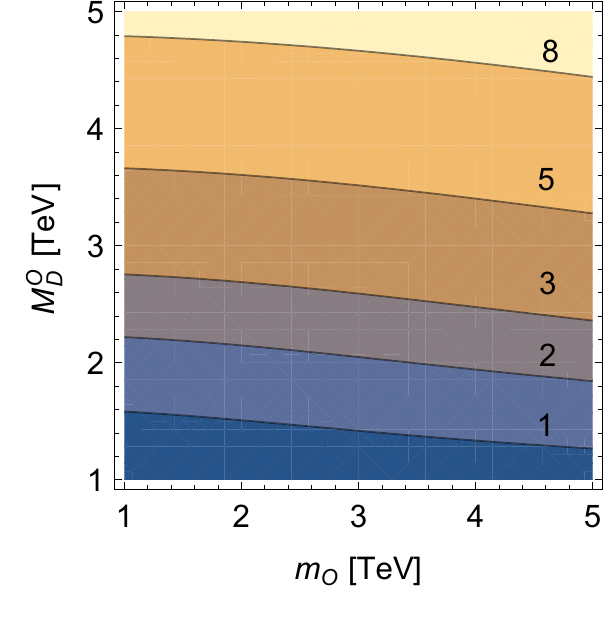}}
  \qquad
    \subfloat[]{\includegraphics[width=0.4\textwidth]{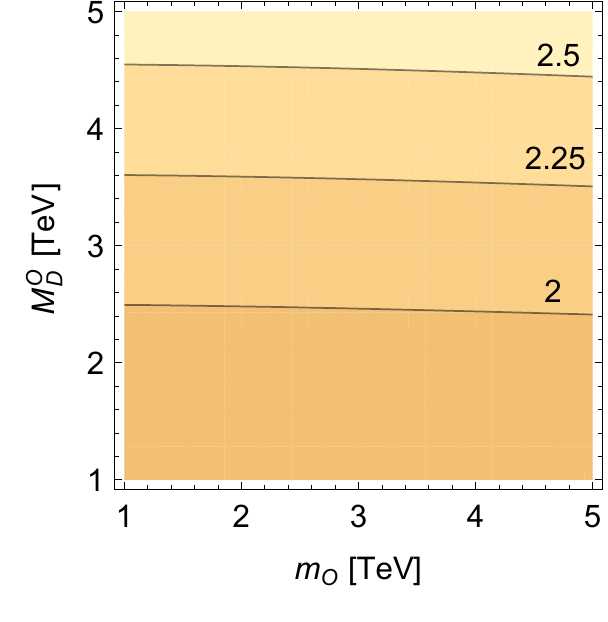}}
  \caption{Contour lines for an approximate two-loop corrections to the Higgs boson mass (in GeV) for $m_{\tilde q} = m_{\tilde u} = 1$ (a) and 5 TeV (b). Threshold corrections to $\alpha_s$ and $Y_t$ are not taken here into account. \label{fig:2loop_size}}
\end{figure}
Instead \Autoref{fig:2loop_size} shows the size of numerical result calculated assuming that the lightest Higgs is a pure $\phi_u$ state, using formula
\begin{equation}
  m_h^{\text{ref}} \sqrt{1+\frac{\delta m_h^2}{(m_h^{\text{ref}})^2}}
\end{equation}
for $m_h^{\text{ref}} = 125$ GeV and without threshold corrections to $\alpha_s$ and $Y_t$. 
$g_s$ and $Y_t$ are fixed to tentative values of 1, while $v_u = 246$ GeV (these also fixes running top mass as $m_t = \frac{1}{\sqrt{2}} Y_t v_u$).
The plot is done for 2 different choices of squark masses:  $m_{\tilde q} = m_{\tilde u} = 1$ TeV (a) and 5 TeV (b).
As was shown in Ref.~\cite{Diessner:2015yna}, in the MSSM for the case when stop squarks do not mix the contribution from gluino turns negative for large values of the gluino mass.
Since this contribution is identical in the MRSSM, one could expect a similar result.
It turns out though, that positive contribution from the diagram with sgluon exchange overpowers negative contribution from gluino diagram as it grows faster with $M_O^D$.

In total, the contribution from these diagrams is around +1 GeV for TeV squarks.

%
%
%

%
%

\section{Numerical results}
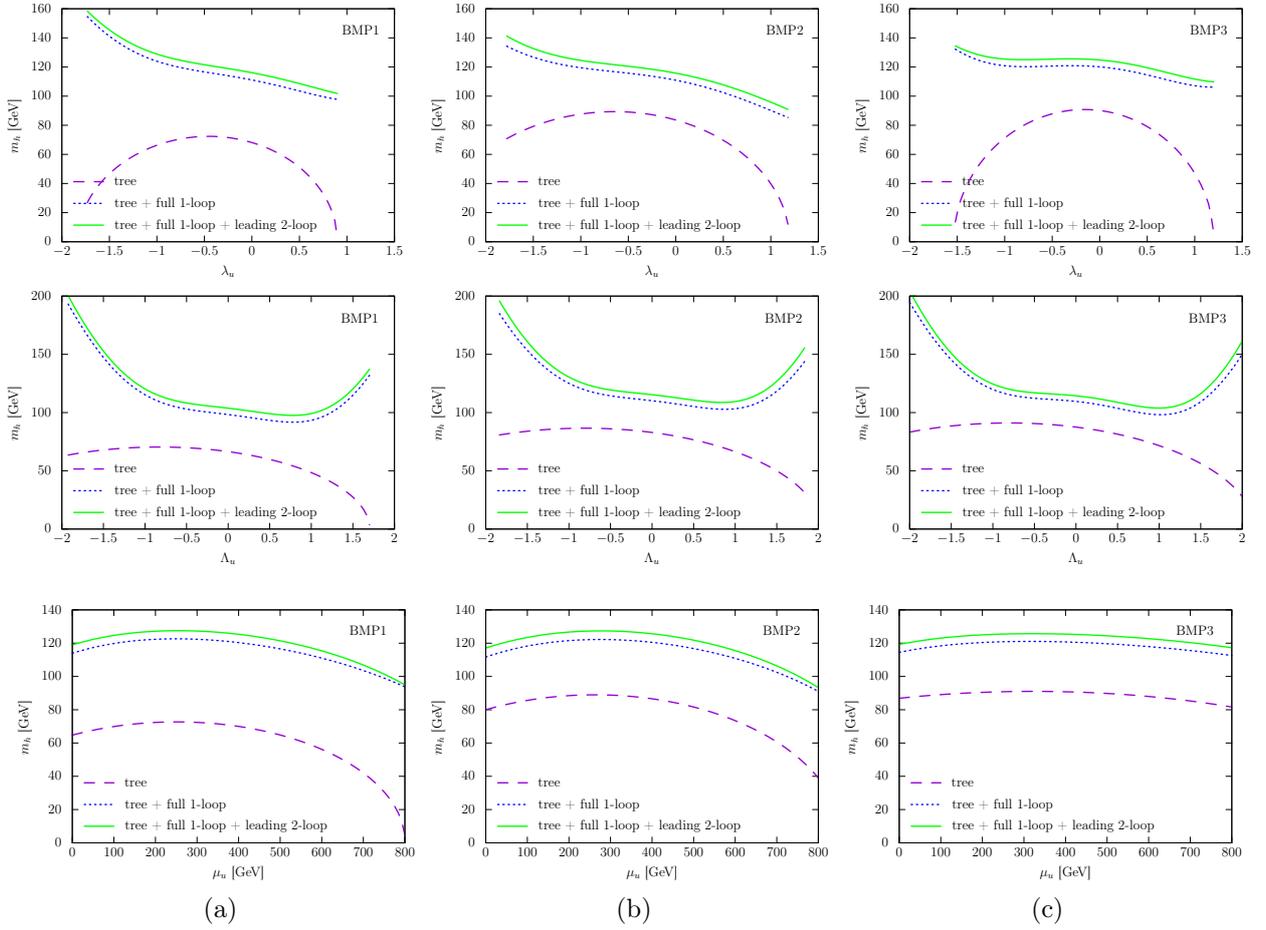
\begin{figure}
  \centering
  \resizebox{0.32\textwidth}{!}{\begin{tikzpicture}[gnuplot]
\path (0.000,0.000) rectangle (12.500,8.750);
\gpcolor{color=gp lt color border}
\gpsetlinetype{gp lt border}
\gpsetdashtype{gp dt solid}
\gpsetlinewidth{1.00}
\draw[gp path] (1.688,1.139)--(1.868,1.139);
\draw[gp path] (11.947,1.139)--(11.767,1.139);
\node[gp node right] at (1.504,1.139) {$0$};
\draw[gp path] (1.688,2.044)--(1.868,2.044);
\draw[gp path] (11.947,2.044)--(11.767,2.044);
\node[gp node right] at (1.504,2.044) {$20$};
\draw[gp path] (1.688,2.950)--(1.868,2.950);
\draw[gp path] (11.947,2.950)--(11.767,2.950);
\node[gp node right] at (1.504,2.950) {$40$};
\draw[gp path] (1.688,3.855)--(1.868,3.855);
\draw[gp path] (11.947,3.855)--(11.767,3.855);
\node[gp node right] at (1.504,3.855) {$60$};
\draw[gp path] (1.688,4.760)--(1.868,4.760);
\draw[gp path] (11.947,4.760)--(11.767,4.760);
\node[gp node right] at (1.504,4.760) {$80$};
\draw[gp path] (1.688,5.665)--(1.868,5.665);
\draw[gp path] (11.947,5.665)--(11.767,5.665);
\node[gp node right] at (1.504,5.665) {$100$};
\draw[gp path] (1.688,6.571)--(1.868,6.571);
\draw[gp path] (11.947,6.571)--(11.767,6.571);
\node[gp node right] at (1.504,6.571) {$120$};
\draw[gp path] (1.688,7.476)--(1.868,7.476);
\draw[gp path] (11.947,7.476)--(11.767,7.476);
\node[gp node right] at (1.504,7.476) {$140$};
\draw[gp path] (1.688,8.381)--(1.868,8.381);
\draw[gp path] (11.947,8.381)--(11.767,8.381);
\node[gp node right] at (1.504,8.381) {$160$};
\draw[gp path] (1.688,1.139)--(1.688,1.319);
\draw[gp path] (1.688,8.381)--(1.688,8.201);
\node[gp node center] at (1.688,0.831) {$-2$};
\draw[gp path] (3.154,1.139)--(3.154,1.319);
\draw[gp path] (3.154,8.381)--(3.154,8.201);
\node[gp node center] at (3.154,0.831) {$-1.5$};
\draw[gp path] (4.619,1.139)--(4.619,1.319);
\draw[gp path] (4.619,8.381)--(4.619,8.201);
\node[gp node center] at (4.619,0.831) {$-1$};
\draw[gp path] (6.085,1.139)--(6.085,1.319);
\draw[gp path] (6.085,8.381)--(6.085,8.201);
\node[gp node center] at (6.085,0.831) {$-0.5$};
\draw[gp path] (7.550,1.139)--(7.550,1.319);
\draw[gp path] (7.550,8.381)--(7.550,8.201);
\node[gp node center] at (7.550,0.831) {$0$};
\draw[gp path] (9.016,1.139)--(9.016,1.319);
\draw[gp path] (9.016,8.381)--(9.016,8.201);
\node[gp node center] at (9.016,0.831) {$0.5$};
\draw[gp path] (10.481,1.139)--(10.481,1.319);
\draw[gp path] (10.481,8.381)--(10.481,8.201);
\node[gp node center] at (10.481,0.831) {$1$};
\draw[gp path] (11.947,1.139)--(11.947,1.319);
\draw[gp path] (11.947,8.381)--(11.947,8.201);
\node[gp node center] at (11.947,0.831) {$1.5$};
\draw[gp path] (1.688,8.381)--(1.688,1.139)--(11.947,1.139)--(11.947,8.381)--cycle;
\node[gp node left] at (10.188,7.702) {BMP1};
\node[gp node center,rotate=-270] at (0.246,4.760) {$m_h$ [GeV]};
\node[gp node center] at (6.817,0.215) {$\lambda_u$};
\node[gp node left] at (3.156,3.006) {tree};
\gpcolor{rgb color={0.580,0.000,0.827}}
\gpsetdashtype{gp dt 2}
\gpsetlinewidth{3.00}
\draw[gp path] (2.056,3.006)--(2.972,3.006);
\draw[gp path] (2.474,2.358)--(2.551,2.500)--(2.629,2.627)--(2.707,2.742)--(2.785,2.847)%
  --(2.863,2.944)--(2.941,3.034)--(3.019,3.119)--(3.097,3.198)--(3.175,3.272)--(3.253,3.343)%
  --(3.331,3.409)--(3.409,3.473)--(3.487,3.533)--(3.565,3.590)--(3.642,3.644)--(3.720,3.695)%
  --(3.798,3.745)--(3.876,3.791)--(3.954,3.836)--(4.032,3.878)--(4.110,3.919)--(4.188,3.957)%
  --(4.266,3.994)--(4.344,4.029)--(4.422,4.062)--(4.500,4.093)--(4.578,4.123)--(4.656,4.151)%
  --(4.733,4.177)--(4.811,4.202)--(4.889,4.226)--(4.967,4.248)--(5.045,4.268)--(5.123,4.287)%
  --(5.201,4.305)--(5.279,4.322)--(5.357,4.337)--(5.435,4.350)--(5.513,4.363)--(5.591,4.374)%
  --(5.669,4.383)--(5.746,4.392)--(5.824,4.399)--(5.902,4.405)--(5.980,4.409)--(6.058,4.412)%
  --(6.136,4.414)--(6.214,4.415)--(6.292,4.415)--(6.370,4.413)--(6.448,4.410)--(6.526,4.405)%
  --(6.604,4.399)--(6.682,4.392)--(6.760,4.384)--(6.837,4.375)--(6.915,4.364)--(6.993,4.351)%
  --(7.071,4.338)--(7.149,4.323)--(7.227,4.306)--(7.305,4.288)--(7.383,4.269)--(7.461,4.248)%
  --(7.539,4.226)--(7.617,4.202)--(7.695,4.177)--(7.773,4.150)--(7.851,4.122)--(7.928,4.091)%
  --(8.006,4.059)--(8.084,4.026)--(8.162,3.990)--(8.240,3.953)--(8.318,3.913)--(8.396,3.871)%
  --(8.474,3.828)--(8.552,3.782)--(8.630,3.733)--(8.708,3.683)--(8.786,3.629)--(8.864,3.573)%
  --(8.941,3.513)--(9.019,3.451)--(9.097,3.385)--(9.175,3.315)--(9.253,3.241)--(9.331,3.162)%
  --(9.409,3.078)--(9.487,2.988)--(9.565,2.891)--(9.643,2.787)--(9.721,2.672)--(9.799,2.546)%
  --(9.877,2.404)--(9.955,2.241)--(10.032,2.046)--(10.110,1.788)--(10.188,1.253);
\gpcolor{color=gp lt color border}
\node[gp node left] at (3.156,2.331) {tree + full 1-loop};
\gpcolor{rgb color={0.000,0.000,1.000}}
\gpsetdashtype{gp dt 4}
\draw[gp path] (2.056,2.331)--(2.972,2.331);
\draw[gp path] (2.474,8.139)--(2.551,8.060)--(2.629,7.983)--(2.707,7.909)--(2.785,7.837)%
  --(2.863,7.768)--(2.941,7.701)--(3.019,7.636)--(3.097,7.573)--(3.175,7.513)--(3.253,7.455)%
  --(3.331,7.399)--(3.409,7.346)--(3.487,7.294)--(3.565,7.244)--(3.642,7.197)--(3.720,7.151)%
  --(3.798,7.107)--(3.876,7.065)--(3.954,7.025)--(4.032,6.987)--(4.110,6.950)--(4.188,6.915)%
  --(4.266,6.882)--(4.344,6.850)--(4.422,6.820)--(4.500,6.791)--(4.578,6.764)--(4.656,6.738)%
  --(4.733,6.713)--(4.811,6.690)--(4.889,6.667)--(4.967,6.646)--(5.045,6.625)--(5.123,6.606)%
  --(5.201,6.587)--(5.279,6.569)--(5.357,6.552)--(5.435,6.536)--(5.513,6.520)--(5.591,6.505)%
  --(5.669,6.490)--(5.746,6.476)--(5.824,6.462)--(5.902,6.448)--(5.980,6.435)--(6.058,6.422)%
  --(6.136,6.409)--(6.214,6.397)--(6.292,6.384)--(6.370,6.372)--(6.448,6.359)--(6.526,6.347)%
  --(6.604,6.334)--(6.682,6.322)--(6.760,6.309)--(6.837,6.297)--(6.915,6.284)--(6.993,6.270)%
  --(7.071,6.257)--(7.149,6.243)--(7.227,6.229)--(7.305,6.215)--(7.383,6.201)--(7.461,6.186)%
  --(7.539,6.171)--(7.617,6.155)--(7.695,6.140)--(7.773,6.123)--(7.851,6.107)--(7.928,6.090)%
  --(8.006,6.073)--(8.084,6.055)--(8.162,6.037)--(8.240,6.019)--(8.318,6.001)--(8.396,5.982)%
  --(8.474,5.963)--(8.552,5.944)--(8.630,5.924)--(8.708,5.905)--(8.786,5.885)--(8.864,5.865)%
  --(8.941,5.845)--(9.019,5.825)--(9.097,5.805)--(9.175,5.785)--(9.253,5.766)--(9.331,5.746)%
  --(9.409,5.727)--(9.487,5.708)--(9.565,5.689)--(9.643,5.671)--(9.721,5.654)--(9.799,5.637)%
  --(9.877,5.620)--(9.955,5.605)--(10.032,5.590)--(10.110,5.577)--(10.188,5.566);
\gpcolor{color=gp lt color border}
\node[gp node left] at (3.156,1.656) {tree + full 1-loop + leading 2-loop};
\gpcolor{rgb color={0.000,1.000,0.000}}
\gpsetdashtype{gp dt solid}
\draw[gp path] (2.056,1.656)--(2.972,1.656);
\draw[gp path] (2.474,8.308)--(2.551,8.232)--(2.629,8.159)--(2.707,8.088)--(2.785,8.019)%
  --(2.863,7.952)--(2.941,7.888)--(3.019,7.826)--(3.097,7.766)--(3.175,7.708)--(3.253,7.653)%
  --(3.331,7.599)--(3.409,7.548)--(3.487,7.498)--(3.565,7.451)--(3.642,7.405)--(3.720,7.361)%
  --(3.798,7.319)--(3.876,7.279)--(3.954,7.240)--(4.032,7.203)--(4.110,7.168)--(4.188,7.134)%
  --(4.266,7.102)--(4.344,7.071)--(4.422,7.041)--(4.500,7.013)--(4.578,6.986)--(4.656,6.961)%
  --(4.733,6.937)--(4.811,6.914)--(4.889,6.892)--(4.967,6.871)--(5.045,6.850)--(5.123,6.831)%
  --(5.201,6.812)--(5.279,6.795)--(5.357,6.778)--(5.435,6.761)--(5.513,6.745)--(5.591,6.730)%
  --(5.669,6.715)--(5.746,6.700)--(5.824,6.686)--(5.902,6.673)--(5.980,6.659)--(6.058,6.646)%
  --(6.136,6.633)--(6.214,6.620)--(6.292,6.608)--(6.370,6.595)--(6.448,6.582)--(6.526,6.570)%
  --(6.604,6.557)--(6.682,6.545)--(6.760,6.532)--(6.837,6.519)--(6.915,6.507)--(6.993,6.494)%
  --(7.071,6.480)--(7.149,6.467)--(7.227,6.453)--(7.305,6.439)--(7.383,6.425)--(7.461,6.411)%
  --(7.539,6.396)--(7.617,6.381)--(7.695,6.366)--(7.773,6.350)--(7.851,6.334)--(7.928,6.318)%
  --(8.006,6.301)--(8.084,6.284)--(8.162,6.267)--(8.240,6.249)--(8.318,6.231)--(8.396,6.212)%
  --(8.474,6.194)--(8.552,6.175)--(8.630,6.155)--(8.708,6.135)--(8.786,6.116)--(8.864,6.096)%
  --(8.941,6.076)--(9.019,6.055)--(9.097,6.035)--(9.175,6.014)--(9.253,5.993)--(9.331,5.972)%
  --(9.409,5.951)--(9.487,5.930)--(9.565,5.909)--(9.643,5.888)--(9.721,5.867)--(9.799,5.846)%
  --(9.877,5.825)--(9.955,5.805)--(10.032,5.785)--(10.110,5.766)--(10.188,5.749);
\gpcolor{color=gp lt color border}
\gpsetlinewidth{1.00}
\draw[gp path] (1.688,8.381)--(1.688,1.139)--(11.947,1.139)--(11.947,8.381)--cycle;
\gpdefrectangularnode{gp plot 1}{\pgfpoint{1.688cm}{1.139cm}}{\pgfpoint{11.947cm}{8.381cm}}
\end{tikzpicture}
}
  \resizebox{0.32\textwidth}{!}{\begin{tikzpicture}[gnuplot]
\path (0.000,0.000) rectangle (12.500,8.750);
\gpcolor{color=gp lt color border}
\gpsetlinetype{gp lt border}
\gpsetdashtype{gp dt solid}
\gpsetlinewidth{1.00}
\draw[gp path] (1.688,1.139)--(1.868,1.139);
\draw[gp path] (11.947,1.139)--(11.767,1.139);
\node[gp node right] at (1.504,1.139) {$0$};
\draw[gp path] (1.688,2.044)--(1.868,2.044);
\draw[gp path] (11.947,2.044)--(11.767,2.044);
\node[gp node right] at (1.504,2.044) {$20$};
\draw[gp path] (1.688,2.950)--(1.868,2.950);
\draw[gp path] (11.947,2.950)--(11.767,2.950);
\node[gp node right] at (1.504,2.950) {$40$};
\draw[gp path] (1.688,3.855)--(1.868,3.855);
\draw[gp path] (11.947,3.855)--(11.767,3.855);
\node[gp node right] at (1.504,3.855) {$60$};
\draw[gp path] (1.688,4.760)--(1.868,4.760);
\draw[gp path] (11.947,4.760)--(11.767,4.760);
\node[gp node right] at (1.504,4.760) {$80$};
\draw[gp path] (1.688,5.665)--(1.868,5.665);
\draw[gp path] (11.947,5.665)--(11.767,5.665);
\node[gp node right] at (1.504,5.665) {$100$};
\draw[gp path] (1.688,6.571)--(1.868,6.571);
\draw[gp path] (11.947,6.571)--(11.767,6.571);
\node[gp node right] at (1.504,6.571) {$120$};
\draw[gp path] (1.688,7.476)--(1.868,7.476);
\draw[gp path] (11.947,7.476)--(11.767,7.476);
\node[gp node right] at (1.504,7.476) {$140$};
\draw[gp path] (1.688,8.381)--(1.868,8.381);
\draw[gp path] (11.947,8.381)--(11.767,8.381);
\node[gp node right] at (1.504,8.381) {$160$};
\draw[gp path] (1.688,1.139)--(1.688,1.319);
\draw[gp path] (1.688,8.381)--(1.688,8.201);
\node[gp node center] at (1.688,0.831) {$-2$};
\draw[gp path] (3.154,1.139)--(3.154,1.319);
\draw[gp path] (3.154,8.381)--(3.154,8.201);
\node[gp node center] at (3.154,0.831) {$-1.5$};
\draw[gp path] (4.619,1.139)--(4.619,1.319);
\draw[gp path] (4.619,8.381)--(4.619,8.201);
\node[gp node center] at (4.619,0.831) {$-1$};
\draw[gp path] (6.085,1.139)--(6.085,1.319);
\draw[gp path] (6.085,8.381)--(6.085,8.201);
\node[gp node center] at (6.085,0.831) {$-0.5$};
\draw[gp path] (7.550,1.139)--(7.550,1.319);
\draw[gp path] (7.550,8.381)--(7.550,8.201);
\node[gp node center] at (7.550,0.831) {$0$};
\draw[gp path] (9.016,1.139)--(9.016,1.319);
\draw[gp path] (9.016,8.381)--(9.016,8.201);
\node[gp node center] at (9.016,0.831) {$0.5$};
\draw[gp path] (10.481,1.139)--(10.481,1.319);
\draw[gp path] (10.481,8.381)--(10.481,8.201);
\node[gp node center] at (10.481,0.831) {$1$};
\draw[gp path] (11.947,1.139)--(11.947,1.319);
\draw[gp path] (11.947,8.381)--(11.947,8.201);
\node[gp node center] at (11.947,0.831) {$1.5$};
\draw[gp path] (1.688,8.381)--(1.688,1.139)--(11.947,1.139)--(11.947,8.381)--cycle;
\node[gp node left] at (10.188,7.702) {BMP2};
\node[gp node center,rotate=-270] at (0.246,4.760) {$m_h$ [GeV]};
\node[gp node center] at (6.817,0.215) {$\lambda_u$};
\node[gp node left] at (3.156,3.006) {tree};
\gpcolor{rgb color={0.580,0.000,0.827}}
\gpsetdashtype{gp dt 2}
\gpsetlinewidth{3.00}
\draw[gp path] (2.056,3.006)--(2.972,3.006);
\draw[gp path] (2.333,4.342)--(2.420,4.390)--(2.508,4.436)--(2.596,4.481)--(2.683,4.523)%
  --(2.771,4.564)--(2.859,4.604)--(2.946,4.642)--(3.034,4.678)--(3.122,4.713)--(3.209,4.746)%
  --(3.297,4.778)--(3.385,4.809)--(3.472,4.838)--(3.560,4.866)--(3.647,4.893)--(3.735,4.918)%
  --(3.823,4.942)--(3.910,4.965)--(3.998,4.987)--(4.086,5.007)--(4.173,5.027)--(4.261,5.045)%
  --(4.349,5.062)--(4.436,5.078)--(4.524,5.092)--(4.611,5.106)--(4.699,5.118)--(4.787,5.130)%
  --(4.874,5.140)--(4.962,5.149)--(5.050,5.157)--(5.137,5.164)--(5.225,5.170)--(5.313,5.175)%
  --(5.400,5.179)--(5.488,5.181)--(5.575,5.183)--(5.663,5.183)--(5.751,5.183)--(5.838,5.181)%
  --(5.926,5.178)--(6.014,5.175)--(6.101,5.170)--(6.189,5.164)--(6.277,5.156)--(6.364,5.148)%
  --(6.452,5.139)--(6.539,5.128)--(6.627,5.117)--(6.715,5.104)--(6.802,5.090)--(6.890,5.075)%
  --(6.978,5.059)--(7.065,5.041)--(7.153,5.023)--(7.241,5.003)--(7.328,4.982)--(7.416,4.959)%
  --(7.504,4.936)--(7.591,4.911)--(7.679,4.884)--(7.766,4.857)--(7.854,4.828)--(7.942,4.797)%
  --(8.029,4.765)--(8.117,4.732)--(8.205,4.697)--(8.292,4.660)--(8.380,4.622)--(8.468,4.582)%
  --(8.555,4.541)--(8.643,4.498)--(8.730,4.453)--(8.818,4.406)--(8.906,4.357)--(8.993,4.305)%
  --(9.081,4.252)--(9.169,4.197)--(9.256,4.139)--(9.344,4.078)--(9.432,4.015)--(9.519,3.949)%
  --(9.607,3.880)--(9.694,3.807)--(9.782,3.731)--(9.870,3.651)--(9.957,3.567)--(10.045,3.478)%
  --(10.133,3.384)--(10.220,3.284)--(10.308,3.177)--(10.396,3.062)--(10.483,2.937)--(10.571,2.801)%
  --(10.658,2.651)--(10.746,2.481)--(10.834,2.282)--(10.921,2.039)--(11.009,1.686);
\gpcolor{color=gp lt color border}
\node[gp node left] at (3.156,2.331) {tree + full 1-loop};
\gpcolor{rgb color={0.000,0.000,1.000}}
\gpsetdashtype{gp dt 4}
\draw[gp path] (2.056,2.331)--(2.972,2.331);
\draw[gp path] (2.333,7.220)--(2.420,7.176)--(2.508,7.134)--(2.596,7.094)--(2.683,7.055)%
  --(2.771,7.018)--(2.859,6.983)--(2.946,6.949)--(3.034,6.917)--(3.122,6.887)--(3.209,6.858)%
  --(3.297,6.830)--(3.385,6.804)--(3.472,6.779)--(3.560,6.755)--(3.647,6.732)--(3.735,6.711)%
  --(3.823,6.691)--(3.910,6.671)--(3.998,6.653)--(4.086,6.636)--(4.173,6.620)--(4.261,6.604)%
  --(4.349,6.589)--(4.436,6.575)--(4.524,6.562)--(4.611,6.549)--(4.699,6.537)--(4.787,6.525)%
  --(4.874,6.514)--(4.962,6.503)--(5.050,6.492)--(5.137,6.482)--(5.225,6.472)--(5.313,6.462)%
  --(5.400,6.453)--(5.488,6.443)--(5.575,6.434)--(5.663,6.424)--(5.751,6.415)--(5.838,6.405)%
  --(5.926,6.396)--(6.014,6.386)--(6.101,6.376)--(6.189,6.366)--(6.277,6.356)--(6.364,6.345)%
  --(6.452,6.334)--(6.539,6.323)--(6.627,6.311)--(6.715,6.299)--(6.802,6.287)--(6.890,6.274)%
  --(6.978,6.260)--(7.065,6.246)--(7.153,6.232)--(7.241,6.217)--(7.328,6.201)--(7.416,6.185)%
  --(7.504,6.169)--(7.591,6.151)--(7.679,6.133)--(7.766,6.115)--(7.854,6.095)--(7.942,6.076)%
  --(8.029,6.055)--(8.117,6.034)--(8.205,6.012)--(8.292,5.989)--(8.380,5.966)--(8.468,5.942)%
  --(8.555,5.917)--(8.643,5.891)--(8.730,5.865)--(8.818,5.839)--(8.906,5.811)--(8.993,5.783)%
  --(9.081,5.754)--(9.169,5.725)--(9.256,5.694)--(9.344,5.664)--(9.432,5.632)--(9.519,5.600)%
  --(9.607,5.568)--(9.694,5.535)--(9.782,5.501)--(9.870,5.467)--(9.957,5.433)--(10.045,5.398)%
  --(10.133,5.363)--(10.220,5.327)--(10.308,5.291)--(10.396,5.255)--(10.483,5.219)--(10.571,5.183)%
  --(10.658,5.147)--(10.746,5.110)--(10.834,5.074)--(10.921,5.038)--(11.009,5.003);
\gpcolor{color=gp lt color border}
\node[gp node left] at (3.156,1.656) {tree + full 1-loop + leading 2-loop};
\gpcolor{rgb color={0.000,1.000,0.000}}
\gpsetdashtype{gp dt solid}
\draw[gp path] (2.056,1.656)--(2.972,1.656);
\draw[gp path] (2.333,7.535)--(2.420,7.484)--(2.508,7.437)--(2.596,7.392)--(2.683,7.348)%
  --(2.771,7.306)--(2.859,7.266)--(2.946,7.228)--(3.034,7.192)--(3.122,7.157)--(3.209,7.125)%
  --(3.297,7.093)--(3.385,7.064)--(3.472,7.036)--(3.560,7.009)--(3.647,6.984)--(3.735,6.960)%
  --(3.823,6.937)--(3.910,6.915)--(3.998,6.895)--(4.086,6.875)--(4.173,6.857)--(4.261,6.840)%
  --(4.349,6.823)--(4.436,6.807)--(4.524,6.792)--(4.611,6.778)--(4.699,6.764)--(4.787,6.751)%
  --(4.874,6.739)--(4.962,6.727)--(5.050,6.715)--(5.137,6.704)--(5.225,6.693)--(5.313,6.683)%
  --(5.400,6.672)--(5.488,6.662)--(5.575,6.652)--(5.663,6.642)--(5.751,6.632)--(5.838,6.622)%
  --(5.926,6.612)--(6.014,6.602)--(6.101,6.592)--(6.189,6.581)--(6.277,6.571)--(6.364,6.560)%
  --(6.452,6.549)--(6.539,6.538)--(6.627,6.526)--(6.715,6.514)--(6.802,6.502)--(6.890,6.489)%
  --(6.978,6.476)--(7.065,6.463)--(7.153,6.449)--(7.241,6.434)--(7.328,6.419)--(7.416,6.404)%
  --(7.504,6.388)--(7.591,6.371)--(7.679,6.354)--(7.766,6.336)--(7.854,6.318)--(7.942,6.299)%
  --(8.029,6.280)--(8.117,6.259)--(8.205,6.239)--(8.292,6.217)--(8.380,6.195)--(8.468,6.172)%
  --(8.555,6.149)--(8.643,6.125)--(8.730,6.100)--(8.818,6.075)--(8.906,6.049)--(8.993,6.022)%
  --(9.081,5.995)--(9.169,5.967)--(9.256,5.938)--(9.344,5.909)--(9.432,5.879)--(9.519,5.849)%
  --(9.607,5.818)--(9.694,5.786)--(9.782,5.754)--(9.870,5.721)--(9.957,5.687)--(10.045,5.653)%
  --(10.133,5.619)--(10.220,5.584)--(10.308,5.549)--(10.396,5.513)--(10.483,5.477)--(10.571,5.440)%
  --(10.658,5.403)--(10.746,5.366)--(10.834,5.329)--(10.921,5.291)--(11.009,5.253);
\gpcolor{color=gp lt color border}
\gpsetlinewidth{1.00}
\draw[gp path] (1.688,8.381)--(1.688,1.139)--(11.947,1.139)--(11.947,8.381)--cycle;
\gpdefrectangularnode{gp plot 1}{\pgfpoint{1.688cm}{1.139cm}}{\pgfpoint{11.947cm}{8.381cm}}
\end{tikzpicture}
}
  \resizebox{0.32\textwidth}{!}{\begin{tikzpicture}[gnuplot]
\path (0.000,0.000) rectangle (12.500,8.750);
\gpcolor{color=gp lt color border}
\gpsetlinetype{gp lt border}
\gpsetdashtype{gp dt solid}
\gpsetlinewidth{1.00}
\draw[gp path] (1.688,1.139)--(1.868,1.139);
\draw[gp path] (11.947,1.139)--(11.767,1.139);
\node[gp node right] at (1.504,1.139) {$0$};
\draw[gp path] (1.688,2.044)--(1.868,2.044);
\draw[gp path] (11.947,2.044)--(11.767,2.044);
\node[gp node right] at (1.504,2.044) {$20$};
\draw[gp path] (1.688,2.950)--(1.868,2.950);
\draw[gp path] (11.947,2.950)--(11.767,2.950);
\node[gp node right] at (1.504,2.950) {$40$};
\draw[gp path] (1.688,3.855)--(1.868,3.855);
\draw[gp path] (11.947,3.855)--(11.767,3.855);
\node[gp node right] at (1.504,3.855) {$60$};
\draw[gp path] (1.688,4.760)--(1.868,4.760);
\draw[gp path] (11.947,4.760)--(11.767,4.760);
\node[gp node right] at (1.504,4.760) {$80$};
\draw[gp path] (1.688,5.665)--(1.868,5.665);
\draw[gp path] (11.947,5.665)--(11.767,5.665);
\node[gp node right] at (1.504,5.665) {$100$};
\draw[gp path] (1.688,6.571)--(1.868,6.571);
\draw[gp path] (11.947,6.571)--(11.767,6.571);
\node[gp node right] at (1.504,6.571) {$120$};
\draw[gp path] (1.688,7.476)--(1.868,7.476);
\draw[gp path] (11.947,7.476)--(11.767,7.476);
\node[gp node right] at (1.504,7.476) {$140$};
\draw[gp path] (1.688,8.381)--(1.868,8.381);
\draw[gp path] (11.947,8.381)--(11.767,8.381);
\node[gp node right] at (1.504,8.381) {$160$};
\draw[gp path] (1.688,1.139)--(1.688,1.319);
\draw[gp path] (1.688,8.381)--(1.688,8.201);
\node[gp node center] at (1.688,0.831) {$-2$};
\draw[gp path] (3.154,1.139)--(3.154,1.319);
\draw[gp path] (3.154,8.381)--(3.154,8.201);
\node[gp node center] at (3.154,0.831) {$-1.5$};
\draw[gp path] (4.619,1.139)--(4.619,1.319);
\draw[gp path] (4.619,8.381)--(4.619,8.201);
\node[gp node center] at (4.619,0.831) {$-1$};
\draw[gp path] (6.085,1.139)--(6.085,1.319);
\draw[gp path] (6.085,8.381)--(6.085,8.201);
\node[gp node center] at (6.085,0.831) {$-0.5$};
\draw[gp path] (7.550,1.139)--(7.550,1.319);
\draw[gp path] (7.550,8.381)--(7.550,8.201);
\node[gp node center] at (7.550,0.831) {$0$};
\draw[gp path] (9.016,1.139)--(9.016,1.319);
\draw[gp path] (9.016,8.381)--(9.016,8.201);
\node[gp node center] at (9.016,0.831) {$0.5$};
\draw[gp path] (10.481,1.139)--(10.481,1.319);
\draw[gp path] (10.481,8.381)--(10.481,8.201);
\node[gp node center] at (10.481,0.831) {$1$};
\draw[gp path] (11.947,1.139)--(11.947,1.319);
\draw[gp path] (11.947,8.381)--(11.947,8.201);
\node[gp node center] at (11.947,0.831) {$1.5$};
\draw[gp path] (1.688,8.381)--(1.688,1.139)--(11.947,1.139)--(11.947,8.381)--cycle;
\node[gp node left] at (10.188,7.702) {BMP3};
\node[gp node center,rotate=-270] at (0.246,4.760) {$m_h$ [GeV]};
\node[gp node center] at (6.817,0.215) {$\lambda_u$};
\node[gp node left] at (3.156,3.006) {tree};
\gpcolor{rgb color={0.580,0.000,0.827}}
\gpsetdashtype{gp dt 2}
\gpsetlinewidth{3.00}
\draw[gp path] (2.056,3.006)--(2.972,3.006);
\draw[gp path] (3.095,1.756)--(3.175,2.130)--(3.256,2.422)--(3.337,2.635)--(3.417,2.822)%
  --(3.498,2.990)--(3.578,3.139)--(3.659,3.275)--(3.739,3.401)--(3.820,3.518)--(3.900,3.627)%
  --(3.981,3.729)--(4.061,3.825)--(4.142,3.916)--(4.222,4.002)--(4.303,4.083)--(4.383,4.160)%
  --(4.464,4.233)--(4.545,4.302)--(4.625,4.368)--(4.706,4.431)--(4.786,4.491)--(4.867,4.548)%
  --(4.947,4.602)--(5.028,4.653)--(5.108,4.702)--(5.189,4.748)--(5.269,4.792)--(5.350,4.833)%
  --(5.430,4.873)--(5.511,4.910)--(5.591,4.945)--(5.672,4.978)--(5.753,5.008)--(5.833,5.037)%
  --(5.914,5.064)--(5.994,5.089)--(6.075,5.112)--(6.155,5.133)--(6.236,5.153)--(6.316,5.170)%
  --(6.397,5.186)--(6.477,5.200)--(6.558,5.212)--(6.638,5.223)--(6.719,5.231)--(6.799,5.238)%
  --(6.880,5.243)--(6.961,5.247)--(7.041,5.249)--(7.122,5.249)--(7.202,5.247)--(7.283,5.243)%
  --(7.363,5.238)--(7.444,5.231)--(7.524,5.222)--(7.605,5.212)--(7.685,5.200)--(7.766,5.186)%
  --(7.846,5.170)--(7.927,5.152)--(8.007,5.133)--(8.088,5.111)--(8.168,5.088)--(8.249,5.063)%
  --(8.330,5.036)--(8.410,5.006)--(8.491,4.975)--(8.571,4.942)--(8.652,4.906)--(8.732,4.869)%
  --(8.813,4.829)--(8.893,4.787)--(8.974,4.742)--(9.054,4.695)--(9.135,4.646)--(9.215,4.593)%
  --(9.296,4.538)--(9.376,4.481)--(9.457,4.420)--(9.538,4.356)--(9.618,4.288)--(9.699,4.217)%
  --(9.779,4.142)--(9.860,4.063)--(9.940,3.980)--(10.021,3.891)--(10.101,3.798)--(10.182,3.698)%
  --(10.262,3.593)--(10.343,3.479)--(10.423,3.358)--(10.504,3.226)--(10.584,3.082)--(10.665,2.925)%
  --(10.746,2.744)--(10.826,2.541)--(10.907,2.317)--(10.987,1.936)--(11.068,1.385);
\gpcolor{color=gp lt color border}
\node[gp node left] at (3.156,2.331) {tree + full 1-loop};
\gpcolor{rgb color={0.000,0.000,1.000}}
\gpsetdashtype{gp dt 4}
\draw[gp path] (2.056,2.331)--(2.972,2.331);
\draw[gp path] (3.095,7.126)--(3.175,7.077)--(3.256,7.030)--(3.337,6.986)--(3.417,6.945)%
  --(3.498,6.907)--(3.578,6.871)--(3.659,6.838)--(3.739,6.808)--(3.820,6.779)--(3.900,6.753)%
  --(3.981,6.729)--(4.061,6.708)--(4.142,6.688)--(4.222,6.670)--(4.303,6.655)--(4.383,6.641)%
  --(4.464,6.629)--(4.545,6.619)--(4.625,6.610)--(4.706,6.602)--(4.786,6.596)--(4.867,6.591)%
  --(4.947,6.587)--(5.028,6.584)--(5.108,6.582)--(5.189,6.581)--(5.269,6.581)--(5.350,6.581)%
  --(5.430,6.582)--(5.511,6.583)--(5.591,6.585)--(5.672,6.587)--(5.753,6.589)--(5.833,6.592)%
  --(5.914,6.594)--(5.994,6.597)--(6.075,6.599)--(6.155,6.602)--(6.236,6.604)--(6.316,6.606)%
  --(6.397,6.608)--(6.477,6.609)--(6.558,6.610)--(6.638,6.611)--(6.719,6.611)--(6.799,6.611)%
  --(6.880,6.610)--(6.961,6.608)--(7.041,6.606)--(7.122,6.603)--(7.202,6.600)--(7.283,6.595)%
  --(7.363,6.590)--(7.444,6.585)--(7.524,6.578)--(7.605,6.571)--(7.685,6.563)--(7.766,6.554)%
  --(7.846,6.544)--(7.927,6.534)--(8.007,6.522)--(8.088,6.510)--(8.168,6.498)--(8.249,6.484)%
  --(8.330,6.470)--(8.410,6.455)--(8.491,6.440)--(8.571,6.423)--(8.652,6.406)--(8.732,6.389)%
  --(8.813,6.371)--(8.893,6.353)--(8.974,6.334)--(9.054,6.314)--(9.135,6.295)--(9.215,6.275)%
  --(9.296,6.255)--(9.376,6.235)--(9.457,6.214)--(9.538,6.194)--(9.618,6.174)--(9.699,6.154)%
  --(9.779,6.134)--(9.860,6.115)--(9.940,6.096)--(10.021,6.077)--(10.101,6.060)--(10.182,6.043)%
  --(10.262,6.027)--(10.343,6.012)--(10.423,5.998)--(10.504,5.986)--(10.584,5.975)--(10.665,5.966)%
  --(10.746,5.958)--(10.826,5.952)--(10.907,5.947)--(10.987,5.946)--(11.068,5.947);
\gpcolor{color=gp lt color border}
\node[gp node left] at (3.156,1.656) {tree + full 1-loop + leading 2-loop};
\gpcolor{rgb color={0.000,1.000,0.000}}
\gpsetdashtype{gp dt solid}
\draw[gp path] (2.056,1.656)--(2.972,1.656);
\draw[gp path] (3.095,7.229)--(3.175,7.189)--(3.256,7.151)--(3.337,7.115)--(3.417,7.082)%
  --(3.498,7.052)--(3.578,7.023)--(3.659,6.997)--(3.739,6.972)--(3.820,6.950)--(3.900,6.929)%
  --(3.981,6.910)--(4.061,6.893)--(4.142,6.877)--(4.222,6.863)--(4.303,6.851)--(4.383,6.841)%
  --(4.464,6.832)--(4.545,6.824)--(4.625,6.817)--(4.706,6.812)--(4.786,6.807)--(4.867,6.804)%
  --(4.947,6.801)--(5.028,6.799)--(5.108,6.798)--(5.189,6.798)--(5.269,6.798)--(5.350,6.799)%
  --(5.430,6.800)--(5.511,6.801)--(5.591,6.803)--(5.672,6.805)--(5.753,6.807)--(5.833,6.809)%
  --(5.914,6.812)--(5.994,6.814)--(6.075,6.816)--(6.155,6.818)--(6.236,6.820)--(6.316,6.822)%
  --(6.397,6.823)--(6.477,6.824)--(6.558,6.825)--(6.638,6.825)--(6.719,6.825)--(6.799,6.824)%
  --(6.880,6.823)--(6.961,6.821)--(7.041,6.819)--(7.122,6.816)--(7.202,6.812)--(7.283,6.808)%
  --(7.363,6.803)--(7.444,6.797)--(7.524,6.791)--(7.605,6.784)--(7.685,6.776)--(7.766,6.768)%
  --(7.846,6.758)--(7.927,6.749)--(8.007,6.738)--(8.088,6.727)--(8.168,6.715)--(8.249,6.702)%
  --(8.330,6.688)--(8.410,6.674)--(8.491,6.659)--(8.571,6.644)--(8.652,6.628)--(8.732,6.611)%
  --(8.813,6.594)--(8.893,6.576)--(8.974,6.558)--(9.054,6.540)--(9.135,6.520)--(9.215,6.501)%
  --(9.296,6.481)--(9.376,6.461)--(9.457,6.441)--(9.538,6.421)--(9.618,6.400)--(9.699,6.380)%
  --(9.779,6.359)--(9.860,6.339)--(9.940,6.319)--(10.021,6.299)--(10.101,6.279)--(10.182,6.260)%
  --(10.262,6.242)--(10.343,6.224)--(10.423,6.207)--(10.504,6.191)--(10.584,6.176)--(10.665,6.161)%
  --(10.746,6.148)--(10.826,6.136)--(10.907,6.125)--(10.987,6.116)--(11.068,6.110);
\gpcolor{color=gp lt color border}
\gpsetlinewidth{1.00}
\draw[gp path] (1.688,8.381)--(1.688,1.139)--(11.947,1.139)--(11.947,8.381)--cycle;
\gpdefrectangularnode{gp plot 1}{\pgfpoint{1.688cm}{1.139cm}}{\pgfpoint{11.947cm}{8.381cm}}
\end{tikzpicture}
}\\
  \resizebox{0.32\textwidth}{!}{\begin{tikzpicture}[gnuplot]
\path (0.000,0.000) rectangle (12.500,8.750);
\gpcolor{color=gp lt color border}
\gpsetlinetype{gp lt border}
\gpsetdashtype{gp dt solid}
\gpsetlinewidth{1.00}
\draw[gp path] (1.688,1.139)--(1.868,1.139);
\draw[gp path] (11.947,1.139)--(11.767,1.139);
\node[gp node right] at (1.504,1.139) {$0$};
\draw[gp path] (1.688,2.950)--(1.868,2.950);
\draw[gp path] (11.947,2.950)--(11.767,2.950);
\node[gp node right] at (1.504,2.950) {$50$};
\draw[gp path] (1.688,4.760)--(1.868,4.760);
\draw[gp path] (11.947,4.760)--(11.767,4.760);
\node[gp node right] at (1.504,4.760) {$100$};
\draw[gp path] (1.688,6.571)--(1.868,6.571);
\draw[gp path] (11.947,6.571)--(11.767,6.571);
\node[gp node right] at (1.504,6.571) {$150$};
\draw[gp path] (1.688,8.381)--(1.868,8.381);
\draw[gp path] (11.947,8.381)--(11.767,8.381);
\node[gp node right] at (1.504,8.381) {$200$};
\draw[gp path] (1.688,1.139)--(1.688,1.319);
\draw[gp path] (1.688,8.381)--(1.688,8.201);
\node[gp node center] at (1.688,0.831) {$-2$};
\draw[gp path] (2.970,1.139)--(2.970,1.319);
\draw[gp path] (2.970,8.381)--(2.970,8.201);
\node[gp node center] at (2.970,0.831) {$-1.5$};
\draw[gp path] (4.253,1.139)--(4.253,1.319);
\draw[gp path] (4.253,8.381)--(4.253,8.201);
\node[gp node center] at (4.253,0.831) {$-1$};
\draw[gp path] (5.535,1.139)--(5.535,1.319);
\draw[gp path] (5.535,8.381)--(5.535,8.201);
\node[gp node center] at (5.535,0.831) {$-0.5$};
\draw[gp path] (6.818,1.139)--(6.818,1.319);
\draw[gp path] (6.818,8.381)--(6.818,8.201);
\node[gp node center] at (6.818,0.831) {$0$};
\draw[gp path] (8.100,1.139)--(8.100,1.319);
\draw[gp path] (8.100,8.381)--(8.100,8.201);
\node[gp node center] at (8.100,0.831) {$0.5$};
\draw[gp path] (9.382,1.139)--(9.382,1.319);
\draw[gp path] (9.382,8.381)--(9.382,8.201);
\node[gp node center] at (9.382,0.831) {$1$};
\draw[gp path] (10.665,1.139)--(10.665,1.319);
\draw[gp path] (10.665,8.381)--(10.665,8.201);
\node[gp node center] at (10.665,0.831) {$1.5$};
\draw[gp path] (11.947,1.139)--(11.947,1.319);
\draw[gp path] (11.947,8.381)--(11.947,8.201);
\node[gp node center] at (11.947,0.831) {$2$};
\draw[gp path] (1.688,8.381)--(1.688,1.139)--(11.947,1.139)--(11.947,8.381)--cycle;
\node[gp node left] at (10.152,7.657) {BMP1};
\node[gp node center,rotate=-270] at (0.246,4.760) {$m_h$ [GeV]};
\node[gp node center] at (6.817,0.215) {$\Lambda_u$};
\node[gp node left] at (3.156,3.006) {tree};
\gpcolor{rgb color={0.580,0.000,0.827}}
\gpsetdashtype{gp dt 2}
\gpsetlinewidth{3.00}
\draw[gp path] (2.056,3.006)--(2.972,3.006);
\draw[gp path] (1.883,3.439)--(1.977,3.456)--(2.071,3.473)--(2.165,3.488)--(2.258,3.503)%
  --(2.352,3.518)--(2.446,3.532)--(2.540,3.545)--(2.634,3.557)--(2.728,3.569)--(2.822,3.581)%
  --(2.916,3.591)--(3.010,3.601)--(3.103,3.611)--(3.197,3.620)--(3.291,3.628)--(3.385,3.636)%
  --(3.479,3.644)--(3.573,3.650)--(3.667,3.657)--(3.761,3.662)--(3.855,3.667)--(3.948,3.672)%
  --(4.042,3.676)--(4.136,3.679)--(4.230,3.682)--(4.324,3.684)--(4.418,3.686)--(4.512,3.688)%
  --(4.606,3.688)--(4.699,3.689)--(4.793,3.688)--(4.887,3.687)--(4.981,3.686)--(5.075,3.684)%
  --(5.169,3.682)--(5.263,3.679)--(5.357,3.675)--(5.451,3.671)--(5.544,3.666)--(5.638,3.661)%
  --(5.732,3.655)--(5.826,3.649)--(5.920,3.642)--(6.014,3.634)--(6.108,3.626)--(6.202,3.618)%
  --(6.296,3.608)--(6.389,3.599)--(6.483,3.588)--(6.577,3.577)--(6.671,3.565)--(6.765,3.553)%
  --(6.859,3.540)--(6.953,3.527)--(7.047,3.512)--(7.141,3.498)--(7.234,3.482)--(7.328,3.466)%
  --(7.422,3.449)--(7.516,3.431)--(7.610,3.413)--(7.704,3.393)--(7.798,3.374)--(7.892,3.353)%
  --(7.985,3.331)--(8.079,3.309)--(8.173,3.286)--(8.267,3.261)--(8.361,3.236)--(8.455,3.210)%
  --(8.549,3.183)--(8.643,3.155)--(8.737,3.126)--(8.830,3.096)--(8.924,3.065)--(9.018,3.032)%
  --(9.112,2.998)--(9.206,2.963)--(9.300,2.927)--(9.394,2.889)--(9.488,2.849)--(9.582,2.807)%
  --(9.675,2.764)--(9.769,2.719)--(9.863,2.671)--(9.957,2.622)--(10.051,2.569)--(10.145,2.514)%
  --(10.239,2.455)--(10.333,2.393)--(10.426,2.326)--(10.520,2.254)--(10.614,2.176)--(10.708,2.091)%
  --(10.802,1.995)--(10.896,1.886)--(10.990,1.755)--(11.084,1.586)--(11.178,1.271);
\gpcolor{color=gp lt color border}
\node[gp node left] at (3.156,2.331) {tree + full 1-loop};
\gpcolor{rgb color={0.000,0.000,1.000}}
\gpsetdashtype{gp dt 4}
\draw[gp path] (2.056,2.331)--(2.972,2.331);
\draw[gp path] (1.883,8.134)--(1.977,7.968)--(2.071,7.807)--(2.165,7.650)--(2.258,7.496)%
  --(2.352,7.347)--(2.446,7.202)--(2.540,7.061)--(2.634,6.925)--(2.728,6.794)--(2.822,6.666)%
  --(2.916,6.544)--(3.010,6.426)--(3.103,6.313)--(3.197,6.205)--(3.291,6.102)--(3.385,6.003)%
  --(3.479,5.910)--(3.573,5.821)--(3.667,5.737)--(3.761,5.658)--(3.855,5.583)--(3.948,5.513)%
  --(4.042,5.447)--(4.136,5.385)--(4.230,5.327)--(4.324,5.274)--(4.418,5.224)--(4.512,5.178)%
  --(4.606,5.136)--(4.699,5.097)--(4.793,5.061)--(4.887,5.029)--(4.981,4.999)--(5.075,4.971)%
  --(5.169,4.946)--(5.263,4.924)--(5.357,4.903)--(5.451,4.884)--(5.544,4.866)--(5.638,4.850)%
  --(5.732,4.835)--(5.826,4.821)--(5.920,4.808)--(6.014,4.795)--(6.108,4.784)--(6.202,4.772)%
  --(6.296,4.761)--(6.389,4.749)--(6.483,4.738)--(6.577,4.727)--(6.671,4.716)--(6.765,4.704)%
  --(6.859,4.692)--(6.953,4.680)--(7.047,4.668)--(7.141,4.655)--(7.234,4.642)--(7.328,4.628)%
  --(7.422,4.615)--(7.516,4.601)--(7.610,4.587)--(7.704,4.573)--(7.798,4.559)--(7.892,4.546)%
  --(7.985,4.532)--(8.079,4.520)--(8.173,4.508)--(8.267,4.496)--(8.361,4.486)--(8.455,4.478)%
  --(8.549,4.470)--(8.643,4.465)--(8.737,4.462)--(8.830,4.461)--(8.924,4.462)--(9.018,4.467)%
  --(9.112,4.475)--(9.206,4.486)--(9.300,4.502)--(9.394,4.521)--(9.488,4.545)--(9.582,4.573)%
  --(9.675,4.607)--(9.769,4.645)--(9.863,4.689)--(9.957,4.739)--(10.051,4.794)--(10.145,4.854)%
  --(10.239,4.921)--(10.333,4.993)--(10.426,5.072)--(10.520,5.157)--(10.614,5.247)--(10.708,5.344)%
  --(10.802,5.446)--(10.896,5.554)--(10.990,5.669)--(11.084,5.789)--(11.178,5.916);
\gpcolor{color=gp lt color border}
\node[gp node left] at (3.156,1.656) {tree + full 1-loop + leading 2-loop};
\gpcolor{rgb color={0.000,1.000,0.000}}
\gpsetdashtype{gp dt solid}
\draw[gp path] (2.056,1.656)--(2.972,1.656);
\draw[gp path] (1.890,8.381)--(1.977,8.221)--(2.071,8.051)--(2.165,7.886)--(2.258,7.726)%
  --(2.352,7.571)--(2.446,7.420)--(2.540,7.274)--(2.634,7.133)--(2.728,6.997)--(2.822,6.866)%
  --(2.916,6.741)--(3.010,6.619)--(3.103,6.504)--(3.197,6.393)--(3.291,6.288)--(3.385,6.188)%
  --(3.479,6.093)--(3.573,6.003)--(3.667,5.918)--(3.761,5.839)--(3.855,5.763)--(3.948,5.693)%
  --(4.042,5.627)--(4.136,5.566)--(4.230,5.508)--(4.324,5.455)--(4.418,5.406)--(4.512,5.361)%
  --(4.606,5.319)--(4.699,5.281)--(4.793,5.246)--(4.887,5.214)--(4.981,5.185)--(5.075,5.158)%
  --(5.169,5.134)--(5.263,5.112)--(5.357,5.091)--(5.451,5.073)--(5.544,5.056)--(5.638,5.041)%
  --(5.732,5.026)--(5.826,5.013)--(5.920,5.001)--(6.014,4.989)--(6.108,4.977)--(6.202,4.966)%
  --(6.296,4.956)--(6.389,4.945)--(6.483,4.935)--(6.577,4.924)--(6.671,4.913)--(6.765,4.902)%
  --(6.859,4.891)--(6.953,4.880)--(7.047,4.868)--(7.141,4.856)--(7.234,4.843)--(7.328,4.831)%
  --(7.422,4.818)--(7.516,4.805)--(7.610,4.792)--(7.704,4.778)--(7.798,4.765)--(7.892,4.752)%
  --(7.985,4.740)--(8.079,4.728)--(8.173,4.716)--(8.267,4.706)--(8.361,4.696)--(8.455,4.688)%
  --(8.549,4.681)--(8.643,4.676)--(8.737,4.673)--(8.830,4.672)--(8.924,4.674)--(9.018,4.679)%
  --(9.112,4.686)--(9.206,4.697)--(9.300,4.712)--(9.394,4.731)--(9.488,4.754)--(9.582,4.782)%
  --(9.675,4.815)--(9.769,4.852)--(9.863,4.895)--(9.957,4.943)--(10.051,4.997)--(10.145,5.057)%
  --(10.239,5.122)--(10.333,5.193)--(10.426,5.271)--(10.520,5.354)--(10.614,5.443)--(10.708,5.539)%
  --(10.802,5.640)--(10.896,5.748)--(10.990,5.862)--(11.084,5.981)--(11.178,6.108);
\gpcolor{color=gp lt color border}
\gpsetlinewidth{1.00}
\draw[gp path] (1.688,8.381)--(1.688,1.139)--(11.947,1.139)--(11.947,8.381)--cycle;
\gpdefrectangularnode{gp plot 1}{\pgfpoint{1.688cm}{1.139cm}}{\pgfpoint{11.947cm}{8.381cm}}
\end{tikzpicture}
}
  \resizebox{0.32\textwidth}{!}{\begin{tikzpicture}[gnuplot]
\path (0.000,0.000) rectangle (12.500,8.750);
\gpcolor{color=gp lt color border}
\gpsetlinetype{gp lt border}
\gpsetdashtype{gp dt solid}
\gpsetlinewidth{1.00}
\draw[gp path] (1.688,1.139)--(1.868,1.139);
\draw[gp path] (11.947,1.139)--(11.767,1.139);
\node[gp node right] at (1.504,1.139) {$0$};
\draw[gp path] (1.688,2.950)--(1.868,2.950);
\draw[gp path] (11.947,2.950)--(11.767,2.950);
\node[gp node right] at (1.504,2.950) {$50$};
\draw[gp path] (1.688,4.760)--(1.868,4.760);
\draw[gp path] (11.947,4.760)--(11.767,4.760);
\node[gp node right] at (1.504,4.760) {$100$};
\draw[gp path] (1.688,6.571)--(1.868,6.571);
\draw[gp path] (11.947,6.571)--(11.767,6.571);
\node[gp node right] at (1.504,6.571) {$150$};
\draw[gp path] (1.688,8.381)--(1.868,8.381);
\draw[gp path] (11.947,8.381)--(11.767,8.381);
\node[gp node right] at (1.504,8.381) {$200$};
\draw[gp path] (1.688,1.139)--(1.688,1.319);
\draw[gp path] (1.688,8.381)--(1.688,8.201);
\node[gp node center] at (1.688,0.831) {$-2$};
\draw[gp path] (2.970,1.139)--(2.970,1.319);
\draw[gp path] (2.970,8.381)--(2.970,8.201);
\node[gp node center] at (2.970,0.831) {$-1.5$};
\draw[gp path] (4.253,1.139)--(4.253,1.319);
\draw[gp path] (4.253,8.381)--(4.253,8.201);
\node[gp node center] at (4.253,0.831) {$-1$};
\draw[gp path] (5.535,1.139)--(5.535,1.319);
\draw[gp path] (5.535,8.381)--(5.535,8.201);
\node[gp node center] at (5.535,0.831) {$-0.5$};
\draw[gp path] (6.818,1.139)--(6.818,1.319);
\draw[gp path] (6.818,8.381)--(6.818,8.201);
\node[gp node center] at (6.818,0.831) {$0$};
\draw[gp path] (8.100,1.139)--(8.100,1.319);
\draw[gp path] (8.100,8.381)--(8.100,8.201);
\node[gp node center] at (8.100,0.831) {$0.5$};
\draw[gp path] (9.382,1.139)--(9.382,1.319);
\draw[gp path] (9.382,8.381)--(9.382,8.201);
\node[gp node center] at (9.382,0.831) {$1$};
\draw[gp path] (10.665,1.139)--(10.665,1.319);
\draw[gp path] (10.665,8.381)--(10.665,8.201);
\node[gp node center] at (10.665,0.831) {$1.5$};
\draw[gp path] (11.947,1.139)--(11.947,1.319);
\draw[gp path] (11.947,8.381)--(11.947,8.201);
\node[gp node center] at (11.947,0.831) {$2$};
\draw[gp path] (1.688,8.381)--(1.688,1.139)--(11.947,1.139)--(11.947,8.381)--cycle;
\node[gp node left] at (10.152,7.657) {BMP2};
\node[gp node center,rotate=-270] at (0.246,4.760) {$m_h$ [GeV]};
\node[gp node center] at (6.817,0.215) {$\Lambda_u$};
\node[gp node left] at (3.156,3.006) {tree};
\gpcolor{rgb color={0.580,0.000,0.827}}
\gpsetdashtype{gp dt 2}
\gpsetlinewidth{3.00}
\draw[gp path] (2.056,3.006)--(2.972,3.006);
\draw[gp path] (2.109,4.063)--(2.204,4.078)--(2.299,4.093)--(2.394,4.107)--(2.489,4.120)%
  --(2.584,4.133)--(2.679,4.145)--(2.775,4.157)--(2.870,4.168)--(2.965,4.179)--(3.060,4.189)%
  --(3.155,4.198)--(3.250,4.207)--(3.345,4.216)--(3.440,4.223)--(3.536,4.231)--(3.631,4.237)%
  --(3.726,4.243)--(3.821,4.249)--(3.916,4.254)--(4.011,4.258)--(4.106,4.262)--(4.201,4.265)%
  --(4.297,4.268)--(4.392,4.271)--(4.487,4.272)--(4.582,4.273)--(4.677,4.274)--(4.772,4.274)%
  --(4.867,4.274)--(4.962,4.273)--(5.058,4.271)--(5.153,4.269)--(5.248,4.266)--(5.343,4.263)%
  --(5.438,4.259)--(5.533,4.255)--(5.628,4.250)--(5.724,4.245)--(5.819,4.239)--(5.914,4.232)%
  --(6.009,4.225)--(6.104,4.218)--(6.199,4.209)--(6.294,4.200)--(6.389,4.191)--(6.485,4.181)%
  --(6.580,4.170)--(6.675,4.159)--(6.770,4.147)--(6.865,4.135)--(6.960,4.122)--(7.055,4.108)%
  --(7.150,4.094)--(7.246,4.079)--(7.341,4.063)--(7.436,4.047)--(7.531,4.030)--(7.626,4.012)%
  --(7.721,3.994)--(7.816,3.975)--(7.911,3.955)--(8.007,3.934)--(8.102,3.913)--(8.197,3.891)%
  --(8.292,3.868)--(8.387,3.845)--(8.482,3.820)--(8.577,3.795)--(8.673,3.768)--(8.768,3.741)%
  --(8.863,3.713)--(8.958,3.684)--(9.053,3.654)--(9.148,3.623)--(9.243,3.591)--(9.338,3.558)%
  --(9.434,3.524)--(9.529,3.489)--(9.624,3.452)--(9.719,3.414)--(9.814,3.375)--(9.909,3.334)%
  --(10.004,3.292)--(10.099,3.248)--(10.195,3.202)--(10.290,3.155)--(10.385,3.106)--(10.480,3.055)%
  --(10.575,3.001)--(10.670,2.945)--(10.765,2.887)--(10.860,2.825)--(10.956,2.761)--(11.051,2.692)%
  --(11.146,2.620)--(11.241,2.543)--(11.336,2.460)--(11.431,2.370)--(11.526,2.272);
\gpcolor{color=gp lt color border}
\node[gp node left] at (3.156,2.331) {tree + full 1-loop};
\gpcolor{rgb color={0.000,0.000,1.000}}
\gpsetdashtype{gp dt 4}
\draw[gp path] (2.056,2.331)--(2.972,2.331);
\draw[gp path] (2.109,7.836)--(2.204,7.693)--(2.299,7.553)--(2.394,7.418)--(2.489,7.287)%
  --(2.584,7.160)--(2.679,7.038)--(2.775,6.921)--(2.870,6.808)--(2.965,6.700)--(3.060,6.596)%
  --(3.155,6.496)--(3.250,6.401)--(3.345,6.311)--(3.440,6.226)--(3.536,6.144)--(3.631,6.068)%
  --(3.726,5.995)--(3.821,5.927)--(3.916,5.863)--(4.011,5.803)--(4.106,5.747)--(4.201,5.695)%
  --(4.297,5.647)--(4.392,5.602)--(4.487,5.561)--(4.582,5.523)--(4.677,5.488)--(4.772,5.456)%
  --(4.867,5.426)--(4.962,5.400)--(5.058,5.375)--(5.153,5.353)--(5.248,5.332)--(5.343,5.314)%
  --(5.438,5.297)--(5.533,5.281)--(5.628,5.267)--(5.724,5.253)--(5.819,5.241)--(5.914,5.229)%
  --(6.009,5.218)--(6.104,5.207)--(6.199,5.196)--(6.294,5.186)--(6.389,5.176)--(6.485,5.165)%
  --(6.580,5.155)--(6.675,5.144)--(6.770,5.133)--(6.865,5.122)--(6.960,5.111)--(7.055,5.099)%
  --(7.150,5.087)--(7.246,5.074)--(7.341,5.061)--(7.436,5.047)--(7.531,5.034)--(7.626,5.020)%
  --(7.721,5.006)--(7.816,4.991)--(7.911,4.977)--(8.007,4.963)--(8.102,4.949)--(8.197,4.935)%
  --(8.292,4.922)--(8.387,4.910)--(8.482,4.899)--(8.577,4.888)--(8.673,4.879)--(8.768,4.872)%
  --(8.863,4.866)--(8.958,4.862)--(9.053,4.861)--(9.148,4.862)--(9.243,4.866)--(9.338,4.873)%
  --(9.434,4.884)--(9.529,4.898)--(9.624,4.916)--(9.719,4.938)--(9.814,4.964)--(9.909,4.995)%
  --(10.004,5.031)--(10.099,5.072)--(10.195,5.118)--(10.290,5.169)--(10.385,5.226)--(10.480,5.288)%
  --(10.575,5.356)--(10.670,5.430)--(10.765,5.510)--(10.860,5.595)--(10.956,5.686)--(11.051,5.782)%
  --(11.146,5.885)--(11.241,5.993)--(11.336,6.106)--(11.431,6.225)--(11.526,6.349);
\gpcolor{color=gp lt color border}
\node[gp node left] at (3.156,1.656) {tree + full 1-loop + leading 2-loop};
\gpcolor{rgb color={0.000,1.000,0.000}}
\gpsetdashtype{gp dt solid}
\draw[gp path] (2.056,1.656)--(2.972,1.656);
\draw[gp path] (2.109,8.233)--(2.204,8.071)--(2.299,7.916)--(2.394,7.765)--(2.489,7.619)%
  --(2.584,7.479)--(2.679,7.343)--(2.775,7.213)--(2.870,7.089)--(2.965,6.971)--(3.060,6.858)%
  --(3.155,6.750)--(3.250,6.646)--(3.345,6.549)--(3.440,6.456)--(3.536,6.369)--(3.631,6.286)%
  --(3.726,6.209)--(3.821,6.136)--(3.916,6.068)--(4.011,6.005)--(4.106,5.946)--(4.201,5.891)%
  --(4.297,5.841)--(4.392,5.794)--(4.487,5.751)--(4.582,5.711)--(4.677,5.675)--(4.772,5.642)%
  --(4.867,5.611)--(4.962,5.584)--(5.058,5.559)--(5.153,5.536)--(5.248,5.515)--(5.343,5.496)%
  --(5.438,5.479)--(5.533,5.464)--(5.628,5.449)--(5.724,5.436)--(5.819,5.424)--(5.914,5.412)%
  --(6.009,5.401)--(6.104,5.390)--(6.199,5.380)--(6.294,5.370)--(6.389,5.360)--(6.485,5.350)%
  --(6.580,5.340)--(6.675,5.330)--(6.770,5.320)--(6.865,5.309)--(6.960,5.298)--(7.055,5.287)%
  --(7.150,5.275)--(7.246,5.263)--(7.341,5.251)--(7.436,5.238)--(7.531,5.225)--(7.626,5.212)%
  --(7.721,5.199)--(7.816,5.185)--(7.911,5.172)--(8.007,5.159)--(8.102,5.146)--(8.197,5.134)%
  --(8.292,5.122)--(8.387,5.111)--(8.482,5.101)--(8.577,5.092)--(8.673,5.085)--(8.768,5.080)%
  --(8.863,5.076)--(8.958,5.075)--(9.053,5.076)--(9.148,5.079)--(9.243,5.086)--(9.338,5.097)%
  --(9.434,5.110)--(9.529,5.128)--(9.624,5.150)--(9.719,5.176)--(9.814,5.208)--(9.909,5.244)%
  --(10.004,5.285)--(10.099,5.332)--(10.195,5.385)--(10.290,5.443)--(10.385,5.508)--(10.480,5.578)%
  --(10.575,5.655)--(10.670,5.739)--(10.765,5.828)--(10.860,5.924)--(10.956,6.026)--(11.051,6.135)%
  --(11.146,6.250)--(11.241,6.372)--(11.336,6.499)--(11.431,6.634)--(11.526,6.774);
\gpcolor{color=gp lt color border}
\gpsetlinewidth{1.00}
\draw[gp path] (1.688,8.381)--(1.688,1.139)--(11.947,1.139)--(11.947,8.381)--cycle;
\gpdefrectangularnode{gp plot 1}{\pgfpoint{1.688cm}{1.139cm}}{\pgfpoint{11.947cm}{8.381cm}}
\end{tikzpicture}
}
  \resizebox{0.32\textwidth}{!}{\begin{tikzpicture}[gnuplot]
\path (0.000,0.000) rectangle (12.500,8.750);
\gpcolor{color=gp lt color border}
\gpsetlinetype{gp lt border}
\gpsetdashtype{gp dt solid}
\gpsetlinewidth{1.00}
\draw[gp path] (1.688,1.139)--(1.868,1.139);
\draw[gp path] (11.947,1.139)--(11.767,1.139);
\node[gp node right] at (1.504,1.139) {$0$};
\draw[gp path] (1.688,2.950)--(1.868,2.950);
\draw[gp path] (11.947,2.950)--(11.767,2.950);
\node[gp node right] at (1.504,2.950) {$50$};
\draw[gp path] (1.688,4.760)--(1.868,4.760);
\draw[gp path] (11.947,4.760)--(11.767,4.760);
\node[gp node right] at (1.504,4.760) {$100$};
\draw[gp path] (1.688,6.571)--(1.868,6.571);
\draw[gp path] (11.947,6.571)--(11.767,6.571);
\node[gp node right] at (1.504,6.571) {$150$};
\draw[gp path] (1.688,8.381)--(1.868,8.381);
\draw[gp path] (11.947,8.381)--(11.767,8.381);
\node[gp node right] at (1.504,8.381) {$200$};
\draw[gp path] (1.688,1.139)--(1.688,1.319);
\draw[gp path] (1.688,8.381)--(1.688,8.201);
\node[gp node center] at (1.688,0.831) {$-2$};
\draw[gp path] (2.970,1.139)--(2.970,1.319);
\draw[gp path] (2.970,8.381)--(2.970,8.201);
\node[gp node center] at (2.970,0.831) {$-1.5$};
\draw[gp path] (4.253,1.139)--(4.253,1.319);
\draw[gp path] (4.253,8.381)--(4.253,8.201);
\node[gp node center] at (4.253,0.831) {$-1$};
\draw[gp path] (5.535,1.139)--(5.535,1.319);
\draw[gp path] (5.535,8.381)--(5.535,8.201);
\node[gp node center] at (5.535,0.831) {$-0.5$};
\draw[gp path] (6.818,1.139)--(6.818,1.319);
\draw[gp path] (6.818,8.381)--(6.818,8.201);
\node[gp node center] at (6.818,0.831) {$0$};
\draw[gp path] (8.100,1.139)--(8.100,1.319);
\draw[gp path] (8.100,8.381)--(8.100,8.201);
\node[gp node center] at (8.100,0.831) {$0.5$};
\draw[gp path] (9.382,1.139)--(9.382,1.319);
\draw[gp path] (9.382,8.381)--(9.382,8.201);
\node[gp node center] at (9.382,0.831) {$1$};
\draw[gp path] (10.665,1.139)--(10.665,1.319);
\draw[gp path] (10.665,8.381)--(10.665,8.201);
\node[gp node center] at (10.665,0.831) {$1.5$};
\draw[gp path] (11.947,1.139)--(11.947,1.319);
\draw[gp path] (11.947,8.381)--(11.947,8.201);
\node[gp node center] at (11.947,0.831) {$2$};
\draw[gp path] (1.688,8.381)--(1.688,1.139)--(11.947,1.139)--(11.947,8.381)--cycle;
\node[gp node left] at (10.152,7.657) {BMP3};
\node[gp node center,rotate=-270] at (0.246,4.760) {$m_h$ [GeV]};
\node[gp node center] at (6.817,0.215) {$\Lambda_u$};
\node[gp node left] at (3.156,3.006) {tree};
\gpcolor{rgb color={0.580,0.000,0.827}}
\gpsetdashtype{gp dt 2}
\gpsetlinewidth{3.00}
\draw[gp path] (2.056,3.006)--(2.972,3.006);
\draw[gp path] (1.688,4.153)--(1.792,4.172)--(1.895,4.191)--(1.999,4.209)--(2.103,4.226)%
  --(2.206,4.242)--(2.310,4.258)--(2.413,4.273)--(2.517,4.287)--(2.621,4.301)--(2.724,4.314)%
  --(2.828,4.326)--(2.932,4.338)--(3.035,4.349)--(3.139,4.359)--(3.242,4.369)--(3.346,4.378)%
  --(3.450,4.386)--(3.553,4.394)--(3.657,4.401)--(3.761,4.407)--(3.864,4.413)--(3.968,4.418)%
  --(4.071,4.423)--(4.175,4.427)--(4.279,4.430)--(4.382,4.433)--(4.486,4.434)--(4.590,4.436)%
  --(4.693,4.437)--(4.797,4.437)--(4.900,4.436)--(5.004,4.435)--(5.108,4.433)--(5.211,4.431)%
  --(5.315,4.427)--(5.419,4.424)--(5.522,4.419)--(5.626,4.414)--(5.729,4.409)--(5.833,4.402)%
  --(5.937,4.395)--(6.040,4.388)--(6.144,4.379)--(6.248,4.370)--(6.351,4.361)--(6.455,4.350)%
  --(6.558,4.339)--(6.662,4.328)--(6.766,4.315)--(6.869,4.302)--(6.973,4.288)--(7.077,4.274)%
  --(7.180,4.258)--(7.284,4.242)--(7.387,4.225)--(7.491,4.208)--(7.595,4.189)--(7.698,4.170)%
  --(7.802,4.150)--(7.906,4.130)--(8.009,4.108)--(8.113,4.085)--(8.216,4.062)--(8.320,4.038)%
  --(8.424,4.013)--(8.527,3.987)--(8.631,3.960)--(8.735,3.932)--(8.838,3.903)--(8.942,3.872)%
  --(9.045,3.841)--(9.149,3.809)--(9.253,3.776)--(9.356,3.741)--(9.460,3.705)--(9.564,3.668)%
  --(9.667,3.630)--(9.771,3.590)--(9.874,3.549)--(9.978,3.506)--(10.082,3.461)--(10.185,3.415)%
  --(10.289,3.367)--(10.393,3.318)--(10.496,3.266)--(10.600,3.212)--(10.703,3.155)--(10.807,3.096)%
  --(10.911,3.035)--(11.014,2.970)--(11.118,2.902)--(11.222,2.829)--(11.325,2.753)--(11.429,2.672)%
  --(11.532,2.585)--(11.636,2.491)--(11.740,2.389)--(11.843,2.277)--(11.947,2.150);
\gpcolor{color=gp lt color border}
\node[gp node left] at (3.156,2.331) {tree + full 1-loop};
\gpcolor{rgb color={0.000,0.000,1.000}}
\gpsetdashtype{gp dt 4}
\draw[gp path] (2.056,2.331)--(2.972,2.331);
\draw[gp path] (1.688,8.185)--(1.792,8.011)--(1.895,7.842)--(1.999,7.678)--(2.103,7.519)%
  --(2.206,7.364)--(2.310,7.215)--(2.413,7.073)--(2.517,6.936)--(2.621,6.804)--(2.724,6.678)%
  --(2.828,6.557)--(2.932,6.442)--(3.035,6.334)--(3.139,6.231)--(3.242,6.134)--(3.346,6.042)%
  --(3.450,5.957)--(3.553,5.877)--(3.657,5.803)--(3.761,5.734)--(3.864,5.671)--(3.968,5.613)%
  --(4.071,5.559)--(4.175,5.511)--(4.279,5.467)--(4.382,5.428)--(4.486,5.393)--(4.590,5.361)%
  --(4.693,5.333)--(4.797,5.309)--(4.900,5.287)--(5.004,5.269)--(5.108,5.252)--(5.211,5.238)%
  --(5.315,5.226)--(5.419,5.215)--(5.522,5.205)--(5.626,5.197)--(5.729,5.189)--(5.833,5.182)%
  --(5.937,5.176)--(6.040,5.169)--(6.144,5.162)--(6.248,5.155)--(6.351,5.148)--(6.455,5.140)%
  --(6.558,5.132)--(6.662,5.123)--(6.766,5.112)--(6.869,5.101)--(6.973,5.089)--(7.077,5.076)%
  --(7.180,5.062)--(7.284,5.046)--(7.387,5.030)--(7.491,5.012)--(7.595,4.994)--(7.698,4.974)%
  --(7.802,4.954)--(7.906,4.934)--(8.009,4.912)--(8.113,4.891)--(8.216,4.869)--(8.320,4.847)%
  --(8.424,4.825)--(8.527,4.804)--(8.631,4.784)--(8.735,4.765)--(8.838,4.747)--(8.942,4.732)%
  --(9.045,4.718)--(9.149,4.707)--(9.253,4.699)--(9.356,4.695)--(9.460,4.694)--(9.564,4.697)%
  --(9.667,4.705)--(9.771,4.719)--(9.874,4.737)--(9.978,4.762)--(10.082,4.793)--(10.185,4.830)%
  --(10.289,4.874)--(10.393,4.926)--(10.496,4.984)--(10.600,5.049)--(10.703,5.122)--(10.807,5.203)%
  --(10.911,5.292)--(11.014,5.388)--(11.118,5.491)--(11.222,5.602)--(11.325,5.720)--(11.429,5.845)%
  --(11.532,5.977)--(11.636,6.116)--(11.740,6.261)--(11.843,6.413)--(11.947,6.572);
\gpcolor{color=gp lt color border}
\node[gp node left] at (3.156,1.656) {tree + full 1-loop + leading 2-loop};
\gpcolor{rgb color={0.000,1.000,0.000}}
\gpsetdashtype{gp dt solid}
\draw[gp path] (2.056,1.656)--(2.972,1.656);
\draw[gp path] (1.772,8.381)--(1.792,8.344)--(1.895,8.158)--(1.999,7.977)--(2.103,7.803)%
  --(2.206,7.635)--(2.310,7.473)--(2.413,7.318)--(2.517,7.170)--(2.621,7.030)--(2.724,6.896)%
  --(2.828,6.768)--(2.932,6.646)--(3.035,6.531)--(3.139,6.423)--(3.242,6.321)--(3.346,6.226)%
  --(3.450,6.138)--(3.553,6.055)--(3.657,5.979)--(3.761,5.908)--(3.864,5.843)--(3.968,5.784)%
  --(4.071,5.730)--(4.175,5.681)--(4.279,5.637)--(4.382,5.597)--(4.486,5.562)--(4.590,5.530)%
  --(4.693,5.502)--(4.797,5.478)--(4.900,5.456)--(5.004,5.438)--(5.108,5.422)--(5.211,5.408)%
  --(5.315,5.395)--(5.419,5.385)--(5.522,5.375)--(5.626,5.367)--(5.729,5.360)--(5.833,5.353)%
  --(5.937,5.346)--(6.040,5.340)--(6.144,5.334)--(6.248,5.327)--(6.351,5.320)--(6.455,5.312)%
  --(6.558,5.304)--(6.662,5.295)--(6.766,5.286)--(6.869,5.275)--(6.973,5.263)--(7.077,5.251)%
  --(7.180,5.237)--(7.284,5.223)--(7.387,5.207)--(7.491,5.190)--(7.595,5.173)--(7.698,5.155)%
  --(7.802,5.136)--(7.906,5.116)--(8.009,5.096)--(8.113,5.075)--(8.216,5.055)--(8.320,5.034)%
  --(8.424,5.014)--(8.527,4.995)--(8.631,4.976)--(8.735,4.959)--(8.838,4.943)--(8.942,4.929)%
  --(9.045,4.917)--(9.149,4.908)--(9.253,4.902)--(9.356,4.900)--(9.460,4.901)--(9.564,4.907)%
  --(9.667,4.918)--(9.771,4.934)--(9.874,4.956)--(9.978,4.984)--(10.082,5.018)--(10.185,5.059)%
  --(10.289,5.108)--(10.393,5.164)--(10.496,5.227)--(10.600,5.299)--(10.703,5.379)--(10.807,5.467)%
  --(10.911,5.564)--(11.014,5.669)--(11.118,5.782)--(11.222,5.904)--(11.325,6.034)--(11.429,6.172)%
  --(11.532,6.318)--(11.636,6.472)--(11.740,6.635)--(11.843,6.805)--(11.947,6.982);
\gpcolor{color=gp lt color border}
\gpsetlinewidth{1.00}
\draw[gp path] (1.688,8.381)--(1.688,1.139)--(11.947,1.139)--(11.947,8.381)--cycle;
\gpdefrectangularnode{gp plot 1}{\pgfpoint{1.688cm}{1.139cm}}{\pgfpoint{11.947cm}{8.381cm}}
\end{tikzpicture}
}\\
  \subfloat[]{\resizebox{0.32\textwidth}{!}{\begin{tikzpicture}[gnuplot]
\path (0.000,0.000) rectangle (12.500,8.750);
\gpcolor{color=gp lt color border}
\gpsetlinetype{gp lt border}
\gpsetdashtype{gp dt solid}
\gpsetlinewidth{1.00}
\draw[gp path] (1.688,1.139)--(1.868,1.139);
\draw[gp path] (11.947,1.139)--(11.767,1.139);
\node[gp node right] at (1.504,1.139) {$0$};
\draw[gp path] (1.688,2.174)--(1.868,2.174);
\draw[gp path] (11.947,2.174)--(11.767,2.174);
\node[gp node right] at (1.504,2.174) {$20$};
\draw[gp path] (1.688,3.208)--(1.868,3.208);
\draw[gp path] (11.947,3.208)--(11.767,3.208);
\node[gp node right] at (1.504,3.208) {$40$};
\draw[gp path] (1.688,4.243)--(1.868,4.243);
\draw[gp path] (11.947,4.243)--(11.767,4.243);
\node[gp node right] at (1.504,4.243) {$60$};
\draw[gp path] (1.688,5.277)--(1.868,5.277);
\draw[gp path] (11.947,5.277)--(11.767,5.277);
\node[gp node right] at (1.504,5.277) {$80$};
\draw[gp path] (1.688,6.312)--(1.868,6.312);
\draw[gp path] (11.947,6.312)--(11.767,6.312);
\node[gp node right] at (1.504,6.312) {$100$};
\draw[gp path] (1.688,7.346)--(1.868,7.346);
\draw[gp path] (11.947,7.346)--(11.767,7.346);
\node[gp node right] at (1.504,7.346) {$120$};
\draw[gp path] (1.688,8.381)--(1.868,8.381);
\draw[gp path] (11.947,8.381)--(11.767,8.381);
\node[gp node right] at (1.504,8.381) {$140$};
\draw[gp path] (1.688,1.139)--(1.688,1.319);
\draw[gp path] (1.688,8.381)--(1.688,8.201);
\node[gp node center] at (1.688,0.831) {$0$};
\draw[gp path] (2.970,1.139)--(2.970,1.319);
\draw[gp path] (2.970,8.381)--(2.970,8.201);
\node[gp node center] at (2.970,0.831) {$100$};
\draw[gp path] (4.253,1.139)--(4.253,1.319);
\draw[gp path] (4.253,8.381)--(4.253,8.201);
\node[gp node center] at (4.253,0.831) {$200$};
\draw[gp path] (5.535,1.139)--(5.535,1.319);
\draw[gp path] (5.535,8.381)--(5.535,8.201);
\node[gp node center] at (5.535,0.831) {$300$};
\draw[gp path] (6.818,1.139)--(6.818,1.319);
\draw[gp path] (6.818,8.381)--(6.818,8.201);
\node[gp node center] at (6.818,0.831) {$400$};
\draw[gp path] (8.100,1.139)--(8.100,1.319);
\draw[gp path] (8.100,8.381)--(8.100,8.201);
\node[gp node center] at (8.100,0.831) {$500$};
\draw[gp path] (9.382,1.139)--(9.382,1.319);
\draw[gp path] (9.382,8.381)--(9.382,8.201);
\node[gp node center] at (9.382,0.831) {$600$};
\draw[gp path] (10.665,1.139)--(10.665,1.319);
\draw[gp path] (10.665,8.381)--(10.665,8.201);
\node[gp node center] at (10.665,0.831) {$700$};
\draw[gp path] (11.947,1.139)--(11.947,1.319);
\draw[gp path] (11.947,8.381)--(11.947,8.201);
\node[gp node center] at (11.947,0.831) {$800$};
\draw[gp path] (1.688,8.381)--(1.688,1.139)--(11.947,1.139)--(11.947,8.381)--cycle;
\node[gp node left] at (10.088,7.709) {BMP1};
\node[gp node center,rotate=-270] at (0.246,4.760) {$m_h$ [GeV]};
\node[gp node center] at (6.817,0.215) {$\mu_u$ [GeV]};
\node[gp node left] at (3.156,3.006) {tree};
\gpcolor{rgb color={0.580,0.000,0.827}}
\gpsetdashtype{gp dt 2}
\gpsetlinewidth{3.00}
\draw[gp path] (2.056,3.006)--(2.972,3.006);
\draw[gp path] (1.688,4.487)--(1.792,4.515)--(1.895,4.541)--(1.999,4.566)--(2.103,4.590)%
  --(2.206,4.613)--(2.310,4.635)--(2.413,4.657)--(2.517,4.677)--(2.621,4.696)--(2.724,4.715)%
  --(2.828,4.732)--(2.932,4.748)--(3.035,4.764)--(3.139,4.779)--(3.242,4.793)--(3.346,4.805)%
  --(3.450,4.818)--(3.553,4.829)--(3.657,4.839)--(3.761,4.849)--(3.864,4.857)--(3.968,4.865)%
  --(4.071,4.872)--(4.175,4.878)--(4.279,4.884)--(4.382,4.888)--(4.486,4.892)--(4.590,4.895)%
  --(4.693,4.897)--(4.797,4.898)--(4.900,4.899)--(5.004,4.898)--(5.108,4.897)--(5.211,4.895)%
  --(5.315,4.892)--(5.419,4.889)--(5.522,4.884)--(5.626,4.879)--(5.729,4.873)--(5.833,4.866)%
  --(5.937,4.859)--(6.040,4.850)--(6.144,4.841)--(6.248,4.830)--(6.351,4.819)--(6.455,4.807)%
  --(6.558,4.795)--(6.662,4.781)--(6.766,4.766)--(6.869,4.751)--(6.973,4.734)--(7.077,4.717)%
  --(7.180,4.699)--(7.284,4.679)--(7.387,4.659)--(7.491,4.638)--(7.595,4.616)--(7.698,4.592)%
  --(7.802,4.568)--(7.906,4.543)--(8.009,4.516)--(8.113,4.489)--(8.216,4.460)--(8.320,4.430)%
  --(8.424,4.399)--(8.527,4.367)--(8.631,4.333)--(8.735,4.298)--(8.838,4.262)--(8.942,4.224)%
  --(9.045,4.185)--(9.149,4.144)--(9.253,4.101)--(9.356,4.057)--(9.460,4.012)--(9.564,3.964)%
  --(9.667,3.914)--(9.771,3.863)--(9.874,3.809)--(9.978,3.753)--(10.082,3.695)--(10.185,3.634)%
  --(10.289,3.570)--(10.393,3.504)--(10.496,3.434)--(10.600,3.360)--(10.703,3.282)--(10.807,3.201)%
  --(10.911,3.114)--(11.014,3.021)--(11.118,2.922)--(11.222,2.816)--(11.325,2.700)--(11.429,2.573)%
  --(11.532,2.431)--(11.636,2.268)--(11.740,2.075)--(11.843,1.826)--(11.947,1.384);
\gpcolor{color=gp lt color border}
\node[gp node left] at (3.156,2.331) {tree + full 1-loop};
\gpcolor{rgb color={0.000,0.000,1.000}}
\gpsetdashtype{gp dt 4}
\draw[gp path] (2.056,2.331)--(2.972,2.331);
\draw[gp path] (1.688,7.033)--(1.792,7.065)--(1.895,7.096)--(1.999,7.124)--(2.103,7.153)%
  --(2.206,7.180)--(2.310,7.205)--(2.413,7.229)--(2.517,7.252)--(2.621,7.273)--(2.724,7.293)%
  --(2.828,7.312)--(2.932,7.330)--(3.035,7.347)--(3.139,7.363)--(3.242,7.377)--(3.346,7.391)%
  --(3.450,7.403)--(3.553,7.414)--(3.657,7.425)--(3.761,7.434)--(3.864,7.443)--(3.968,7.451)%
  --(4.071,7.457)--(4.175,7.463)--(4.279,7.468)--(4.382,7.472)--(4.486,7.476)--(4.590,7.478)%
  --(4.693,7.480)--(4.797,7.481)--(4.900,7.482)--(5.004,7.481)--(5.108,7.480)--(5.211,7.478)%
  --(5.315,7.475)--(5.419,7.472)--(5.522,7.468)--(5.626,7.463)--(5.729,7.458)--(5.833,7.452)%
  --(5.937,7.446)--(6.040,7.438)--(6.144,7.430)--(6.248,7.422)--(6.351,7.413)--(6.455,7.403)%
  --(6.558,7.393)--(6.662,7.382)--(6.766,7.370)--(6.869,7.358)--(6.973,7.345)--(7.077,7.332)%
  --(7.180,7.318)--(7.284,7.304)--(7.387,7.288)--(7.491,7.273)--(7.595,7.256)--(7.698,7.240)%
  --(7.802,7.222)--(7.906,7.204)--(8.009,7.185)--(8.113,7.166)--(8.216,7.146)--(8.320,7.126)%
  --(8.424,7.105)--(8.527,7.083)--(8.631,7.061)--(8.735,7.038)--(8.838,7.015)--(8.942,6.991)%
  --(9.045,6.966)--(9.149,6.941)--(9.253,6.915)--(9.356,6.888)--(9.460,6.861)--(9.564,6.833)%
  --(9.667,6.805)--(9.771,6.776)--(9.874,6.746)--(9.978,6.716)--(10.082,6.685)--(10.185,6.653)%
  --(10.289,6.620)--(10.393,6.587)--(10.496,6.553)--(10.600,6.518)--(10.703,6.482)--(10.807,6.446)%
  --(10.911,6.409)--(11.014,6.371)--(11.118,6.332)--(11.222,6.293)--(11.325,6.252)--(11.429,6.211)%
  --(11.532,6.168)--(11.636,6.125)--(11.740,6.081)--(11.843,6.036)--(11.947,5.991);
\gpcolor{color=gp lt color border}
\node[gp node left] at (3.156,1.656) {tree + full 1-loop + leading 2-loop};
\gpcolor{rgb color={0.000,1.000,0.000}}
\gpsetdashtype{gp dt solid}
\draw[gp path] (2.056,1.656)--(2.972,1.656);
\draw[gp path] (1.688,7.294)--(1.792,7.325)--(1.895,7.354)--(1.999,7.381)--(2.103,7.408)%
  --(2.206,7.435)--(2.310,7.459)--(2.413,7.482)--(2.517,7.504)--(2.621,7.525)--(2.724,7.545)%
  --(2.828,7.563)--(2.932,7.581)--(3.035,7.597)--(3.139,7.613)--(3.242,7.627)--(3.346,7.640)%
  --(3.450,7.653)--(3.553,7.664)--(3.657,7.674)--(3.761,7.684)--(3.864,7.693)--(3.968,7.700)%
  --(4.071,7.707)--(4.175,7.714)--(4.279,7.719)--(4.382,7.723)--(4.486,7.727)--(4.590,7.730)%
  --(4.693,7.732)--(4.797,7.734)--(4.900,7.734)--(5.004,7.734)--(5.108,7.733)--(5.211,7.732)%
  --(5.315,7.730)--(5.419,7.727)--(5.522,7.723)--(5.626,7.719)--(5.729,7.714)--(5.833,7.708)%
  --(5.937,7.702)--(6.040,7.695)--(6.144,7.688)--(6.248,7.679)--(6.351,7.670)--(6.455,7.661)%
  --(6.558,7.650)--(6.662,7.639)--(6.766,7.628)--(6.869,7.616)--(6.973,7.603)--(7.077,7.589)%
  --(7.180,7.575)--(7.284,7.560)--(7.387,7.545)--(7.491,7.528)--(7.595,7.511)--(7.698,7.494)%
  --(7.802,7.476)--(7.906,7.457)--(8.009,7.437)--(8.113,7.417)--(8.216,7.396)--(8.320,7.374)%
  --(8.424,7.351)--(8.527,7.328)--(8.631,7.304)--(8.735,7.279)--(8.838,7.254)--(8.942,7.227)%
  --(9.045,7.201)--(9.149,7.173)--(9.253,7.144)--(9.356,7.115)--(9.460,7.085)--(9.564,7.054)%
  --(9.667,7.022)--(9.771,6.989)--(9.874,6.955)--(9.978,6.921)--(10.082,6.885)--(10.185,6.849)%
  --(10.289,6.811)--(10.393,6.772)--(10.496,6.733)--(10.600,6.692)--(10.703,6.650)--(10.807,6.607)%
  --(10.911,6.563)--(11.014,6.518)--(11.118,6.471)--(11.222,6.424)--(11.325,6.375)--(11.429,6.324)%
  --(11.532,6.272)--(11.636,6.219)--(11.740,6.165)--(11.843,6.109)--(11.947,6.052);
\gpcolor{color=gp lt color border}
\gpsetlinewidth{1.00}
\draw[gp path] (1.688,8.381)--(1.688,1.139)--(11.947,1.139)--(11.947,8.381)--cycle;
\gpdefrectangularnode{gp plot 1}{\pgfpoint{1.688cm}{1.139cm}}{\pgfpoint{11.947cm}{8.381cm}}
\end{tikzpicture}
}}
  \subfloat[]{\resizebox{0.32\textwidth}{!}{\begin{tikzpicture}[gnuplot]
\path (0.000,0.000) rectangle (12.500,8.750);
\gpcolor{color=gp lt color border}
\gpsetlinetype{gp lt border}
\gpsetdashtype{gp dt solid}
\gpsetlinewidth{1.00}
\draw[gp path] (1.688,1.139)--(1.868,1.139);
\draw[gp path] (11.947,1.139)--(11.767,1.139);
\node[gp node right] at (1.504,1.139) {$0$};
\draw[gp path] (1.688,2.174)--(1.868,2.174);
\draw[gp path] (11.947,2.174)--(11.767,2.174);
\node[gp node right] at (1.504,2.174) {$20$};
\draw[gp path] (1.688,3.208)--(1.868,3.208);
\draw[gp path] (11.947,3.208)--(11.767,3.208);
\node[gp node right] at (1.504,3.208) {$40$};
\draw[gp path] (1.688,4.243)--(1.868,4.243);
\draw[gp path] (11.947,4.243)--(11.767,4.243);
\node[gp node right] at (1.504,4.243) {$60$};
\draw[gp path] (1.688,5.277)--(1.868,5.277);
\draw[gp path] (11.947,5.277)--(11.767,5.277);
\node[gp node right] at (1.504,5.277) {$80$};
\draw[gp path] (1.688,6.312)--(1.868,6.312);
\draw[gp path] (11.947,6.312)--(11.767,6.312);
\node[gp node right] at (1.504,6.312) {$100$};
\draw[gp path] (1.688,7.346)--(1.868,7.346);
\draw[gp path] (11.947,7.346)--(11.767,7.346);
\node[gp node right] at (1.504,7.346) {$120$};
\draw[gp path] (1.688,8.381)--(1.868,8.381);
\draw[gp path] (11.947,8.381)--(11.767,8.381);
\node[gp node right] at (1.504,8.381) {$140$};
\draw[gp path] (1.688,1.139)--(1.688,1.319);
\draw[gp path] (1.688,8.381)--(1.688,8.201);
\node[gp node center] at (1.688,0.831) {$0$};
\draw[gp path] (2.970,1.139)--(2.970,1.319);
\draw[gp path] (2.970,8.381)--(2.970,8.201);
\node[gp node center] at (2.970,0.831) {$100$};
\draw[gp path] (4.253,1.139)--(4.253,1.319);
\draw[gp path] (4.253,8.381)--(4.253,8.201);
\node[gp node center] at (4.253,0.831) {$200$};
\draw[gp path] (5.535,1.139)--(5.535,1.319);
\draw[gp path] (5.535,8.381)--(5.535,8.201);
\node[gp node center] at (5.535,0.831) {$300$};
\draw[gp path] (6.818,1.139)--(6.818,1.319);
\draw[gp path] (6.818,8.381)--(6.818,8.201);
\node[gp node center] at (6.818,0.831) {$400$};
\draw[gp path] (8.100,1.139)--(8.100,1.319);
\draw[gp path] (8.100,8.381)--(8.100,8.201);
\node[gp node center] at (8.100,0.831) {$500$};
\draw[gp path] (9.382,1.139)--(9.382,1.319);
\draw[gp path] (9.382,8.381)--(9.382,8.201);
\node[gp node center] at (9.382,0.831) {$600$};
\draw[gp path] (10.665,1.139)--(10.665,1.319);
\draw[gp path] (10.665,8.381)--(10.665,8.201);
\node[gp node center] at (10.665,0.831) {$700$};
\draw[gp path] (11.947,1.139)--(11.947,1.319);
\draw[gp path] (11.947,8.381)--(11.947,8.201);
\node[gp node center] at (11.947,0.831) {$800$};
\draw[gp path] (1.688,8.381)--(1.688,1.139)--(11.947,1.139)--(11.947,8.381)--cycle;
\node[gp node left] at (10.088,7.709) {BMP2};
\node[gp node center,rotate=-270] at (0.246,4.760) {$m_h$ [GeV]};
\node[gp node center] at (6.817,0.215) {$\mu_u$ [GeV]};
\node[gp node left] at (3.156,3.006) {tree};
\gpcolor{rgb color={0.580,0.000,0.827}}
\gpsetdashtype{gp dt 2}
\gpsetlinewidth{3.00}
\draw[gp path] (2.056,3.006)--(2.972,3.006);
\draw[gp path] (1.688,5.277)--(1.792,5.307)--(1.895,5.335)--(1.999,5.362)--(2.103,5.388)%
  --(2.206,5.413)--(2.310,5.437)--(2.413,5.460)--(2.517,5.482)--(2.621,5.504)--(2.724,5.524)%
  --(2.828,5.543)--(2.932,5.561)--(3.035,5.579)--(3.139,5.595)--(3.242,5.611)--(3.346,5.625)%
  --(3.450,5.639)--(3.553,5.652)--(3.657,5.664)--(3.761,5.675)--(3.864,5.685)--(3.968,5.694)%
  --(4.071,5.703)--(4.175,5.710)--(4.279,5.717)--(4.382,5.723)--(4.486,5.728)--(4.590,5.732)%
  --(4.693,5.735)--(4.797,5.738)--(4.900,5.740)--(5.004,5.740)--(5.108,5.740)--(5.211,5.740)%
  --(5.315,5.738)--(5.419,5.735)--(5.522,5.732)--(5.626,5.728)--(5.729,5.723)--(5.833,5.717)%
  --(5.937,5.710)--(6.040,5.702)--(6.144,5.694)--(6.248,5.685)--(6.351,5.674)--(6.455,5.663)%
  --(6.558,5.651)--(6.662,5.638)--(6.766,5.625)--(6.869,5.610)--(6.973,5.594)--(7.077,5.578)%
  --(7.180,5.560)--(7.284,5.542)--(7.387,5.523)--(7.491,5.502)--(7.595,5.481)--(7.698,5.459)%
  --(7.802,5.435)--(7.906,5.411)--(8.009,5.385)--(8.113,5.359)--(8.216,5.331)--(8.320,5.303)%
  --(8.424,5.273)--(8.527,5.242)--(8.631,5.210)--(8.735,5.176)--(8.838,5.141)--(8.942,5.106)%
  --(9.045,5.068)--(9.149,5.030)--(9.253,4.990)--(9.356,4.948)--(9.460,4.905)--(9.564,4.861)%
  --(9.667,4.815)--(9.771,4.767)--(9.874,4.718)--(9.978,4.667)--(10.082,4.614)--(10.185,4.559)%
  --(10.289,4.502)--(10.393,4.443)--(10.496,4.381)--(10.600,4.317)--(10.703,4.251)--(10.807,4.182)%
  --(10.911,4.110)--(11.014,4.035)--(11.118,3.956)--(11.222,3.874)--(11.325,3.788)--(11.429,3.697)%
  --(11.532,3.602)--(11.636,3.501)--(11.740,3.394)--(11.843,3.279)--(11.947,3.157);
\gpcolor{color=gp lt color border}
\node[gp node left] at (3.156,2.331) {tree + full 1-loop};
\gpcolor{rgb color={0.000,0.000,1.000}}
\gpsetdashtype{gp dt 4}
\draw[gp path] (2.056,2.331)--(2.972,2.331);
\draw[gp path] (1.688,6.914)--(1.792,6.950)--(1.895,6.984)--(1.999,7.015)--(2.103,7.046)%
  --(2.206,7.076)--(2.310,7.104)--(2.413,7.131)--(2.517,7.157)--(2.621,7.182)--(2.724,7.205)%
  --(2.828,7.227)--(2.932,7.249)--(3.035,7.269)--(3.139,7.287)--(3.242,7.305)--(3.346,7.322)%
  --(3.450,7.338)--(3.553,7.352)--(3.657,7.366)--(3.761,7.378)--(3.864,7.390)--(3.968,7.401)%
  --(4.071,7.411)--(4.175,7.419)--(4.279,7.427)--(4.382,7.434)--(4.486,7.441)--(4.590,7.446)%
  --(4.693,7.450)--(4.797,7.454)--(4.900,7.457)--(5.004,7.459)--(5.108,7.460)--(5.211,7.461)%
  --(5.315,7.460)--(5.419,7.459)--(5.522,7.457)--(5.626,7.455)--(5.729,7.451)--(5.833,7.447)%
  --(5.937,7.442)--(6.040,7.437)--(6.144,7.430)--(6.248,7.423)--(6.351,7.416)--(6.455,7.407)%
  --(6.558,7.398)--(6.662,7.388)--(6.766,7.378)--(6.869,7.366)--(6.973,7.354)--(7.077,7.342)%
  --(7.180,7.328)--(7.284,7.314)--(7.387,7.299)--(7.491,7.284)--(7.595,7.268)--(7.698,7.251)%
  --(7.802,7.233)--(7.906,7.215)--(8.009,7.196)--(8.113,7.176)--(8.216,7.156)--(8.320,7.134)%
  --(8.424,7.112)--(8.527,7.090)--(8.631,7.066)--(8.735,7.042)--(8.838,7.017)--(8.942,6.991)%
  --(9.045,6.965)--(9.149,6.938)--(9.253,6.910)--(9.356,6.881)--(9.460,6.851)--(9.564,6.820)%
  --(9.667,6.789)--(9.771,6.757)--(9.874,6.724)--(9.978,6.690)--(10.082,6.655)--(10.185,6.619)%
  --(10.289,6.582)--(10.393,6.545)--(10.496,6.506)--(10.600,6.466)--(10.703,6.426)--(10.807,6.384)%
  --(10.911,6.341)--(11.014,6.297)--(11.118,6.252)--(11.222,6.206)--(11.325,6.158)--(11.429,6.110)%
  --(11.532,6.060)--(11.636,6.009)--(11.740,5.957)--(11.843,5.903)--(11.947,5.848);
\gpcolor{color=gp lt color border}
\node[gp node left] at (3.156,1.656) {tree + full 1-loop + leading 2-loop};
\gpcolor{rgb color={0.000,1.000,0.000}}
\gpsetdashtype{gp dt solid}
\draw[gp path] (2.056,1.656)--(2.972,1.656);
\draw[gp path] (1.688,7.195)--(1.792,7.229)--(1.895,7.261)--(1.999,7.291)--(2.103,7.321)%
  --(2.206,7.350)--(2.310,7.377)--(2.413,7.403)--(2.517,7.428)--(2.621,7.452)--(2.724,7.475)%
  --(2.828,7.496)--(2.932,7.517)--(3.035,7.536)--(3.139,7.555)--(3.242,7.572)--(3.346,7.589)%
  --(3.450,7.604)--(3.553,7.618)--(3.657,7.632)--(3.761,7.644)--(3.864,7.656)--(3.968,7.667)%
  --(4.071,7.676)--(4.175,7.685)--(4.279,7.693)--(4.382,7.700)--(4.486,7.707)--(4.590,7.712)%
  --(4.693,7.717)--(4.797,7.720)--(4.900,7.723)--(5.004,7.726)--(5.108,7.727)--(5.211,7.728)%
  --(5.315,7.727)--(5.419,7.726)--(5.522,7.725)--(5.626,7.722)--(5.729,7.719)--(5.833,7.715)%
  --(5.937,7.710)--(6.040,7.705)--(6.144,7.698)--(6.248,7.691)--(6.351,7.684)--(6.455,7.675)%
  --(6.558,7.666)--(6.662,7.656)--(6.766,7.645)--(6.869,7.634)--(6.973,7.622)--(7.077,7.609)%
  --(7.180,7.595)--(7.284,7.581)--(7.387,7.565)--(7.491,7.549)--(7.595,7.533)--(7.698,7.515)%
  --(7.802,7.497)--(7.906,7.478)--(8.009,7.458)--(8.113,7.437)--(8.216,7.416)--(8.320,7.393)%
  --(8.424,7.370)--(8.527,7.346)--(8.631,7.322)--(8.735,7.296)--(8.838,7.269)--(8.942,7.242)%
  --(9.045,7.214)--(9.149,7.184)--(9.253,7.154)--(9.356,7.123)--(9.460,7.091)--(9.564,7.058)%
  --(9.667,7.024)--(9.771,6.989)--(9.874,6.953)--(9.978,6.916)--(10.082,6.877)--(10.185,6.838)%
  --(10.289,6.797)--(10.393,6.756)--(10.496,6.713)--(10.600,6.669)--(10.703,6.623)--(10.807,6.577)%
  --(10.911,6.529)--(11.014,6.479)--(11.118,6.429)--(11.222,6.376)--(11.325,6.322)--(11.429,6.267)%
  --(11.532,6.210)--(11.636,6.152)--(11.740,6.091)--(11.843,6.029)--(11.947,5.965);
\gpcolor{color=gp lt color border}
\gpsetlinewidth{1.00}
\draw[gp path] (1.688,8.381)--(1.688,1.139)--(11.947,1.139)--(11.947,8.381)--cycle;
\gpdefrectangularnode{gp plot 1}{\pgfpoint{1.688cm}{1.139cm}}{\pgfpoint{11.947cm}{8.381cm}}
\end{tikzpicture}
}}
  \subfloat[]{\resizebox{0.32\textwidth}{!}{\begin{tikzpicture}[gnuplot]
\path (0.000,0.000) rectangle (12.500,8.750);
\gpcolor{color=gp lt color border}
\gpsetlinetype{gp lt border}
\gpsetdashtype{gp dt solid}
\gpsetlinewidth{1.00}
\draw[gp path] (1.688,1.139)--(1.868,1.139);
\draw[gp path] (11.947,1.139)--(11.767,1.139);
\node[gp node right] at (1.504,1.139) {$0$};
\draw[gp path] (1.688,2.174)--(1.868,2.174);
\draw[gp path] (11.947,2.174)--(11.767,2.174);
\node[gp node right] at (1.504,2.174) {$20$};
\draw[gp path] (1.688,3.208)--(1.868,3.208);
\draw[gp path] (11.947,3.208)--(11.767,3.208);
\node[gp node right] at (1.504,3.208) {$40$};
\draw[gp path] (1.688,4.243)--(1.868,4.243);
\draw[gp path] (11.947,4.243)--(11.767,4.243);
\node[gp node right] at (1.504,4.243) {$60$};
\draw[gp path] (1.688,5.277)--(1.868,5.277);
\draw[gp path] (11.947,5.277)--(11.767,5.277);
\node[gp node right] at (1.504,5.277) {$80$};
\draw[gp path] (1.688,6.312)--(1.868,6.312);
\draw[gp path] (11.947,6.312)--(11.767,6.312);
\node[gp node right] at (1.504,6.312) {$100$};
\draw[gp path] (1.688,7.346)--(1.868,7.346);
\draw[gp path] (11.947,7.346)--(11.767,7.346);
\node[gp node right] at (1.504,7.346) {$120$};
\draw[gp path] (1.688,8.381)--(1.868,8.381);
\draw[gp path] (11.947,8.381)--(11.767,8.381);
\node[gp node right] at (1.504,8.381) {$140$};
\draw[gp path] (1.688,1.139)--(1.688,1.319);
\draw[gp path] (1.688,8.381)--(1.688,8.201);
\node[gp node center] at (1.688,0.831) {$0$};
\draw[gp path] (2.970,1.139)--(2.970,1.319);
\draw[gp path] (2.970,8.381)--(2.970,8.201);
\node[gp node center] at (2.970,0.831) {$100$};
\draw[gp path] (4.253,1.139)--(4.253,1.319);
\draw[gp path] (4.253,8.381)--(4.253,8.201);
\node[gp node center] at (4.253,0.831) {$200$};
\draw[gp path] (5.535,1.139)--(5.535,1.319);
\draw[gp path] (5.535,8.381)--(5.535,8.201);
\node[gp node center] at (5.535,0.831) {$300$};
\draw[gp path] (6.818,1.139)--(6.818,1.319);
\draw[gp path] (6.818,8.381)--(6.818,8.201);
\node[gp node center] at (6.818,0.831) {$400$};
\draw[gp path] (8.100,1.139)--(8.100,1.319);
\draw[gp path] (8.100,8.381)--(8.100,8.201);
\node[gp node center] at (8.100,0.831) {$500$};
\draw[gp path] (9.382,1.139)--(9.382,1.319);
\draw[gp path] (9.382,8.381)--(9.382,8.201);
\node[gp node center] at (9.382,0.831) {$600$};
\draw[gp path] (10.665,1.139)--(10.665,1.319);
\draw[gp path] (10.665,8.381)--(10.665,8.201);
\node[gp node center] at (10.665,0.831) {$700$};
\draw[gp path] (11.947,1.139)--(11.947,1.319);
\draw[gp path] (11.947,8.381)--(11.947,8.201);
\node[gp node center] at (11.947,0.831) {$800$};
\draw[gp path] (1.688,8.381)--(1.688,1.139)--(11.947,1.139)--(11.947,8.381)--cycle;
\node[gp node left] at (10.088,7.709) {BMP3};
\node[gp node center,rotate=-270] at (0.246,4.760) {$m_h$ [GeV]};
\node[gp node center] at (6.817,0.215) {$\mu_u$ [GeV]};
\node[gp node left] at (3.156,3.006) {tree};
\gpcolor{rgb color={0.580,0.000,0.827}}
\gpsetdashtype{gp dt 2}
\gpsetlinewidth{3.00}
\draw[gp path] (2.056,3.006)--(2.972,3.006);
\draw[gp path] (1.688,5.633)--(1.792,5.644)--(1.895,5.654)--(1.999,5.665)--(2.103,5.675)%
  --(2.206,5.684)--(2.310,5.694)--(2.413,5.703)--(2.517,5.712)--(2.621,5.720)--(2.724,5.728)%
  --(2.828,5.736)--(2.932,5.744)--(3.035,5.751)--(3.139,5.759)--(3.242,5.765)--(3.346,5.772)%
  --(3.450,5.778)--(3.553,5.784)--(3.657,5.790)--(3.761,5.795)--(3.864,5.800)--(3.968,5.805)%
  --(4.071,5.810)--(4.175,5.814)--(4.279,5.818)--(4.382,5.822)--(4.486,5.826)--(4.590,5.829)%
  --(4.693,5.832)--(4.797,5.835)--(4.900,5.837)--(5.004,5.840)--(5.108,5.842)--(5.211,5.843)%
  --(5.315,5.845)--(5.419,5.846)--(5.522,5.847)--(5.626,5.847)--(5.729,5.848)--(5.833,5.848)%
  --(5.937,5.848)--(6.040,5.847)--(6.144,5.847)--(6.248,5.846)--(6.351,5.844)--(6.455,5.843)%
  --(6.558,5.841)--(6.662,5.839)--(6.766,5.837)--(6.869,5.834)--(6.973,5.832)--(7.077,5.829)%
  --(7.180,5.825)--(7.284,5.822)--(7.387,5.818)--(7.491,5.814)--(7.595,5.809)--(7.698,5.804)%
  --(7.802,5.799)--(7.906,5.794)--(8.009,5.789)--(8.113,5.783)--(8.216,5.777)--(8.320,5.771)%
  --(8.424,5.764)--(8.527,5.757)--(8.631,5.750)--(8.735,5.742)--(8.838,5.735)--(8.942,5.727)%
  --(9.045,5.718)--(9.149,5.710)--(9.253,5.701)--(9.356,5.692)--(9.460,5.682)--(9.564,5.672)%
  --(9.667,5.662)--(9.771,5.652)--(9.874,5.641)--(9.978,5.630)--(10.082,5.619)--(10.185,5.607)%
  --(10.289,5.596)--(10.393,5.583)--(10.496,5.571)--(10.600,5.558)--(10.703,5.545)--(10.807,5.531)%
  --(10.911,5.518)--(11.014,5.503)--(11.118,5.489)--(11.222,5.474)--(11.325,5.459)--(11.429,5.444)%
  --(11.532,5.428)--(11.636,5.412)--(11.740,5.395)--(11.843,5.378)--(11.947,5.361);
\gpcolor{color=gp lt color border}
\node[gp node left] at (3.156,2.331) {tree + full 1-loop};
\gpcolor{rgb color={0.000,0.000,1.000}}
\gpsetdashtype{gp dt 4}
\draw[gp path] (2.056,2.331)--(2.972,2.331);
\draw[gp path] (1.688,7.060)--(1.792,7.082)--(1.895,7.102)--(1.999,7.120)--(2.103,7.139)%
  --(2.206,7.157)--(2.310,7.173)--(2.413,7.189)--(2.517,7.205)--(2.621,7.219)--(2.724,7.233)%
  --(2.828,7.246)--(2.932,7.259)--(3.035,7.270)--(3.139,7.282)--(3.242,7.292)--(3.346,7.302)%
  --(3.450,7.312)--(3.553,7.320)--(3.657,7.329)--(3.761,7.336)--(3.864,7.344)--(3.968,7.350)%
  --(4.071,7.357)--(4.175,7.362)--(4.279,7.368)--(4.382,7.373)--(4.486,7.377)--(4.590,7.381)%
  --(4.693,7.385)--(4.797,7.388)--(4.900,7.391)--(5.004,7.393)--(5.108,7.395)--(5.211,7.397)%
  --(5.315,7.398)--(5.419,7.400)--(5.522,7.400)--(5.626,7.401)--(5.729,7.401)--(5.833,7.401)%
  --(5.937,7.400)--(6.040,7.399)--(6.144,7.398)--(6.248,7.397)--(6.351,7.395)--(6.455,7.393)%
  --(6.558,7.391)--(6.662,7.388)--(6.766,7.386)--(6.869,7.383)--(6.973,7.380)--(7.077,7.376)%
  --(7.180,7.372)--(7.284,7.368)--(7.387,7.364)--(7.491,7.360)--(7.595,7.355)--(7.698,7.350)%
  --(7.802,7.345)--(7.906,7.340)--(8.009,7.334)--(8.113,7.328)--(8.216,7.322)--(8.320,7.316)%
  --(8.424,7.309)--(8.527,7.303)--(8.631,7.296)--(8.735,7.289)--(8.838,7.281)--(8.942,7.274)%
  --(9.045,7.266)--(9.149,7.258)--(9.253,7.250)--(9.356,7.241)--(9.460,7.233)--(9.564,7.224)%
  --(9.667,7.215)--(9.771,7.206)--(9.874,7.196)--(9.978,7.187)--(10.082,7.177)--(10.185,7.167)%
  --(10.289,7.156)--(10.393,7.146)--(10.496,7.135)--(10.600,7.124)--(10.703,7.113)--(10.807,7.102)%
  --(10.911,7.090)--(11.014,7.078)--(11.118,7.066)--(11.222,7.054)--(11.325,7.042)--(11.429,7.029)%
  --(11.532,7.016)--(11.636,7.003)--(11.740,6.990)--(11.843,6.976)--(11.947,6.963);
\gpcolor{color=gp lt color border}
\node[gp node left] at (3.156,1.656) {tree + full 1-loop + leading 2-loop};
\gpcolor{rgb color={0.000,1.000,0.000}}
\gpsetdashtype{gp dt solid}
\draw[gp path] (2.056,1.656)--(2.972,1.656);
\draw[gp path] (1.688,7.307)--(1.792,7.328)--(1.895,7.347)--(1.999,7.364)--(2.103,7.382)%
  --(2.206,7.400)--(2.310,7.416)--(2.413,7.432)--(2.517,7.446)--(2.621,7.461)--(2.724,7.474)%
  --(2.828,7.487)--(2.932,7.499)--(3.035,7.510)--(3.139,7.521)--(3.242,7.532)--(3.346,7.541)%
  --(3.450,7.551)--(3.553,7.559)--(3.657,7.567)--(3.761,7.575)--(3.864,7.582)--(3.968,7.589)%
  --(4.071,7.595)--(4.175,7.601)--(4.279,7.606)--(4.382,7.611)--(4.486,7.616)--(4.590,7.620)%
  --(4.693,7.623)--(4.797,7.627)--(4.900,7.630)--(5.004,7.632)--(5.108,7.635)--(5.211,7.636)%
  --(5.315,7.638)--(5.419,7.639)--(5.522,7.640)--(5.626,7.641)--(5.729,7.641)--(5.833,7.641)%
  --(5.937,7.641)--(6.040,7.640)--(6.144,7.639)--(6.248,7.638)--(6.351,7.637)--(6.455,7.635)%
  --(6.558,7.633)--(6.662,7.631)--(6.766,7.628)--(6.869,7.626)--(6.973,7.623)--(7.077,7.619)%
  --(7.180,7.616)--(7.284,7.612)--(7.387,7.608)--(7.491,7.604)--(7.595,7.600)--(7.698,7.595)%
  --(7.802,7.590)--(7.906,7.585)--(8.009,7.579)--(8.113,7.574)--(8.216,7.568)--(8.320,7.562)%
  --(8.424,7.555)--(8.527,7.549)--(8.631,7.542)--(8.735,7.535)--(8.838,7.528)--(8.942,7.521)%
  --(9.045,7.513)--(9.149,7.505)--(9.253,7.497)--(9.356,7.489)--(9.460,7.480)--(9.564,7.471)%
  --(9.667,7.462)--(9.771,7.453)--(9.874,7.443)--(9.978,7.434)--(10.082,7.424)--(10.185,7.414)%
  --(10.289,7.403)--(10.393,7.393)--(10.496,7.382)--(10.600,7.371)--(10.703,7.359)--(10.807,7.348)%
  --(10.911,7.336)--(11.014,7.324)--(11.118,7.312)--(11.222,7.299)--(11.325,7.286)--(11.429,7.273)%
  --(11.532,7.260)--(11.636,7.246)--(11.740,7.233)--(11.843,7.219)--(11.947,7.204);
\gpcolor{color=gp lt color border}
\gpsetlinewidth{1.00}
\draw[gp path] (1.688,8.381)--(1.688,1.139)--(11.947,1.139)--(11.947,8.381)--cycle;
\gpdefrectangularnode{gp plot 1}{\pgfpoint{1.688cm}{1.139cm}}{\pgfpoint{11.947cm}{8.381cm}}
\end{tikzpicture}
}}
  \caption{Higgs mass at tree, one-loop and two-loop levels as a function of superpotential parameters $\lambda_u$, $\Lambda_u$ and $\mu_u$. 
  All other parameters are fixed  as for BMP1 in column (a), BMP2 in (b) and BMP3 in (c).
  \label{fig:mh_1d}}
\end{figure}

\Autoref{fig:mh_1d} shows the lightest Higgs boson mass at tree, one-loop and two-loop levels as a function of superpotential couplings $\lambda_{u,d}$, $\Lambda_{u,d}$ and $\mu_u$.
The plots are done varying these parameters while keeping others fixed to their respective benchmark point values - that is BMP1 in column (a), BMP2 in (b) and BMP3 in (c).
The masses are calculated using \texttt{SARAH} generated \texttt{SPheno} so,
contrary to the previous section, corrections to all elements of the Higgs mass matrix and from all possible particles are taken into account in full and threshold corrections to quark Yukawas from strongly interacting particles were included as well.\footnote{Technically, \texttt{SARAH} implements 3 methods for calculating two-loop corrections (controlled by the value of flag 8 in \texttt{SphenoInput}): numerical (value 1) and semi-analytical (2) methods based on the effective potential and a diagrammatic approach (3) \cite{Goodsell:2015ira}.
    Numerically, for BMPs of \autoref{tab:BMP} the latter two give the lightest Higgs mass which differ by less than 10 MeV, while the numerical one gives the value which is less than 100 MeV different   for BMP1 compared to diagrammatic or semi-analytical approach.
    The diagrammatic approach allows also to calculate two-loop corrections to the mass of the pseudoscalar Higgs bosons as explained in Ref.~\cite{Goodsell:2016udb}.
    }
    
In \Autoref{fig:mh_1d} the tree-level results are denoted by dashed lines. 
They exhibit, as discussed earlier, a quadratic dependence on $\Lambda$ and $\lambda$  due to the mixing with the triplet and the singlet states that reduces the predicted Higgs boson mass well below the MSSM upper limit of $m_Z |\cos 2\beta|$. 
The dotted lines show the one-loop corrected mass. 
Due to the large and positive corrections from top Yukawa and $\Lambda/\lambda$ couplings the dependence on $\Lambda/\lambda$ parameters is reversed and  allows to reach values in the 125 GeV range. 
Finally, solid lines show the predicted Higgs boson mass with leading two-loop corrections which approximately amount to a constant upward shift. 
The two-loop corrections are moderate, of order a few GeV, which in spite of large one-loop corrections, give us confidence in the perturbative expansion. 

\section{Conclusions}

This chapter discussed the calculation of the Higgs boson mass in the MRSSM.
In the considered setup, where the soft masses $|m_S|$ and $|m_T|$ are in the few TeV range, the tree-level Higgs mass is bounded from above by $m_Z |\cos 2\beta|$ like in the MSSM.
Due to mixing with singlet and triplet states the mass is additionally reduced, though.
For MRSSM to be phenomenologically viable, large radiative corrections must be invoked.
This was shown to be possible already at  one-loop level in Ref.~\cite{Diessner:2014ksa}.
Enlarged MRSSM EW sector gives sufficiently large contributions to Higgs mass even without left-right mixing in the stop sector and at a moderate values of the stop squark masses, i.e. below 1 TeV.
One-loop contributions from different particle groups were given in this chapter, together with overall dependence of one-loop contribution on the most important superpotential parameters.

After Ref.~\cite{Diessner:2014ksa}, the leading two-loop corrections to the Higgs boson mass in the MRSSM became also known.
Benchmark points in this thesis are characterized by values of $\Lambda_u$ smaller than in that work, chosen to reproduce the correct Higgs boson mass at the two-loop level.
The two-loop corrections turned out to be moderate, increasing its mass by around +5 GeV for considered benchmark points.

This chapter discussed also the setup of this calculation and gave analytic formulas for leading MRSSM specific contributions. 
The formulas were re-derived following the original Ref.~\cite{Martin:2001vx}, and revealed a mistake in the original implementation of diagram (a) in \autoref{fig:2L_diagrams} in \texttt{SARAH}. 

With two-loop augmented calculation of the Higgs boson mass, in the next chapter predictions for $m_W$ and $m_h$ are confronted and their interdependence on superpotential parameters is discussed.

\chapter{Interdependence of Higgs mass and other electroweak observables \label{sec:mhmw} }

\begin{figure}
  \centering
  \subfloat[]{\includegraphics[width=0.49\textwidth]{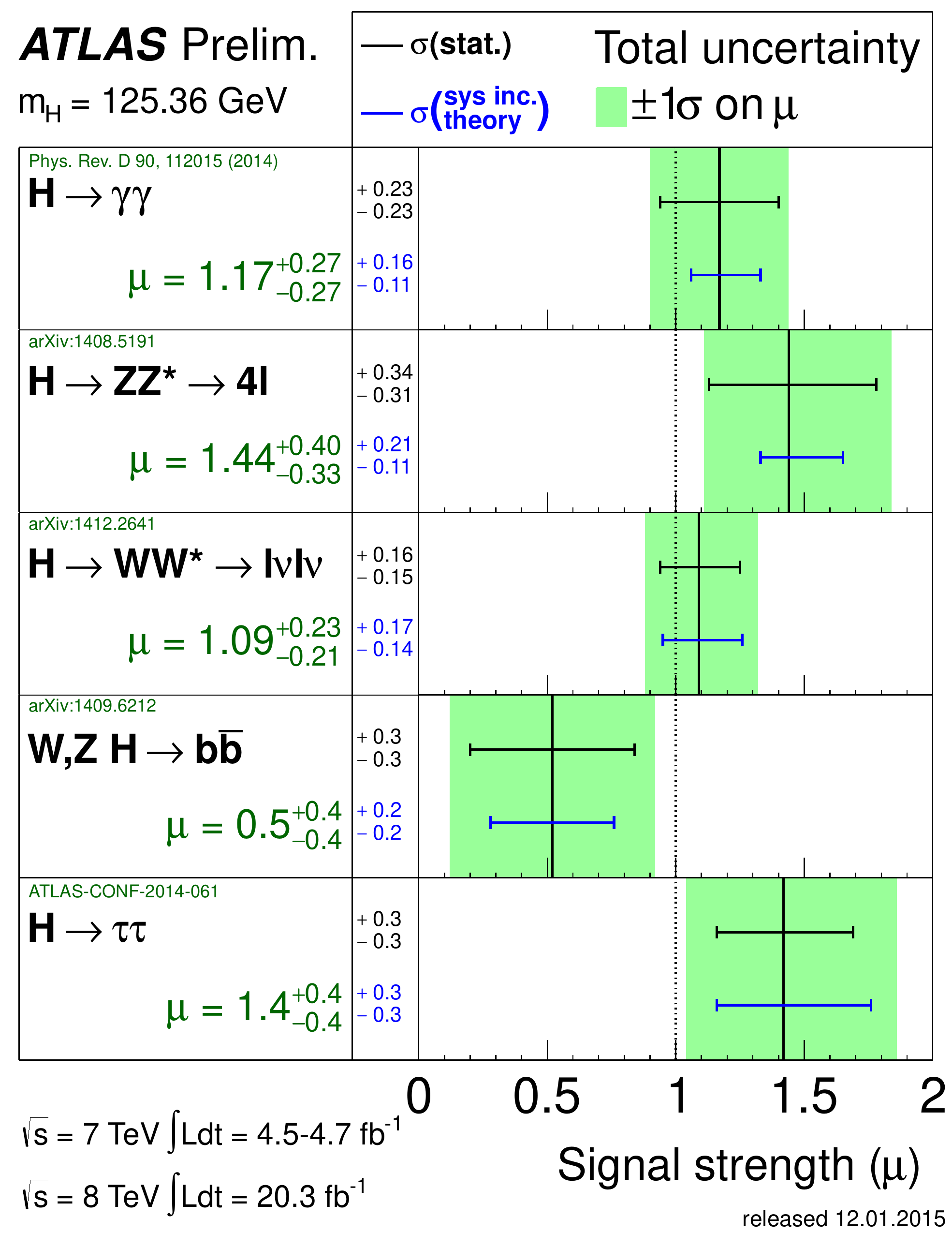}}
  \subfloat[]{
  \raisebox{+1.25cm}{\includegraphics[width=0.49\textwidth]{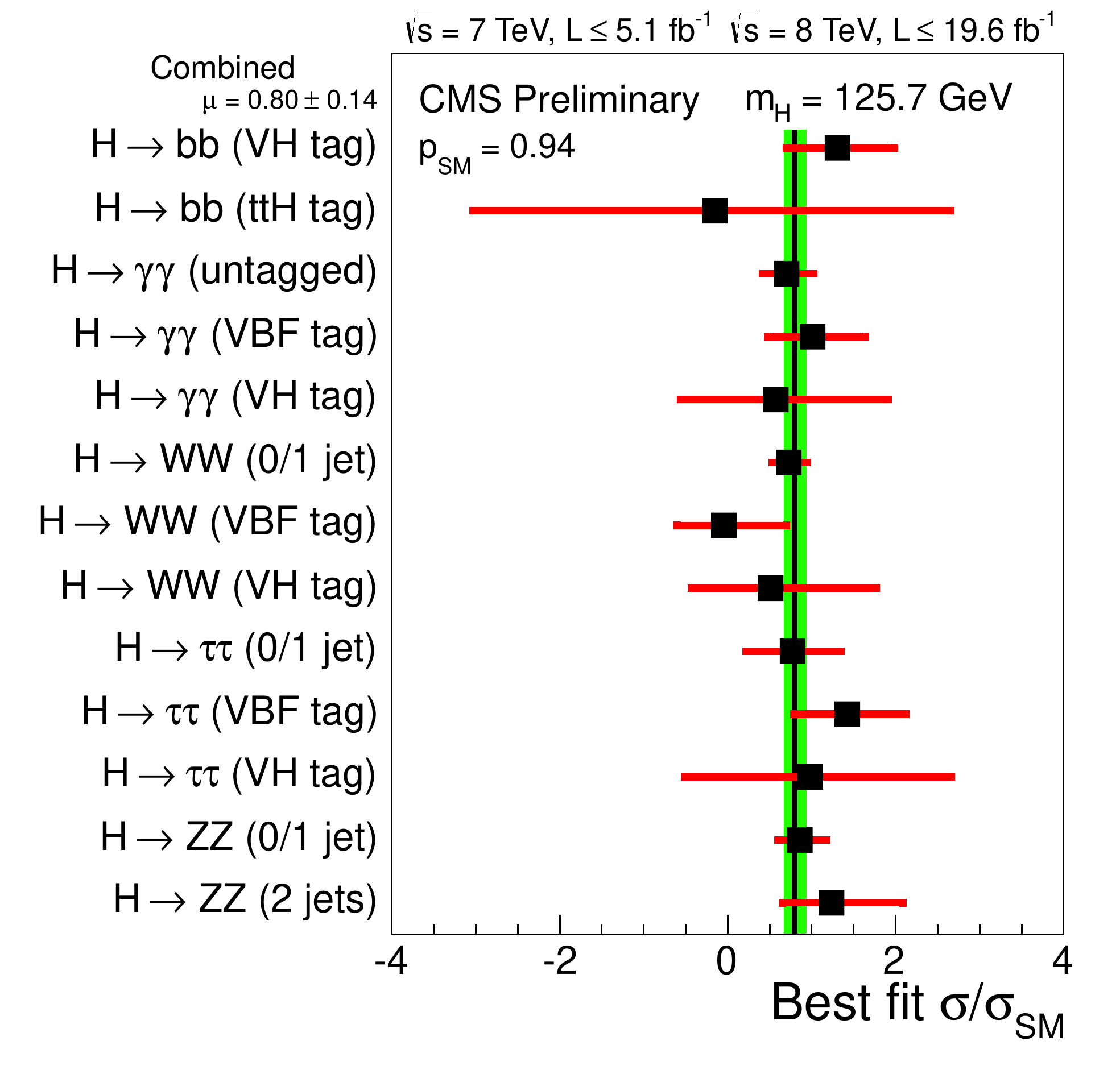}}}
  \caption{ 
  The measured production strengths for a Higgs boson from ATLAS (a) and CMS (b), normalized to the SM expectations.
  The best-fit values for a given channel are shown by the solid vertical lines on (a) and points on (b).
  Dotted vertical line on (a) and vertical line on (b) show the overall best-fit value.
  \label{fig:higgs_signal_strength_ATLAS}
  }
\end{figure}

Previous two chapters were devoted to describing the setup for calculation of loop corrections to the Higgs and $W$ boson masses.
This included both full one-loop corrections for both observables and dominant two-loop corrections to CP-even Higgs boson mass.
Although these chapters showed that both observables can be explained within MRSSM, their interdependence has not yet been discussed.
To fill this gap, this chapter will present a set of benchmark points in agreement with both of them.
Since we know quite precisely from Run 1 of the LHC  not only the Higgs boson mass, but also its main decay patterns (see for example \autoref{fig:higgs_signal_strength_ATLAS}), observables such as Higgs signal strength and negative results from Higgs searches must be taken into account when checking validity of the benchmark points.

The benchmark points will also be shown to pass basic tests such as vacuum stability and selected $b$-physics observables.  
Around these points, valid regions of parameter space will be identified, both by $2d$ parameters scans and by random scatter scans.

The chapter is structured as follows. 
The next section presents used numeric tools and mass spectra plots.
It also gives numerical values for masses of selected particles.
The philosophy behind checking the Higgs sector against experimental data is explained, and \ac{LO} couplings of the Higgs boson to SM particles are given.
\Autoref{sec:contour_plots} shows $2d$ scans around a selected benchmark point, while \Autoref{sec:scatter_scans} does the same in a higher number of parameters.
The chapter ends with conclusions in \autoref{sec:mwmh_conclusions}.

\section{Properties of the benchmark points}
Following the discussion of the computational setup done in \autoref{sec:MRSSM}, \ref{sec:ew} and \ref{sec:higgs_chapter} this section presents 
\begin{figure}
\centering
\subfloat[]{\includegraphics[width=0.49\textwidth]{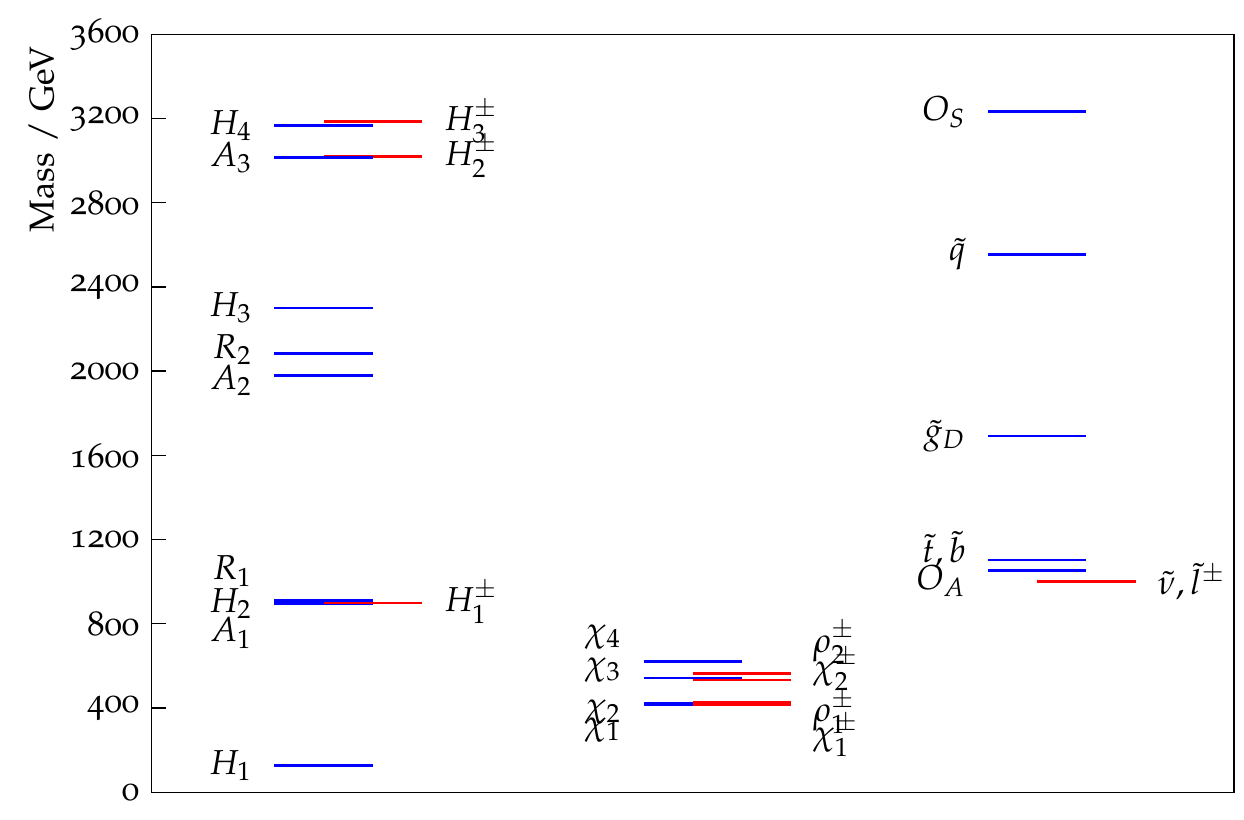}}
\subfloat[]{\includegraphics[width=0.49\textwidth]{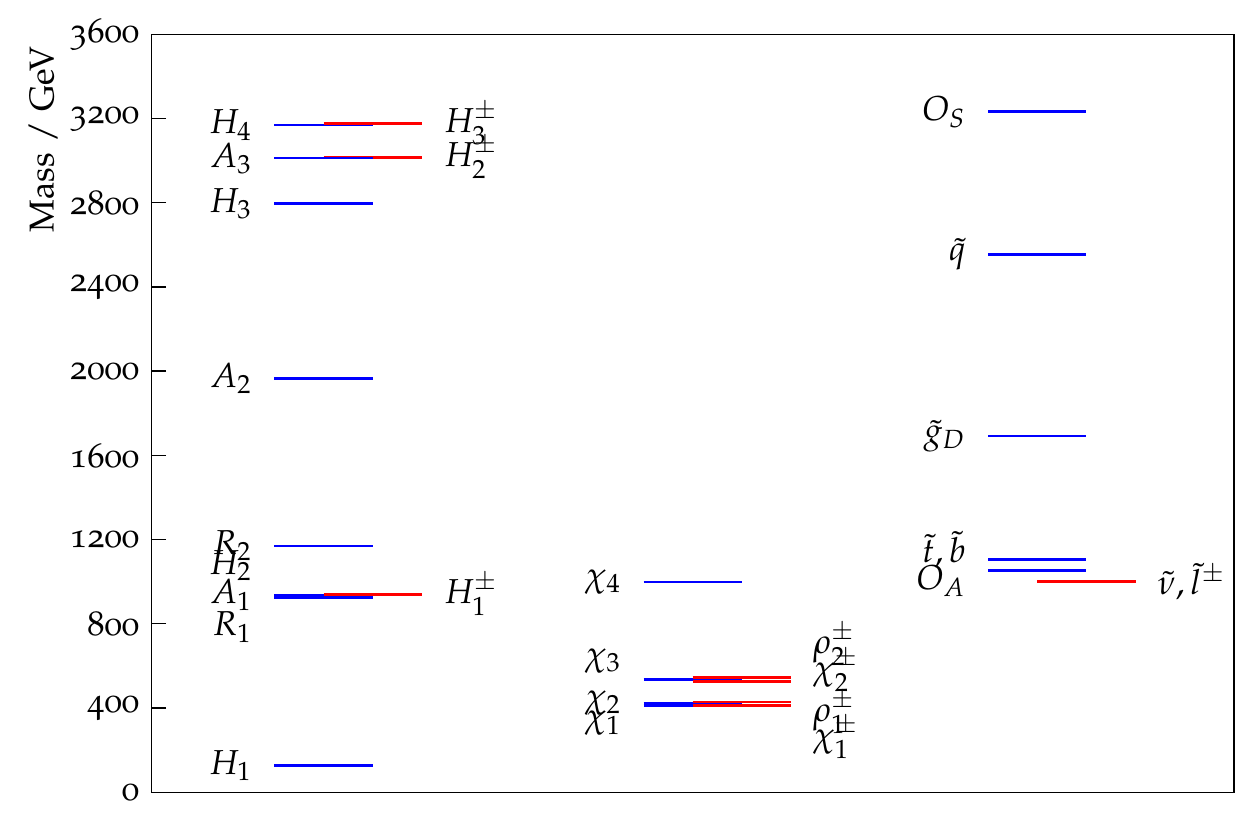}}\\
\subfloat[]{\includegraphics[width=0.49\textwidth]{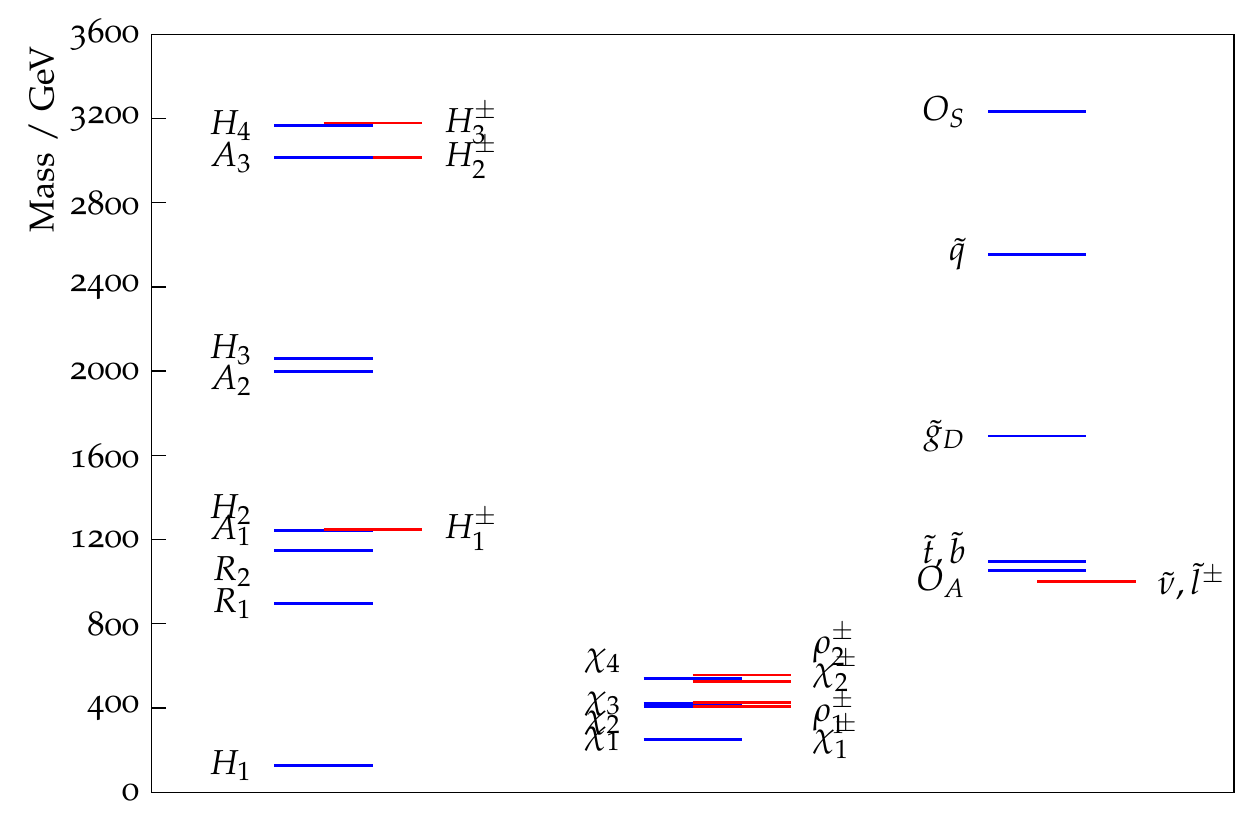}}
\caption{
  Particle mass spectra for $\tan \beta = 3$ (a), $\tan \beta = 10$ (b) and $\tan \beta = 40$ (c). Since all 1st- and 2nd-generation squarks are almost degenerate in mass they they are commonly denoted as $\tilde{q}$.
  Also, since left/right mass splitting for stops (sbottoms) is small, they are commonly denoted as $\tilde t$ ($\tilde b$).
  Plots were done using \texttt{PySLHA}~\cite{Buckley:2013jua}. \label{fig:mass_spectrum} 
}
\end{figure}
complete mass spectra for the benchmark points in \autoref{tab:BMP}.
The particle masses were calculated using \texttt{SARAH} generated \texttt{SPheno} code and visualized in \autoref{fig:mass_spectrum}.
As discussed in previous chapters, all masses are calculated at full one-loop level, with the exception of scalar and pseudocalar Higgs masses, which are calculated with leading two-loop corrections.\footnote{
  Since Ref.~\cite{Goodsell:2016udb} two-loop corrections to psedudoscalar Higgs masses are also available.
    Since they are not essential to this analysis and for BMPs of \autoref{tab:BMP} are roughly 0, they are not discussed here further.
}
The left column of every spectrum plot contains only "Higgs" states (that is, including R-Higgses), the middle one contains only EW-inos, while the right one shows strongly interacting particles and sleptons.
The spectra reach up to ~3 TeV, where the heaviest states are mainly composed of scalar triplet or octet.
These masses are driven by the large value of the soft SUSY breaking triplet mass parameter $m_T^2$, which must be large to guarantee a small triplet VEV.
Masses of the Higgs bosons lying around 2 TeV could in principle be lowered without changing conclusions of this chapter.
Since both the pseudoscalar sgluon and gluino must be at least in the 1 TeV range, this implies that scalar sgluon will also be heavier than 2 TeV.
Masses of squarks of 1st and 2nd generation are constrained by flavor observables.
For quantitative analysis, \autoref{tab:mass_spectrum} contains masses of selected particles (the mass of the lightest Higgs boson is shown in \autoref{tab:higgs_properties}).

\begin{table}
\centering
\begin{tabular}{l|ccccc|cccc|cc|c}
& $H_2$ & $A_1$ & $H_1^\pm$ & $R_1$ & $R_1^\pm$ & $\chi_1$ & $\chi_2$ & $\chi_1^\pm$ & $\rho_1^\pm$ & $\tilde{t}_1$ & $\tilde{b}_1$ & $\tilde{\nu}$ \\
\hline
BMP1 & 897 & 896  & 899  & 912 & 906 & 415 & 420 & 416 & 427 & 1059 & 1061 & 1002 \\
BMP2 & 937 & 937  & 940  & 926 & 921 & 413 & 423 & 413 & 429 & 1061 & 1062 & 1003 \\
BMP3 & 1245 & 1245 & 1248 & 896 & 891 & 251 & 408 & 408 & 424 & 1060 & 1056 & 1000
\end{tabular}
\caption{ Masses of selected particles (in GeV) for three benchmark points.  \label{tab:mass_spectrum} }
\end{table}

For every BMP, the Higgs sector was checked against \ac{LEP} and hadron colliders (both Tevatron and LHC) data using \texttt{HiggsBounds}~\cite{Bechtle:2008jh,Bechtle:2011sb,Bechtle:2013gu,Bechtle:2013wla,Bechtle:2015pma}~\textit{v4.3.1} and \texttt{HiggsSignals}~\cite{Bechtle:2013xfa,Stal:2013hwa,Bechtle:2014ewa}~\textit{v1.4.0}.
Both programs are run in the \texttt{effc} mode, in which the input to the analysis comprises:
\begin{itemize}
  \item the masses of neutral and singly charged Higgs bosons;
  \item their total decay widths;
  \item the squared normalized scalar (subscript $s$) and pseudoscalar ($p$) effective Higgs couplings to fermions\\
$(g^{\text{MRSSM}}_{s, h_k (\text{OP})}/g^{\text{SM}}_{H (\text{OP})})^2$,
$(g^{\text{MRSSM}}_{p, h_k (\text{OP})}/g^{\text{SM}}_{H (\text{OP})})^2$, where OP = $\{ s\bar s, c \bar c, b \bar b, t \bar t, \mu^+ \mu^-, \tau^+ \tau^- \}$;
  \item the squared normalized effective Higgs couplings to bosons\footnote{\texttt{HiggsBounds} is also capable of checking Higgs-$Z\gamma$ and Higgs-$Zgg$ couplings but since for \texttt{SARAH}~\textit{v4.8.6} these couplings are not calculated, those channels were removed from the list.
  }\\
$(g^{\text{MRSSM}}_{h_i h_j Z}/g^{\text{ref}}_{H' H Z})^2$, 
$(g^{\text{MRSSM}}_{s, h_k (\text{OP})}/g^{\text{SM}}_{H (\text{OP})})^2$,
where OP = $\{ W^+ W^-, ZZ, \gamma \gamma, gg \}$;
  \item the neutral Higgs branching ratios without SM equivalents\\
  $\text{BR}_\text{MRSSM}(h_k \to h_i h_j)$, $\text{BR}_\text{MRSSM}(h_k \to \text{invisible})$;
  \item  the charged Higgs branching ratios to SM particles\\
    $\text{BR}_\text{MRSSM}(H_j^+ \to SM)$, where SM = $\{ c\bar{s}, c\bar b, \tau^+ v_\tau \}$.
\end{itemize}
The reference value $g^{\text{ref}}_{H' H Z}$, which does not have its  counterpart in the SM, is defined as $(g^{\text{ref}}_{H' H Z})^2 = \equiv e^2/(4 s_w^2 c_w^2)$.
All these quantities are calculated automatically by \texttt{SPheno}.
\Autoref{tab:higgs_coupling_strength} gives values of selected effective Higgs couplings for the benchmark points.
\begin{table}
\centering
\begin{tabular}{l|ccccccc}
& $\gamma \gamma$ & gg & $W^\pm W^{\mp*}$ & $Z Z^*$ & $\tau^+ \tau^-$ & $c \bar{c}$ & $b \bar{b}$ \\
\hline
BMP1 & 0.914 & 1.005 & 0.999 & 0.999 & 1.196 & 0.981 & 1.196 \\
BMP2 & 0.928 & 1.024 & 1 & 1 & 1.17 & 0.998  & 1.17 \\
BMP3 & 0.915 & 1.028 & 1     & 1     & 1.109 & 1  & 1.109  
\end{tabular}
\caption{
  Squared effective couplings of the lightest Higgs boson to the gauge boson and fermion pairs normalized to the SM values.
  Couplings are calculated at the leading order.
\label{tab:higgs_coupling_strength} 
}
\end{table}

To use both LEP and hadron collider data in \texttt{HiggsBounds}, the option \texttt{whichanalyses} must be set to \texttt{LandH}.
\texttt{HiggsSignals} is run using the \texttt{latestresults} flag with the peak-centered $\chi^2$ method and Gaussian Higgs mass uncertainty. 
Both \texttt{HiggsBounds} and \texttt{HiggsSignals} are run with and without theoretical uncertainty on the SM-like Higgs mass, which is assumed to be $\pm 5$ GeV.

\Autoref{tab:higgs_properties} summarizes properties of the Higgs sector.
It gives the values of the two-loop corrected SM-like Higgs boson masses, the \texttt{HiggsBounds}'s \texttt{obsratio}s and \texttt{HiggsSignals} $p$-values (both with and without assumed theoretical uncertainty of 5 GeV on the Higgs mass).
The value of \texttt{obsratio} $\geq 1$ would mean that the parameter point is excluded at 95\% C.L. Likewise, the parameter point is excluded if the \texttt{HiggsSignals} $p$-value < 0.05.

The BMPs were also checked for stability of the EW minimum in the $\{ v_d, v_u, v_S, v_T \}$ space using \texttt{Vevacious}~\textit{v1.2.01}~\cite{Camargo-Molina:2013qva,Camargo-Molina:2014pwa}, and found to be stable.

Selected $b$-physics observables, namely $B \to X_s \gamma$ and $B_{s/d}\to \mu^+ \mu^-$, were calculated using $\texttt{SARAH}'s$ \texttt{FlavourKit} package \cite{Dreiner:2012dh,Porod:2014xia} and found to be in agreement with experiment (see \autoref{tab:b_physics_observables}), as generally expected from the R-symmetric models \cite{Kribs:2007ac}.\footnote{See also Ref.~\cite{Dudas:2013gga} for the discussion of flavor observables in models with Dirac gauginos beyond the mass insertion approximation.}

To summarize, the three benchmark points are consistent with mentioned experimental measurements despite different values of Lagrangian parameters.

\begin{table}
  \centering
  \begin{tabular}{c|ccc}
    & $10^4 \cdot$BR($B \to X_s \gamma$) & BR($B^0_d \to \mu^+ \mu^-$) & $10^9  \cdot$BR($B^0_s \to \mu^+ \mu^-$) 
    \\
    \hline
    BMP1 & $3.47$ & $1.04 \cdot 10^{-10}$ & 3.24
    \\
    BMP2 & $3.45$ & $1.03 \cdot 10^{-10}$ & 3.19
    \\
    BMP3 & $3.35$ & $9.52 \cdot 10^{-11}$ & 2.95
    \\
    \hline
    \multirow{2}{*}{experiment} & 
    \multirow{2}{*}{$3.43 \pm 0.21 \pm 0.07$~\cite{Amhis:2014hma}} &
        \multirow{2}{*}{$< 7.4 \cdot 10^{-10}$~\cite{Aaij:2013aka}} & 
    $2.9^{+1.1}_{-1.0}$~\cite{Aaij:2013aka} \\
    & & & $3.0^{+1.0}_{-0.9}$~\cite{Chatrchyan:2013bka}
  \end{tabular}
  \caption{ 
  Values of selected $b$-physics observables for benchmark points 1-3 compared to experimental results.
  \label{tab:b_physics_observables}
  }
\end{table}
 
\begin{table}
\begin{center}
\begin{tabular}{l|ccc}
  & BMP1 & BMP2 & BMP3 \\
  \hline
$m_h$ & 125.3 GeV  & 125.7 GeV  & 125.4 GeV  \\
$m_W$ & 80.396 GeV & 80.382 GeV & 80.386 GeV \\
\texttt{HiggsBounds}'s \texttt{obsratio} (w/o t.u.) & $0.62$ & $0.68$ & $0.88$\\
\texttt{HiggsBounds}'s \texttt{obsratio} (w/ t.u.) & $0.58$ & $0.57$ & $0.88$\\
 \texttt{HiggsSignals}'s p-value (w/o t.u.) & 0.77 & 0.14 & 0.77 \\
 \texttt{HiggsSignals}'s p-value (w/ t.u.) & 0.79 & 0.81 & 0.82 \\
\end{tabular}
\caption{Collection of different predictions for the benchmark points
defined in \autoref{tab:BMP}. For details see the full text. 
Compared to the original publication \cite{Diessner:2014ksa} the $p$-value for BMP2 without theoretical uncertainty dropped significantly.
This is due to new data, which pulled the Higgs mass closer to 125 GeV. \label{tab:higgs_properties}
 }
\end{center}
\end{table}

\section{Contour plots \label{sec:contour_plots}}

Having discussed the properties of the benchmark points, I now come to the analysis of parameter space around them.
This is mainly to illustrate that there is nothing special about the choice of input parameters in \autoref{tab:BMP}.
To limit the number of plots, this analysis is shown only for BMP3.
Similar plots for other BMPs can be found in Refs.~\cite{Diessner:2014ksa, Diessner:2015yna}.

\begin{figure}
  \centering
  \includegraphics[width=0.99\textwidth]{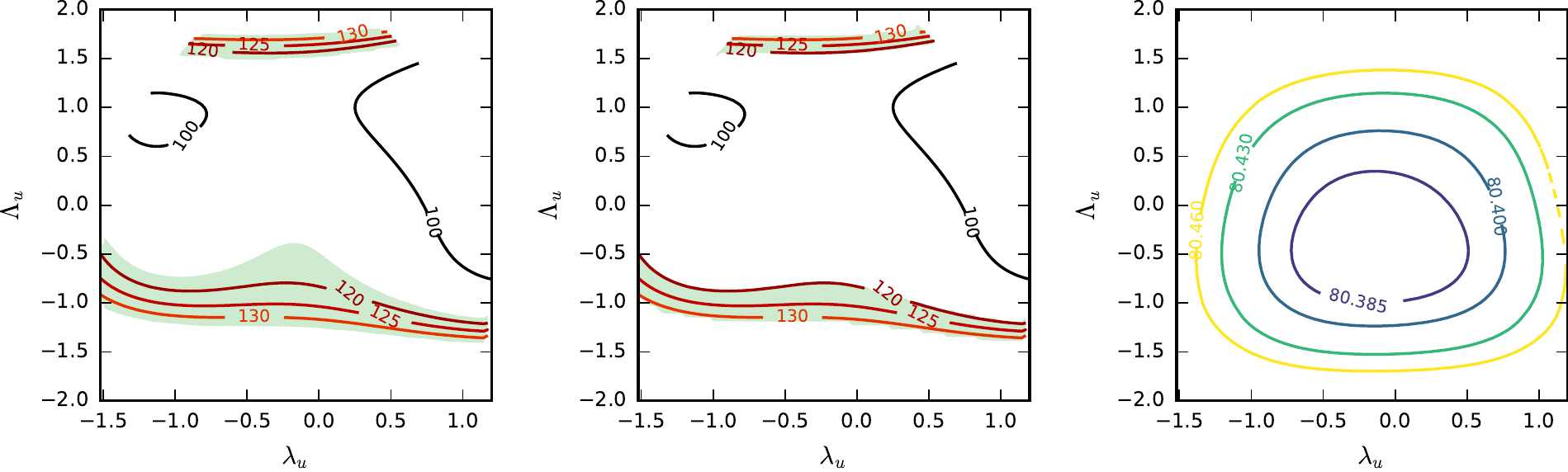}\\
  \includegraphics[width=0.99\textwidth]{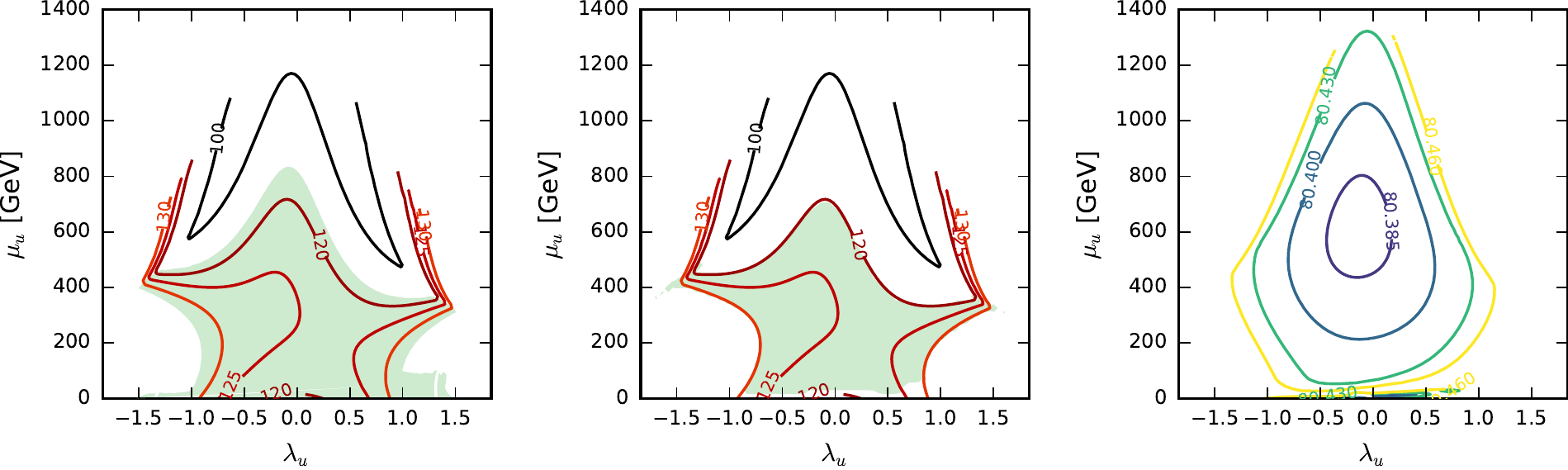}\\
  \includegraphics[width=0.99\textwidth]{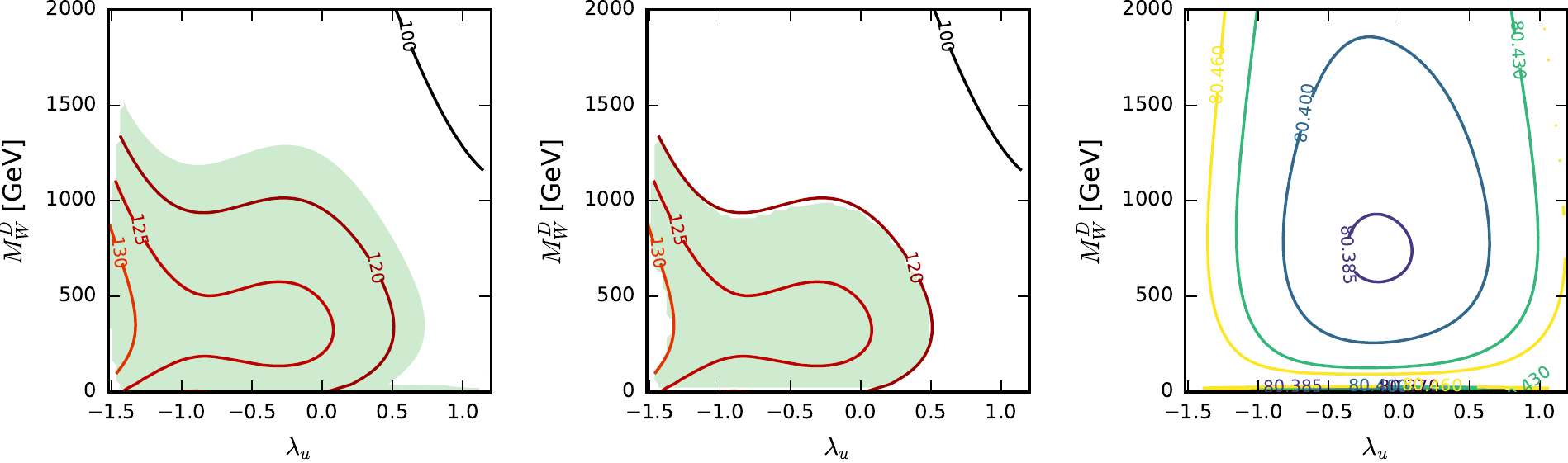}\\
  \includegraphics[width=0.99\textwidth]{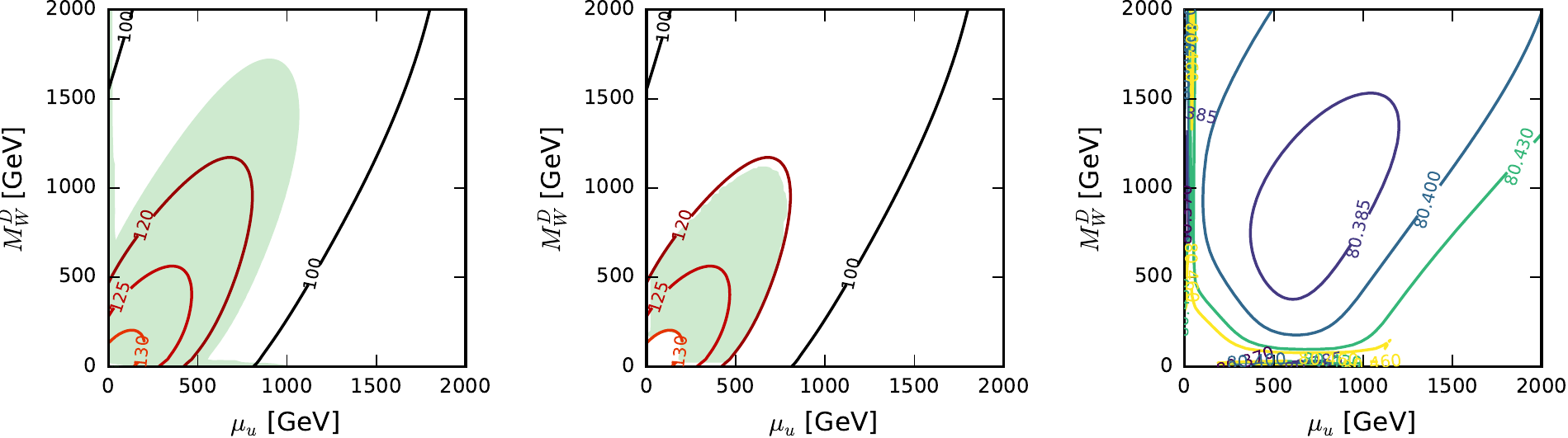}
\caption{
  Plots showing contours of constant Higgs (left and center column) and $W$ masses (right column) in the plane of two parameters as shown, with other parameters fixed as for BMP3. 
  Colored regions in the left column are allowed by \texttt{HiggsBounds}, in the center one by \texttt{HiggsBounds} and \texttt{HiggsSignals}.
  \label{fig:2d_scans}
}
\end{figure}

\autoref{fig:2d_scans} presents contour plots in the plane of two selected Lagrangian parameters while keeping all others fixed to the BMP3 values.
The left column shows the lightest Higgs mass contours for 100, 120, 125, 130 GeV together with the colored region, allowed at 95\% C.L. by \texttt{HiggsBounds}, while the middle column does the same for the \texttt{HiggsSignals}.
The right column shows predictions for the $W$ boson mass in the same parameter space with contours for 80.385, 80.4, 80.43, 80.46 GeV.
As can be seen in the plot, the \texttt{HiggsSignals} non-exclusion bound is more stringent and usually follows the assumed 5 GeV theoretical uncertainty on Higgs mass calculation.
The important feature of the $m_W - m_H$ interdependence is the removal of two-fold degeneration for some parameters such as $\Lambda_u$.
From the point of view of Higgs mass, both positive and negative $\Lambda_u$ are allowed. But since $m_W$ mass ellipses are shifted toward negative $\Lambda_u$, this puts large positive $\Lambda_u$ in conflict with precision EW observables.

\section{Multiparameter scan \label{sec:scatter_scans}}

To conclude the analysis of valid parameter space, this section presents multidimensional scans around BMP3.
The scan is done simultaneously for $\lambda_{u,d}, \Lambda_{u,d}, \mu_{u,d}, M_{1/2}$ and $\tan \beta$.
These parameters are scanned in the following ranges
\begin{eqnarray}
 -1.5  \leq & \lambda_i, \Lambda_i & \leq 1.5\\
  0  \leq &  \mu_{u,d},\, M^D_{S/T} & \leq 1 \text{ TeV}\\
  1.1 \leq & \tan \beta & \leq 60
\end{eqnarray}

\begin{figure}
\centering
\resizebox{0.49\textwidth}{!}{\input{img/mhmW/tanB.tex}} 
\resizebox{0.49\textwidth}{!}{\input{img/mhmW/LamSU.tex}}  
\includegraphics{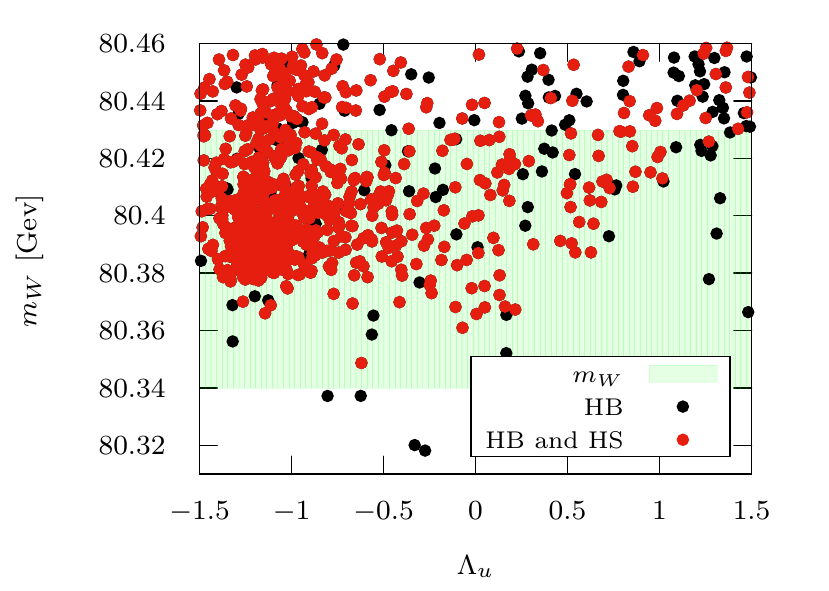}
\resizebox{0.49\textwidth}{!}{\input{img/mhmW/muU.tex}}  
\includegraphics{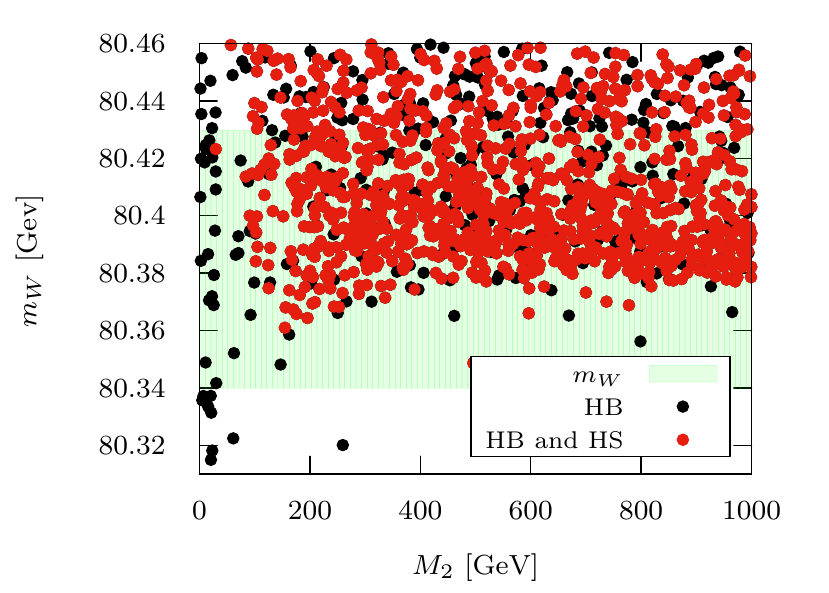} 
\caption{
  Multidimensional scans around BMP3 projected on $2d$ planes. 
  Green band shows 3$\sigma$ region around the measured value of $m_W$, $m_W = 80.385 \pm 3 \cdot 0.015$ GeV. 
  Black points are in agreement with \texttt{HiggsBounds} (HB), red points with both \texttt{HiggsBounds} and \texttt{HiggsSignals} (HS).
  \label{fig:scatter_scans}
}
\end{figure}

The basic requirement for an acceptable parameter point is that it be non-tachionic and have a neutralino or a sneutrino as a lightest supersymmetric particle.
They were then projected on the $m_W$ - $X$ plane, for $X \in \{ \tan \beta, \lambda_u, \Lambda_u, \mu_u, M^D_T \}$, and visualized in \autoref{fig:scatter_scans}.
The green band is the $3\sigma$ region around the measured value of $W$ boson mass.
White points are in agreement with \texttt{HiggsBounds}, while the red ones also pass the \texttt{HiggsSignals} test.
It is clear that there is no preference as far as the value of $\tan \beta$ is concerned.
Also, as discussed in the previous section, $m_W$ tends to prefer negative values of $\lambda_u$ and $\Lambda_u$.
A quite generic feature is that the contributions to the $T$ parameter from \autoref{eq:T-parameter_definition} scale like $T \propto \mu_u^{-2}$ (see Ref.~\cite{philips_phd} and the discussion around Eq.~4.16 in \cite{Diessner:2014ksa}).
This explains why $m_W$ goes out of green band when $\mu_u \to 0$ on 
\autoref{fig:scatter_scans}.

The multiparameter scan proves that there is nothing unique in the selected benchmark points.
Around each BMP exists a large, connected multidimensional parameter space predicting correct Higgs and $W$ boson masses but with other possibly distinct features while in Ref~\cite{Diessner:2015iln} a separate parameter region with the second-to-lightest Higgs as a SM-like Higgs was also identified.

\section{Conclussion \label{sec:mwmh_conclusions}}
This chapter ends the discussion of electroweak observables within the MRSSM.
\Autoref{sec:ew} showed that even despite the presence of $SU(2)_L$ triplet with a non-zero vacuum expectation value and an extended electroweak sector the model can be in agreement with the $W$ boson mass and other EW measurements.
\Autoref{sec:higgs_chapter} dealt with the analysis of the Higgs sector at tree, one-loop and two-loop levels, describing the Higgs boson mass calculation setup and identifying parameter regions where the SM-like Higgs state has a mass of around 125 GeV, even without the left-right mixing in the stop sector.
The lack of this contribution is compensated by new contributions from the extended Higgs sector with Yukawa-like couplings $\lambda, \Lambda$.

In the present chapter, all these results were combined, pointing to a large region of parameter space with interesting phenomenology.
Characteristically, this region features light stops and EW-inos well within the reach of Run 2 of the LHC.
Devised benchmark points were thoroughly tested, especially against data concerning the Higgs sector.
As such, the MRSSM presents a viable alternative to MSSM and NMSSM.

\chapter{Next-to-Leading Order QCD corrections to sgluon pair production \label{sec:sgluon_analytic}}

\begin{figure}[h!]
  \centering
  \includegraphics[width=0.4\textwidth]{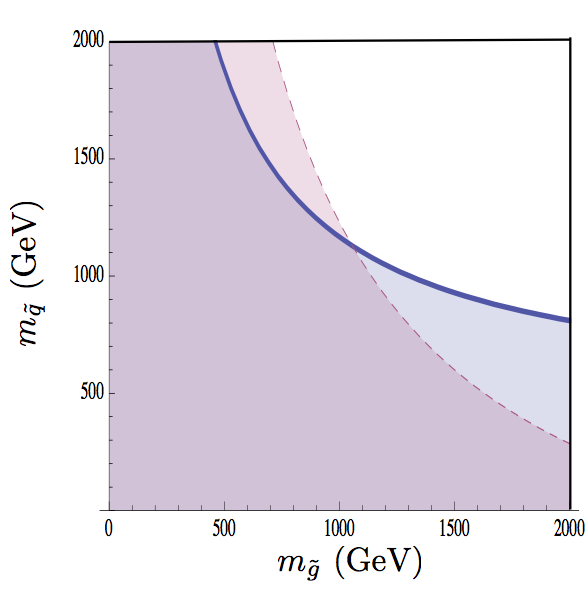}
  \caption{
  ATLAS exclusion limit \cite{Aad:2011ib} in the gluino-squark mass plane for the MSSM (solid line).
  In the model with Dirac gluinos, like MRSSM, an increased cross section for the production of gluinos would make the analysis more sensitive in the low gluino mass region, while the reduction of squark cross section, because of lack of certain squark production channels, would make squark exclusion weaker (see also Ref. \cite{Choi:2008pi}).
  This is represented by the dashed line.
  Figure comes from Ref. \cite{Heikinheimo:2011fk}.
  \label{fig:sanz}
  }
\end{figure}

After the discussion of the electroweak sector I now turn to the analysis of strongly interacting particles of the MRSSM.
As in the previous chapter, their phenomenology is different than in the MSSM.
As was shown in Refs.~\cite{Choi:2008pi,Heikinheimo:2011fk}, LHC exclusion limits for squarks (and gluinos) are altered in the MRSSM due to the Dirac nature of the gluino (see \autoref{fig:sanz}).
As suggested by high scale models of R-symmetric gauge mediation \cite{Amigo:2008rc}, Dirac gauginos might be significantly heavier than squarks, which makes them kinematically not accessible at the LHC.
In that case one is left only with squarks and a pair of (real) color octet scalars.
Moreover, since the masses of sgluons are split by the $D$-term contribution as discussed in \autoref{sec:particle_spectrum}, one of them will be substantially heavier than the other if the gluino is heavy.
The other one, whose mass is controlled at tree level only by a soft-breaking mass parameter $m_O$, might be light and produced copiously at the LHC through its coupling to gluons.\footnote{See the next chapter for the discussion of experimental limits on the sgluon pair production.}
Moreover, as mentioned in \autoref{sec:field_content}, sgluons are uncharged under R-symmetry and can be produced by, and decay directly to, SM particles.
It is therefore important to analyze their production at the LHC, as the discovery of a colored resonance decaying to SM particles without an LSP would be a smoking gun for the MRSSM.
In what follows the discussion is limited only to the sgluon sector.
For a more thorough discussion of the strongly interacting sector of the MRSSM, I refer to the forthcoming publication \cite{tobepublished}.

\section{Sgluons pair production at leading order \label{sec:sgluon_pair_production_@lo}}
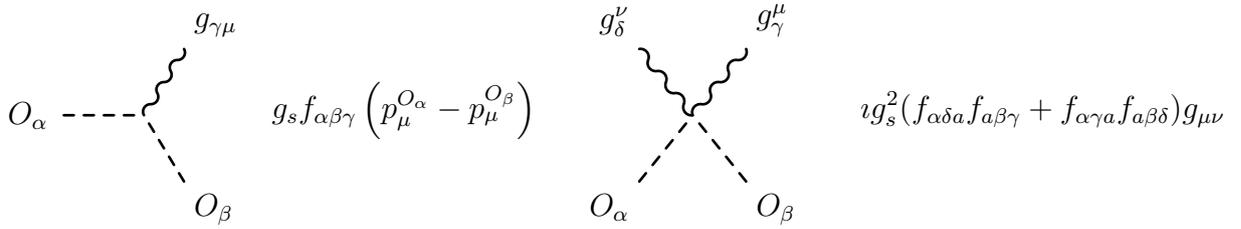
\begin{figure}
\begin{align*}
    \adjustbox{raise=-8.5ex}{
\begin{fmffile}{img/feynman_rules/OOg} 
\fmfframe(20,20)(20,20){ 
\begin{fmfgraph*}(50,50) 
\fmfleft{l1}
\fmfright{r1,r2}
\fmf{dashes}{l1,v1}
\fmf{dashes}{r1,v1}
\fmf{wiggly}{r2,v1}
\fmflabel{$O_{{\alpha}}$}{l1}
\fmflabel{$O_{{\beta}}$}{r1}
\fmflabel{$g_{{\gamma \mu}}$}{r2}
\end{fmfgraph*}} 
\end{fmffile} 
}
&
g_s f_{\alpha\beta\gamma} \left( p^{O_{{\alpha}}}_{\mu} - p^{O_{{\beta}}}_{\mu}\right)
\adjustbox{raise=-8.5ex}{
\begin{fmffile}{img/feynman_rules/OOgg} 
\fmfframe(20,20)(20,20){ 
\begin{fmfgraph*}(50,50) 
\fmfleft{l1,l2}
\fmfright{r1,r2}
\fmf{dashes}{l1,v1}
\fmf{dashes}{r1,v1}
\fmf{wiggly}{r2,v1}
\fmf{wiggly}{l2,v1}
\fmflabel{$O_{{\alpha}}$}{l1}
\fmflabel{$O_{{\beta}}$}{r1}
\fmflabel{$g_\gamma^\mu$}{r2}
\fmflabel{$g_\delta^\nu$}{l2}
\end{fmfgraph*}} 
\end{fmffile} 
}
&
\imath g_s^2 (f_{\alpha \delta a} f_{a \beta \gamma}  + f_{\alpha \gamma a} f_{a \beta \delta}) g_{\mu \nu}
\end{align*}
\caption{Non-SM Feynman rules for the simplified model given by Lagrangian from \autoref{eq:simp_lag}. \label{fig:simp_lag}}
\end{figure}

\begin{figure}
  \centering
  \includegraphics[width=0.76\textwidth]{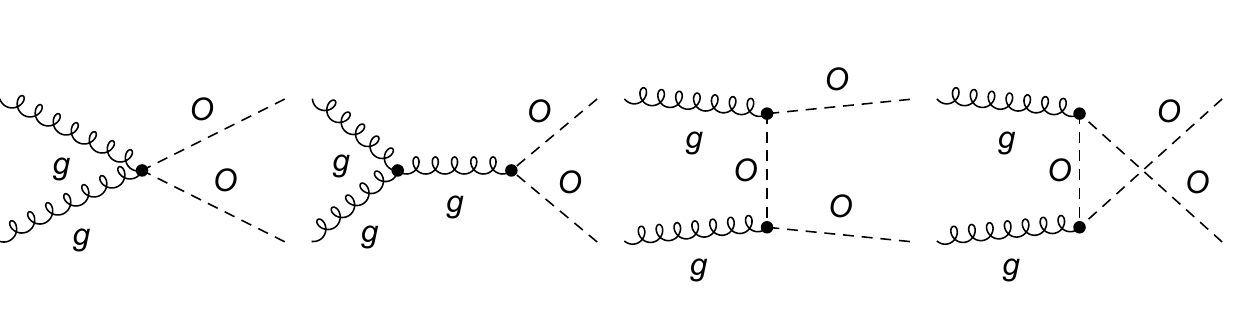}
  \includegraphics[width=0.19\textwidth]{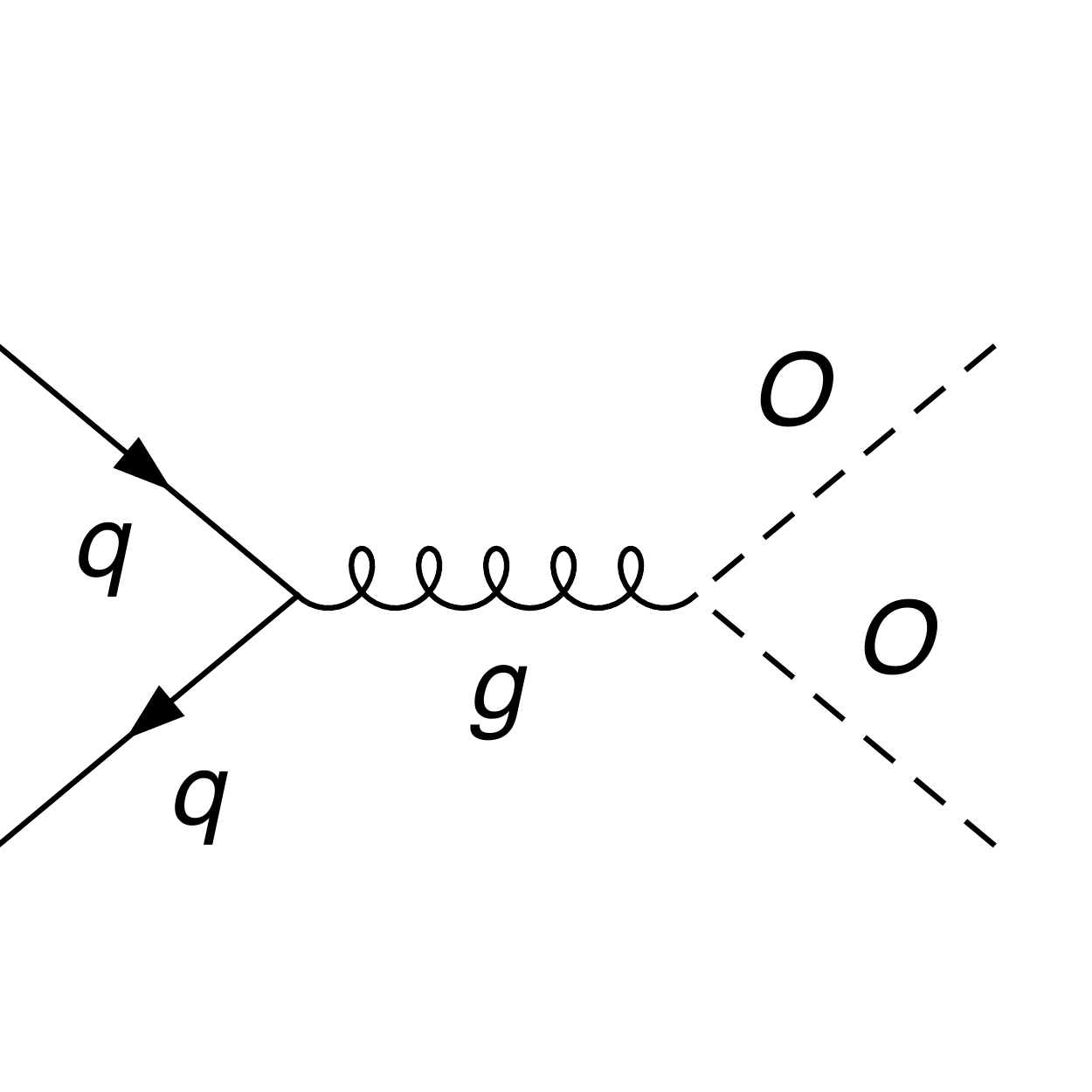}
  \caption{Diagrams for sgluon pair production at the LO. \label{fig:sgluon_feynman_diagrams_@LO}}
\end{figure}

At the LHC, sgluons can be produced in pairs via tree-level couplings to gluons (a model-dependent single sgluon production via loop-induced coupling to $gg$ or $q\bar q$ might become competitive only for heavy states).
If squarks and gluinos are heavy, as suggested by the LHC, sgluon pair production can be described by a simplified model given by the Lagrangian
\begin{equation}
\label{eq:simp_lag}
\mathcal{L} = \mathcal{L}_{\text{SM}} + \frac{1}{2} D^\mu O D_\mu O - \frac{1}{2} m_O^2 O^2 , 
\end{equation}
with the corresponding Feynman rules given in~\autoref{fig:simp_lag}.
For further analysis, a \texttt{FeynArts} add-on model (for the SM model file) with the Lagrangian of \autoref{eq:simp_lag} was created.
Sgluons are produced in $q\bar{q}$ or $gg$ channels through diagrams shown in \autoref{fig:sgluon_feynman_diagrams_@LO} with partonic cross sections

\begin{align}
\label{eq:lo_crosssection1}
  \hat{\sigma}_{q \bar{q}}^B & = \frac{2\pi\alpha_s^2}{9\hat{s}} \beta^3 ,\\
\label{eq:lo_crosssection2}
  \hat{\sigma}_{g g}^B & = \frac{3 \pi \alpha_s^2}{32 \hat{s}} \left ( 27 \beta - 17 \beta^3 +6 (-3 + 2\beta^2 + \beta^4 )\arctanh \beta \right  ),
\end{align}
where $\hat s=(p_q + p_{\bar{q}})^2$ or $(p_g + p_{g'})^2$ and $\beta$ is sgluon's velocity in the center of mass system of colliding partons.\footnote{\Autoref{eq:lo_crosssection1} and \autoref{eq:lo_crosssection2} differ by a factor of 1/2 compared to \cite{Choi:2008ub}, where authors assumed that sgluons are complex scalars.} 
\begin{figure}
  \centering
  \includegraphics{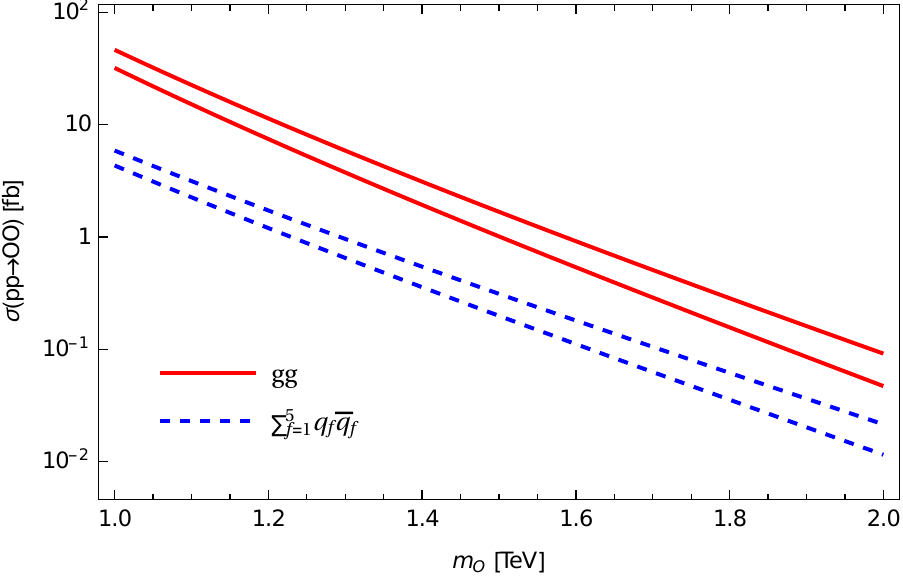}
  \caption{
  Cross sections per production channel for sgluon pair production at LO in $\alpha_s$. 
  The upper line of each type is provided for LHC's energy $\sqrt{S} = 14$ TeV, the lower 13 TeV.
  Renormalization and factorization scales were set equal to the sgluon mass, $\mu_F = \mu_R = m_O$. 
  \label{fig:s8_xsec_lo}
  }
\end{figure}

\begin{table}
  \centering
  \begin{tabular}{c|cc}
    \multirow{2}{*}{sgluon mass [TeV]} & 
    \multicolumn{2}{c}{cross section [fb]} \\
    & 13 TeV & 14 TeV \\
    \hline
    1  & $36.2^{+43.9\% + 8.5\%}_{-28.4\%-5.2\%}$ & $52.11^{+42.9\%+7.9\%}_{-28\%-4.9\%}$ \\
    1.25 & $6.14^{+45.7\%+10.4\%}_{-29.2\%-6.2\%}$ & $9.396^{+44.6\%+9.7\%}_{-28.7\%-5.8\%}$ \\
    1.5 & $1.208^{+47.4\%+12.5\%}_{-29.9\%-7.1\%}$ & $1.983^{+46.2\%+11.5\%}_{-29.4\%-6.6\%}$\\
    1.75 & $0.2602^{+49\%+14.7\%}_{-30.6\%-8\%}$ & $0.4606^{+47.6\%+13.5\%}_{-30\%-7.5\%}$\\
    2 & $0.05868^{+50.3\%+16.9\%}_{-31.1\%-8.9\%}$ & $0.1131^{+49\%+15.5\%}_{-30.6\%-8.3\%}$\\
  \end{tabular}
  \caption{
    Cross sections for sgluon pair production for $\sqrt{S}=$13 and 14 TeV LHC as a function of the sgluon mass.
    Cross sections are calculated using 5-flavor sgluon UFO model~\cite{Degrande:2011ua} from \texttt{FeynRules}~\cite{Alloul:2013bka} and \texttt{MadGraph5\_aMC@NLO}~\cite{Alwall:2014hca} at LO in $\alpha_s$ using \texttt{MMHT2014} LO PDF fit from \texttt{LHAPDF6}~\cite{Buckley:2014ana}.
    The first error comes from the scale variation, the second represents the PDF uncertainty (see main text for more details).
    Relative statistical errors are of the order of $10^{-3}$ and not shown here.
    \label{tab:sgluon_xsection_LO}
    }
\end{table}

The total hadronic cross section as a function of hadrons CMS energy $\sqrt S$ is given by a convolution of partonic cross sections and proton parton distribution functions (PDF) $f(x, \mu_F)$ as
\begin{equation}
  \sigma = \sum_{ij} \int_{4 m_O^2/S}^1 d x_1 \int_{4 m_O^2/(x_1 S)}^1 d x_2 \, f_i(x_1, \mu_F) \, f_j(x_2, \mu_F) \,\hat{\sigma}_{ij} (\sqrt{\hat{s}}, \mu_R^2) + \mathcal{O}(\Lambda_{QCD}^2/\hat{s}),
\end{equation}
where the last term represents non-factorizable (but $\hat s$-suppressed) contributions.
\Autoref{fig:s8_xsec_lo} shows the cross section plot as a function of a sgluon mass obtained using the \texttt{MMHT2014} \cite{Harland-Lang:2014zoa} baseline LO \ac{PDF} fit ($\alpha_s(m_Z) = 0.135$ and up to 5 active flavors).
As shown in \autoref{tab:sgluon_xsection_LO}, these cross sections exhibit large renormalization and factorization scale dependence (with much smaller PDF uncertainty).
Scale uncertainty is calculated as an envelope over a grid where renormalization and factorization scales are varied independently by a factor of 2 up and down.
PDF uncertainty comes from 50 PDF eigenvectors plus a central set, calculated using hessian method.

By an analogy with squark pair production \cite{Beenakker:1996ch}, NLO QCD corrections are expected to be large.
They will also reduce the scale dependence.
It therefore calls for further study.

\section{QCD corrections to sgluon pair production \label{sec:sgluon_pair_production_@nlo}}
I now turn to the calculation of NLO \ac{QCD} correction to sgluon pair production.
\footnote{
  The NLO corrections to the cross section for sgluon pair pair production for the complex version of Lagrangian of \autoref{eq:QCD_lag} are known since \cite{GoncalvesNetto:2012nt}, that calculation was based on \texttt{MadGolem} framework with Catani-Seymour dipole subtraction \cite{Catani:1996vz} (although authors say the result for virtual part was cross-checked against \texttt{FeynArts} + \texttt{FormCalc} + \texttt{LoopTools}~\cite{Hahn:1998yk}). 
  In this respect my calculation is more explicit in the treatment of IR singularities, extracting them using the so-called two cut method and giving compact expressions for (UV and IR) divergent parts of the cross section. It is also a first step in the direction of full NLO analysis of SQCD sector of the MRSSM, which is beyond the scope of this thesis.
  }
For this purpose, Lagrangian of \autoref{eq:simp_lag} needs to be supplemented by the counter-term Lagrangian
\begin{align}
  \label{eq:QCD_lag}
  \mathcal{L} = & \frac{1}{2} D^\mu O D_\mu O - \frac{1}{2} m_O^2 O^2 + \frac{1}{2} \delta Z_O \partial^\mu O \partial_\mu O - \frac{1}{2} \left( m_O^2 \delta Z_O + \delta m^2\right) O^2 
   \\
  & + \frac{1}{2} \left(\delta g_s + \delta Z_O + \frac{1}{2} \delta Z_g \right ) f^{abc} \left[ (\partial_\mu O_b) O_c - O_b  (\partial_\mu O_c) \right] g_a^\mu
  \nonumber \\
  & + \frac{1}{2} \left (2\delta g_s + \delta Z_O + \delta Z_g \right) (f_{\alpha \delta a} f_{a \beta \gamma}  + f_{\alpha \gamma a} f_{a \beta \delta}) O_{c} O_{e} g^\mu_{b} g_\mu^{d} \nonumber,
\end{align}
where $O$, $g$, $g_s$ and $m_O$ are the renormalized sgluon and gluon fields, QCD coupling and sgluon mass, respectively.
Renormalized parameters are connected to the bare ones in the usual way
\begin{align}
  O_0 = &\sqrt{Z_O} O \approx \left( 1+ \frac{1}{2} \delta Z_O \right) O, &
  g_0 = \sqrt{Z_g} g \approx \left (1 + \frac{1}{2} \delta Z_g \right) g ,
\end{align}
\begin{align}
   g_{s,0} & = g_s + \delta g_s ,\\
   m_{O,0}^2 & = m^2_O + \delta m^2_O .
\end{align}
The resulting counter-term Feynman diagrams do not differ in their structure from the Born ones and therefore are not shown here.

The rest of the section is structured as follows. Firstly, I discus the virtual amplitude and its renormalization.
Then I turn to the calculation of real-emission contribution, proving the cancelation of IR divergences.
The section finishes with combining results and giving final numbers for the NLO cross section for sgluon pair production for a few selected sgluon masses. 

\subsection{Virtual correction}

\begin{figure}
  \centering
  \includegraphics[width=0.95\textwidth]{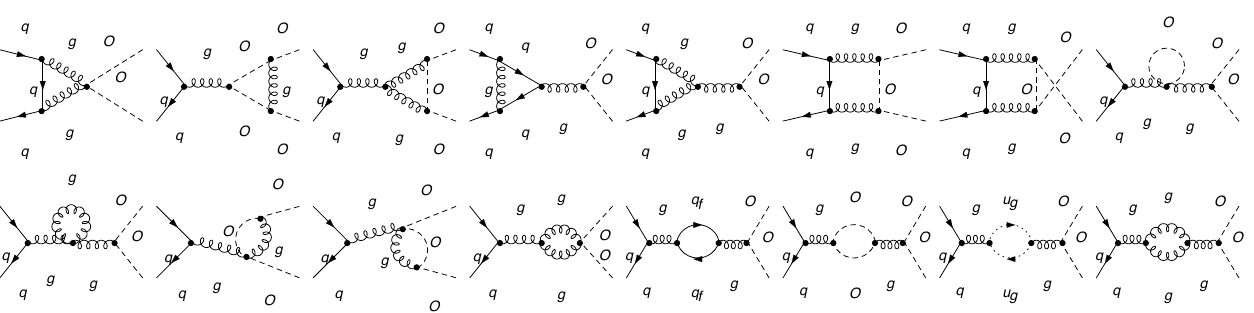}  \caption{
  Pure QCD corrections to the $q\bar q$ initiated-process (without wave-function corrections). 
  $q$ represents a quark with the flavor from \textit{up} to \textit{bottom}.
  A closed quark loop is summed over $f$, with $f$ from 1 to 6.  \label{fig:qqbar_to_s8s8_@NLO}}
\end{figure}

\begin{figure}
  \centering
  \includegraphics[width=0.95\textwidth]{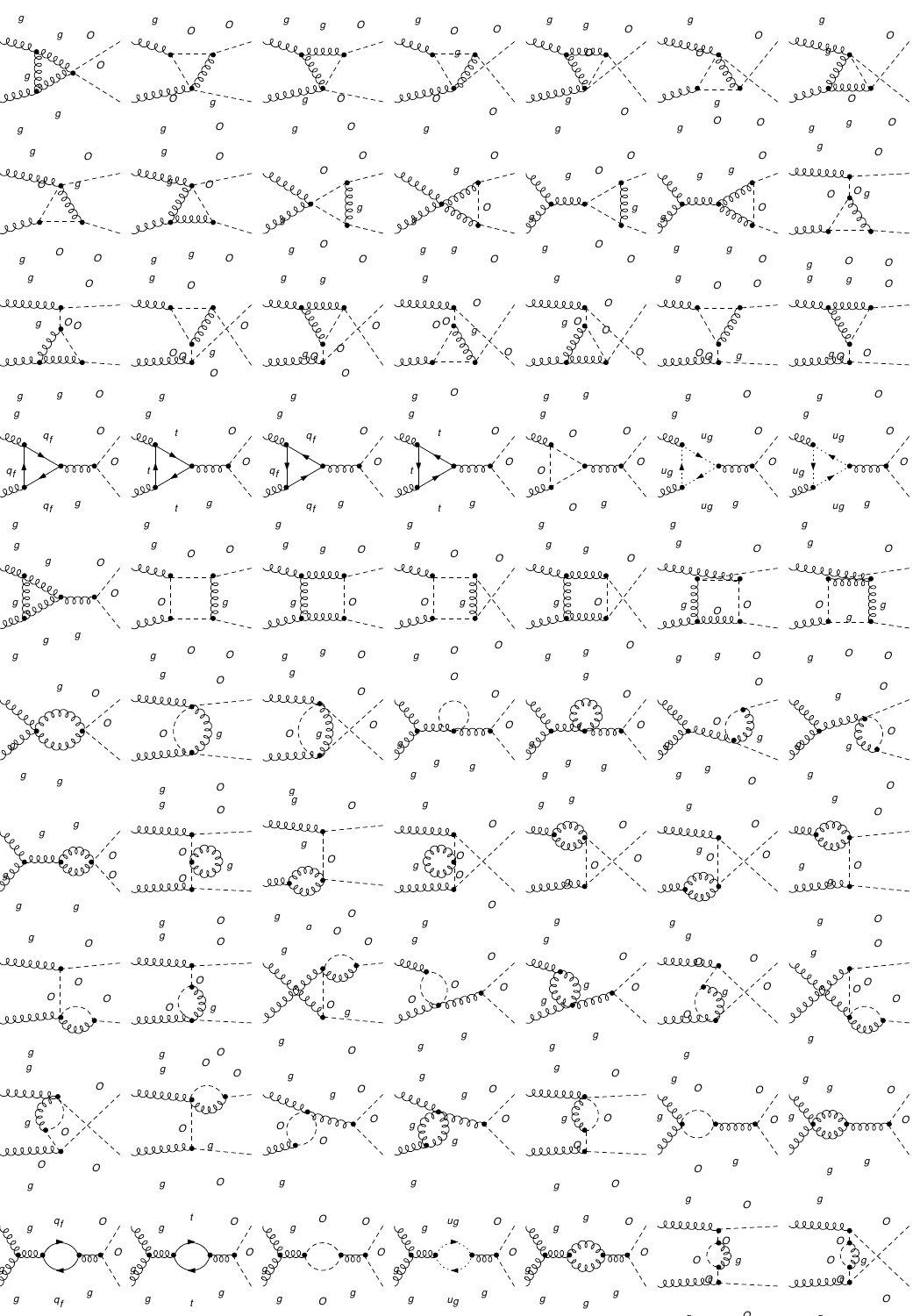}  \caption{
  Pure QCD corrections to the $gg$-initiated process (without external wave function corrections). Closed quark loops are summed over $q_f$ with $f$ from 1 to 5.  \label{fig:gg_to_s8s8_@NLO}}
\end{figure}

Loop diagrams contributing to sgluon pair production in $q\bar{q}$ and $gg$ processes are shown in \autoref{fig:qqbar_to_s8s8_@NLO} and \autoref{fig:gg_to_s8s8_@NLO}, respectively. 
Analytic expression for them was generated using \texttt{FeynArts} model and processed by \texttt{FormCalc}.

\begin{figure}
  \centering
  \subfloat[]{\includegraphics[width=0.95\textwidth]{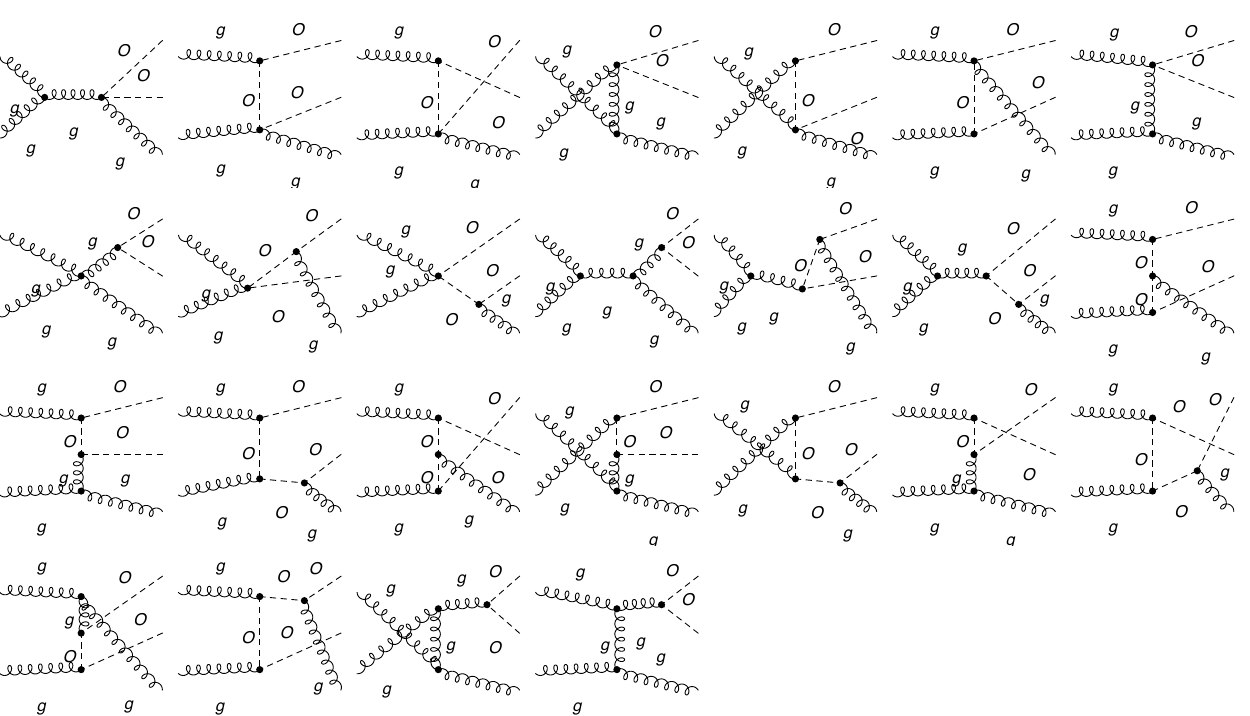}}\\
  \subfloat[]{\includegraphics[width=0.95\textwidth]{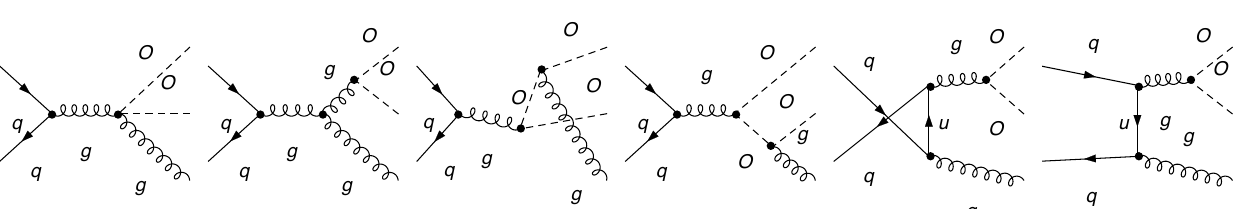}}
  \caption{Real emission contribution to the $gg$ (a) and $q\bar q$ (b) channels. \label{fig:pure_QCD_real_cor_to_s8s8}}
\end{figure}

Sgluons are renormalized \textit{on-shell}
\begin{align}
  \delta Z_O & = -\Re \left . \frac{\partial \Sigma_{OO}}{\partial k^2}\right|_{k^2=m_O^2}, & \delta m_O^2 & = \Re (\Sigma_{OO} (m_O^2)),
\end{align}

\begin{figure}
  \centering
  \subfloat[]{\includegraphics[width=0.35\textwidth]{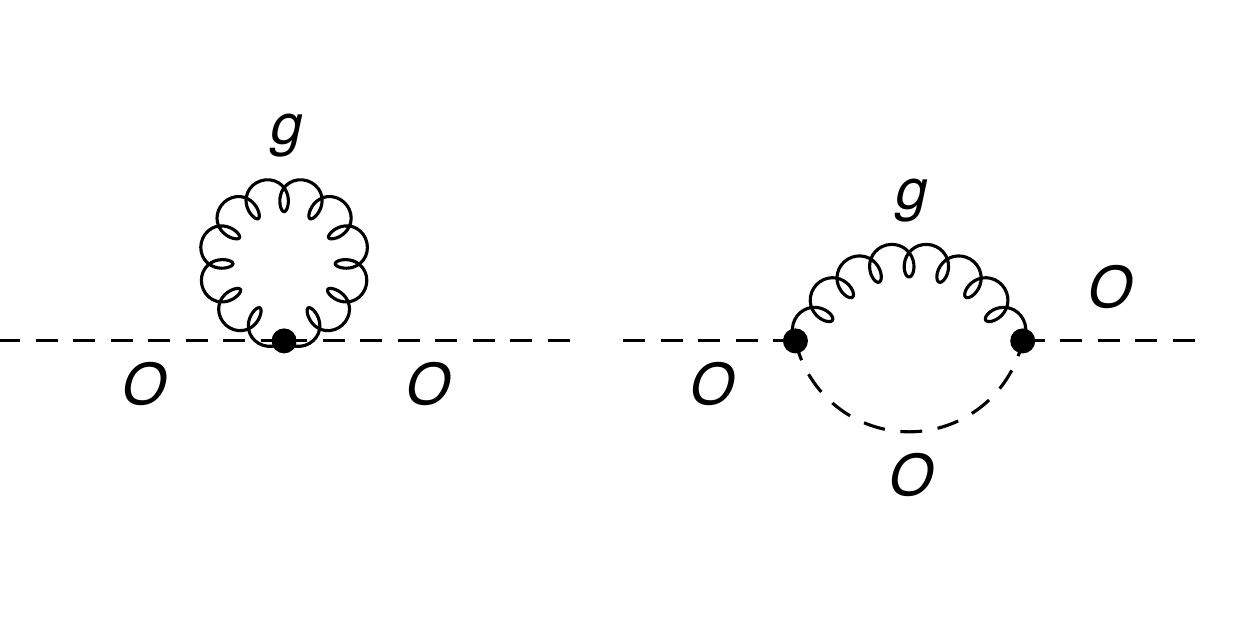}}
   \subfloat[]{\includegraphics[width=0.35\textwidth]{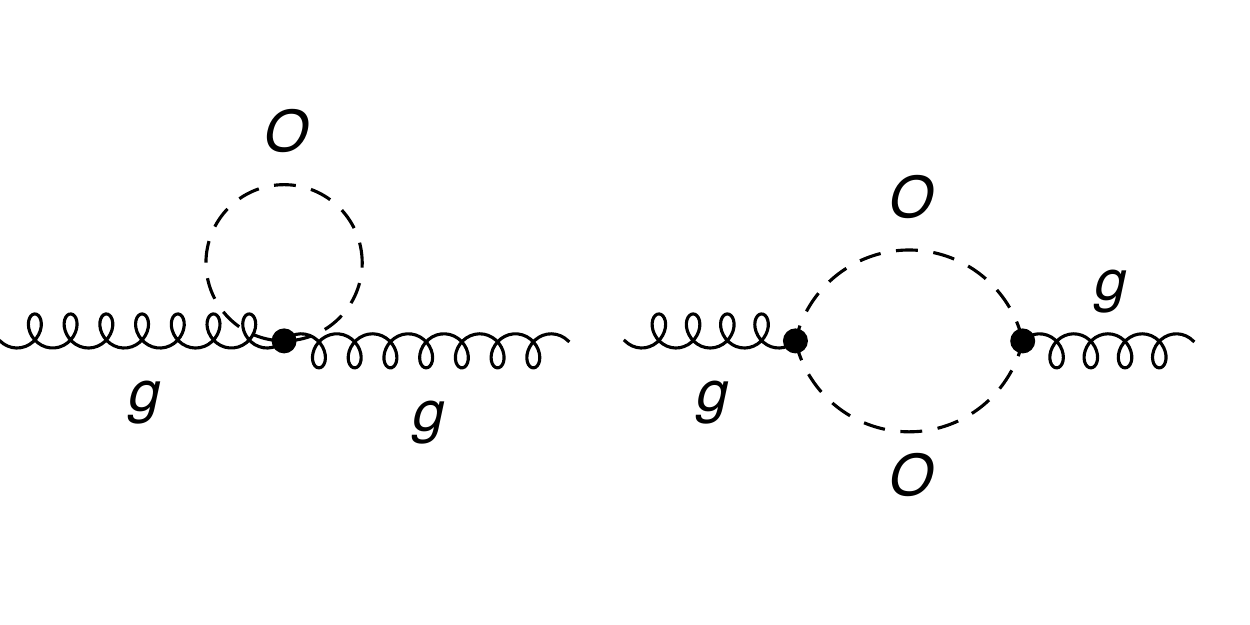}}
  \caption{Diagrams giving the wave function and mass renormalization constants for sgluon (a) and gluon (b). For gluon, only non-SM diagrams are drawn.
  \label{fig:wave_func_ren}
  }
\end{figure}

giving
\begin{eqnarray}
  \delta Z_O &=& \frac{3\alpha_s}{2\pi} (\Delta_{\text{UV}}-\Delta_{\text{IR}}),\\
  \delta m_O^2 & = &-\frac{3\alpha_s}{2\pi} m^2_O \Re \left [2 B_0(m^2_O,0,m^2_O) + B_1(m^2_O, 0, m^2_O) \right] \\
  & = &-\frac{3 \alpha_s}{4\pi} m_O^2 \left( 3\Delta_{\text{UV}} + 7 - 3 \log \frac{m_O^2}{\mu_R^2}  \right) ,
\end{eqnarray}
where $\Delta_i \equiv 1/\bar \epsilon_\text{i} \equiv 1/ \epsilon_i - \gamma + \ln 4\pi$ and where $\gamma \approx 0.577$ is the Euler-Mascheroni constant.
The expression for mass renormalization constant agrees with (B6) of \cite{GoncalvesNetto:2012nt}, but the wave function renormalization constant does not. 
They both agree with \cite{Degrande:2014sta}, though.

Since sgluon interacts at the tree-level only with gluons, it does not contribute to $\delta Z_1$ or $\delta Z_2$ counter-terms at one-loop level, giving SM result $\delta Z_1 - \delta Z_2 = - C_A \frac{\alpha_s}{4\pi} \Delta_{\text{UV}}$.
Contributions from heavy particles to the gluon self-energy, depicted in \autoref{fig:wave_func_ren}, are renormalized in zero-momentum subtraction scheme
\begin{equation}
\label{eq:MOM}
\left .  \frac{d \Sigma^{\text{light}}_{gg}(p^2)}{d p^2} \right |_{p^2=0,\,\Delta_{\text{UV}}} + 
\left.\frac{d \Sigma^{\text{heavy}}_{gg}(p^2)}{d p^2}\right|_{p^2=0} + \delta Z_g= 0 ,
\end{equation}
which yields
\begin{equation}
  \delta Z_g = \frac{\alpha_s}{4\pi} \left (\frac{5}{3} C_A - \frac{N_f-1}{2} C_F \right ) \frac{1}{\bar \epsilon_{\text{UV}}}
  - \frac{\alpha_s}{6\pi} \left (\frac{1}{\bar \epsilon_{\text{UV}}} - \log \frac{m_t^2}{\mu_R^2} \right )
  - \frac{\alpha_s}{8\pi} \left (\frac{1}{\bar \epsilon_{\text{UV}}} - \log \frac{m_O^2}{\mu_R^2} \right ) .
\end{equation}
The remaining wave-function correction is pure IR divergence\footnote{This correction enters with a minus sign multiplying the Born cross section.}
\begin{equation}
  \left.\frac{d \Sigma^{\text{total}}_{gg}(p^2)}{d p^2}\right|_{p^2=0} + \delta Z_g =\frac{5\alpha_s}{6\pi} \frac{1}{\bar \epsilon_{\text{IR}}} .
\end{equation}
Using the vertex counterterm to determine $\delta g_s$, one finds 
\begin{multline}
  \label{eq:dgs}
  \frac{\delta g_s}{g_s} = \delta Z_1 - \delta Z_2 -\frac{1}{2} \delta Z_g = \frac{\alpha_s}{4\pi} \left ( \frac{N_f - 1}{3} - \frac{11}{6} C_A\right) \frac{1}{\bar \epsilon_{\text{UV}}} \\
  + \frac{\alpha_s}{12 \pi} \left ( \frac{1}{\bar \epsilon_{\text{UV}}} - \log \frac{m_t^2}{\mu_R^2}\right) 
  + \frac{\alpha_s}{16 \pi} \left ( \frac{1}{\bar \epsilon_{\text{UV}}} - \log \frac{m_O^2}{\mu_R^2}\right) . 
\end{multline}
The QCD $\beta$ function can then be obtained in a standard way as
\begin{equation}
  \mu \frac{d}{d \mu} g_s^0 = \mu \frac{d}{d \mu} \mu^\epsilon g_s Z_{g_s} = \mu^\epsilon g_s Z_{g_s} \left (\epsilon + \frac{1}{g_s} \beta + \frac{\mu}{Z_{g_s}} \frac{d Z_{g_s}}{d \mu} \right ) ,
\end{equation}
where I defined $Z_{g_s} = 1-\delta g_s/g_s$, which evaluates to 
\begin{equation}
  \mu \frac{d Z_{g_s}}{d \mu} = - \frac{\alpha_s}{2\pi} \left ( \frac{N_f -1}{3} - \frac{11}{6} C_A \right ) .
\end{equation}
In the last equation, terms coming from top quark and sgluon drop out.
Combined with the previous equation, this finally gives the evolution of a strong coupling constant in this scheme
\begin{equation}
  \mu \frac{d \alpha_s}{d\mu} = - \frac{\alpha_s^2}{2\pi} \left ( \frac{11}{3} C_A - 2\frac{N_f-1}{3} \right ).
\end{equation}
$N_f = 6$, which means that the strong coupling constant evolution is given by 5-flavor $\beta$ function.
This motivates the use of the zero-momentum subtraction for heavy particles, as it allows easily to consistently combine the calculation with used PDFs, in which $\alpha_s$ also evolves with 5 flavors.

UV-finiteness of the final result was checked using \texttt{FormCalc} function \texttt{UVDivergentPart}, which replaces the Passarino-Veltman loop functions with their UV-divergent parts.
The result for the renormalized virtual part was cross-checked against a standalone \texttt{MadGraph5\_aMC@NLO} output, using NLO-capable 5-flavor UFO model generated with \texttt{FeynRules}~\cite{Degrande:2014vpa,Degrande:2014sta}, for a few phase space points showing perfect agreement. 

Even after UV renormalization, the result still contains infrared poles.
Their treatment is described in the next section.
%
%

\subsubsection{Note about the \texttt{LoopTools} convention}
In the \texttt{LoopTools} convention, the loop integrals are defined as 
\begin{equation}
  T^N_{\mu_1 \ldots \mu_P} = \frac{\mu^{4-D}}{\imath \pi^{D/2} r_\Gamma} \int d^D q \frac{q_{\mu_1} \ldots q_{\mu_P}}{[q^2-m_1^2][(q+k_1)^2-m_2^2]\ldots [(q+k_{N-1})^2-m_N^2]},
\end{equation}
where 
\begin{equation}
  r_\Gamma =  \frac{\Gamma^2(1-\epsilon) \Gamma(1+\epsilon)}{\Gamma(1-2\epsilon)} \text{ and } D= 4-2\epsilon .
\end{equation}
Compared to the standard definition of the loop integral measure, which is
\begin{equation}
  \frac{\mu^{2\epsilon} \text{d}^D k}{(2\pi)^D} = 
  \frac{\imath}{16 \pi^2} \cdot
  (4\pi)^\epsilon \cdot 
  \frac{ \mu^{2\epsilon} \text{d}^D k}{\imath \pi^{D/2}} =
  \frac{\imath}{16 \pi^2} \cdot 
  (4\pi)^\epsilon r_\Gamma \cdot
  \frac{\mu^{2\epsilon}}{\imath \pi^{D/2} r_\Gamma},
  \end{equation}
one misses the factor of
\begin{equation}
    (4\pi)^\epsilon r_\Gamma = (4\pi)^{\epsilon} \frac{\Gamma^2(1-\epsilon)\Gamma(1+\epsilon)}{\Gamma(1-2\epsilon)} = \frac{(4\pi)^\epsilon}{\Gamma(1-\epsilon)} + \mathcal{O}(\epsilon^3),
\end{equation}
since \texttt{LoopTools} includes prefactor $-\imath (2\pi)^{-4 \cdot \texttt{loop number}}$ in the final result.\footnote{The global minus sign in this prefactor does not matter as long as one considers squared amplitudes.}
It should be noted that not including this factor for UV-poles corresponds to working in the $\overline{\text{MS}}$-scheme, as
\begin{equation}
  \frac{(4\pi)^\epsilon}{\Gamma(1-\epsilon)} = 1 + (- \gamma + \ln 4\pi ) \epsilon + \mathcal{O}(\epsilon^2).
\end{equation}
This factor does matter, though, for IR-divergent contributions, where \texttt{LoopTools} results are combined with real emissions. 

\subsection{Real-emission corrections}

Even after the renormalization, the virtual matrix element still contains singularities.
These are of soft and/or collinear origin. 
Soft singularity is canceled between virtual and real emission corrections.
This is known as the Kinoshita-Lee-Nauenberg theorem \cite{Kinoshita:1962ur, PhysRev.133.B1549}. 
Initial state collinear singularities are removed by mass factorization~\cite{Collins:1985ue,Bodwin:1984hc}.

The treatment of infrared singularities follows the two-cut phase space slicing method, as documented in Ref.~\cite{Harris:2001sx}. 
The main steps are briefly outlined below.

For the considered process, 3-body real emission phase space is decomposed into soft and/or collinear parts with respect to the additional parton. 
This can be schematically written out for the (formally infinite) real emission cross section $\sigma_R$ as

\begin{equation}
  \sigma_R = \int d \sigma_R = \int_S d \sigma_R + \int_H d \sigma_R = \int_S d \sigma_R + \int_{HC} d \sigma_R + \int_{H\overline{C}} d \sigma_R,
\end{equation}
where in the last step the collinear singular part HC was extracted from the hard part H, leaving finite hard non-collinear reminder H$\overline{\text{C}}$. 

\subsubsection{Soft emissions} 
The soft phase space is defined by the condition that the energy of an outgoing gluon $E_5$ in the rest frame of colliding partons fulfills
\begin{equation}
  \label{eq:gluon_en_soft_lim}
  E_5 < \delta_s \frac{\sqrt{\hat s}}{2},
\end{equation}
where $\hat s$ represents the Mandelstam variable for the incoming partons. 
$\delta_s$ is an arbitrary parameter, of which the final result should be independent.
In the soft limit and $4-2\epsilon$ dimensions the $2 \to 3$ amplitude can be written as
\begin{equation}
  M_{3}^{\text{soft}} = g_s \mu^{\epsilon} \epsilon_\mu(p_5) \,\textbf{J}^{\mu} (p_5) \cdot \textbf{M}_2 + \text{finite terms},
  \label{eq:eik}
\end{equation}
where $p_5$ is the gluon 4-momentum and
\begin{equation}
  \textbf{J}^{\mu} (p_5) = \sum_{f=1}^4 \textbf{T}_f \frac{p_f^\mu}{p_f \cdot p_5}
\end{equation}
is the non-abelian eikonal current (with the sum over particles except for the final-state gluon), which is color-connected with the $2 \to 2$ process $M_2$ through generator $T_f$ of $SU(3)_C$ representation for particle $f$. 
"Finite terms" denotes sub-amplitudes, which are not singular in the soft limit.
After taking the modulus squared of \autoref{eq:eik}, one obtains the general expression of the form 
\begin{equation}
  |\mathcal{M}|^2 \sim \sum_{f,f'=1}^4 \frac{p_f \cdot p_{f'}}{(p_f \cdot p_g)(p_{f'}\cdot p_g)} + \text{less singular terms} .
  \label{eq:eik2}
\end{equation}
In the soft limit the 3-body phase space factorizes as
\begin{equation}
  d\Phi_3^{\text{soft}} = d\Phi_2 \cdot \left(\frac{4\pi}{\hat s}\right)^\epsilon \frac{\Gamma(1-\epsilon)}{\Gamma(1-2\epsilon)} \frac{1}{2(2\pi)^2} dS,
  \label{eq:eik4}
\end{equation}
where 
\begin{equation}
  dS = \frac{1}{\pi} \left (\frac{\hat s}{4}\right)^{\epsilon} E_5^{1-2\epsilon} \, d E_5 \, \sin^{1-2\epsilon} \theta_1 \, d \theta_1 \sin^{-2\epsilon} \theta_2 \, d \theta_2,
\end{equation}
and the phase space is integrated over $E_5$ from 0 to $\delta_s \sqrt{\hat s}/2$.
\Autoref{eq:eik2}, as far as the dependence on $E_5$ is concerned, produces 
\begin{equation}
  \left (\frac{\hat s}{4}\right)^\epsilon \int_0^{\delta_s \sqrt{\hat s}/2} d E_5 E_5^{1-2\epsilon - n} = \frac{2^{-2+n}}{2-n-2\epsilon} \hat s^{1-\frac{1}{2}n} \delta_s^{2-n-2\epsilon}
  \label{eq:eik3}
\end{equation}
for $n \leq 2$ and $\epsilon < 0$. 
For $n=2$, this gives
\begin{equation}
  - \frac{1}{2\epsilon} \delta_s^{-2\epsilon} = - \frac{1}{2\epsilon} + \ln \delta_s + \mathcal{O}(\epsilon^2),
\end{equation}
while for $n = 1$, which corresponds to the "less singular terms" of \autoref{eq:eik2}, is regular in $\epsilon$ and goes to 0 as $\delta_s \to 0$.
Since the phase space volume also goes to 0 as $\delta_s \to 0$, this justifies neglecting this part as long as $\delta_s \ll 1$.

The necessary angular integrals are listed in \autoref{a:ang_integrals}.

Following this discussion, the singular parts of soft cross sections for the partonic channels $\sigma_{q\bar{q}}$ and $\sigma_{gg}$ are given by
\begin{align}
  \label{eq:qq_channel_soft}
  \sigma_{q\bar{q}}^{soft} = &\frac{8 \alpha_s^3 \beta^3}{27 s} \frac{1}{\epsilon^2} + 
  \frac{\alpha_s^3}{81 s} \bigg [ -27\beta + 2(83-24\gamma)\beta^3 + 27(-1+4\beta^2+\beta^4)\arctanh \beta \\
&  \left .- 48 \beta^3 \log \left ( \frac{\hat s \beta \delta_s}{4\pi \mu_R^2} \right)
   \right ] \frac{1}{\epsilon} 
   + \text{finite} \nonumber, \\
  \sigma_{gg}^{soft} = & \frac{9 \alpha_s^3}{32 s} \left(27 \beta - 17 \beta^3 + 6(-3+2\beta^2+\beta^4) \arctanh \beta \right) \frac{1}{\epsilon^2} + \mathcal{O}(1/\epsilon).
\end{align} 
The finite parts of both expressions, as well as the single pole part for the $gg$ channel, are lengthy and not given here explicitly.
The double-pole terms are proportional to the 4-dimensional Born cross sections in \autoref{eq:lo_crosssection1}
\begin{align}
 \left . \sigma_{q\bar{q}}^{soft}\right |_{\text{double pole}} = & 
 \sigma_{q\bar{q}}^{B} \cdot 2 \frac{\alpha_s}{2\pi} C_F \cdot \frac{1}{\epsilon^2} , \\
\left.  \sigma_{gg}^{soft} \right |_{\text{double pole}}= & 
\sigma_{gg}^{B} \cdot 2 \frac{\alpha_s}{2\pi} C_A \cdot \frac{1}{\epsilon^2} .
\end{align}

The cancelation of double poles between the soft and virtual parts was checked numerically, giving no less than 14 digits of relative accuracy (depending on the partonic channel and phase space point).

The single-pole coefficient is not canceled completely between virtual and soft contribution.
Remaining terms have the form
\begin{align}
\label{eq:cos1}
 \left . \sigma_{q\bar{q}}^{soft}\right |_{\text{soft-collinear remainder}} = & 
 - \frac{1}{\epsilon} \,\sigma_{q\bar{q}}^{B} \cdot 2 \, \frac{\alpha_s}{2\pi} C_F (3/2 + 2\log \delta_s),
 \\
 \label{eq:cos2}
\left.  \sigma_{gg}^{soft} \right |_{\text{soft-collinear remainder}}= & - \frac{1}{\epsilon} \,\sigma_{gg}^{B} \cdot 2 \, \frac{\alpha_s}{2\pi} [2N \log \delta_s + (11 N -2 N_f)/6 ] .
\end{align}
They come from the region of the phase space where the gluon is collinear with an incoming parton, but its energy is non-zero.
These terms will not cancel out with the virtual contribution as they have different kinematics.
As discussed in the next subsection, these are the terms that will cancel out with the soft-collinear pieces of the initial state factorization counterterms.

\subsubsection{Collinear emissions}
In the collinear limit, the real-emission cross section factorizes also at the level of matrix element squared.
The double differential hadronic cross section is given by
\begin{multline}
\label{eq:PDFconv2}
\frac{d \sigma^{HC}}{d x_1 d x_2}  =  \sum_{ij} \frac{2}{1+\delta_{ij}} \hat \sigma^B_{ij} \frac{\alpha_s}{2 \pi} \frac{\Gamma(1-\epsilon)}{\Gamma(1-2\epsilon)} \left(\frac{4 \pi \mu_R^2}{\hat s} \right)^\epsilon \left ( - \frac{1}{\epsilon} \right )\delta_c^{-\epsilon} \\ 
 \sum_k \left ( \int_{x_1}^{1-\delta_s \delta_{ik}} \frac{d z}{z} f_{k/p} \left( \frac{x_1}{z}\right ) f_{j/p} \left( x_2\right ) 
P_{ik}(z, \epsilon)  \left [\frac{1-z}{z} \right]^{-\epsilon} + \right.
\\ 
\left. \int_{x_2}^{1-\delta_s \delta_{jk}} \frac{d z}{z} f_{k/p} \left( x_1 \right ) f_{j/p} \left( \frac{x_2}{z} \right ) 
P_{ik}(z, \epsilon)  \left [\frac{1-z}{z} \right]^{-\epsilon} \right).
\end{multline}
The factor $2/(1+\delta_{ij})$ accounts for two possible ways in which $q\bar q$ in the initial state can be obtained from the proton-proton system.
$P_{ik}(z, \epsilon)$ are the $D$-dimensional unregulated Altarelli-Parisi splitting kernels \cite{Altarelli:1977zs}
\begin{align}
  P_{qq}(z, \epsilon) = & C_F \left[ \frac{1+z^2}{1-z} - \epsilon(1-z)\right],\\
  P_{gg}(z, \epsilon) = & 2 N \left ( \frac{z}{1-z} + \frac{1-z}{z} + z(1-z) \right) .
\end{align}
$\delta_{ik}$ in the integration boundaries of \autoref{eq:PDFconv2} ensures that for kernels which are singular as $z \to 1$ (so $P_{qq}$ and $P_{gg}$), the integral is taken up to $z = 1 - \delta_s$.

The Bjorken variable in $f_{k/p}$ is rescaled so that the Born configuration $\hat \sigma_B$ is taken at $\hat{s} = x_1 x_2 S$.\footnote{For quark radiation, the integral will be taken up to 1 as there is no soft singularity in this case.}
Collinear singularities will cancel out with the renormalized PDFs. 
The first-order correction to $i$-th flavor PDF in the $\overline{\text{MS}}$ prescription is given by

\begin{equation}
\label{eq:sdPDFs}
  f_{i/p} (x, \mu_F) \equiv f_{i/p}(x) - \frac{1}{\epsilon} \left [ \frac{\alpha_s}{2 \pi} \frac{\Gamma(1-\epsilon)}{\Gamma(1-2\epsilon)} \left(\frac{4 \pi \mu_R^2}{\mu_F^2} \right)^\epsilon \right ] \sum_j \int_x^1 \frac{d z}{z} P_{ij}^+(z) f_{j/p} (x/z),
\end{equation}
where $P_{ij}^+(z)$ are the '+' regulated Altarelli-Parisi splitting kernels
\begin{align}
  P_{qq}^+(z) = & C_F \left ( \frac{1+z^2}{(1-z)_+} + \frac{3}{2} \delta(1-z)\right),\\
  P_{gg}^+(z) = & 2 N \left ( \frac{z}{(1-z)_+} + \frac{1-z}{z} + z(1-z) \right) + (11 N - 2 N_f)/6 \, \delta(1-z),
\end{align}
where '+' prescription is defined as 
\begin{equation}
  \int_x^1 \frac{f(z)}{(1-z)_+} \equiv \int_x^1 \frac{f(z)-f(1)}{1-z}.
  \end{equation}
For partonic processes which have soft singularity there is a mismatch in the $z$ integration boundary between \autoref{eq:PDFconv2} and \autoref{eq:sdPDFs}.
\Autoref{eq:sdPDFs} can be rewritten as
\begin{align}
\label{eq:sdPDFs2}
  f_{i/p} (x, \mu_F) \approx & f_{i/p}(x)\left [ 1  - \frac{1}{\epsilon}  \frac{\alpha_s}{2 \pi} \frac{\Gamma(1-\epsilon)}{\Gamma(1-2\epsilon)} \left(\frac{4 \pi \mu_R^2}{\mu_F^2} \right)^\epsilon A_1^{sc} \right] 
  \nonumber \\
  & - \frac{1}{\epsilon} \frac{\alpha_s}{2 \pi} \frac{\Gamma(1-\epsilon)}{\Gamma(1-2\epsilon)} \left(\frac{4 \pi \mu_R^2}{\mu_F^2} \right)^\epsilon \int_x^{1-\delta_s} \frac{d z}{z} P_{ij}(z) f_{j/p} (x/z),
\end{align}
where now the unregularized AP functions appear and soft-collinear factors $A_1^\text{sc}$ for the splittings with soft gluon $(g)$ are given by
\begin{align}
  A_1^\text{sc}(q \to q (g)) & = C_F ( 2 \ln \delta_s + 3/2 ) ,\\
  A_1^\text{sc}(g \to g (g)) & = 2 N \ln \delta_s + (11 N + 2 N_f)/6 .
\end{align}

Solving \autoref{eq:sdPDFs} for $f(x)$ in the lowest order in $\alpha_s$ and convoluting with the Born cross section gives 
\begin{multline}
\label{eq:PDFconv}
\frac{d \sigma_{ij}}{d x_1 d x_2} = \sum_{ij} \frac{2}{1+\delta_{ij}} \sigma^B_{ij} \left \{ f_{i/p} (x_1, \mu_F) f_{j/p} (x_2, \mu_F) \left (1 
+ \frac{\alpha_s}{2 \pi} \frac{\Gamma(1-\epsilon)}{\Gamma(1-2\epsilon)} \left(\frac{4 \pi \mu_R^2}{\mu_F^2} \right)^\epsilon \frac{1}{\epsilon} A^\text{sc} \right) 
  \right. \\ 
 + \sum_k 
\frac{\alpha_s}{2 \pi} \frac{\Gamma(1-\epsilon)}{\Gamma(1-2\epsilon)} \left(\frac{4 \pi \mu_R^2}{\mu_F^2} \right)^\epsilon \frac{1}{\epsilon}
\left [ \int_{x_1}^{1-\delta_s \delta_{ik}} \frac{d z}{z}  f_{k/p} \left( \frac{x_1}{z}, \mu_F \right )  f_{j/p} \left( x_2, \mu_F \right ) 
P_{ik}(z) \right. \\
 \left. \left. + \int_{x_2}^{1-\delta_s \delta_{ij}}
f_{i/p} \left( x_1, \mu_F \right ) f_{k/p} \left( \frac{x_2}{z}, \mu_F \right )   
P_{jk}(z)
 \right ]  \right \}.
\end{multline}
The first term is just the Born partonic cross section convoluted with the scale-dependent PDFs.
The term $A_{\text{sc}}$, coming from the part of the integration of '+' regularized AP splitting functions, cancels out with \autoref{eq:cos1} and \autoref{eq:cos2}.
In other words, it is absorbed into the renormalized PDFs.
Adding now \autoref{eq:sdPDFs2} and \autoref{eq:PDFconv} gives the final result for the hard-collinear part
\begin{multline}
   \frac{d \sigma}{d x_1 d x_2} = \sum_{ij} \frac{2}{1+\delta_{ij}}  
   \hat \sigma^B_{ij} \left \{ \frac{\alpha_s}{2 \pi} \frac{\Gamma(1-\epsilon)}{\Gamma(1-2\epsilon)} \left (\frac{4 \pi \mu_R^2}{\hat s} \right)^\epsilon  \frac{1}{\epsilon} \right.
   \\ 
   \int_{x_1}^{1-\delta_s} \frac{dz}{z} \left[ \left( \left(\frac{\hat s}{\mu_F^2} \right)^\epsilon 
   P_{ik}(z) - 
   \delta_c^{-\epsilon} P_{ik}(z, \epsilon)  \left [\frac{1-z}{z} \right]^{-\epsilon} \right) f_{i/p} (x_1/z, \mu_F) f_{j/p} (x_2, \mu_F) + x_1 \leftrightarrow x_2 \right) \\
   = \sum_{ij} \frac{2}{1+\delta_{ij}} 
   \hat \sigma^B_{ij} \{ f_{i/p} (x_1, \mu_F) f_{j/p} (x_2, \mu_F) \left (1    + 2 \frac{1}{\epsilon}  \frac{\alpha_s}{2 \pi} \frac{\Gamma(1-\epsilon)}{\Gamma(1-2\epsilon)} \left(\frac{4 \pi \mu_R^2}{\mu_F^2} \right)^\epsilon A_1^{sc} \right )\\
   + \frac{\alpha_s}{2\pi} \int_{x_1}^{1-\delta_s} \frac{dz}{z} \left [  \left(P_{ik}(z) \ln \left( \delta_c \frac{1-z}{z} \frac{\hat s}{\mu_F^2} \right) - P_{ik}'(z)\right) f_{i/p} (x_1/z, \mu_F) f_{j/p} (x_2, \mu_F) + x_1 \leftrightarrow x_2 \right ] .
\end{multline} 

It was checked numerically that adding the virtual, soft- and hard-collinear contributions cancels the single-pole coefficient with the relative accuracy of at least 10 digits.
\subsubsection{Hard non-collinear part}
After adding virtual, soft- and hard-collinear contributions the result is finite, although still dependent on the regulators $\delta_s$ and $\delta_c$. 
This dependence vanishes for inclusive observables, i.e. after adding hard non-collinear part of the real-emission contribution.
Fortunately, this part does not pose any conceptual difficulty as it is finite and can be evaluated from the start in four dimensions using numerical integration.

A $\texttt{C++}$ library containing $2 \to 3$ matrix element was created using \texttt{MadGraph5\_aMC@NLO} and used in the standalone $\texttt{C++}$ code developed for this work.
The code was linked against the \texttt{rk} library~\cite{RK}, which was used to perform the Euler rotations and \texttt{LHAPDF6} to access \texttt{MMHT2014} PDF grids.

\subsection{Numeric results for the total cross section}
Numerical calculation was divided into two parts.
The soft, collinear and virtual pieces were evaluated in \textit{Mathematica}~\cite{mathematica} while the hard non-collinear part was evaluated in the standalone $\texttt{C++}$ code.
Numerical results were obtained using the \texttt{MMHT2014} NLO fit with $\alpha_s(m_Z) = 0.12$ and $N_f = 5$.
PDFs were accessed either through \textit{Mathematica} interface to \texttt{MMHT2014} or through \texttt{LHAPDF6} with the value of the $\alpha_s$ extracted from the PDFs.
The integrations over the phase space, both in the case of \textit{Mathematica} and $\texttt{C++}$ codes were done using the \texttt{Cuba} library \cite{Hahn:2004fe}.
In \textit{Mathematica}, \texttt{Cuhre} and \texttt{Divonne} integrators were used.
In $\texttt{C++}$ part, \texttt{Vegas} turned out to offer the highest precision.
Where advantageous, parallelized versions of \texttt{Cuba} routines were used \cite{Hahn:2014fua}.
Integrals evaluated with \texttt{Cuba} were maximally 7-dimensional, two of the dimensions coming always from the integration over the Bjorken variables $x$ in the range $\frac{4 m_O^2}{S} \leq x_1 \leq 1$ and $ \frac{4 m_O^2}{x_1 S} \leq x_2 \leq 1$. 
Apart of the sgluon mass, the cross section depends also on top quark mass, which was set to $m_t = 173$ GeV.

\subsubsection{Cancelation of $\delta_s$ and $\delta_c$ dependencies}
\begin{figure}
\centering
\subfloat[]{\includegraphics[width=0.495\textwidth]{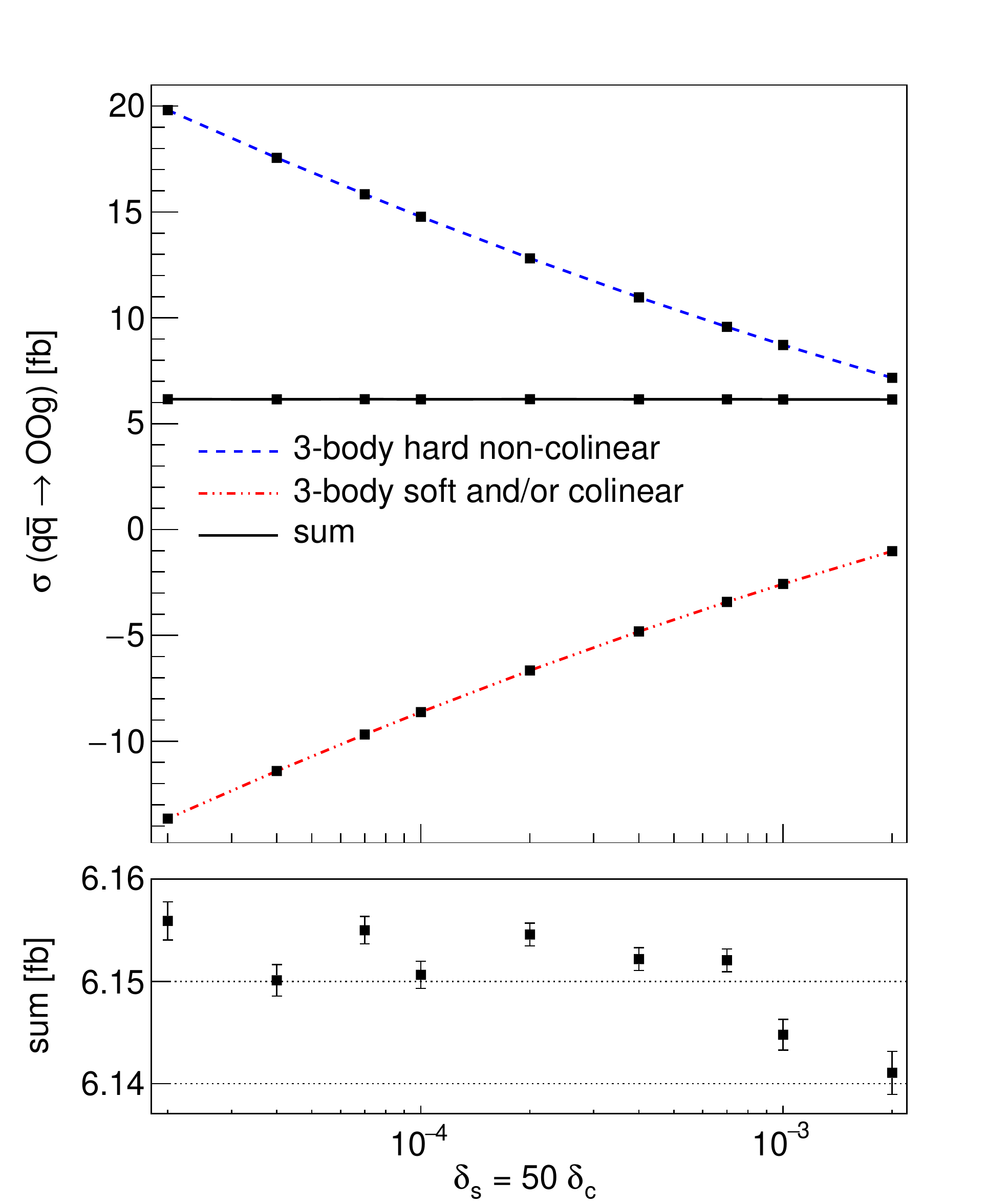}}
\subfloat[]{\includegraphics[width=0.495\textwidth]{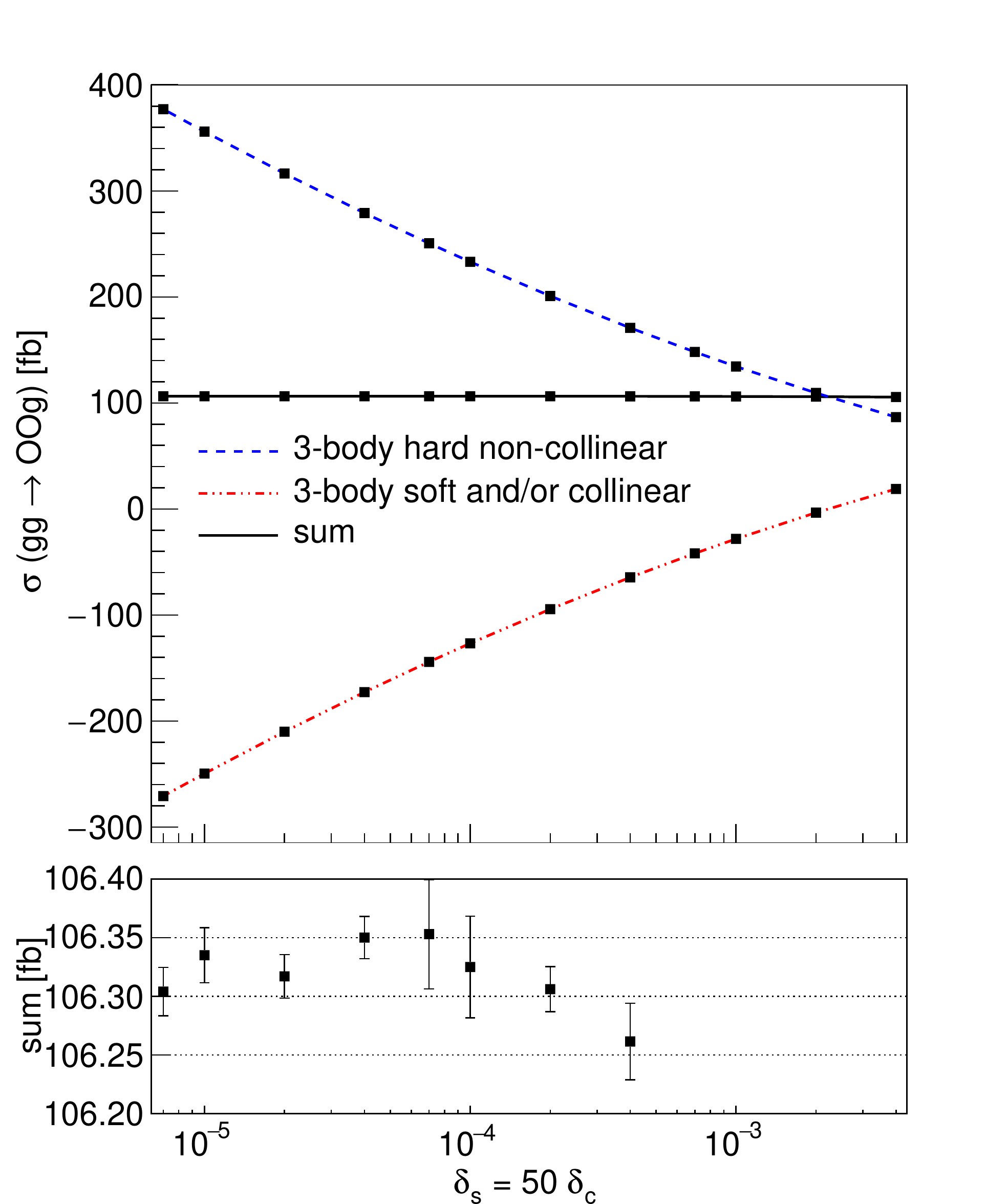}}
\\
\subfloat[]{\includegraphics[width=0.495\textwidth]{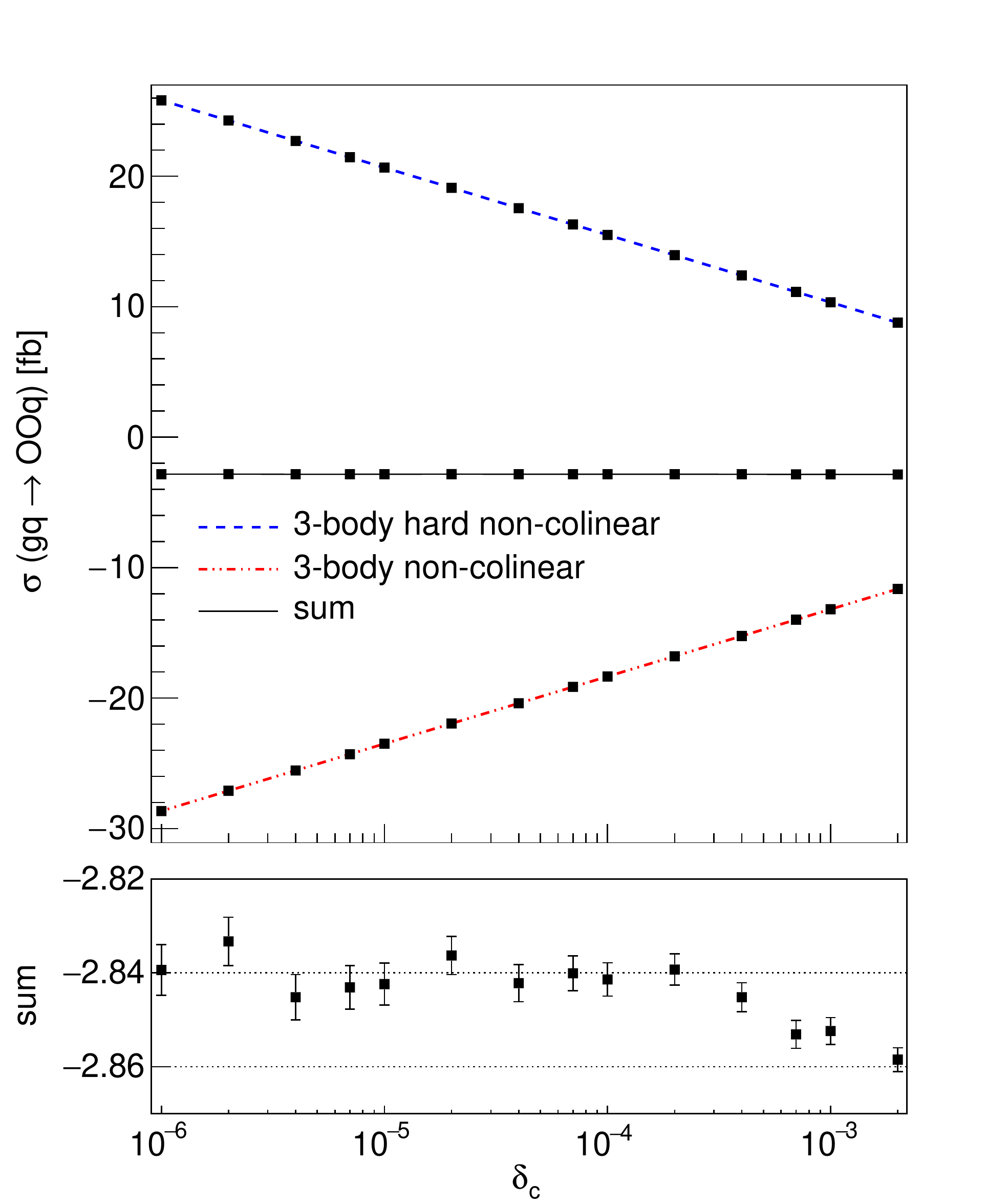}}
  \caption{Cancelation of $\delta_s$ and $\delta_c$ dependence for $gg$ (a), $q\bar q$ (b) and $q g/\bar{q} g$ (c) initiated processes.  Channels involving quarks are implicitly summed in 5-flavor scheme. 
  Plots are done for LHC at $\sqrt{S} = 13$ TeV and $\mu_R = \mu_F = m_O = 1$ TeV using \texttt{MMHT2014} baseline NLO fit with $\alpha_s (m_Z) = 0.12$.\label{fig:cancel_dS}}
\end{figure}

\Autoref{fig:cancel_dS} shows the dependence of the total cross section for partonic channels $gg$, $gq/g\bar q$ and $q\bar q$ on the phase space cut $\delta_s$ for $\delta_c = \frac{1}{50} \delta_s$ (in the case of channels $gg$ and $q\bar q$).
Plots are done for $\sqrt{S} = 13$ TeV, the sgluon mass of 1 TeV and the 
factorization and renormalization scales set to the sgluon mass.

As can be seen in \autoref{fig:cancel_dS}, for $\delta_s \lesssim 10^{-3}$ the dependence becomes flat.
Judging from this I fix $\delta_s = 50 \, \delta_c = 2 \cdot 10^{-4}$ in the case of $gg$, $\delta_s = 50 \, \delta_c = 2 \cdot 10^{-4}$ in the case of $q\bar{q}$ channels and $\delta_c = 10^{-4}$ in the case of $qg/\bar{q}g$ one.
Choosing smaller $\delta_s$ is disadvantageous as it increases both the integration time and integration error.

\begin{figure}
  \centering
  \includegraphics[width=0.6\textwidth]{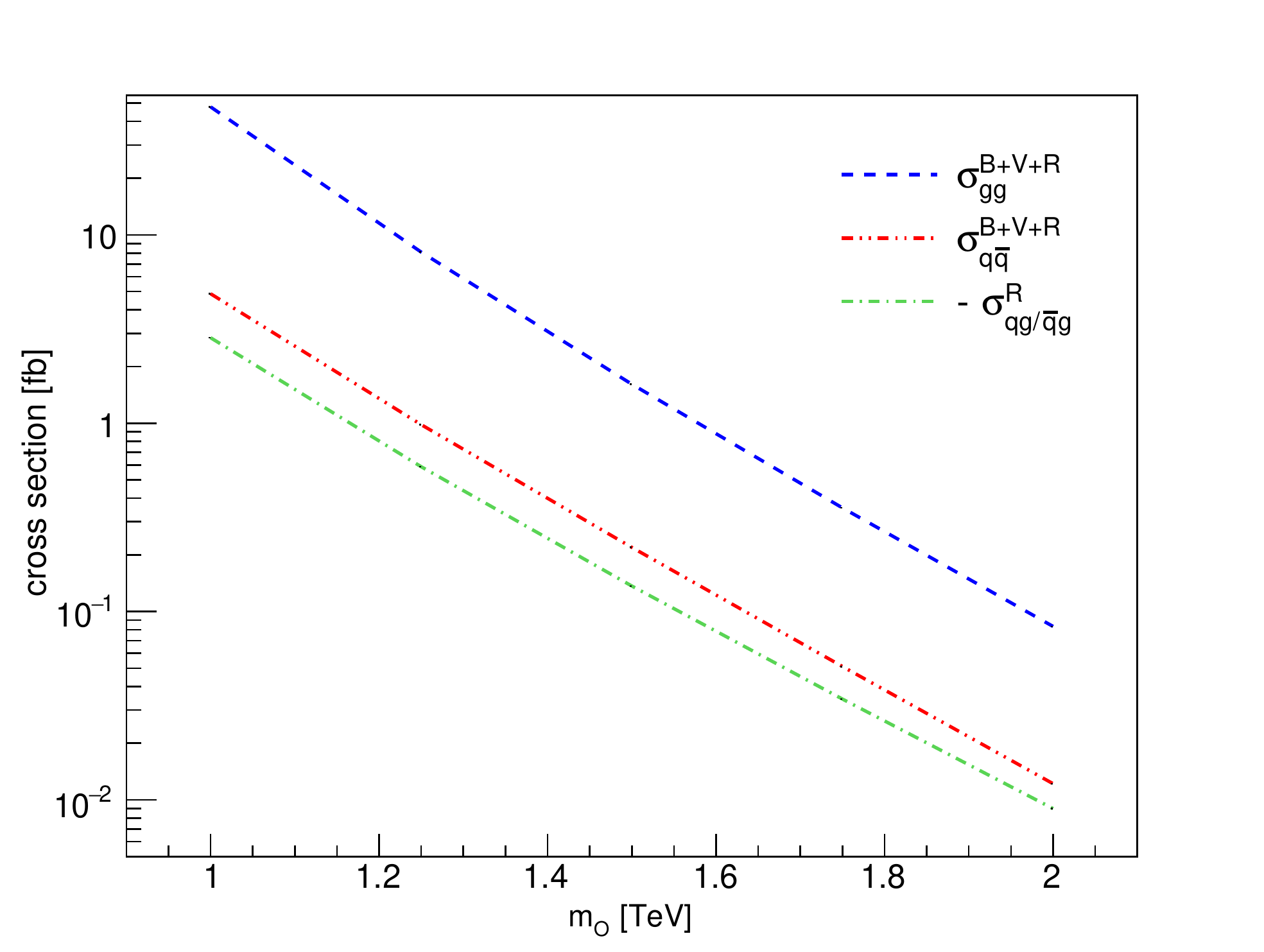}
  \caption{
  Decomposition of the NLO cross section into contributions from different (initial state) partonic channels. 
  Channels involving quarks are implicitly summed (in a 5-flavor scheme). 
  Plot was done for $\sqrt{S} = 13$ TeV and $\mu_R = \mu_F = m_O$ using the \texttt{MMHT2014} baseline NLO fit with $\alpha_s (m_Z) = 0.12$. 
  Real emission corrections are evaluated for $\delta_s = 50\,\delta_c = 2 \cdot 10^{-4}$ in the case of $gg$ channels, $\delta_s = 50\,\delta_c =  10^{-4}$ for $q\bar{q}$ and for $\delta_c = 2 \cdot 10^{-4}$ in the case of $qg/\bar{q}g$.
      As $qg/\bar{q}g$ contribution is negative, it is plotted with reversed sign for readability (see legend).
  \label{fig:nlo_contributions}}
\end{figure}
\begin{figure}
  \centering
  \input{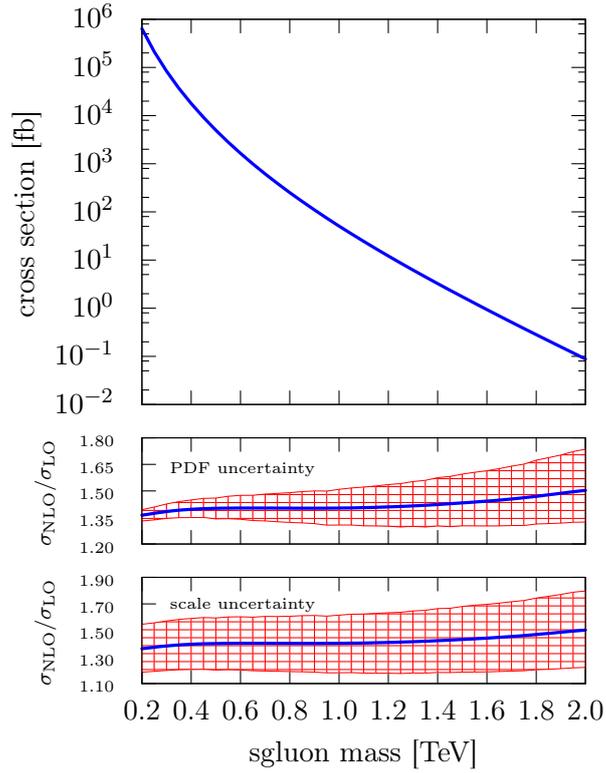}
  \caption{NLO cross section for sgluon pair production in function of their mass.
  Middle subfigure shows K-factor together with an uncertainty coming from the PDFs.
  Lower one does the same for uncertainty coming from scale variation.
  }
\end{figure}
The virtual part does not depend on the value of $\delta_s$ or $\delta_c$ cuts and as such its contribution is not included in the figures.

\subsubsection{Final results and discussion}

Putting everything together one can evaluate the total cross section for the sgluon pair production.
\Autoref{fig:nlo_contributions} shows contributions from different components of the \ac{NLO} cross section as a function of the sgluon mass (different quark channel are implicitly summed). 
To cross-check the result, final numbers where compared with 5-flavor version of the sgluon model described in Ref. \cite{Degrande:2014sta}.
As in the case of 4-flavor version, 5-flavor one was obtained using the \texttt{NLOCT} package \cite{Degrande:2014vpa}.
The results for 13 TeV LHC and 5 selected sgluon masses given in \Autoref{tab:sgluon_xsection_NLO} differed by at most 3.7\textperthousand.
It should be noted that \texttt{MadGraph5\_aMC@NLO} does not use the phase space slicing. 
Instead it uses a subtraction scheme (precisely the FKS subtraction \cite{Frixione:1995ms,Frixione:1997np}).
The differences between phase space slicing and subtraction schemes was analyzed in \cite{Dittmaier:1999mb,Eynck:2001en}.
Phase space slicing neglects terms which go to 0 as $\delta_s$ or $\delta_c \to 0$.
This simplifies computation but renders the result cut dependent.
It is not always possible to choose the cut parameters $\delta_s$ and $\delta_c \to 0$ small enough to get sub-permille accuracy due to numerical instabilities.
Compared to this, the subtraction methods are 'exact' and allow for much higher accuracy.
That said, achieved accuracy is enough to claim correctness of my result.

Mentioned \autoref{tab:sgluon_xsection_NLO} gives cross sections values for selected sgluon masses at 13 and 14 TeV runs of the LHC, together with systematical uncertainties coming from variation of $\mu_F$ and $\mu_R$ and PDF uncertainty.
Uncertainties are calculated as for the case of LO calculation in \autoref{sec:sgluon_pair_production_@lo}.

It was also checked that the results are in agreement the the ones obtained from \texttt{NNPDF3.0} set~(\texttt{LHAPDF} id 260000) \cite{Ball:2014uwa} within the PDF uncertainty.
The table also contains global K-factors, calculated with respect to values from \autoref{tab:sgluon_xsection_LO}.

The calculation setup described in this chapter is the first step in an effort  to calculate NLO (S)QCD corrections to pair production of strongly interacting MRSSM particles.
Written code were designed with extensibility in mind and at the time of writing they were also successfully applied to study of $p p \to \tilde u_L \tilde u_R$ process.
Therefore the content of this chapter should be treated not only as a cross-check of results already available in the literature, but also as a important step in this direction.

In the next chapter, I discuss phenomenological significance of results presented here.

\begin{table}
  \centering
  \begin{tabular}{c||cc||cc}
    sgluon mass [TeV] & cross section at 13 TeV [fb] & $K$ & cross section at 14 TeV [fb] & $K$\\
    \hline
    1  & $50.79^{+15.3\%+7.7\%}_{-15.7\%-6.7\%}$ & 1.40 & $71.41^{+14.1\%+7.2\%}_{-15\%-6.3\%}$ & 1.37\\
    1.25 & $8.656^{+16.3\%+9.5\%}_{-16.5\%-7.9\%}$ & 1.38 & $12.91^{+14.9\%+8.8\%}_{-15.7\%-7.4\%}$ & 1.41 \\
    1.5 & $1.726^{+17.3\%+11.3\%}_{-17.2\%-9.1\%}$ & 1.40 & $2.752^{+15.8\%+10.5\%}_{-16.3\%-8.5\%}$ & 1.39\\
    1.75 & $0.3797^{+18.4\%+13.3\%}_{-17.9\%-10.5\%}$ & 1.46 & $0.6482^{+16.7\%+12.3\%}_{-17\%-9.7\%}$ & 1.41 \\
    2 & $0.08832^{+19.7\%+15.5\%}_{-18.8\%-11.9\%}$ & 1.47 & $0.1635^{+17.8\%+14.2\%}_{-16.5\%-11\%}$ & 1.45 \\
  \end{tabular}
  \caption{
    Cross sections for sgluon pair production for 13 and 14 TeV LHC as a function of the sgluon mass.
    Cross sections are calculated using 5-flavor NLO capable sgluon UFO model from \texttt{FeynRuls} and \texttt{MadGraph5\_aMC@NLO} at NLO in $\alpha_s$ using \texttt{MMHT2014} NLO PDF fit from \texttt{LHAPDF6} (see main text for more details).
    First error comes from the scale variation, second is the PDF uncertainty.
    Relative statistical errors are below $10^{-3}$ and not shown here.
    Global $K$-factors $K$ are calculated with respect to \autoref{tab:sgluon_xsection_LO}.
    \label{tab:sgluon_xsection_NLO}
    }
\end{table}

\chapter{Collider phenomenology of sgluon pair production\label{sec:sgluon_mc}}

\section{Introduction}

In the previous section next-to-leading order QCD corrections to the sgluon pair production were calculated.
This process is of high phenomenological importance since sgluons can be one of the lightest color-charged MRSSM particles, copiously produced at the Large Hadron Collider.
Their LHC phenomenology was previously investigated in the context of \textit{R}-symmetric/$\mathcal{N}=2$/Dirac gaugino SUSY models, hyper-pions in vector-like confinement gauge theories and universal extra dimensions~\cite{Choi:2008ub,Plehn:2008ae,Calvet:2012rk,Kilic:2008ub,Schumann:2011ji,Chen:2014haa,Beck:2015cga,Degrande:2014sta,Dobrescu:2007yp,Kilic:2010et,Burdman:2006gy,Benakli:2016ybe}.

Produced, sgluons are expected to decay, depending on their mass, mainly into pairs of gluons or quarks (assuming that squarks and gluinos are too heavy).
Since due to chirality the $O \to q\bar{q}$ coupling is suppressed by the quark mass, above the top quark threshold there are only two decay channels: $O \to t\bar{t}$ and $O \to gg$.
\begin{figure}
  \centering
    \subfloat[]{  \raisebox{+0.75cm}{\includegraphics[width=0.49\textwidth]{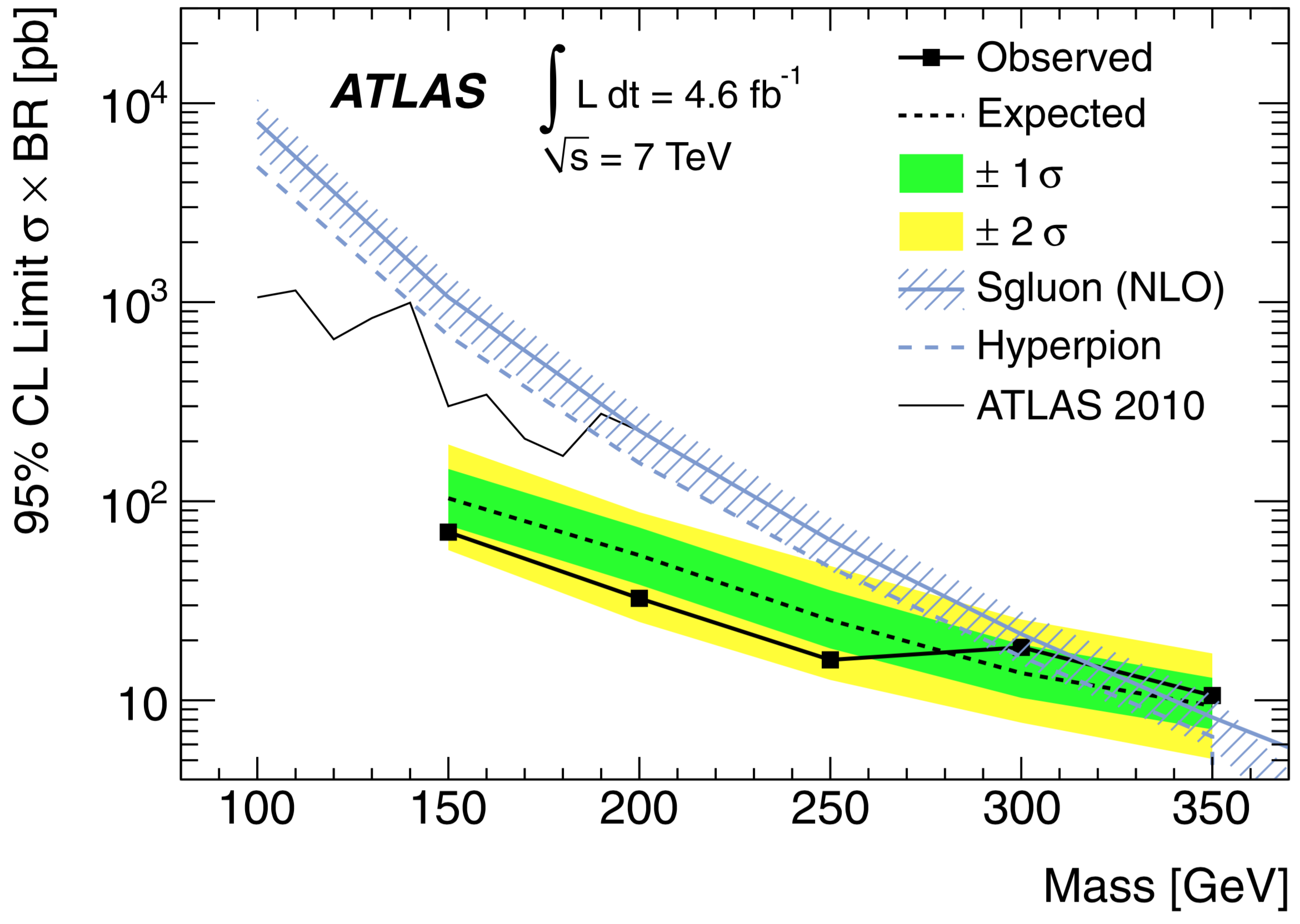}}}
\subfloat[]{  \includegraphics[width=0.49\textwidth]{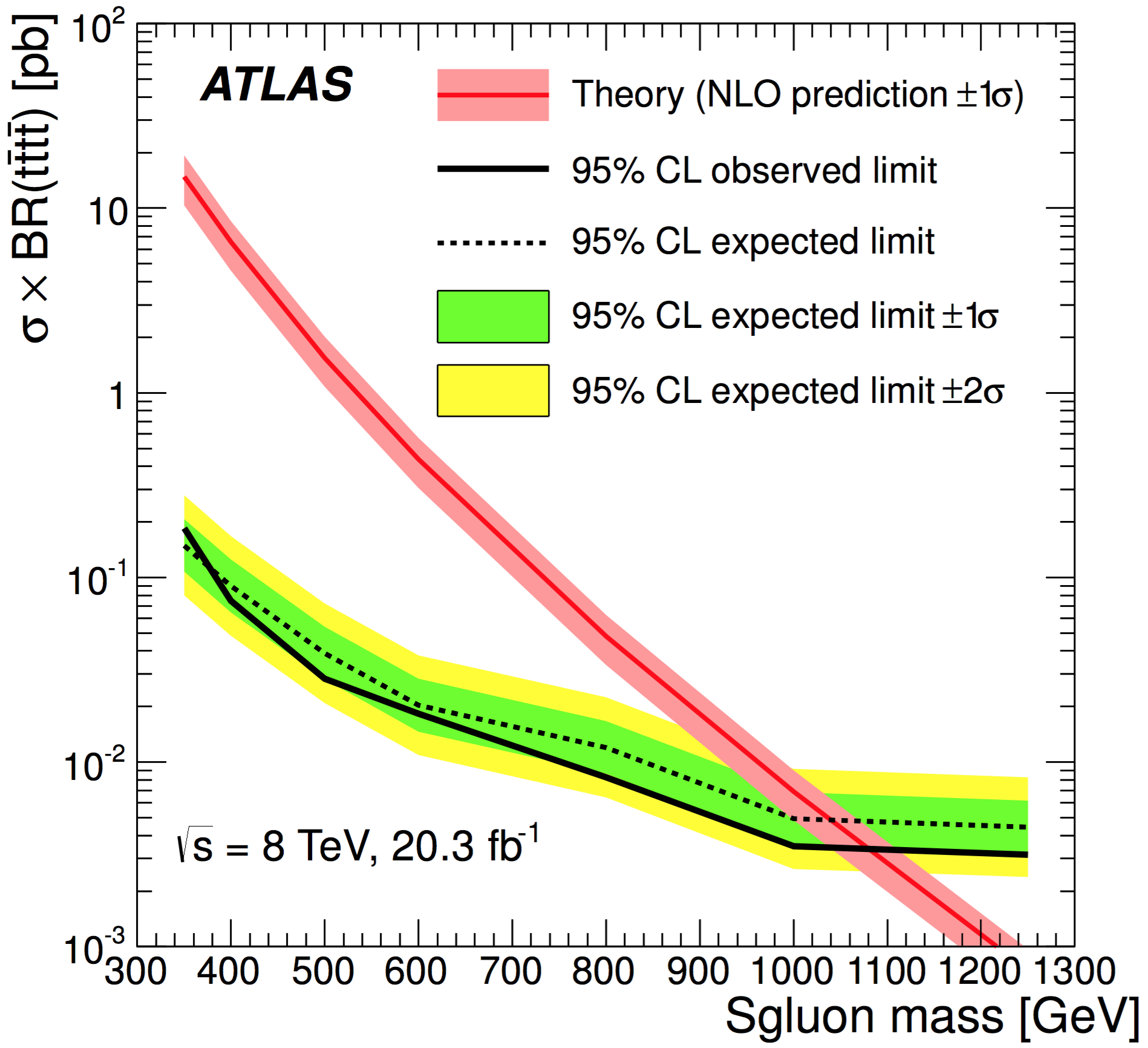}}
  \caption{Sgluon cross section exclusion limit as determined by ATLAS in the gg (a) \cite{ATLAS:2012ds} and $t\bar{t}$ (b) \cite{Aad:2015kqa} decay modes. Theory predictions (dashed blue band in (a) and red band in (b)) comes from the model with a complex sgluon \cite{GoncalvesNetto:2012nt}. \label{fig:sgluon_exclusion_plot}}
\end{figure}
This kind of signatures, in both channels, was searched for by the experimental collaborations.
ATLAS excludes at 95\% CL pair produced, complex sgluons decaying (with branching fraction 1) to gluon pairs in the mass range from 100 to 287 GeV \cite{ATLAS:2012ds} (see \autoref{fig:sgluon_exclusion_plot} a).
For $t\bar{t}$ decay mode, sgluons are exluded at 95\% CL limit to 1.06 TeV  \cite{Aad:2015kqa} (see \autoref{fig:sgluon_exclusion_plot} b).
Two things should be noted, though.
First, those exclusions are based on a simplified model with complex sgluon from Ref.~\cite{GoncalvesNetto:2012nt} while in the MRSSM the cross section is roughly 2 times smaller.
Second, ATLAS analysis does not specify form of the used sgluon - top coupling.

At the time of writing there are no 13 TeV analyses addressing directly sgluon pair production (although similar final state topologies were searched for).
Therefore, all mentioned exclusions come from Run 1.
This makes any projections for final Run 2 exclusion limits difficult.  
To fill this gap, this section is devoted to recasting current ATLAS limits from search of SUSY in the 4-top quark final state in Ref.~\cite{Aad:2016tuk} to sgluon pair production.
Since for phenomenologically viable Dirac gluino masses pseudoscalar sgluons are significantly lighter than the scalar ones (see \autoref{eq:sgluons_mass_splitting1} and \ref{eq:sgluons_mass_splitting2}), they could be discovered much sooner.
Therefore in this chapter a simplified model with a sgluon coupled to $t\bar{t}$ pair through the pure-axial coupling is used.

The effective Lagrangian for the model consists of the SM and a BSM part given by
\begin{equation}
\mathcal{L}^{\text{BMS}} = \frac{1}{2} D_\mu O^a D^\mu O^a - \frac{1}{2} m^2_{O} O^2 - \imath c \bar{t} \gamma^5 T^a t O^a ,
\label{eq:lagrangian}
\end{equation}
with the sum over a color index $a$. 
As explained in \autoref{sec:sgluon_decays}, if pseudoscalar sgluon mass is smaller than twice the mass of the Dirac gluino it will decay almost exclusively to $t\bar t$ pairs.
This justifies the form of \autoref{eq:lagrangian}.
Also, the precise value of the loop-induced coupling $c$ is not important, as the branching ratio is almost 1.\footnote{Since coupling $c$ is loop-induced, it is expected to be small so that off-shell effects in sgluon pair production are also negligible.}

For the phenomenological studies, a few sgluon masses between 0.9 and 1.5 TeV were selected, with the 0.9 TeV one being already at the border of an exclusion with currently available Run 1 data.

This chapter is structured as follows.
The next section describes a Monte Carlo simulation setup.
Although ATLAS analysis specifies both measured and expected number of background events in the relevant signal region, simulation of associated production of a gauge boson with a $t\bar{t}$ pair is used to validate encoded analysis and detector "simulation" setup.
The same technique is then applied to simulated signal events.
\Autoref{sec:sgluons_mc_analysis} describes the analysis and detector parametrization setup together with final results.
The chapter finishes with conclusions and prospects for expected $\geq 100~\text{fb}^{-1}$ data sample of Run 2.

\section{Simulation setup}
Due to technical reasons, samples for signal and background were generated using two different methods.
Both for signal and background simulation the following values of gauge boson masses were used: $m_W = 80.385$~GeV, $m_Z = 91.1876$~GeV. 
Top quark mass was set to 173.21 GeV while other quarks were assumed massless in matrix elements (although in the parton shower $c$ and $b$ quark masses were kept).
The CKM matrix was set to identity.

All samples were generated using the \texttt{MMTH2014} baseline (5-flavor) NLO fit \cite{Harland-Lang:2014zoa} interfaced through \texttt{LHAPDF6}~\cite{Buckley:2014ana}.

\subsection{Signal}
Signal events were generated using \texttt{MadGraph5\_aMC@NLO} and an  NLO capable UFO \cite{Degrande:2011ua} model generated with \texttt{FeynRules}~\cite{Alloul:2013bka,Degrande:2014vpa,Degrande:2014sta}. The original NLO capable sgluon model, available at 
    \url{https://feynrules.irmp.ucl.ac.be/wiki/NLOModels}, 
    does not allow for the pseudoscalar $Ot\bar{t}$ coupling as in \autoref{eq:lagrangian}.
Because of this, a new model with Lagrangian \autoref{eq:lagrangian} was generated (changing also the original 4-flavor scheme to 5-flavor one in the process).
    
Three sgluon masses, 1, 1.25 and 1.5 TeV, were considered. 
The production cross sections for these masses are given in \autoref{tab:sgluon_xsection_NLO}.
Sgluons were then decayed into $t\bar t$ pairs (and further) using \texttt{MadSpin}~\cite{Artoisenet:2012st} as\\
\-\qquad \texttt{decay sig8 > t t\~{}, (t > w+ b, w+ > mu+ vm), (t\~{} > w- b\~{}, w- > all all)}\\
or, after defining multiparticle container \texttt{no\_mu} as\\
\-\qquad \texttt{define no\_mu = u d s c u\~{} d\~{} s\~{} c\~{} e+ e- ta+ ta- ve ve\~{} vt vt\~{}},\\
decayed as\\
\-\qquad \texttt{decay sig8 > t t\~{}, (t > w+ b, w+ > no\_mu no\_mu), (t\~{} > w- b\~{}, w- > mu- vm\~{})}\\
\-\qquad \texttt{decay sig8 > t t\~{}, (t > w+ b, w+ > all all), (t\~{} > w- b\~{}, w- > mu- vm\~{})}\\
generating all configurations that give two same-sign muons.
Total branching ratio into these channels is given by 
$\text{BR}^2(W \to \mu \nu) (2-\text{BR}^2(W \to \mu \nu))$, where $\text{BR}(W \to \mu \nu) \approx 11$\%.
Partonic events were matched to parton shower using \texttt{MC@NLO}~\cite{Frixione:2002ik} prescription and \texttt{Pythia8}~\textit{v219}~\cite{Sjostrand:2014zea}.

By default the final state shower algorithm in \texttt{Pythia8} is based on the dipole-style recoils. 
As stated in \texttt{Pythia8} manual, for \texttt{MC@NLO} where a full analytic knowledge of the shower radiation pattern in needed one has to switch to global recoil approach which does not contain color coherence phenomena (and hence factorizes).
A minimal set of settings needed to consistently shower \texttt{MC@NLO} events is then given by\footnote{See the \texttt{Pythia8} manual at \url{http://home.thep.lu.se/~torbjorn/pythia82html/Welcome.html}, section \textit{Link to Other Programs $\to$ Matching and Merging $\to$  aMC@NLO Matching}. See also the discussion in Ref.~\cite{ATL-PHYS-PUB-2016-005}.}\\
\-\qquad \texttt{SpaceShower:pTmaxMatch = 1}\\
\-\qquad \texttt{SpaceShower:pTmaxFudge = 1.}\\
\-\qquad \texttt{SpaceShower:MEcorrections = off}\\
\-\qquad \texttt{TimeShower:pTmaxMatch = 1}\\
\-\qquad \texttt{TimeShower:pTmaxFudge = 1.}\\
\-\qquad \texttt{TimeShower:MEcorrections = off}\\
\-\qquad \texttt{TimeShower:globalRecoil = on}\\
\-\qquad \texttt{TimeShower:weightGluonToQuark = 1}\\
Those settings cannot be modified.
What can be chosen, though, is when to return from the global recoil mode to the dipole recoil.
Since color coherence phenomena are very important (see for example \cite{Chatrchyan:2013fha}), it is advantageous to switch back to dipole recoils already after the first emission. 
This can be done in two ways, setting \texttt{TimeShower:globalRecoilMode = 1} or \texttt{2}. 
Option 2 applies global recoil only if the first branching in the evolution is a timelike splitting of a parton in an event with Born-like kinematics (the so called $\mathbb{S}$-events in the \texttt{MC@NLO} language), while for option 1 this is done both for Born-like ($\mathbb{S}$) and real-emission events ($\mathbb{H}$-events).
With option 2 the impact of global recoil should be minimal.
For options 1 and 2 a maximal number of splittings in the timelike shower with global recoil strategy should be set to 1 through \texttt{TimeShower:nMaxGlobalBranch} flag.
Also, to distinguish between $\mathbb{S}$ and $\mathbb{H}$ events, the number of color-charged particles for Born-like configurations must be given through \texttt{TimeShower:nPartonsInBorn} option.
The \texttt{MC@NLO} matching is done at the level of the hard process.
To that end, \texttt{Pythia8} removes decay chains generated by \texttt{MadSpin}
by traversing the event tree and identifying intermediate particles with status code \texttt{ISTUP}=$\pm 2$~\cite{Boos:2001cv} which have a single parent. 
\texttt{TimeShower:nPartonsInBorn}  then counts the number of remaining color-charged particles.
For the sgluon pair production I therefore set:
\\
\-\qquad \texttt{TimeShower:globalRecoilMode = 2}\\
\-\qquad \texttt{TimeShower:nMaxGlobalBranch = 1}\\
\-\qquad \texttt{TimeShower:nPartonsInBorn = 2}\\
\-\qquad \texttt{TimeShower:limitPTmaxGlobal = on}\\

Since there are no genuine underlying event NLO tunes in \texttt{Pythia8}, the default LO tune is used.
This degrades quality of low $p_T$ predictions and is a known issue with NLO+PS simulations (see discussion in sec. 3.3 of Ref.~\cite{Frederix:2015eii}).
This can be seen for example in the few first bins in \autoref{fig:z_validation} (a), where the MC simulation fails to correctly describe $Z$-boson transverse momentum spectrum for $p_T$ between 1 and 5 GeV.

\subsection{Background validation}

\begin{figure}
  \centering
  \subfloat[]{\includegraphics[width=0.42\textwidth]{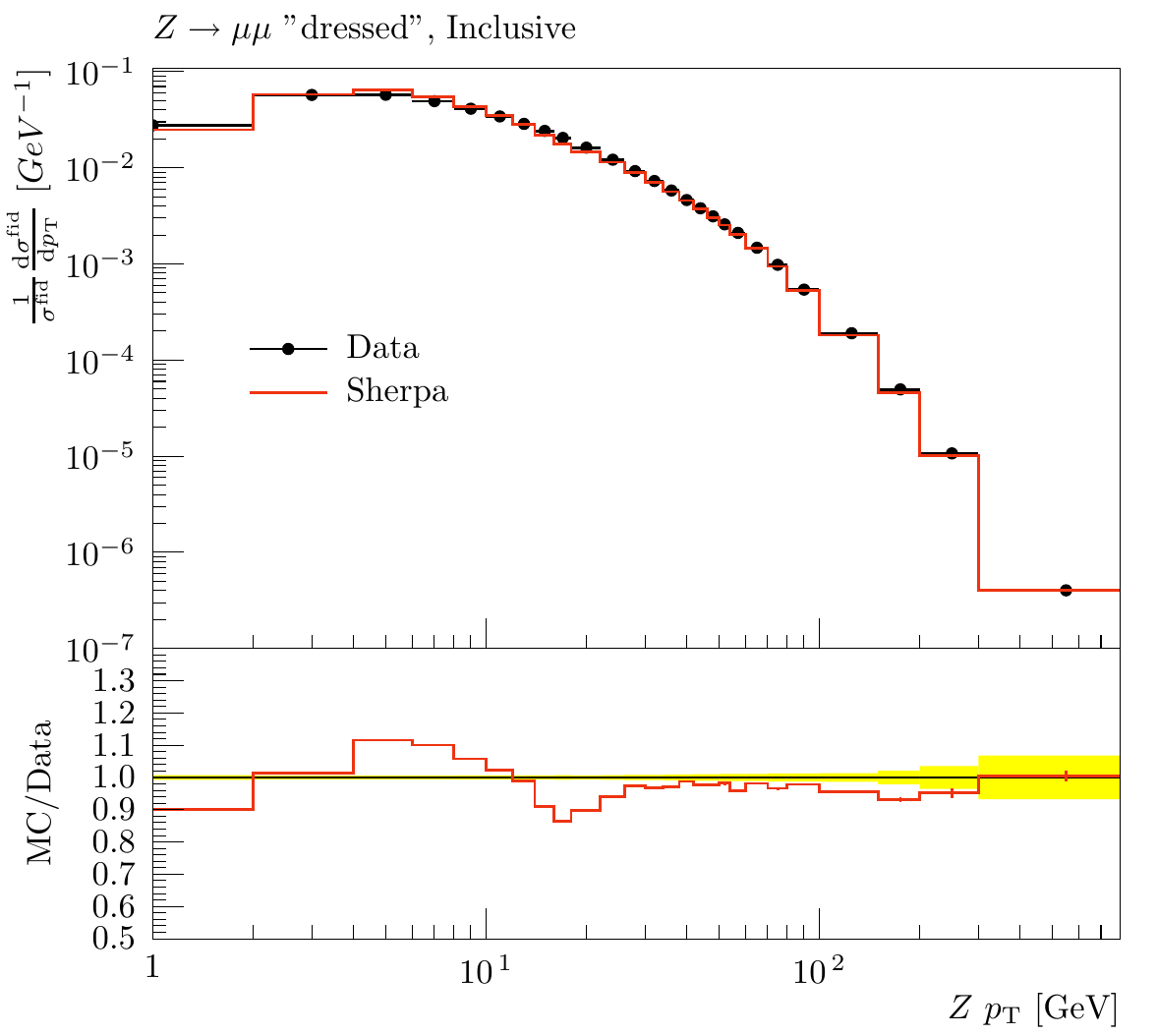}}
  \subfloat[]{\includegraphics[width=0.42\textwidth]{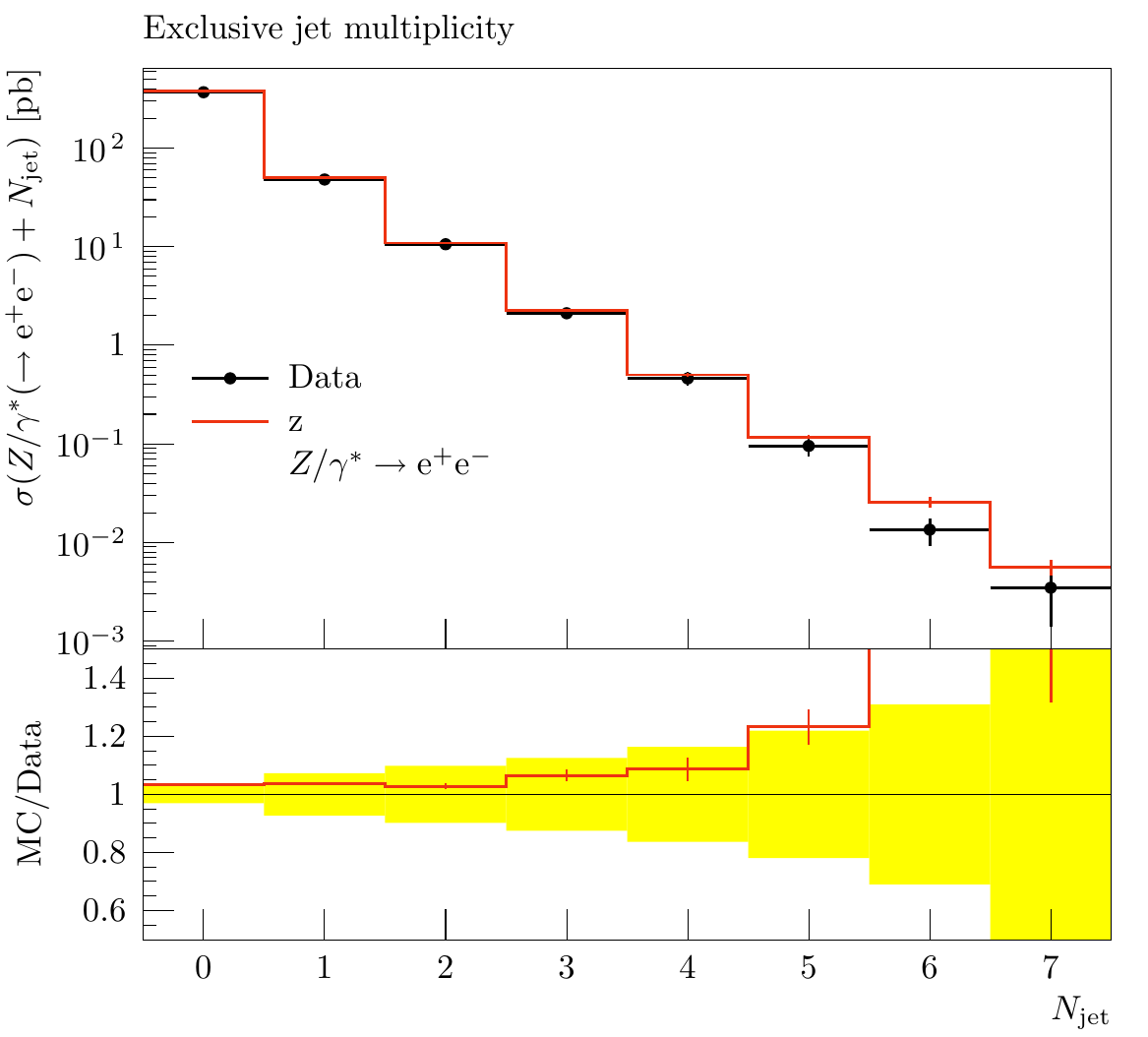}}\\
  \subfloat[]{\includegraphics[width=0.42\textwidth]{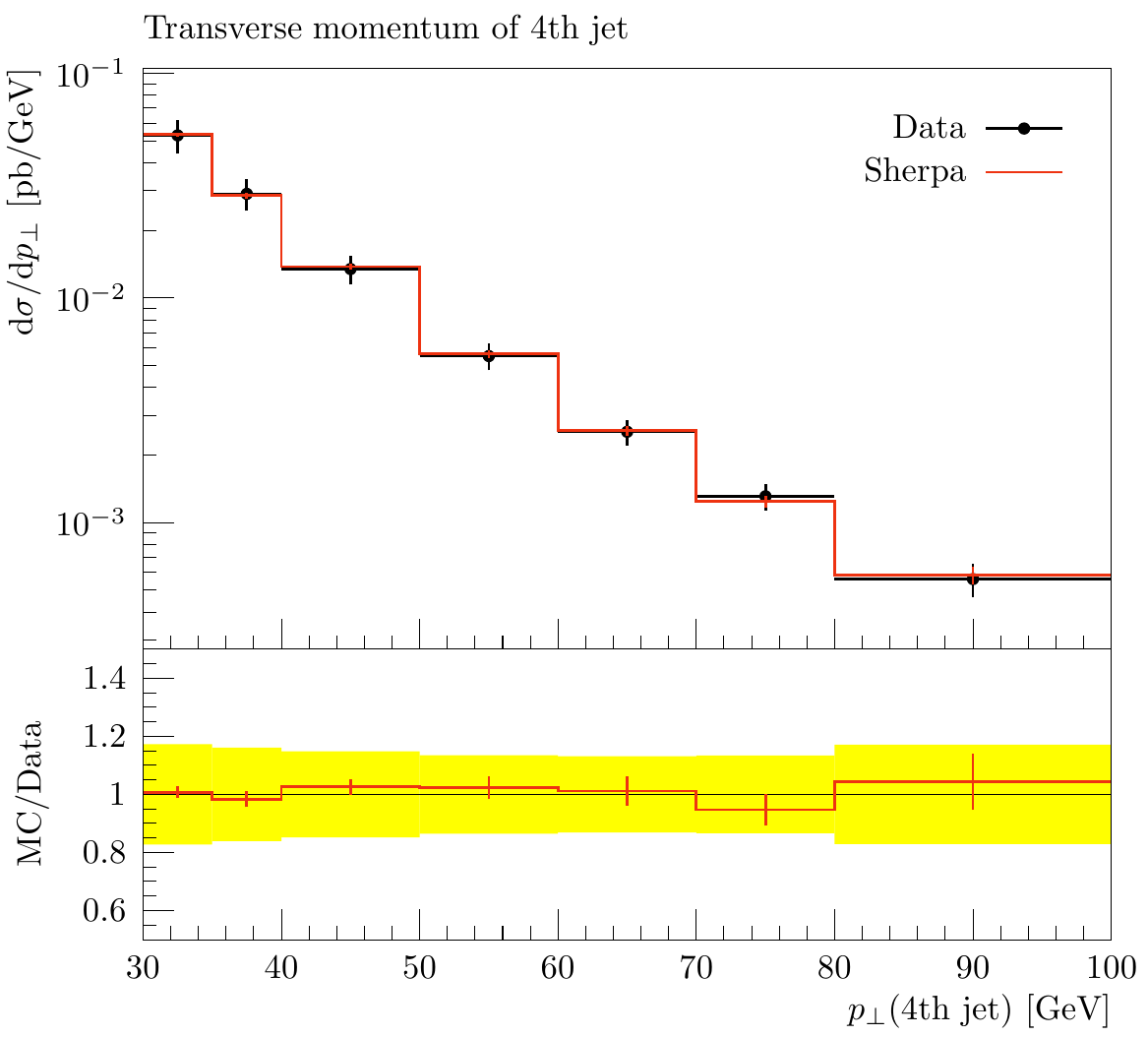}}
  \subfloat[]{\includegraphics[width=0.42\textwidth]{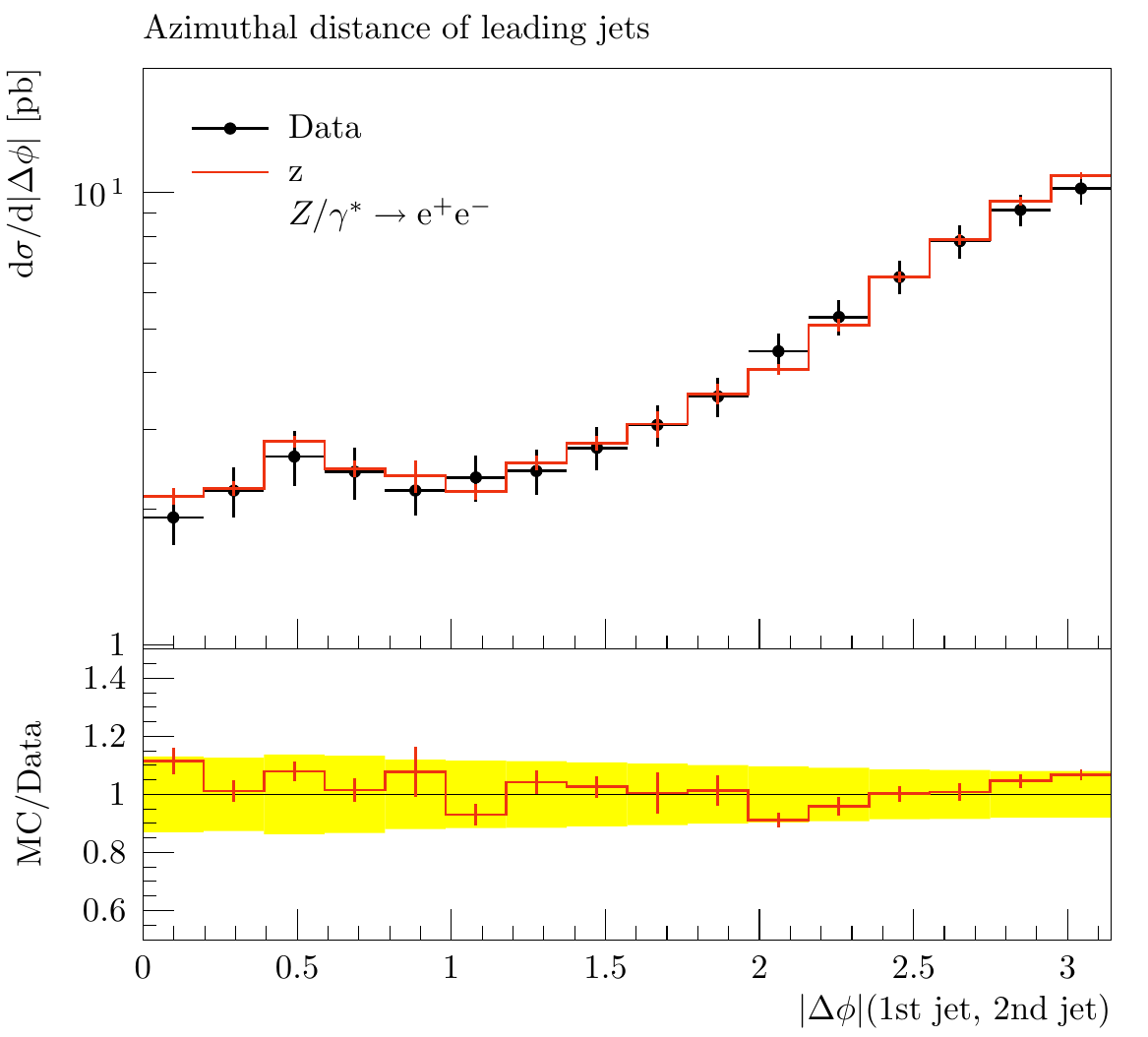}}\\
  \subfloat[]{\includegraphics[width=0.42\textwidth]{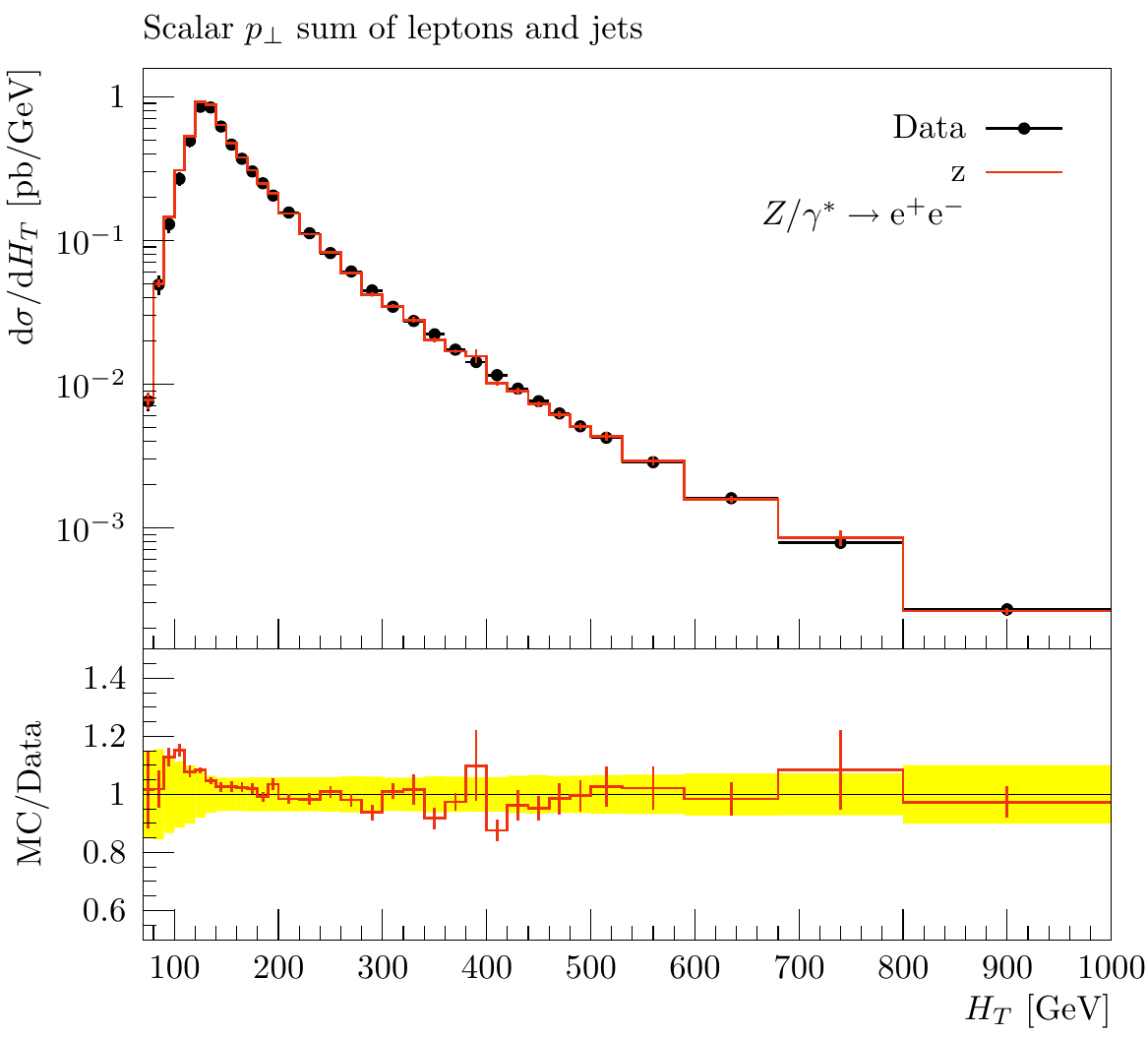}}
  \subfloat[]{\includegraphics[width=0.42\textwidth]{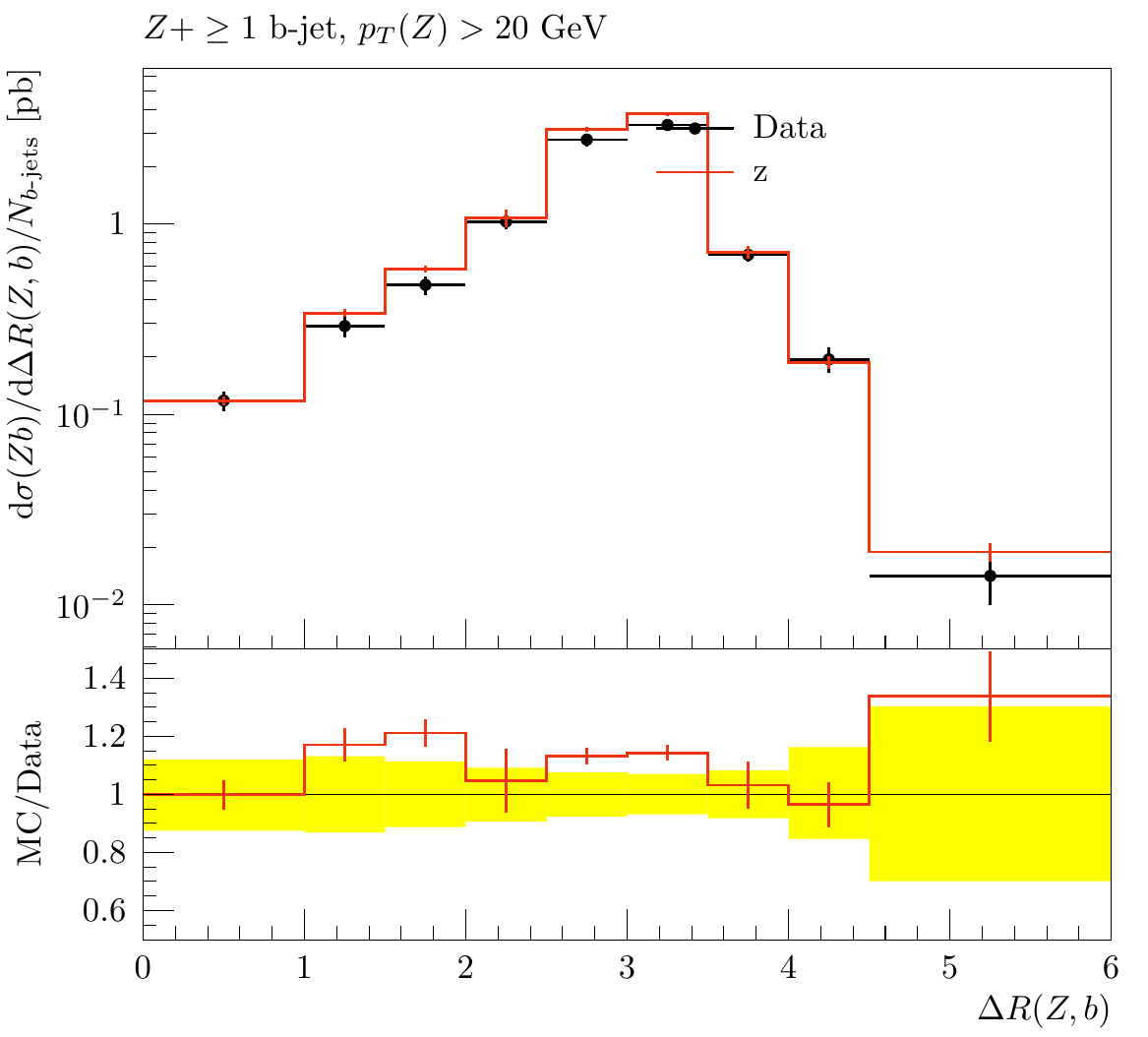}}
  \caption{Validation plots for the Drell-Yan process using \texttt{Rivet} analyses ATLAS\_2014\_I1300647 (a), ATLAS\_2013\_I1230812\_EL (b-e), ATLAS\_2014\_I1306294\_EL~(f) and 7 TeV LHC data from analyses \cite{Aad:2014xaa}~(a), \cite{Aad:2013ysa}~(b-e) and \cite{Aad:2014dvb}~(f).}
  \label{fig:z_validation}
\end{figure}
\begin{figure}
  \centering
  \subfloat[]{ \includegraphics[width=0.49\textwidth]{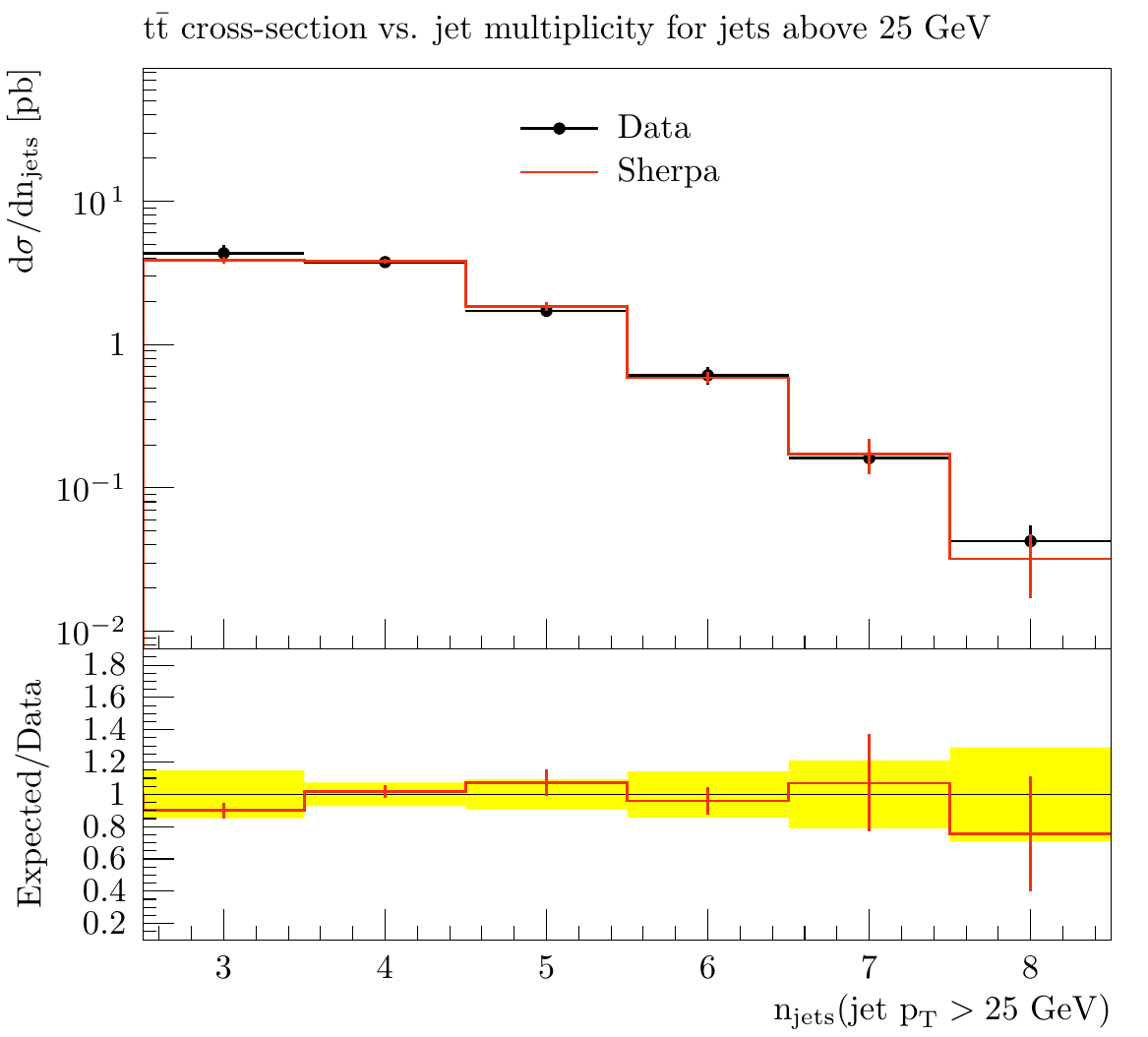}}
  \subfloat[]{ \includegraphics[width=0.49\textwidth]{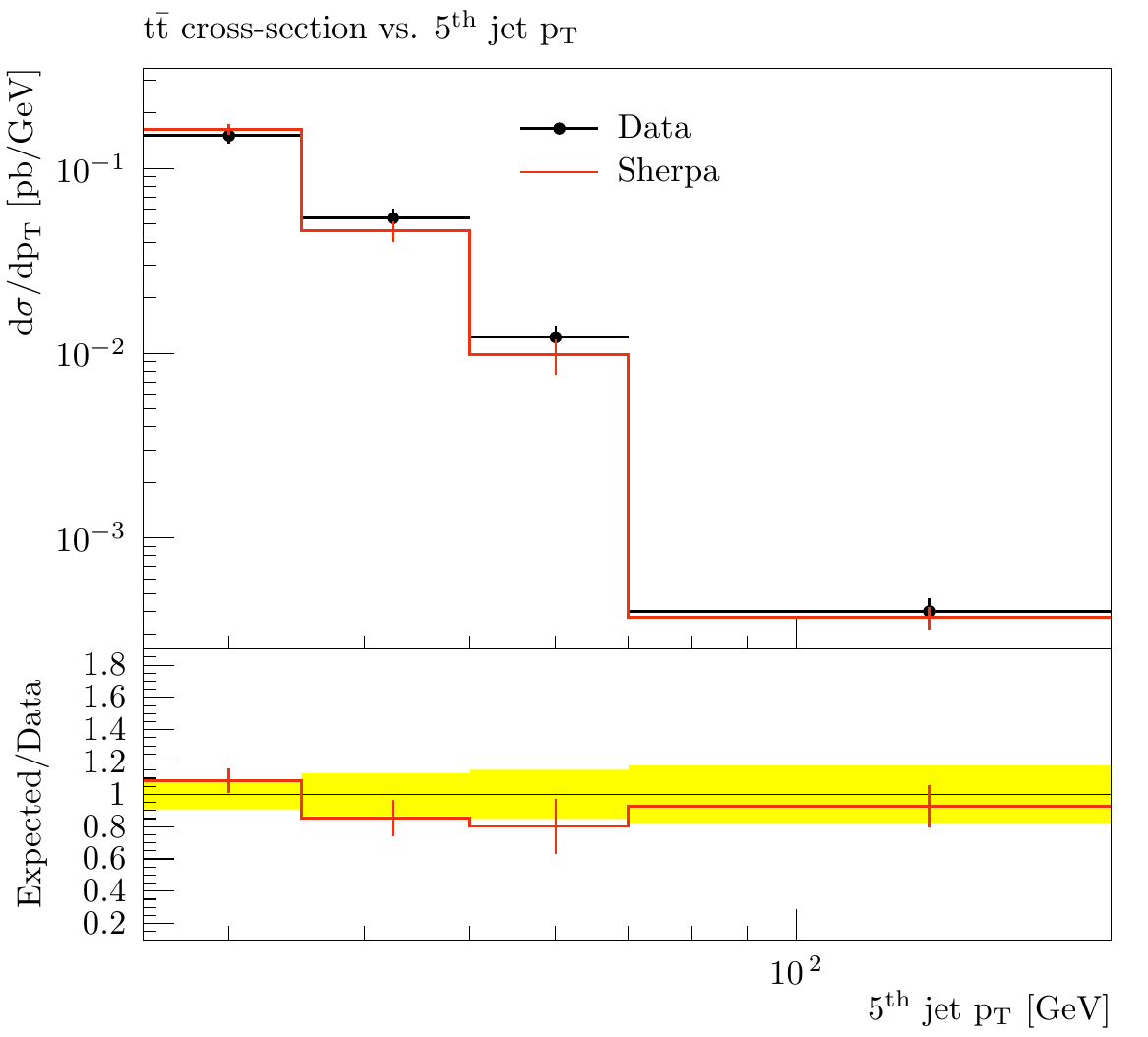}}\\
  \subfloat[]{ \includegraphics[width=0.49\textwidth]{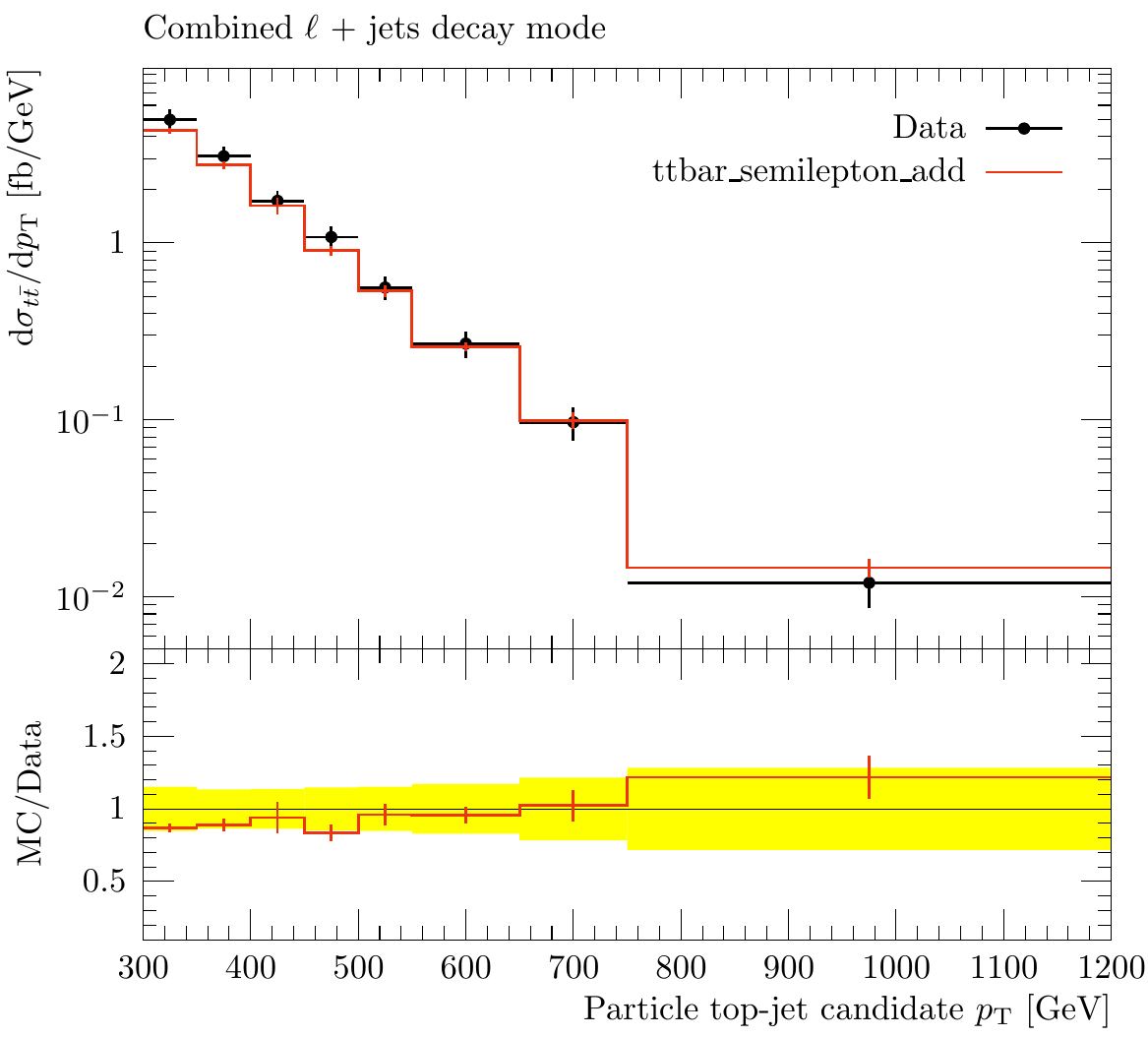}}
  \subfloat[]{ \includegraphics[width=0.49\textwidth]{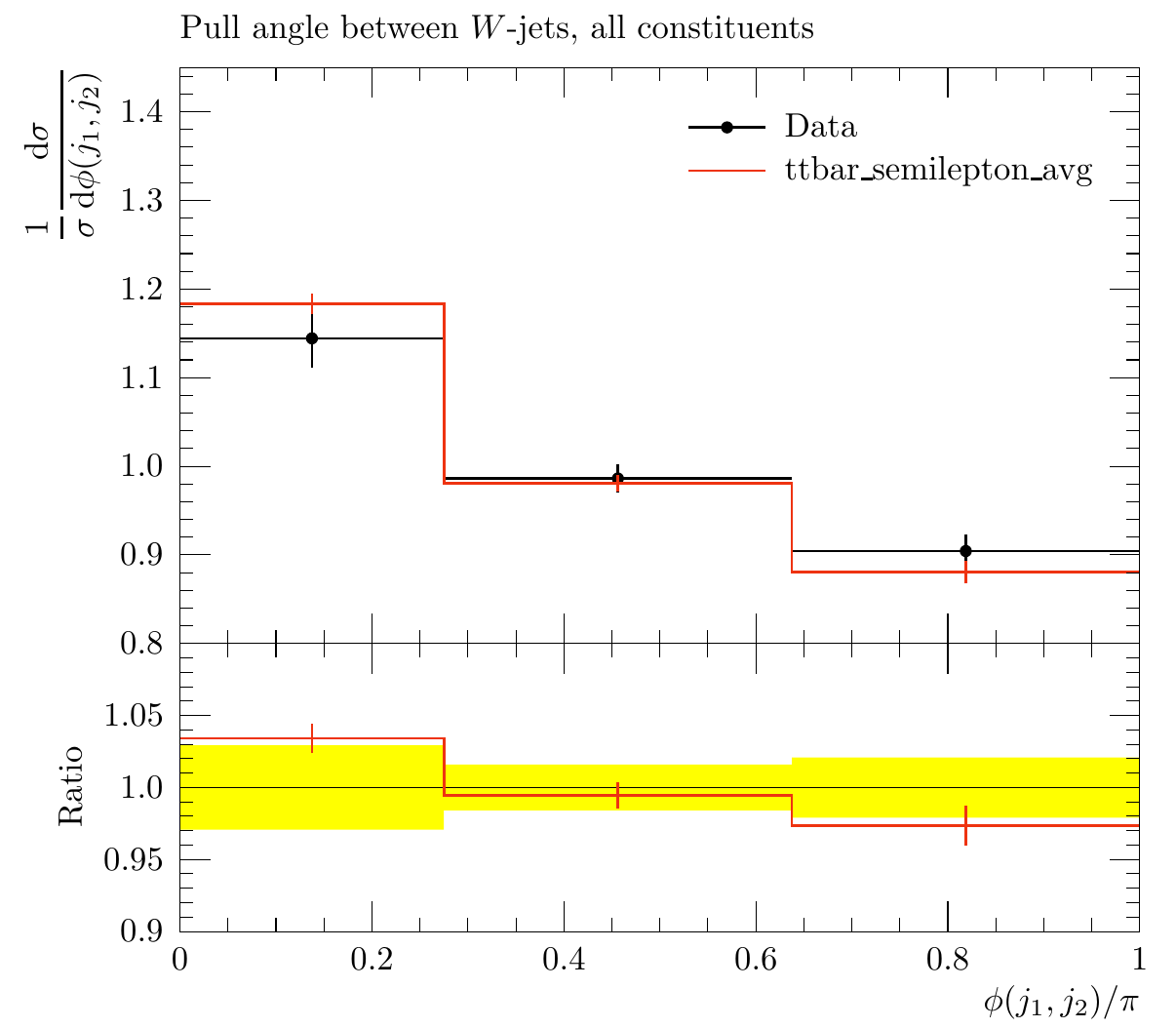}}\\
  \caption{Validation plots for the $t\bar t$ production using \texttt{Rivet} analyses ATLAS\_2014\_I1304688 (a,b), 
ATLAS\_2015\_I1397637~(c) and ATLAS\_2015\_I1376945~(d)
  and 7 TeV data from analysis \cite{Aad:2014iaa}~(a,b) and 8 TeV data from analyses \cite{Aad:2015hna} (c)  \cite{Aad:2015lxa}~(d).
}
  \label{fig:ttbar_validation}
\end{figure}

Background samples were generated using \texttt{Sherpa}~\textit{v.2.2}~\cite{Gleisberg:2008ta}, with virtual matrix elements provided by \texttt{OpenLoops}~\textit{v1.3.1}~\cite{Cascioli:2011va} and evaluated using \texttt{CutTools}~\cite{Ossola:2006us,Ossola:2007ax} or \texttt{COLLIER}~\cite{Denner:2002ii,Denner:2005nn,Denner:2010tr,Denner:2016kdg}. 
$t\bar{t} \mu \nu_\mu$ was generated with up to 1 additional jet at NLO order and 3 jets at LO, while for $t\bar{t} \mu^+ \mu^-$ up to 1 and 2 jets, respectively, were generated.
Different multiplicities were merged using MEPS@NLO technique \cite{Hoeche:2012yf,Gehrmann:2012yg}.
In the case of $t\bar{t} \mu^+ \mu^-$ a generation cut on an invariant mass of  muon pair  $m_{\mu^+ \mu^-} > 20$ GeV was applied.
The inclusive cross sections for those samples (including appropriate top-quarks decays) are 7.77 and 5.43 fb.\footnote{Contrary to UNLOPS~\cite{Lonnblad:2012ix}, MEPS@NLO inclusive cross section depends (slightly) on the number of additional jets.}
These predictions agree within (still very large) experimental uncertainties with the LHC measurements \cite{ATLAS-CONF-2016-003,CMS-PAS-TOP-16-009}.
Top quarks were then decayed in all possible ways that ensure two same-sign muons in the final state.

The setup of \texttt{Sherpa} mostly follows standard settings. 
Here only the most important ones are mentioned. 
Samples were generated with \texttt{EXCLUSIVE\_CLUSTER\_MODE = 1} setting (meaning that only QCD splittings are considered when reconstructing parton shower history) to ensure that $t\bar{t}V$ is always identified as the core process.
Since ATLAS analysis uses jets with $p_T > 20$ GeV, the shower starting scale was set to 15 GeV.
Also, a default scale definition for the core process was used.

%

The setup for the background simulation was thoroughly tested on Drell-Yan and $t\bar{t}$ data from 7 and 8 TeV LHC runs and then compared with experimental analyses encoded in \texttt{Rivet}~\textit{v2.4.2} \cite{Buckley:2010ar}.
A sample of validation plots is shown in \autoref{fig:z_validation} for Drell-Yan and \autoref{fig:ttbar_validation} for $t\bar t$.\footnote{More validation plots can be found under \url{www.fuw.edu.pl/~wkotlarski/MC-validation}.}
The result of background simulation using \texttt{Sherpa} was also compared with samples obtained using the same method as used for signal simulation. 
They agreed within theoretical uncertainties.

\section{Recasting current ATLAS 13 TeV analysis \label{sec:sgluons_mc_analysis}}

\begin{table}
  \centering
  \begin{tabular}{c||ccc|c}
    & SS muon pair & \# b-jets $\geq 3$ & $m_{\text{eff}} > 650$ GeV & $E_T^{\texttt{miss}} > 125$ GeV\\
    \hline
    \hline
    $t\bar{t} \mu \nu$ & 3.1876 & 0.0899 & 0.0198 & $0.0117 \pm 0.0006$\\
    $t\bar{t} \mu^+ \mu^-$ & 2.850 & 0.102 & 0.028 & $0.010 \pm 0.001$\\
    \hline
    $m_O = 0.90$ TeV & 1.352 & 0.707 & 0.629 & $0.424 \pm 0.002$\\
    $m_O = 1.00$ TeV & 0.6410 & 0.3324 & 0.3081 & $0.2172 \pm 0.0007$ \\
    $m_O = 1.25$ TeV & 0.1144 & 0.0569 & 0.0552 & $0.0426 \pm 0.0001$ \\
    $m_O = 1.50$ TeV & 0.02365 & 0.01109 & 0.01094 & $0.00897 \pm 0.00003$
  \end{tabular}
  \caption{Cut-flow analysis summary (numbers in fb).
      For brevity's sake, errors only for the final results are given.
      Errors are only statistical.
  \label{tab:cut-flow}}
\end{table}
\definecolor{Gray}{gray}{0.9}
\begin{table}
  \centering
  \begin{tabular}{c||c|c}
    & this analysis & ATLAS \\
    \hline
    \hline
    $t\bar{t} \mu \nu$ & $0.149 \pm 0.007$ & $0.10 \pm 0.05$\\
    $t\bar{t} \mu^+ \mu^-$ & $0.12 \pm 0.02$ & $0.14 \pm 0.06$\\
    \hline
    $m_O = 0.90$ TeV & $5.42 \pm 0.02$ & \cellcolor{Gray}\\
    $m_O = 1.00$ TeV & $2.781 \pm 0.009$ & \cellcolor{Gray}\\
    $m_O = 1.25$ TeV & $0.546 \pm 0.002$ & \cellcolor{Gray}\\
    $m_O = 1.50$ TeV & $0.1148 \pm 0.0003$ & \cellcolor{Gray} 
  \end{tabular}
  \caption{Final result of the analysis (last column of \autoref{tab:cut-flow}) after multiplying by 3.2 fb$^{-1}$ of integrated luminosity and roughly a factor of 4 to account for all possible leptonic channels taken into account in the ATLAS analysis \cite{Aad:2016tuk} compared to column SRb3 of Tab.~5 of that analysis.
  \label{tab:my_cut-flow_vs_atlas}}
\end{table}

The ATLAS analysis \cite{Aad:2016tuk} targeted topologies with 2 same-sign leptons or 3 leptons, looking at 4 different signal regions.
In the case of sgluon decaying to top quark pairs, the interesting signal region is SR3b, defined in Table 1 of \cite{Aad:2016tuk}.
To match experimental data as closely as possible, the detector response was parametrized using \texttt{Delphes} \cite{deFavereau:2013fsa} \textit{v3.3.2}. 
Events were passed using \texttt{HepMC} interface \cite{Dobbs:2001ck}.

\begin{figure}
  \centering
  \includegraphics[width=0.6\textwidth]{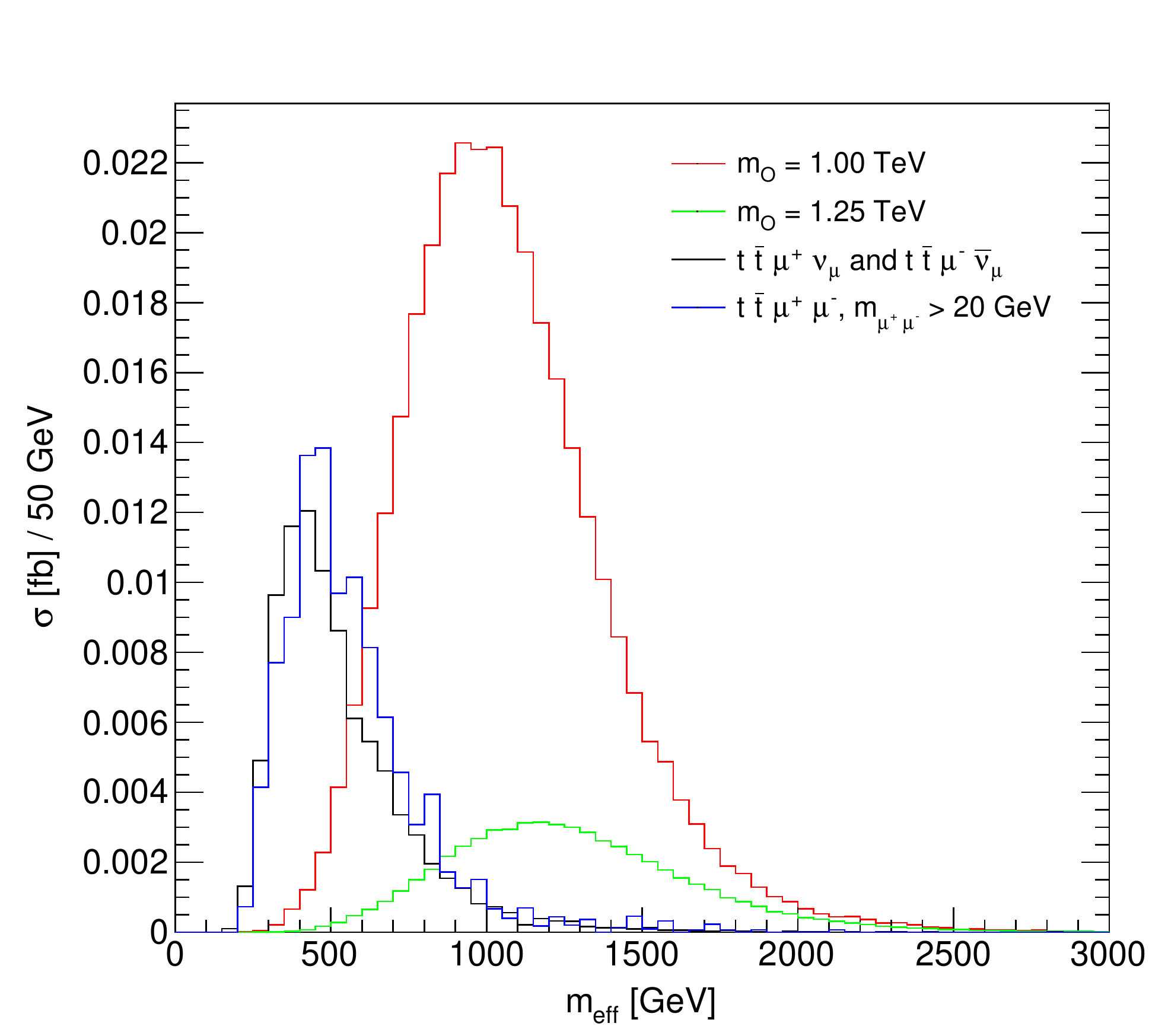}
  \caption{Effective mass spectrum after requiring 2 same-sign leptons and at least 3 $b$-tagged jets (see text for details) for the signal from 1 TeV sgluon pair and for the background from $t\bar t \mu^+ \nu_\mu$/$t\bar t \mu^- \bar \nu_\mu$ and $t\bar t \mu^+ \mu^-$.}
  \label{fig:eff_mass}
\end{figure}

\begin{figure}
  \centering
  \includegraphics[width=0.49\textwidth]{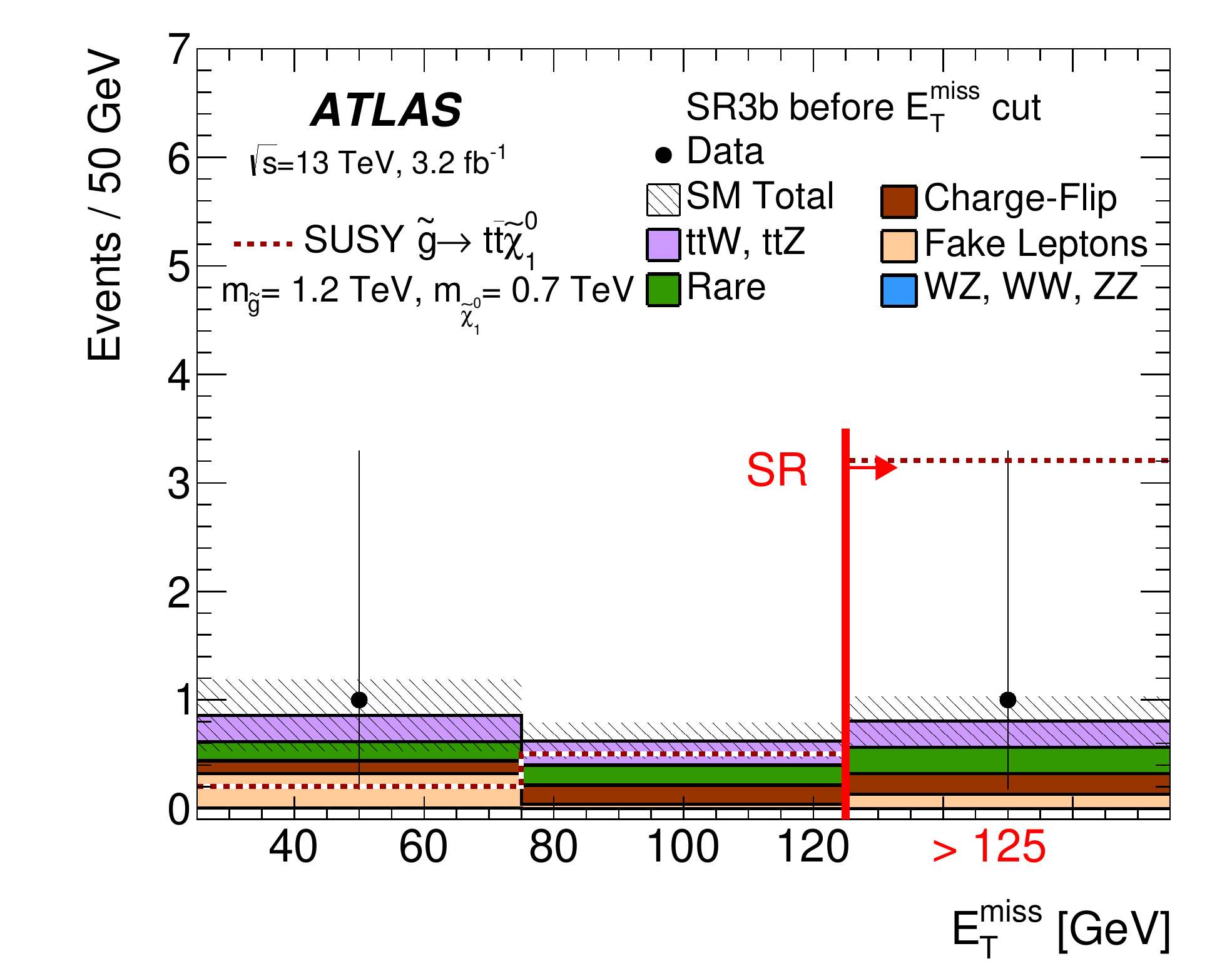}
  \includegraphics[width=0.49\textwidth]{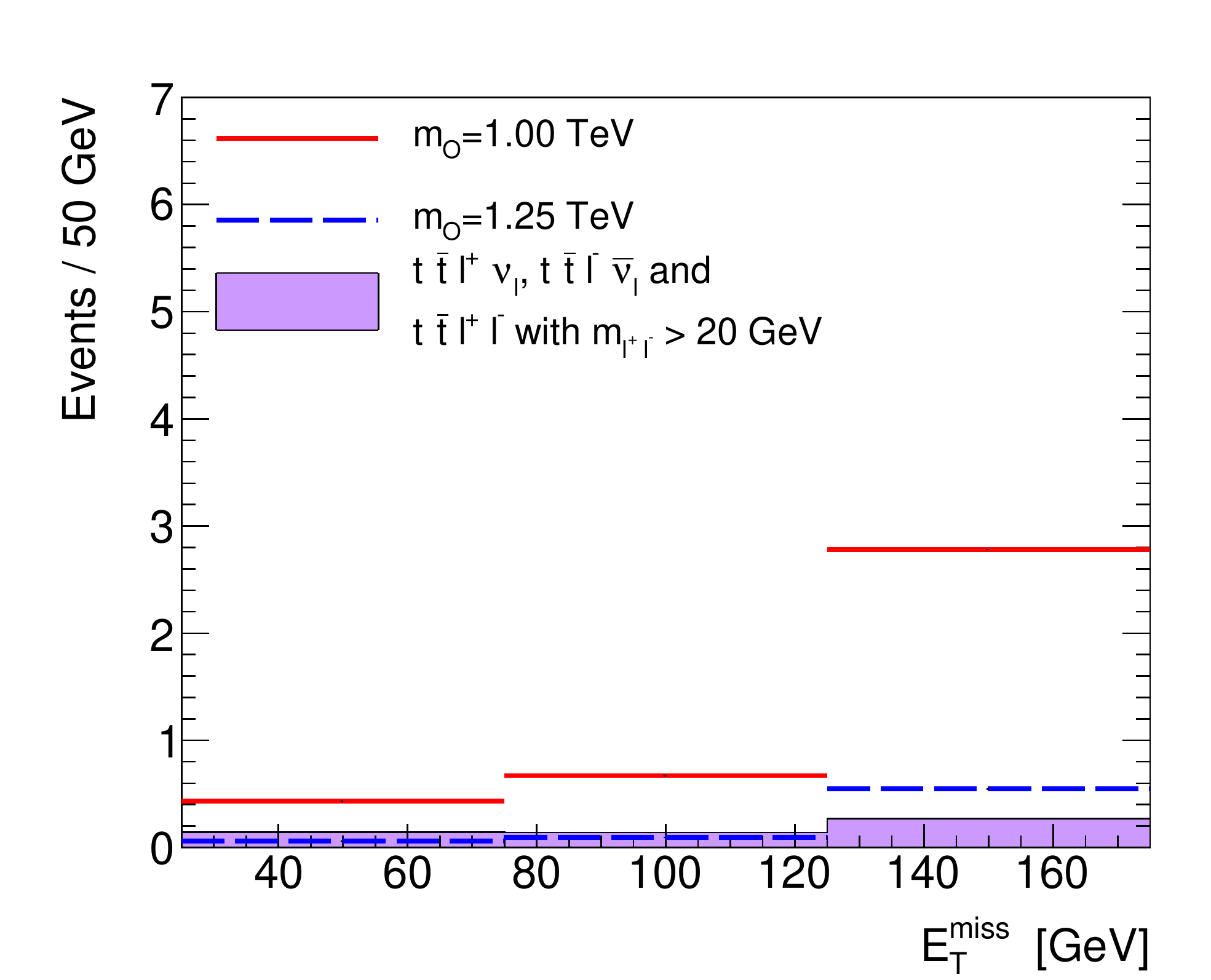}
  \caption{Spectrum of $E_T^{\text{miss}}$ by ATLAS~\cite{Aad:2016tuk} before applying the cut on it (a).
  Right panel (b) shows the analogous plot for $t\bar{t} \mu \nu_\mu$ and $t\bar{t} \mu^+ \mu^-$ backgrounds based on the Monte Carlo simulation used in this work. 
  Red line in right panel b shows the superimposed signal from 1 TeV sgluon production, dashed blue one from 1.25 TeV sgluon. 
  \label{fig:atlas}}
\end{figure}

The following list gives a summary of \texttt{Delphes} detector card settings and applied cuts:
\begin{itemize}

\item[1] Muons are identified with an efficiency of 95\% if they have $p_T > 10$ GeV and $|\eta| < 1.5$ and 85\% if $1.5 < |\eta| < 2.7$.
Candidate muons are required to have $p_T > 20$ GeV and $|\eta| < 2.5$.
Candidate muons must also be isolated, that is have the scalar sum of the $p_T$ of tracks within a variable-size cone around the lepton, excluding its own track, less than 6\% of the muon $p_T$.
The isolation cone size is taken to be the smaller of 10 GeV/$p_T$ and 0.3 (where $p_T$ denotes the muon's transverse momentum).\footnote{\texttt{Delphes} Isolation module was modified to allow for a variable isolation cone size.}

\item[2] At least 3 b-tagged jets  reconstructed using anti-kt algorithm \cite{Cacciari:2008gp} from \texttt{FastJet} with $p_T > 20$ GeV  are required.
Jets are $b$-tagged if they are within $\Delta R_{jb} < 0.3$ of a $b$-quark which had $p_T^b > 5$ GeV and $|\eta_b| < 2.5$ with an efficiency \cite{ATL-PHYS-PUB-2015-022}
\begin{equation}
b\text{-tagging efficient} = \frac{24 \tanh(0.003 \cdot p_T)}{1+0.086 \cdot p_T} 
\end{equation}
Jet energy scale correction is applied according to the formula\footnote{JES is applied \textit{before} the requirement of $p_T > 20$ GeV.} 
\begin{equation}
 E_j \to  \sqrt{1 + (3 - 0.2 |\eta|)^2/p_T} \cdot E_j
\end{equation}
\item[3] Effective mass $m_{\text{eff}}$ of the event, defined as a scalar sum of $p_T$ of signal leptons, b-jets and missing $E_T$, must satisfy $m_{\text{eff}} > 650$ GeV.
The $m_{\text{eff}}$ spectrum for the signal and $t \bar{t} \mu \nu$, $t \bar{t} \mu^+ \mu^-$ is shown in \autoref{fig:eff_mass}.
\item[4] $E_T^{\texttt{miss}} > 125$ GeV
\end{itemize}
\Autoref{tab:cut-flow} shows the cross sections (in fb) for different processes passing this sequence of cuts (cuts are stacked, that is the $n$-th column means that cuts in previous columns were applied).
\Autoref{tab:my_cut-flow_vs_atlas} shows the final numbers of background events, that is after multiplying the last column of \autoref{tab:cut-flow} by 3.2~fb$^{-1}$ of integrated luminosity and roughly a factor of 4 to account for all possible leptonic channels taken into account in the ATLAS analysis.
For example 0.01 fb for process $t\bar{t}\mu^+\mu^-$ in \autoref{tab:cut-flow} corresponds to $0.01 \cdot 3.2 \cdot 4 \approx 0.128$ event.
\Autoref{tab:my_cut-flow_vs_atlas} gives slightly different number, namely 0.12 event, since numbers in \autoref{tab:cut-flow} are already rounded. 
For comparison the last column of \Autoref{tab:my_cut-flow_vs_atlas} shows the column SRb3 of Tab. 5 of Ref.~\cite{Aad:2016tuk}. 
The fact that the simplified analysis based on \texttt{Delphes} predicts roughly the same number of events for background coming from $t\bar{t} \mu \nu_\mu$ and $t\bar{t} \mu^+ \mu^-$ production as the ATLAS one, is a check of its implementation.
Since a significant contribution to the background comes from elements which cannot be reliably simulated by Monte Carlo, like fake/non-prompt leptons and charge flips, the cuts used in the definition of SR3b could not be adapted.
To check the separating power of those cuts on the sgluon signal a plot after cuts on same-sign muon pair and number of $b$-jets was done.
Figure~\ref{fig:eff_mass} shows the spectrum of the effective mass for two sgluons masses: 1 and 1.25 TeV and backgrounds from $t\bar{t} \mu \nu_\mu$ and $t\bar{t} \mu^+ \mu^-$.
It is clear that cut of $m_{\text{eff}} > 650$ GeV used in the ATLAS analysis does also a good job in separating background from the sgluon signal.
For completeness I also show the numbers for background and signal events for sgluons of mass 1 and 1.25 TeV after the effective mass cut but before the cut on missing $E_T$.
They are compared with the original ATLAS plot in Fig.~\ref{fig:atlas} .

\begin{figure}
  \centering
  \includegraphics{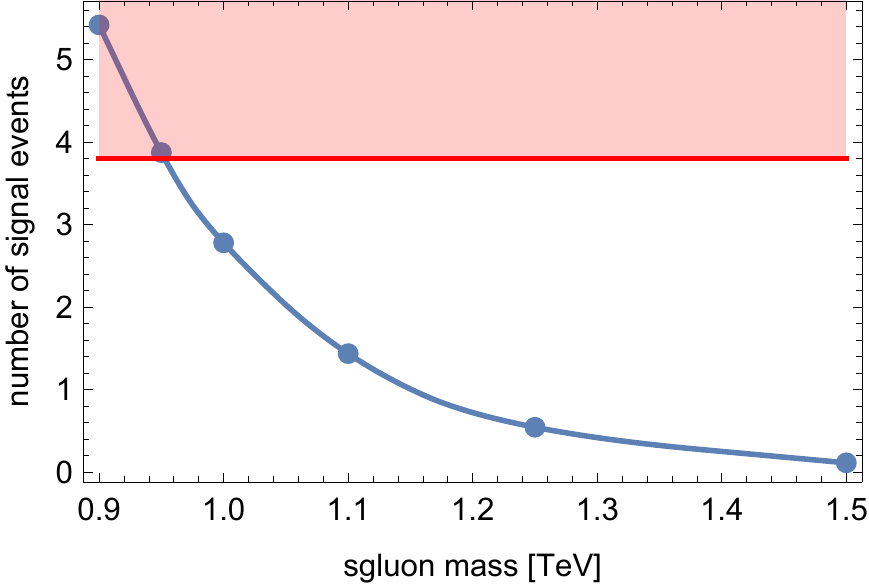}
  \caption{ 
    Predicted number of observed signal events as a function of the sgluon mass (blue points).
    Solid line shows interpolation between these points.
    Red region is excluded by ATLAS for SR3b at 95\% CL.
    Interpreted in the context of sgluon production it corresponds to a lower limit on the sgluon mass $m_O \lesssim 0.95$ TeV.
    \label{fig:mass_exclusion}
  }
\end{figure}

The 95\% CL observed upper limit on the number of signal (BSM) events in the SR3b is 3.8.
The predicted number of signal events for selected sgluon masses are given in Tab.~\ref{tab:my_cut-flow_vs_atlas}.
The ATLAS limit corresponds then to sgluons of mass in the range $0.9 < m_O < 1$ TeV.
To facilitate reading of its precise value, predicted numbers of signal events are plotted in Fig.~\ref{fig:mass_exclusion} together with the interpolation between them.
From this, sgluon masses < 0.95 GeV are excluded at 95\% CL.
This result is already on par with the 8 TeV ATLAS exclusion which was 1.06 TeV for the case of a complex sgluon (i.e. with the cross section greater by a factor of 2).

\section{Conclusions}

The data published at the time of writing this chapter by the Run 2 of the LHC are already competitive with the ones from the Run 1.
With 3.2 fb$^{-1}$ of integrated luminosity they allow to exclude sgluons decaying exclusively to top quarks with mass below $\lesssim$ 0.95 TeV.
By the end of Run 2, the ATLAS experiment is expected to gather $\gtrsim 100~\text{fb}^{-1}$ of integrated luminosity, roughly 30 times more than what is available currently.\footnote{See for example the talk at 2016 Moriond Workshop's Electroweak Interactions and  Unified  Theories session \url{https://indico.in2p3.fr/event/12279/session/7/contribution/118/material/slides/0.pdf} [accessed 9.9.2016]} 
Since statistical significance scales like a square-root of integrated luminosity, numbers in Tab.~\ref{tab:my_cut-flow_vs_atlas} suggest that even without further exploiting event kinematics and adapting cuts it should be possible to exclude (or discover) sgluons with masses up to $\lesssim 1.25$ TeV by the end of Run 2.

With higher statistics one could also think about exploiting hadronic top decays, which have a much larger branching ratio, using one of the well established top-tagging techniques \cite{Seymour:1993mx,Plehn:2011tg,Thaler:2011gf,Soper:2012pb,Almeida:2008yp,Kaplan:2008ie}, as top quarks from decays of heavy sgluons would be highly boosted.
These techniques have already been successfully applied in studies of di-top signatures (like 
the case of $Z'$ boson, see for example \cite{Schaetzel:2013vka,Larkoski:2015yqa}). 
With larger statistics, those techniques could event be applied to 4 top final states like the ones discussed in this chapter.
All of the above improvement will significantly extend the discovery potential of LHC with respect to sgluons at Run 2, well above a conservative estimate of 1.25 TeV done here.

\chapter{Summary \label{sec:conclusions}}

This thesis discussed the Minimal R-symmetric Supersymmetric extension of the Standard Model, proving that the MRSSM is a viable alternative to the MSSM.
It described the motivation for considering non-minimal realizations of supersymmetry together with the construction of the MRSSM.
To allow reasonable comparison with experiments, a calculation of the quantum corrections to most relevant observables was performed.
This thesis presented a set of benchmark points which a posteriori were proven to satisfy constraints such as \textit{W}-boson mass, properties of the Higgs sector, selected $b$-physics observables and the requirement of vacuum stability. 
On top of that, a detailed calculation of NLO QCD corrections to color-octet scalar (sgluon) pair production in a simplified model was presented.
The result of ATLAS search in a similar topology at 13 TeV LHC was interpreted in the context of sgluon production, extracting an exclusion limit for their mass.

In the case of EW observables, quantum corrections included the calculation of one-loop corrections to $W$ and Higgs boson masses.
The latter calculation was supplemented by the leading two-loop corrections in the effective potential approximation.

In connection with the calculation of $W$ mass, the employed calculation procedure was explained.
Since MRSSM contains an $SU(2)_L$ triplet, the treatment of its vacuum expectation value in the setup of Ref.~\cite{Degrassi:1990tu} was described.
A broad region of parameter space, consistent with the measured $W$ mass of $80.385 \pm 0.015$ GeV, was identified.

In the case of Higgs mass calculation, these thesis presented results of tree-level, one- and two-loop calculations.
Although tree-level Higgs mass was lower for selected benchmark points than in the MSSM, radiative corrections from new  MRSSM states were instrumental to raise it to the measured value.
Contributions from different particle sectors were analyzed and emphasis was put on Yukawa-like superpotential parameters $\lambda_u$, $\Lambda_u$.
For $\Lambda_u \approx -1.1$  the lightest Higgs boson with mass at one-loop level of around 120 GeV was obtained.

The calculation of two-loop corrections was explained in detail, and analytic formulas for the most important contributions were given.
At the two-loop level, the Higgs mass becomes for the first time directly sensitive to pure-QCD particles such as sgluons and Dirac gluinos.
These specific contributions were analyzed, showing a non-decoupling behavior.
In general, the leading two-loop corrections for selected benchmark points account for a shift of about +5 GeV to the lightest Higgs mass.
This positive shift is a typical feature of two-loop calculations in the $\overline{\text{DR}}$ scheme.

The constraints resulting from the \textit{W} and Higgs mass calculations were combined, identifying a parameter space in which both of these observables could be accommodated.
The region around the benchmark points, consistent with these two observables, was identified both by $2d$ and random multidimensional scans.

In the case of the QCD sector, the emphasis was put on the production of sgluons. 
These particles are special from the viewpoint of SUSY, as they have R-charge 0 and can decay without an LSP in the final state.
NLO QCD $K$-factors for selected sgluon masses in the range 1-2 TeV at 13 and 14 TeV LHC were calculated.
These \textit{K}-factors range from 1.37 to 1.47. 
This calculation also features a detailed description of the treatment of IR singularities using the two-cut phase space slicing method.
The results of this calculation were then applied to recasting of current ATLAS searches of same-sign lepton production into an exclusion limit on the sgluon mass.
The analysis showed that collected data sample of 3.2/fb was already competitive with the data collected during Run 1, excluding sgluons with masses $\lessapprox 1.1$ TeV.

To conclude, non-minimal supersymmetric extensions of the SM are not only a viable alternative to the MSSM but they also circumvent certain MSSM shortcomings.
It is hoped that this study will encourage experimental groups to perform dedicated searches of signals of the MRSSM.

\part*{Appendices and bibliography}
\appendix


\printacronyms[include-classes=abbrev,name={List of acronyms},heading=chapter]

%
\chapter{Selected mass matrices}

\section{Higgs bosons}

\subsection{Neutral pseudoscalar Higgses \label{sec:pseudohiggs_mass_matrix}}
The pseudoscalar Higgs mass matrix is block diagonal.
In the basis ($\sigma_d$, $\sigma_u$, $\sigma_S$, $\sigma_T$) it is given by
\begin{align}
 m_A^2 =  
 \left(
    \begin{array}{r@{}cc@{}l}
  &  
  \begin{matrix}
    B_\mu \tan \beta & B_\mu \\
    B_\mu & B_\mu \cot \beta
  \end{matrix} 
  & \mbox{\large 0}_{2\times2}  \\  
  &   \mbox{\large 0}_{2\times2} &  
       \begin{matrix}\rule{0pt}{3ex}
        m_S^2 + \frac{1}{2} \left ( \lambda_d^2 v_d^2 + \lambda_u^2 v_u^2 \right ) & 
  \frac{1}{2\sqrt{2}} \left ( \lambda_d \Lambda_d v_d^2 + \lambda_u \Lambda_u v_u^2 \right ) \\
  \frac{1}{2\sqrt{2}} \left ( \lambda_d \Lambda_d v_d^2 + \lambda_u \Lambda_u v_u^2 \right )
  & m_T^2 + \frac{1}{4} \left ( \Lambda_d^2 v_d^2 + \Lambda_u^2 v_u^2 \right )
      \end{matrix}    
    \end{array} 
\right)  \\
\nonumber
+ \, \xi_Z \, m_Z^2 \left(
    \begin{array}{r@{}cc@{}l}
  &  
  \begin{matrix}
    \cos^2 \beta & -\frac{1}{2} \sin 2\beta \\
    -\frac{1}{2} \sin 2\beta & \sin^2 \beta
  \end{matrix} 
  & \mbox{\large 0}_{2\times2}  \\  
  & \mbox{\large 0}_{2\times2} & \mbox{\large 0}_{2\times2}   
    \end{array} 
\right) ,
\end{align}
with $m_Z^2 = \frac{1}{4} (g_1^2+ g_2^2) v^2$ and $O_{2\times2}$ being a 2-by-2 submatrix filled with 0.
The $\sigma_d - \sigma_u$ submatrix is identical to the MSSM one, with eigenstates being the Goldstone boson with mass $m_{G^0}^2 = \xi_Z m_Z^2$ and an MSSM-like pseudoscalar Higgs with mass $m_A^{2,\text{MSSM}} = 2 B_\mu/\sin 2\beta$. 
Since $\sin 2\beta$ is always positive, it implies that with sign convention of \autoref{eq:bmu_sign_convention} $B_\mu > 0$.

\subsection{Neutral scalar Higgses \label{sec:higgs_mass_matrix}}
The scalar Higgs mass matrix in the basis ($\phi_d$, $\phi_u$, $\phi_S$, $\phi_T$) is given by
\begin{align} 
m^2_{H} =  \begin{pmatrix}
\begin{smallmatrix}
\frac{1}{8} (g_1^2 v^2 + 2 \left(g_1^2+g_2^2\right) \cos 2 \beta v^2 \\
+ 4 \lambda_d^2 v_S^2 +2 (2 \mu_d+\Lambda_d v_T)^2 \\
+8 m_{H_d}^2 - 8 \text{g1} M^D_S v_S + 4 \sqrt{2} \\
   \lambda_d v_S (2 \mu_d + \Lambda_d v_T)\\
    +g_2 \left(g_2 v^2+8 M^D_T
   v_T\right)) 
   \end{smallmatrix} 
  & \cdot 
  & \cdot
  & \cdot
   \\ 
   \begin{smallmatrix}
   - B_\mu - \frac{1}{2} m_Z^2 \sin 2 \beta  
   \end{smallmatrix}
    & \begin{smallmatrix}
    \frac{1}{8} (8
   \mu_u^2 + 8 \sqrt{2} \lambda_u v_S \mu_u \\
   -8 \Lambda_u v_T \mu_u + g_1^2 \\
   v^2 + g_2^2 v^2 + 4 \lambda_u^2 v_S^2 \\
   + 2 \Lambda_u^2 v_T^2 + 8 m_{H_u}^2 \\
   + 8 g_1 M^D_S v_S - 8 g_2 M^D_T v_T \\
   - 4 \sqrt{2} \lambda_u \Lambda_u v_S \\
   v_T-2 \left(g_1^2+g_2^2\right) v^2 \cos 2 \beta ) 
       \end{smallmatrix}
   & \cdot
   & \cdot
  \\
   \begin{smallmatrix}
 \frac{1}{2} v (\lambda_d (2 \lambda_d v_S + \\
 \sqrt{2} (2 \mu_d +\Lambda_d v_T)) \\
   -2 g_1 M^D_S ) \cos \beta
   \end{smallmatrix}
   & \begin{smallmatrix}
 \frac{1}{2} v (2 g_1
   M^D_S + \Lambda_u (2 \Lambda_u v_S \\+ \sqrt{2} (2 \mu_u - \Lambda_u
   v_T))) \sin \beta 
   \end{smallmatrix}
   & \begin{smallmatrix} 
   \frac{1}{2} (8 M^{D,2}_S + \lambda_d^2 v^2 \cos ^2 \beta \\
   + \lambda_u^2 v^2 \sin ^2 \beta +2 m_S^2 ) 
   \end{smallmatrix}
   & \cdot
   \\
   \begin{smallmatrix}
   \frac{1}{2} v (2 g_2 M^D_T + \Lambda_d (2 \mu_d \\
   + \sqrt{2} \lambda_d
   v_S + \Lambda_d v_T )) \cos \beta
      \end{smallmatrix}
   & \begin{smallmatrix}
   -\frac{1}{2} v (2 g_2
   M^D_T + \Lambda_u (2 \mu_u \\
   + \sqrt{2} \Lambda_u v_S - \Lambda_u v_T)) \sin \beta  
   \end{smallmatrix}
   & \begin{smallmatrix}  
v^2 (\lambda_d \Lambda_d - \lambda_u \Lambda_u +(\lambda_d \\
   \Lambda_d + \lambda_u \Lambda_u) \cos 2 \beta )\frac{1}{4 \sqrt{2}} 
    \end{smallmatrix}
   & \begin{smallmatrix}
   4 M^{D,2}_T + m_T^2 +\frac{1}{4} v^2 \\
   (\Lambda_d^2 \cos^2 \beta + \Lambda_u^2 \sin^2 \beta ) 
   \end{smallmatrix}
  \end{pmatrix}
\end{align} 
where since the matrix is symmetric and upper-triangle entries were not shown for readability.
Substituting solution of the tadpole equations to the 2-by-2 submatrix in left-top corner gives
\begin{equation}
\centering
 m_H^{2,\text{MSSM}} = \left(\begin{array}{cc}
m_A^{2,\text{MSSM}} \sin^2 \beta + M_Z^2 \cos^2 \beta & -(m_A^{2,\text{MSSM}} + M_Z^2) \sin\beta \cos
\beta\\
-(m_A^{2,\text{MSSM}} + M_Z^2) \sin\beta \cos \beta & m_A^{2,\text{MSSM}} \cos^2 \beta + M_Z^2 \sin^2
\beta\end{array}\right).\label{lll}
\end{equation}
which has the knows MSSM form.

\section{R-Higgs bosons}

\subsection{Neutral R-Higgses}
In the basis $(R_d^0, R_u^0)$, $(R_d^{0,*}, R_u^{0,*})$ the matrix reads
\begin{align}
m_{ R^0 }^2 = 
\begin{pmatrix}
\label{eq:r_higgs_mass_matrix}
\begin{smallmatrix}
m_{R_d}^2 + \frac{1}{8} \left(2 \lambda_d^2 + \Lambda_d^2\right) v^2-\left(g_1^2+ g_2^2-2
   \lambda_d^2-\Lambda_d^2\right) v^2 \cos 2 \beta   \\
   +4 \lambda_d^2 v_S^2 +2 (2
   \mu_d + \Lambda_d v_T)^2 +8 \text{g1} M^D_S v_S - 8 g_2 M^D_T
   v_T \\
   +4 \sqrt{2} \lambda_d v_S (2 \mu_d + \Lambda_d v_T)
   \end{smallmatrix}
   & \frac{1}{8}
   (\Lambda_d \Lambda_u - 2 \lambda_d \lambda_u) v^2 \sin 2 \beta \\
 \frac{1}{8} (\Lambda_d \Lambda_u - 2 \lambda_d \lambda_u) v^2 \sin 2 \beta & 
 \begin{smallmatrix}
 \frac{1}{8} (2
   \left(v^2+2 v_S^2\right) \lambda_u^2-4 \sqrt{2} \Lambda_u v_S v_T \lambda_u+8
   \mu_u^2+8 m_{R_u}^2 \\
   - 8 g_1 M^D_S v_S + 8 g_2 M^D_T v_T + 8 \mu_u \\
   \left(\sqrt{2} \lambda_u v_S - \Lambda_u v_T \right)+ \Lambda_u^2 \left(v^2+2
   v_T^2\right)+
   \\
   \left(g_1^2+g_2^2-2 \Lambda_u^2-\Lambda_u^2\right) v^2 \cos 2 \beta
   )
    \end{smallmatrix}
\end{pmatrix}
\end{align}

\subsection{Charged R-Higgses}
Charged R-Higgses do not mix. Therefore their masses can be given explicitly as
\begin{align}
  \label{eq:rum-higgs_mass}
  m_{R_u^-}^2 =& m_{R_u}^2 + \frac{1}{4} \Lambda_u^2 v^2 + \frac{1}{2} \lambda^2_u v_S^2 + \frac{1}{\sqrt{2}} \lambda_u v_S ( 2 \mu_u + \Lambda_u v_T ) - \frac{1}{2} (g_1 M^D_S v_S + g_2 M^D_T v_T) \\
   & +  \frac{1}{8} \left(g_1^2 - g_2^2 - 2 \Lambda_u^2 \right) v^2 \cos 2 \beta + \frac{1}{4} (2 \mu_u + \Lambda_u v_T)^2 \nonumber \\
  \label{eq:rdp-higgs_mass}
  m_{R_d^+}^2 =& m_{R_d}^2 + \mu_d^2 + g_1 M_S^D v_S + \frac{1}{2} \lambda_d^2 v_S^2 + g_2 M_T^D v_T + \mu_d (\sqrt{2} \lambda_d v_S - \Lambda_d v_T) \\
  & + \frac{1}{4} \Lambda_d(-2\sqrt{2}\lambda_d v_S v_T + \Lambda_d(v^2+v_T^2) ) + \frac{1}{8} \left(g_2^2 - g_1^2 + 2\Lambda_d^2\right) v^2 \cos 2\beta \nonumber
\end{align}

\section{Electroweak gauginos}
\subsection{Charginos \label{sec:chargino_mass_matrix}}
In the basis: ($\tilde T^-$, $\tilde H^-_d$), ($\tilde W^+$, $\tilde R^+_d$) first chargino mass matrix reads
\begin{align}
m_{ \tilde \chi^+} = \left(
  \begin{matrix}
    g_2 v_T + M^D_T & \frac{1}{\sqrt{2}} \Lambda_d v \cos \beta \\
    \frac{1}{\sqrt{2}} g_2 v \cos \beta & -\frac{1}{2} \Lambda_d v_T + \frac{1}{\sqrt 2} \lambda_d v_S + \mu_d 
  \end{matrix}
  \right)
\end{align}
In the basis: ( $\tilde W^-$, $\tilde R^-_u$), ($\tilde T^+$, $\tilde H^+_u$) second chargino mass matrix reads
\begin{align}
m_{ \tilde \rho^-} = \left(
  \begin{matrix}
    -g_2 v_T + M^D_T & \frac{1}{\sqrt{2}} g_2 v \sin \beta \\
    -\frac{1}{\sqrt{2}} \Lambda_u v \sin \beta & -\frac{1}{2} \Lambda_u v_T - \frac{1}{\sqrt 2} \lambda_u v_S + \mu_u
  \end{matrix}
  \right)
\end{align}
These matrices are diagonalized by two independent rotations $U$ and $V$ as
\begin{align}
  \label{eq:diag_rot1}
  U^{1,*} m_{\tilde{\chi}^+} V^{1,\dagger} &= m^{\text{diag}}_{{\chi}^+} \\
    \label{eq:diag_rot2}
  U^{2,*} m_{\tilde{\rho}^-} V^{2,\dagger} & = m^{\text{diag}}_{{\rho}^-}
\end{align}
\subsection{Neutralinos \label{sec:neutralino_mass_matrix} }
In the basis: ($\tilde B$, $\tilde W^0$, $\tilde R^0_d$, $\tilde R^0_u$), ( $\tilde S$, $\tilde T^0$, $\tilde H^0_d$, $\tilde H^0_u$) neutralino mass matrix reads
\begin{equation}
\small
m_{\tilde \chi^0} = \left(
\begin{array}{cccc}
 M^D_S & 0 & -\frac{1}{2} g_1 v \cos \beta & \frac{1}{2} g_1 v \sin \beta \\
 0 & M^D_T & \frac{1}{2} g_2 v \cos \beta & -\frac{1}{2} g_2 v \sin \beta \\
 -\frac{1}{\sqrt{2}} \lambda_d v \cos \beta & -\frac{1}{2} \Lambda_d v \cos \beta &
   -\mu_d - \frac{1}{\sqrt{2}} \lambda_d v_S - \frac{1}{2} \Lambda_d v_T & 0 \\
 \frac{1}{\sqrt{2}} \lambda_u v \sin \beta  & -\frac{1}{2} \Lambda_U v \sin \beta & 0 &
   \mu_u + \frac{1}{\sqrt{2}} \lambda_u v_S - \frac{1}{2} \Lambda_u v_T \\
\end{array}
\right)
\end{equation}
The matrix is diagonalized by two independent rotations $N^{1,2}$ as 
\begin{equation}
  \label{eq:diag_rot3}
  N^{1,*} m_{\tilde \chi^0} N^{2,\dagger} = m^{\text{diag}}_{\tilde \chi^0} .
\end{equation}

\chapter{Selected Feynman rules}

This Appendix presents a selection of MRSSM Feynman rules. 
For brevity's sake, not all Feynman rules used in this thesis are displayed and sfermion mixing was set to 0.
The following abbreviations were also introduced $ \mathbb{P}_L \equiv (1-\gamma_5)/2$ and $ \mathbb{P}_R \equiv (1+\gamma_5)/2$.
Rotation matrices $V, U$ are defined in \autoref{eq:diag_rot1}, \autoref{eq:diag_rot2} and $N$ in \autoref{eq:diag_rot3}.

\section{Electroweak gauginos}
Feynman rules involving charginos and neutralinos used in \autoref{sec:ew}.
\noindent
\begin{longtable}{ll}
\raisebox{-0.475\height}{
\begin{fmffile}{img/feynman_rules/chi0chi+W-} 
\fmfframe(20,20)(20,20){ 
\begin{fmfgraph*}(60,60) 
\fmfleft{l1}
\fmfright{r1,r2}
\fmf{fermion}{v1,l1}
\fmf{fermion}{r1,v1}
\fmf{boson}{r2,v1}
\fmflabel{$\overline{{\chi}_{{i}}}$}{l1}
\fmflabel{${\chi}^+_{{j}}$}{r1}
\fmflabel{$W^-_{{\mu}}$}{r2}
\end{fmfgraph*}} 
\end{fmffile}
}
&
$\frac{\imath}{2} g_2 \gamma_\mu \left [\left(2 V^{1*}_{j 1} N_{{i 2}}^{1}  - \sqrt{2} V^{1*}_{j 2} N_{{i 3}}^{1} \right)\mathbb{P}_L
  +  \left (2 N^{2*}_{i 2} U_{{j 1}}^{1}  + \sqrt{2} N^{2*}_{i 3} U_{{j 2}}^{1} \right )\mathbb{P}_R \right] $\\[3mm]
\raisebox{-0.475\height}{
\begin{fmffile}{img/feynman_rules/chi-chi0W+} 
\fmfframe(20,20)(20,20){ 
\begin{fmfgraph*}(60,60) 
\fmfleft{l1}
\fmfright{r1,r2}
\fmf{fermion}{v1,l1}
\fmf{fermion}{r1,v1}
\fmf{boson}{v1,r2}
\fmflabel{$\overline{{\chi}^+_{{i}}}$}{l1}
\fmflabel{${\chi}_{{j}}$}{r1}
\fmflabel{$W^+_{{\mu}}$}{r2}
\end{fmfgraph*}} 
\end{fmffile} 
}
&
$ \frac{\imath}{2} g_2 \gamma_\mu \left [ \left (2 N^{1*}_{j 2} V_{{i 1}}^{1}  - \sqrt{2} N^{1,*}_{j 3} V_{{i 2}}^{1} \right ) \mathbb{P}_L  
  + \, \left (2 U^{1*}_{i 1} N_{{j 2}}^{2}  + \sqrt{2} U^{1*}_{i 2} N_{{j 3}}^{2} \right ) \mathbb{P}_R \right ]$\\[3mm]
\raisebox{-0.475\height}{
\begin{fmffile}{img/feynman_rules/chi0rho-W+} 
\fmfframe(20,20)(20,20){ 
\begin{fmfgraph*}(60,60) 
\fmfleft{l1}
\fmfright{r1,r2}
\fmf{fermion}{v1,l1}
\fmf{fermion}{r1,v1}
\fmf{boson}{v1,r2}
\fmflabel{$\overline{{{\chi}}_{{i}}}$}{l1}
\fmflabel{${\rho}^-_{{j}}$}{r1}
\fmflabel{$W^+_{{\mu}}$}{r2}
\end{fmfgraph*}} 
\end{fmffile}   
}
&
$-\frac{\imath}{2} g_2 \gamma_\mu \left [\left(2 U^{2*}_{j 1} N_{{i 2}}^{1}  + \sqrt{2} U^{2*}_{j 2} N_{{i 4}}^{1} \right)  \mathbb{P}_L 
   -\left({\sqrt{2}}  N^{2*}_{i 4} V_{{j 2}}^{2}  -2 N^{2*}_{i 2} V_{{j 1}}^{2} \right)  \mathbb{P}_R \right ] $\\[3mm]
\raisebox{-0.475\height}{
\begin{fmffile}{img/feynman_rules/rho+chi0W-} 
\fmfframe(20,20)(20,20){ 
\begin{fmfgraph*}(60,60) 
\fmfleft{l1}
\fmfright{r1,r2}
\fmf{fermion}{v1,l1}
\fmf{fermion}{r1,v1}
\fmf{boson}{r2,v1}
\fmflabel{$\overline{{\rho}^-_{{i}}}$}{l1}
\fmflabel{${\chi}_{{j}}$}{r1}
\fmflabel{$W^-_{{\mu}}$}{r2}
\end{fmfgraph*}} 
\end{fmffile} 
}
&
$-\frac{\imath}{2} g_2 \gamma_\mu \left [ \left (2 N^{1*}_{j 2} U_{{i 1}}^{2}  + \sqrt{2} N^{1*}_{j 4} U_{{i 2}}^{2} \right )  \mathbb{P}_L - \,\left({\sqrt{2}}  V^{2*}_{i 2} N_{{j 4}}^{2}  - V^{2*}_{i 1} N_{{j 2}}^{2} )  \mathbb{P}_R \right)\right]$\\[3mm]
\raisebox{-0.475\height}{
\begin{fmffile}{img/feynman_rules/chi+ellsnu} 
\fmfframe(20,20)(20,20){ 
\begin{fmfgraph*}(60,60) 
\fmfleft{l1}
\fmfright{r1,r2}
\fmf{fermion}{v1,l1}
\fmf{fermion}{r1,v1}
\fmf{scalar}{v1,r2}
\fmflabel{$\overline{{\chi}^-_{{i}}}$}{l1}
\fmflabel{$\ell$}{r1}
\fmflabel{$\tilde{\nu}^*_{{\ell}}$}{r2}
\end{fmfgraph*}} 
\end{fmffile} 
}
&  
$-\imath g_2 V^{1*}_{i 1}    \mathbb{P}_L  
   + \,i Y^*_{\ell}   U_{{i 2}}^{1}  \mathbb{P}_R $\\[3mm]
%
\raisebox{-0.475\height}{
\begin{fmffile}{img/feynman_rules/chi0nusnu} 
\fmfframe(20,20)(20,20){ 
\begin{fmfgraph*}(60,60) 
\fmfleft{l1}
\fmfright{r1,r2}
\fmf{fermion}{v1,l1}
\fmf{fermion}{r1,v1}
\fmf{scalar}{v1,r2}
\fmflabel{$\overline{{\chi}_{{i}}^c}$}{l1}
\fmflabel{${\nu_{{\ell}}}$}{r1}
\fmflabel{$\tilde{\nu}^*_{{\ell}}$}{r2}
\end{fmfgraph*}} 
\end{fmffile} 
}
&$
\frac{\imath}{\sqrt{2}} \left (g_1 N^{1^*}_{i 1}  - g_2 N^{1*}_{i 2} \right ) \mathbb{P}_L $
\\[3mm] 
\raisebox{-0.475\height}{
\begin{fmffile}{img/feynman_rules/rho-nusel} 
\fmfframe(20,20)(20,20){ 
\begin{fmfgraph*}(60,60) 
\fmfleft{l1}
\fmfright{r1,r2}
\fmf{fermion}{v1,l1}
\fmf{fermion}{r1,v1}
\fmf{scalar}{v1,r2}
\fmflabel{$\overline{{\rho}^+_{{i}}}$}{l1}
\fmflabel{$\nu_{{\ell}}$}{r1}
\fmflabel{$\tilde{\ell}^*_{{L}}$}{r2}
\end{fmfgraph*}} 
\end{fmffile} 
}
& 
$ -\imath g_2 U^{2*}_{i 1}    \mathbb{P}_L $
\\[3mm]
\raisebox{-0.475\height}{
\begin{fmffile}{img/feynman_rules/chi0esel} 
\fmfframe(20,20)(20,20){ 
\begin{fmfgraph*}(60,60) 
\fmfleft{l1}
\fmfright{r1,r2}
\fmf{fermion}{v1,l1}
\fmf{fermion}{r1,v1}
\fmf{scalar}{v1,r2}
\fmflabel{$\overline{{\chi}_{{i}}^c}$}{l1}
\fmflabel{${\ell}$}{r1}
\fmflabel{$\tilde{\ell}^*_{{L}}$}{r2}
\end{fmfgraph*}} 
\end{fmffile} 
}
& 
$   \frac{\imath}{\sqrt{2}} \left(g_1 N^{1*}_{i 1}  + g_2 N^{1*}_{i 2} \right)
        \mathbb{P}_L $ 
    $ -\imath Y_{\ell}^* N_{{i 3}}^{2}  \mathbb{P}_R $     
\\[3mm]   
%
\raisebox{-0.475\height}{
\begin{fmffile}{img/feynman_rules/chi0esel2} 
\fmfframe(20,20)(20,20){ 
\begin{fmfgraph*}(60,60) 
\fmfleft{l1}
\fmfright{r1,r2}
\fmf{fermion}{v1,l1}
\fmf{fermion}{r1,v1}
\fmf{scalar}{v1,r2}
\fmflabel{$\overline{{{\chi}}_{{i}}}$}{l1}
\fmflabel{$\ell$}{r1}
\fmflabel{$\tilde{\ell}^*_{{R}}$}{r2}
\end{fmfgraph*}} 
\end{fmffile} 
}
&
$   -i N^{2*}_{i 3} Y_{\ell}     \mathbb{P}_L  
    \,-i \sqrt{2} g_1   N_{{i 1}}^{1}  \mathbb{P}_R $
\\[3mm]
\raisebox{-0.535\height}{
\begin{fmffile}{img/feynman_rules/chi-nusel} 
\fmfframe(20,20)(20,20){ 
\begin{fmfgraph*}(60,60) 
\fmfleft{l1}
\fmfright{r1,r2}
\fmf{fermion}{v1,l1}
\fmf{fermion}{r1,v1}
\fmf{scalar}{v1,r2}
\fmflabel{$\overline{{\chi}^+_{{i}}}$}{l1}
\fmflabel{$\nu_{{\ell}}$}{r1}
\fmflabel{$\tilde{\ell}^*_{{R}}$}{r2}
\end{fmfgraph*}} 
\end{fmffile} 
}
&
$  \imath \, U^{1*}_{i 2} Y_{\ell}\,\mathbb{P}_L $
\end{longtable}

\section{SQCD \label{sec:sqcd_feynman_rules}}
Selection of Feynman rules for the strongly interacting particles.
Below, $\lambda$ denotes Gell-Mann matrices, $f$ are $SU(3)$ structure constants and $\tilde q$ represents squarks (there is no squark mixing). 
\noindent
\begin{longtable}{ll}
\raisebox{-0.465\height}{
\begin{fmffile}{img/feynman_rules/os-gdgdbar} 
\fmfframe(20,20)(20,20){ 
\begin{fmfgraph*}(60,60) 
\fmfleft{l1}
\fmfright{r1,r2}
\fmf{fermion}{v1,l1}
\fmf{fermion}{r1,v1}
\fmf{dashes}{r2,v1}
\fmflabel{$\overline{\tilde{g}_{{\alpha}}}$}{l1}
\fmflabel{$\tilde{g}_{{\beta}}$}{r1}
\fmflabel{$O_{S,\gamma}$}{r2}
\end{fmfgraph*}} 
\end{fmffile} 
} 
& $- g_s f_{\alpha \beta \gamma}$
\\[3mm]

\raisebox{-0.465\height}{ 
\begin{fmffile}{img/feynman_rules/oa-gdgdbar} 
\fmfframe(20,20)(20,20){ 
\begin{fmfgraph*}(60,60) 
\fmfleft{l1}
\fmfright{r1,r2}
\fmf{fermion}{v1,l1}
\fmf{fermion}{r1,v1}
\fmf{dashes}{r2,v1}
\fmflabel{$\overline{\tilde{g}_{{\alpha}}}$}{l1}
\fmflabel{$\tilde{g}_{{\beta}}$}{r1}
\fmflabel{$O_{A, {\gamma}}$}{r2}
\end{fmfgraph*}} 
\end{fmffile} 
} 
 & $- \imath g_s f_{\alpha\beta\gamma} \gamma^5$
\\[3mm]

\raisebox{-0.465\height}{
\begin{fmffile}{img/feynman_rules/os-susu} 
\fmfframe(20,20)(20,20){
\begin{fmfgraph*}(60,60) 
\fmfleft{l1}
\fmfright{r1,r2}
\fmf{dashes}{l1,v1}
\fmf{scalar}{r1,v1}
\fmf{scalar}{v1,r2}
\fmflabel{$O_{S, \alpha}$}{l1}
\fmflabel{$\tilde{q}_{{i, \beta}}$}{r1}
\fmflabel{$\tilde{q}^*_{{i, \gamma}}$}{r2}
\end{fmfgraph*}} 
\end{fmffile} 
}
& $\imath g_s \Re M^D_O \lambda^{\alpha}_{\gamma\beta} (\delta_{iR} - \delta_{iL})$
\\[3mm]

\raisebox{-0.465\height}{
\begin{fmffile}{img/feynman_rules/oa-susu} 
\fmfframe(20,20)(20,20){ 
\begin{fmfgraph*}(60,60) 
\fmfleft{l1}
\fmfright{r1,r2}
\fmf{dashes}{l1,v1}
\fmf{scalar}{r1,v1}
\fmf{scalar}{v1,r2}
\fmflabel{$O_{A,\alpha}$}{l1}
\fmflabel{$\tilde{q}_{{i, \beta}}$}{r1}
\fmflabel{$\tilde{q}^*_{{i, \gamma}}$}{r2}
\end{fmfgraph*}} 
\end{fmffile}
}
& $-\imath g_s \Im M^D_O \lambda^{\alpha}_{\gamma\beta} (\delta_{iR} - \delta_{iL})$
\\[3mm]

\raisebox{-0.465\height}{
\begin{fmffile}{img/feynman_rules/gd-suu} 
\fmfframe(20,20)(20,20){ 
\begin{fmfgraph*}(60,60) 
\fmfleft{l1}
\fmfright{r1,r2}
\fmf{fermion}{v1,l1}
\fmf{fermion}{r1,v1}
\fmf{scalar}{r2,v1}
\fmflabel{$\overline{q_{{i, \alpha}}}$}{l1}
\fmflabel{$\tilde{g}_{{\beta}}$}{r1}
\fmflabel{$\tilde{q}_{{i, \gamma}}$}{r2}
\end{fmfgraph*}} 
\end{fmffile} 
}
& $\frac{\imath}{\sqrt{2}} g_s \lambda^{\beta}_{\alpha\gamma}\mathbb{P}_L$
\\[3mm]

\raisebox{-0.465\height}{
\begin{fmffile}{img/feynman_rules/gd-suu2} 
\fmfframe(20,20)(20,20){ 
\begin{fmfgraph*}(60,60) 
\fmfleft{l1}
\fmfright{r1,r2}
\fmf{fermion}{v1,l1}
\fmf{fermion}{v1,r1}
\fmf{scalar}{r2,v1}
\fmflabel{$\overline{q_{{i, \alpha}}}$}{l1}
\fmflabel{$\overline{\tilde{g}_{{\beta}}}$}{r1}
\fmflabel{$\tilde{q}_{{i, \gamma}}$}{r2}
\end{fmfgraph*}} 
\end{fmffile} 
}
& 
$ - \frac{\imath}{\sqrt{2}} g_s \lambda^{\beta}_{\alpha\gamma} \mathbb{P}_R$
\end{longtable}

\chapter{Kinematics and integrals}

\section{Two-body decay width formula \label{sec:2body_dec}}
The two-body Lorentz-invariant phase space $\Phi_2$ for particles with 4-momenta $p_1$ and $p_2$ is given by
\begin{equation}
  \Phi_2 = \int d \Phi_2 (p_1, p_2) = \int \frac{d^3 p_1}{(2\pi)^3 2 E_1} \frac{d^3 p_2}{(2\pi)^3 2 E_2} (2\pi)^4\delta^4(P-p_1-p_2) = \frac{\beta}{8 \pi} \int \frac{d \Omega}{4\pi},
\end{equation}
where $\Omega$ is the $2d$ solid angle and $\beta$, in the case of particle with mass $M$ decaying in its rest frame into 2 equal mass $m$ particles, is equal to $\sqrt{1- 4m^2/M^2}$.
In case of scalar decay, matrix element $|\mathcal{M}|^2$ cannot depend on angle allowing for trivial integration of differential decay width $d \Gamma$
\begin{equation}
d \Gamma =  \frac{1}{2 M} \overline{|\mathcal{M}|^2} d \Phi_2 \to \Gamma = \frac{1}{2M} \cdot \frac{\beta}{8 \pi} \cdot \overline{|\mathcal{M}|^2},
\end{equation}
where the bar denotes helicity and color sum (average) over final (initial) states.

\section{$2 \to 3$ body eikonal integrals \label{a:ang_integrals} }

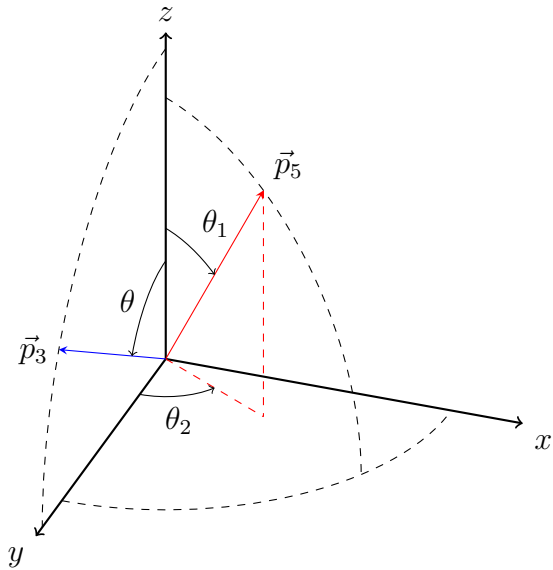
\begin{figure}
\centering

\tdplotsetmaincoords{60}{110}

\pgfmathsetmacro{\rvec}{.8}
\pgfmathsetmacro{\thetavec}{30}
\pgfmathsetmacro{\phivec}{60}

\begin{tikzpicture}[scale=5,tdplot_main_coords]

\coordinate (O) at (0,0,0);

\tdplotsetcoord{P}{\rvec}{\thetavec}{\phivec}


\draw[thick,->] (0,0,0) -- (1,0,0) node[anchor=north east]{$y$};
\draw[thick,->] (0,0,0) -- (0,1,0) node[anchor=north west]{$x$};
\draw[thick,->] (0,0,0) -- (0,0,1) node[anchor=south]{$z$};

\draw[-stealth,color=red] (O) -- (P) node[anchor=south west, color=black]{$\vec{p}_5$};

\draw[dashed, color=red] (O) -- (Pxy);
\draw[dashed, color=red] (P) -- (Pxy);

\tdplotdrawarc[->]{(O)}{0.2}{0}{\phivec}{anchor=north}{$\theta_2$}

\tdplotsetthetaplanecoords{\phivec}

\tdplotdrawarc[tdplot_rotated_coords,->]{(0,0,0)}{0.4}{0}{\thetavec}{anchor=south west}{$\theta_1$}

\draw[dashed,tdplot_rotated_coords] (\rvec,0,0) arc (0:90:\rvec);
\draw[dashed] (\rvec,0,0) arc (0:90:\rvec);


\pgfmathsetmacro{\rvec}{0.95}
\pgfmathsetmacro{\thetavec}{60}
\pgfmathsetmacro{\phivec}{0}
\tdplotsetcoord{P1}{\rvec}{\thetavec}{\phivec}
\draw[-stealth,color=blue] (O) -- (P1) node[anchor=east, color=black]{$\vec{p}_3$};
\tdplotsetthetaplanecoords{\phivec}
\tdplotdrawarc[tdplot_rotated_coords,->]{(0,0,0)}{0.3}{0}{\thetavec}{anchor=east}{$\theta$}

\draw[dashed,tdplot_rotated_coords] (\rvec,0,0) arc (0:90:\rvec);



\end{tikzpicture}
\caption{
  3-body kinematics used in the evaluation of the real-emission cross section in the soft limit.
  It is assumed that the $z$ axis is aligned with the beam axis.
  Momenta lengths are not to scale.
 \label{fig:3body_kinematics}}
\end{figure}

This section collects integrals needed in the calculation of the soft-collinear cross section for the sgluon pair production.
Following the discussion in Ref.~\cite{Harris:2001sx}, the needed integrals have a general form (with the curly bracket collecting terms which may appear in these integrals one-by-one)
\begin{multline}
 I =  \frac{1}{\pi} \left (\frac{\hat s}{4}\right)^{\epsilon} \int_0^{\delta_s \sqrt{\hat s}/2} d E_5 E_5^{1-2\epsilon} \sin^{1-2\epsilon} \theta_1 d \theta_1 \sin^{-2\epsilon} \theta_2 d \theta_2 \\
  \cdot 1/ 
  \{s_{35}^{'} s_{45}^{'}, s_{35}^{'2}, s_{45}^{'2}, t_{15}^{'} t_{25}^{'}, t_{15}^{'2}, t_{25}^{'2}, s_{35}^{'} t_{15}^{'}, s_{45}^{'} t_{25}^{'}, s_{35}^{'} t_{25}^{'}, s_{45}^{'} t_{15}^{'} \},
\end{multline}
where variables $s'_{ij}$, $t'_{ij}$ are related to a generalization of the Mandelstam variables for the case of $2 \to 3$ scattering
\begin{equation}
  s_{ij} \equiv (p_i + p_j)^2, \qquad t_{ij} \equiv (t_i - t_j)^2.
\end{equation}
as, for example, $s'_{35} = s_{35} - m_3^2 - m_5^2$.

Parameterizing the gluon momentum $p_5$ in $D$-dimensions as
\begin{equation}
  p_5 = E_5 (1, \ldots, \sin \theta_1 \sin \theta_2, \sin \theta_1 \cos \theta_2, \cos \theta_1),
\end{equation}
where ellipsis denote $D-4$ unspecified momenta and sgluons momenta $p_3$ and $p_4$ as
\begin{align}
  p_3 =& \frac{\sqrt{s_{12}}}{2} (1, 0, \ldots, 0, \beta \sin \theta, \beta \cos \theta) \\
  p_4 = & \frac{\sqrt{s_{12}}}{2} (1, 0, \ldots, 0, -\beta \sin \theta, -\beta \cos \theta)
\end{align}
where $D-4$ and $x$ components are set to 0 since sgluons momenta are 4 - dimensional and contained in the $yz$ plane.
The kinematics of this final state is shown in \autoref{fig:3body_kinematics}.
The 3-vector $\vec{p}_4$ for the second sgluon is not shown, but in the soft limit is is given by the equality $\vec{p}_4 = - \vec{p}_3$.

The primed Mandelstam variables are then
\begin{align}
  s_{35}^{'} = & \sqrt{\hat s} E_5 (1 - \beta \sin \theta \sin \theta_1 \cos \theta_2 - \beta \cos \theta \cos \theta_1) \\
  s_{45}^{'} = & \sqrt{\hat s} E_5 (1 + \beta \sin \theta \sin \theta_1 \cos \theta_2 + \beta \cos \theta \cos \theta_1) \\
  t_{15} = & -\sqrt{\hat s} E_5 (1 - \cos \theta_1 )\\
  t_{25} = & -\sqrt{\hat s} E_5 (1 + \cos \theta_1 )
\end{align}
The energy integral is universal and gives
\begin{equation}
  \frac{1}{\pi} \left (\frac{\hat s}{4}\right)^{\epsilon} \int_0^{\delta_s \sqrt{\hat s}/2} d E_5 E_5^{1-2\epsilon} = -\frac{\delta_s^{-2\epsilon}}{2\pi \epsilon} \approx -\frac{1}{2\pi \epsilon} + \frac{1}{\pi} \ln \delta_s - \frac{1}{\pi} \epsilon \ln^2 \delta_s ,
\end{equation}
while the remaining angular integrals have a general form of
\begin{equation}
 \int_0^\pi d \theta_1 \frac{\sin^{1-2\epsilon} \theta_1}{(1+a \cos \theta_1)^k} \int_0^\pi d \theta_2 \frac{\sin^{-2\epsilon} \theta_2}{(1+A \cos \theta_1 + B \sin \theta_1 \cos \theta_2)^l}.
\end{equation}

Collinear divergences appear at the boundary of the integration region if $a = \pm 1$ and/or $A^2 + B^2 = 1$.
For massive final state particles  $A^2 + B^2 < 1$, hence collinear singularities appear only in the initial state. 
The angular integrals are symmetric with respect to the change of $\theta_1 \leftrightarrow \pi - \theta_1$ which implies symmetry in simultaneous change of $\{ s_{35}^{'}, t_{15}^{'} \} \leftrightarrow \{ s_{45}^{'}, t_{25}^{'} \}$.  
Large collection of these integrals can be found for example in Refs.~\cite{Beenakker:1988bq,Somogyi:2011ir}.
Below are quoted the ones needed for a considered process of sgluon pair production
\begin{align}
 I( s_{35}^{'} s_{45}^{'} ) = & \frac{1}{\hat{s} \beta} \left ( -\frac{1}{2\epsilon} \log \frac{1+\beta}{1-\beta} - \dilog \frac{2\beta}{1+\beta} - \frac{1}{4} \log^2 \frac{1+\beta}{1-\beta} + \ln \delta_s \ln \frac{1+\beta}{1-\beta}\right ) \\
 I ( s_{35}^{'2} ) = I ( s_{45}^{'2} ) = & \frac{1}{2 m^2} \left ( -\frac{1}{2\epsilon} + \log \delta_s -\frac{1}{2\beta} \log \frac{1+\beta}{1-\beta} \right )\\
 I ( t_{15}^{'} t_{25}^{'} ) = & \frac{1}{2 \hat s} \left ( \frac{1}{\epsilon^2} - \frac{2}{\epsilon} \ln \delta_s + 2 \ln^2 \delta_s\right )\\
 I ( t_{15}^{'2} ) = I ( t_{25}^{'2} )= & \frac{1}{2 \hat s} \left( \frac{1}{\epsilon} - 1 - 2\ln \delta_s \right)& \\
 I ( s_{35}^{'} t_{15}^{'} ) = I ( s_{45}^{'} t_{25}^{'} ) = & - \frac{1}{2(1-\beta \cos \theta) \hat s} \left [ \frac{1}{\epsilon^2} -  \left (2 \ln \delta_s + \ln \frac{(1-\beta \cos\theta)^2}{1-\beta^2} \right ) \frac{1}{\epsilon} \right. \\ 
 & + 2 \ln^2 \delta_s - \frac{1}{2} \ln^2 \frac{1+\beta}{1-\beta} + \ln^2 \frac{1 - \beta \cos \theta}{1-\beta}  + 2 \ln \delta_s \ln \frac{(1-\beta \cos\theta)^2}{1-\beta^2} \nonumber \\
 & \left. + \, 2 \, \dilog \frac{\beta (1-\cos \theta)}{-1 + \beta} - 2 \, \dilog \frac{\beta (1+\cos \theta)}{-1 + \beta \cos \theta} \right ] \nonumber \\
 I ( s_{35}^{'} t_{25}^{'} ) = I ( s_{45}^{'} t_{15}^{'} ) = & - \frac{1}{2(1+ \beta \cos \theta) \hat s} \left [ \frac{1}{\epsilon^2} -  \left (2 \ln \delta_s + \ln \frac{(1+\beta \cos\theta)^2}{1-\beta^2} \right ) \frac{1}{\epsilon} \right. \\ 
 & + 2 \ln^2 \delta_s - \frac{1}{2} \ln^2 \frac{1+\beta}{1-\beta} + \ln^2 \frac{1 + \beta \cos \theta}{1-\beta}  + 2 \ln \delta_s \ln \frac{(1+\beta \cos\theta)^2}{1-\beta^2} \nonumber \\
 & \left. + \, 2 \, \dilog \frac{\beta (1+\cos \theta)}{-1 + \beta} - 2 \, \dilog \frac{\beta (-1+\cos \theta)}{1 + \beta \cos \theta} \right ] \nonumber ,
\end{align}
where $\dilog$ denotes the dilogarithm function $\dilog (x) \equiv \int_x^0 \frac{\ln (1 - t)}{t} dt$.

%
%
%
\chapter{Sgluon decays \label{sec:sgluon_decays}} 
%
%
\section{Scalar sgluons}
Owing to the tree-level relation\footnote{This relation might be modified beyond the LO order.} $m_{O_S}^2 = m_O^2 + 4 (M_D^O)^2$, for scalar sgluon a simple 2-body decay channel to a pair of Dirac gluinos as depicted in one of the Feynman rules in \autoref{sec:sqcd_feynman_rules} is always open with partial decay width (see \autoref{sec:2body_dec})
\begin{equation}
  \label{eq:Os_gluglu_width}
  \Gamma(O_S \to \tilde{g}_D \overline{\tilde{g}_D}) =  \frac{3}{2} \alpha_s m_{O_S} \beta^3_{\tilde{g}_D} ,
\end{equation}
where $\beta_{\tilde g_D} \equiv \sqrt{1 - 4 m_{\tilde g_D}^2/m_{O_S}^2}$ is the absolute value of daughter particle velocity in the rest frame of the sgluon.
If, additionally $m_{O_S} \gtrsim 2 \, m_{\tilde q}$, 2-body decay to pair of squarks is also open with a decay width
\begin{equation}
  \Gamma(O_S \to \tilde{q} \tilde{q}^*) = \frac{1}{2} \alpha_s \frac{|M_O^D|^2}{m_{O_S}} \beta_{\tilde q},
\end{equation}
where $\beta_{\tilde q}$ is defined analogously.

Unless one is interested in a specific final state there is no need to consider loop-induced decay processes for scalar sgluon since at least one decay channel is always open.
\section{Pseudoscalar sgluon}
If the mass of the pseudoscalar sgluon $O_A$ is $\gtrsim 2 m_{\tilde g_D}$, simple 2-body decay channel is open as in the case of scalar sgluon. 
Due to $\gamma^5$ nature of the $O_A \tilde{g}_D \overline{\tilde{g}_D}$ coupling shown in Appendix \ref{sec:sqcd_feynman_rules} it depends linearly on velocity $\beta_{\tilde g_D}$ as (cf. \autoref{eq:Os_gluglu_width})
\begin{eqnarray}
  \Gamma(O_A \to \tilde{g}_D \overline{\tilde{g}_D}) &=&  \frac{3}{2} \alpha_s m_{O_P} \beta_{\tilde g_D}
\end{eqnarray}
No other 2-body decays are allowed at the tree-level.\footnote{This is no longer true if one allows for imaginary part of $M_O^D$. See \autoref{sec:non_real_mdo}.}

If two-body decay channels are closed $O_A$ will decay through loop-induced couplings to SM particles.
In case $\Im M_D^O = 0$, $O_A$ will decay exclusively to $q\bar{q}$ pairs.
\begin{figure}
  \centering
  \includegraphics[width=0.6\textwidth]{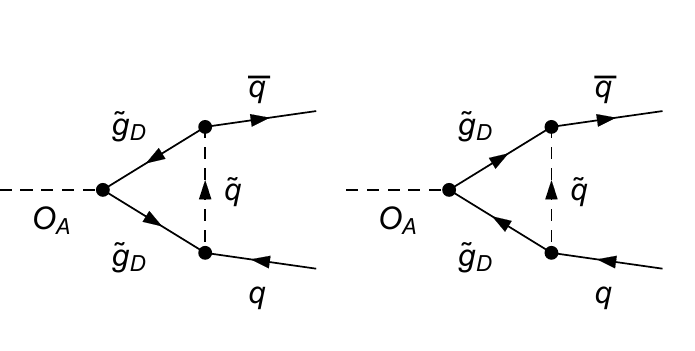}
  \caption{Lowest order diagrams generating (effective) coupling of $O_P$ to quarks.}
  \label{fig:Op2qqbar}
\end{figure}
\Autoref{fig:Op2qqbar} shows diagrams generating effective $O_S q \bar{q}$ coupling. 
The following analytic expression is reads
\begin{align}
\imath V^a_{O_A q \bar{q}} = \frac{3 \imath g_s^3}{16 \pi^2} m_q m_{\tilde{g}_D}  & \left [ C_0 (m_{O_A}^2, m_q^2, m_q^2, m_{\tilde{g}_D}^2, m_{\tilde{g}_D}^2, m_{\tilde{q}_R}^2 ) \right . \nonumber \\
 & \,\,\, \left . - C_0 (m_{O_A}^2, m_q^2, m_q^2, m_{\tilde{g}_D}^2, m_{\tilde{g}_D}^2, m_{\tilde{q}_L}^2 )\right ] T^a \gamma^5  .
\end{align}

Since this amplitude is suppressed by the mass of the quark, above the top threshold $O_A$ decays almost exclusively to top quarks.
General expression for the decay of $O_S \to q\bar{q}$ is then
\begin{align}
  \Gamma (O_A \to q \bar{q} ) = \frac{9\alpha_s^3}{64 \pi^2} \beta_q m_{O_A} m_q^2 m_{\tilde{g}_D}^2 & \left | C_0 (m_{O_A}^2, m_q^2, m_q^2, m_{\tilde{g}_D}^2, m_{\tilde{g}_D}^2, m_{\tilde{q}_R}^2 ) \right . \\
 & \, \left . - C_0 (m_{O_A}^2, m_q^2, m_q^2, m_{\tilde{g}_D}^2, m_{\tilde{g}_D}^2, m_{\tilde{q}_L}^2 )\right |^2 \nonumber 
\end{align}
where $|C_0 (m_{O_P}^2, m_q^2, m_q^2, m_{\tilde{g}_D}^2, m_{\tilde{g}_D}^2, m_{\tilde{q}_R}^2 ) - C_0 (m_{O_P}^2, m_q^2, m_q^2, m_{\tilde{g}_D}^2, m_{\tilde{g}_D}^2, m_{\tilde{q}_L}^2 )|^2 = |\mathcal{I_P}|^2$ of Ref.~\cite{Choi:2008ub}.
\section{Comment on $\Im M_O^D \neq 0$ \label{sec:non_real_mdo}}
\begin{figure}
  \centering
  \includegraphics[width=0.4\textwidth]{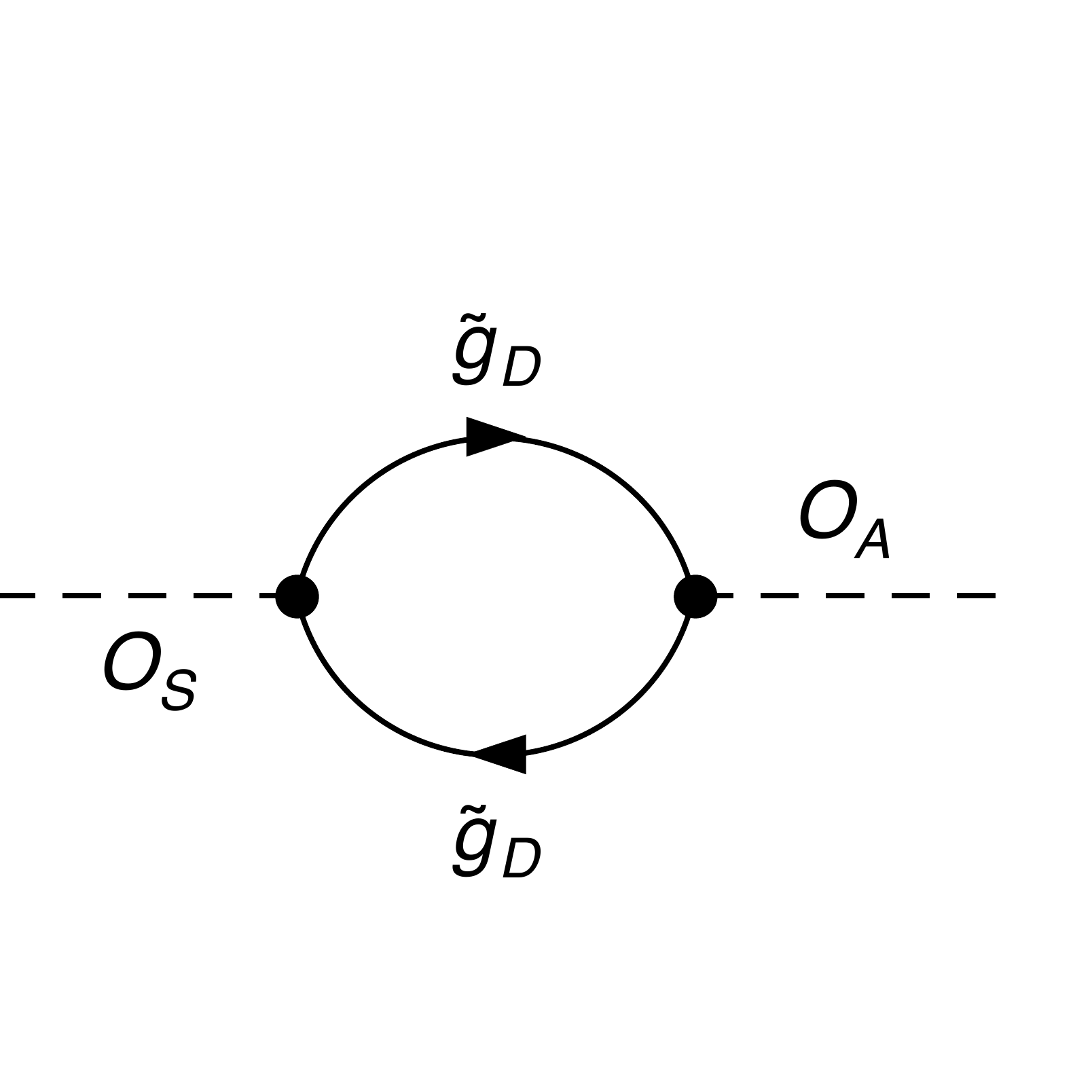}
  \caption{Exemplary diagram generating mixing between CP-odd and CP-even components of the sgluon field if $\Re M_O^D \cdot \Im M_O^D \neq 0$.}
  \label{fig:cp_mixing_of_sgluons}
\end{figure}

Throughout this work it is assumed that all parameters of the MRSSM are real.
If this is not the case and $M_O^D$ has a non-zero imaginary part, pseudoscalar sgluon couples directly to pair of squarks as shown in \autoref{sec:sqcd_feynman_rules}.
So, if kinematically allowed, a 2-body decay to squarks will be occur.
Also, introduction of imaginary part of $M_O^D$ breaks CP in the sgluon sector, mixing scalar and pseudoscalar partners through one-loop diagrams as shown in \autoref{fig:cp_mixing_of_sgluons}, which is proportional to the product of $\Re M_O^D \cdot \Im M_O^D$.

\chapter{\texttt{SARAH} and \texttt{SPheno} setup}

\section{\texttt{SPheno} code generation \label{sec:sarah_mathematica} }
Full custom \texttt{SPheno.m} file generating high scale \texttt{SPheno} module used for the numerical analyses done in this thesis.

\begin{minted}{yaml}

MINPAR = { {3, TanBeta} };

EXTPAR = {	
  {1,   LSDInput},
  {2,   LSUInput},
  {3,   LTDInput},
  {4,   LTUInput},
  {5,   MuDInput},
  {6,   MuUInput},
  {7,   BMuInput},
  {8,   mq2Input},    
  {9,   mq233Input},
  {10,  ml2Input},
  {11,  ml233Input},
  {12,  mu2Input},
  {13,  mu233Input},
  {14,  md2Input},
  {15,  md233Input},
  {16,  me2Input},
  {17,  me233Input},
  {18,  mRu2Input},
  {19,  mRd2Input},
  {20,  mO2Input},
  {21,  M1Input},
  {22,  M2Input},
  {23,  M3Input},
  {24,  mS2Input},
  {25,  mT2Input}
};

RealParameters = {TanBeta};

AssumptionsTadpoleEquations = {conj -> Identity};
ParametersToSolveTadpoles = {mHd2, mHu2, vS, vT};

RenormalizationScaleFirstGuess = 1000^2;
RenormalizationScale =  1000.0^2;

(*  Low-scale R-symmetry  *)

BoundarySUSYScale = {
  {mS2,       mS2Input},
  {mT2,       mT2Input},
  {MDBS,      M1Input},
  {MDWBT,     M2Input},
  {MDGoc,     M3Input},
  {B[\[Mu]],  BMuInput},
  {mq2,       DIAGONAL mq2Input},
  {mq2[3,3],  mq233Input},
  {ml2,       DIAGONAL ml2Input},
  {ml2[3,3],  ml233Input},
  {md2,       DIAGONAL md2Input},
  {md2[3,3],  md233Input},
  {mu2,       DIAGONAL mu2Input},
  {mu2[3,3],  mu233Input},
  {me2,       DIAGONAL me2Input},
  {me2[3,3],  me233Input},
  {LamSD,     LSDInput},
  {LamSU,     LSUInput},
  {LamTD,     LTDInput},
  {LamTU,     LTUInput},
  {MuD,       MuDInput},
  {MuU,       MuUInput},
  {B[MuD],    0},
  {B[MuU],    0},
  {\[Mu],     0},
  {mRu2,      mRu2Input},
  {mRd2,      mRd2Input},
  {moc2,      mO2Input}
};

(*  This is a low-energy model without the GUT-scale so 'HighScale' will not be used. 
    It will be set to an arbitrary value of 10 TeV (so slightly above SUSY scale
    which is fixed to 1 TeV, and contrary to the standard setup, where GUT scale is 
    determined dynamically through ConditionGUTscale variable). To be on the 
    safe side just set B[MuD], B[MuU] and \[Mu] to zero at the GUT and EW scale *)
BoundaryHighScale = {
  {B[MuD],  0},
  {B[MuU],  0},
  {\[Mu],   0}
};

(*  Do not freeze SUSY parameters at the SUSY scale 
but also run them down to the mZ scale  *)
UseBoundarySUSYatEWSB = False

BoundaryEWSBScale = {
  {vd, Sqrt[4 mz2/( g1^2 + g2^2 ) ] * Cos[ ArcTan[TanBeta] ] },
  {vu, Sqrt[4 mz2/( g1^2 + g2^2 ) ] * Sin[ ArcTan[TanBeta] ] },
  {B[MuD],  0},
  {B[MuU],  0},
  {\[Mu],   0}
};

(*  If the Higgs sector of the model is the same as in the MSSM, 
    the original SPheno routines for calculating the two-loop tadpole 
    equations and two-loop self-energies to the the scalar and pseudo 
    scalar Higgses can be activated by setting it to True *)
UseHiggs2LoopMSSM = False;

ListDecayParticles = Automatic;
ListDecayParticles3B = Automatic;

FlagLoopContributions = True

\end{minted}

\section{\texttt{SPheno} run card \label{sec:spheno_card}}
Stering file for \texttt{SPheno}
\begin{minted}{yaml}
Block MODSEL
1   1                 # 1/0: High/low scale input 
6   0                 # Generation Mixing 
Block SMINPUTS        # SM inputs
1   127.94000E+00     # alpha(MZ) SM MSbar
2   1.1663787E-05     # muon decay constant
3   1.1810000E-01     # alpha_s(MZ) SM MSbar 
4   9.1187600E+01     # Z-boson pole mass 
5   4.1800000E+00     # m_b(m_b) SM MSbar 
6   1.7321000E+02     # m_top(pole) 
7   1.7768600E+00     # m_tau(pole) 
Block MINPAR          # SUSY inputs
3   4.0E+01           # TanBeta
Block EXTPAR      
1   1.5E-01           # LSDInput
2   -1.5E-01          # LSUInput
3   -1.0E+00          # LTDInput
4   -1.03E-00         # LTUInput
5   4.0E+02           # MuDInput
6   4.0E+02           # MuUInput
7   4.0E+04           # BmuInput
8   6.25E+06          # mq2Input
9   1.0E+06           # mq233Input
10  1.0E+06           # ml2Input
11  1.0E+06           # ml233Input
12  6.25E+06          # mu2Input
13  1.00E+06          # mu233Input
14  6.25E+06          # md2Input
15  1.0E+06           # md233Input
16  1.0E+06           # me2Input
17  1.0E+06           # me233Input
18  1.0E+06           # mRu2Input
19  4.9E+05           # mRd2Input
20  1.0E+06           # mO2Input
21  2.5E+02           # M1Input
22  5.0E+02           # M2Input
23  1.5E+03           # M3Input
24  4.0E+06           # mS2Input
25  9.0E+06           # mT2Input
Block SPhenoInput     # SPheno settings 
  1  -1               # error level 
  2  1                # fixed SUSY scale = 1 TeV (SPA conventions)
  7  0                # exclude 2-loop corrections to Higgses
  8  3                # use diagrammatic method for 2-loop calculation
 11 1                 # calculate branching ratios 
 13 1                 # include 3-Body decays 
 12 1.000E-04         # write only branching ratios larger than this value 
 31 10.0E+03          # fixed GUT scale (-1: dynamical GUT scale) 
 32 0                 # Strict unification 
 34 1.000E-04         # precision of mass calculation 
 35 40                # maximum number of iterations
 37 1                 # Set Yukawa scheme  
 38 2                 # 1- or 2-loop RGEs 
 50 1                 # Majorana phases: use only positive masses 
 51 0                 # Write Output in CKM basis 
 52 0                 # Write spectrum in case of tachyonic states 
 54 0                 # If Ne1 WriteOutputForNonConvergence=.False.
 55 1                 # Calculate one loop masses 
 57 1                 # Calculate low energy constraints 
 58 1
 60 1                 # Include possible, kinetic mixing 
 65 1                 # Solution tadpole equation 
 75 1                 # Write WHIZARD files 
 76 1                 # Write HiggsBounds file 
 86 0                 # Maximal width to be counted as invisible in Higgs decays;
                      # -1: only LSP 
510 0                 # Write tree level values for tadpole solutions 
515 0                 # Write parameter values at GUT scale 
520 1                 # Write effective Higgs couplings (HiggsBounds blocks) 
525 0                 # Write loop contributions to diphoton decay of Higgs 
530 0                 # Write Blocks for Vevacious 
\end{minted}
\bibliographystyle{jhep.bst}
\bibliography{doktorat}

\end{document}